\documentclass{aa}
\usepackage{graphicx}
\usepackage{txfonts}
\usepackage{subfig}
\usepackage{longtable} 
\usepackage{natbib}
\graphicspath{{./plots/}}
\begin{document} 
\title{Stellar parameters of Be stars observed with X-shooter\thanks{Based on
    observations made with ESO Telescopes at the La Silla Paranal Observatory
    under programme IDs 60.A-9022, 60.A-9024, 077.D-0085, 085.A-0962,
    185.D-0056, 091.B-0900, and 093.D-0415. Tables 6 to 15 are only available
    in electronic form at the CDS via anonymous ftp to cdsarc.u-strasbg.fr
    (130.79.128.5) or via http://cdsweb.u-strasbg.fr/cgi-bin/qcat?J/A+A/} }
\author{A. Shokry \inst{1,2,3} \and Th.~Rivinius\inst{1} \and
  A.~Mehner\inst{1} \and C.~Martayan\inst{1} \and W.~Hummel\inst{4} \and
  R.~H.~D.~Townsend\inst{5} \and A.\ M\'erand\inst{1} \and B.\ Mota\inst{6}
  \and D.~M.\ Faes\inst{6} \and M.~A.~Hamdy\inst{2}\thanks{Deceased} \and
  M.~M.~Beheary\inst{7} \and K.~A.~K~Gadallah\inst{7} \and
  M.~S.~Abo-Elazm\inst{2} }

\institute{ 
ESO - European Organisation for Astronomical Research in the Southern
Hemisphere, Casilla 19001, Santiago, Chile
\and
Astronomy Department, National Research Institute of Astronomy and Geophysics
(NRIAG), 11421 Helwan, Cairo, Egypt.
\and
Kottamia Center of Scientific Excellence in Astronomy and Space Science (KCScE, STDF, ASRT), Cairo, Egypt
\and
ESO - European Organisation for Astronomical Research in the Southern
Hemisphere, Karl-Schwarzschild-Str. 2, Garching Germany.         
\and
Department of Astronomy, University of Wisconsin-Madison, Madison, Wisconsin 53706, USA
\and
Instituto de Astronomia, Geof\'isica e Ci\^encias Atmosf\'ericas, Universidade
de S\~ao Paulo (USP), Rua do Mat\~ao 1226, Cidade Universit\'aria, S\~ao
Paulo, SP - 05508-900, Brazil
\and
Astronomy and Meteorology Department, Faculty of Science, Al-Azhar University, Cairo, Egypt%
}
\date{Received; accepted}
%
% 
%%%%%%%%%%%%%%%%%%%%%%%%%%%%%%%%%%%%%%%%%%%%%%%%%%%%%%%%%%%%%%%%%%%%%%%%%%
%%%%%%%%%%%%%%%%%%%%%%%%%%%%%%%%%%%%%%%%%%%%%%%%%%%%%%%%%%%%%%%%%%%%%%%%%%
\abstract
% context heading (optional)
{}
% aims heading (mandatory) 
{The X-shooter archive of several thousand telluric star spectra was skimmed
  for Be and Be-shell stars to derive the stellar fundamental parameters and
  statistical properties, in particular for the less investigated late type Be
  stars, and the extension of the Be phenomenon into early A stars.}
% methods heading (mandatory)
{An adapted version of the BCD method is used, utilizing the Balmer
  discontinuity parameters to determine effective temperature and surface
  gravity. This method is optimally suited for late B stars.  The projected
  rotational velocity was obtained by profile fitting to the \ion{Mg}{ii}
  lines of the targets, and the spectra were inspected visually for the
  presence of peculiar features such as the infrared \ion{Ca}{ii} triplet or
  the presence of a double Balmer discontinuity. The Balmer line equivalent
  widths were measured, but due to uncertainties in determining the
  photospheric contribution are useful only in a subsample of Be stars for
  determining the pure emission contribution. }
% results heading (mandatory)
{A total of 78 Be stars, mostly late type ones, were identified in the
  X-shooter telluric standard star archive, out of which 48 had not been
  reported before. The general trend of late type Be stars having more tenuous
  disks and being less variable than early type ones is confirmed. The
  relatively large number (48) of relatively bright ($V>8.5$) additional Be
  stars casts some doubt on the statistics of late type Be stars; they are
  more common than currently thought: The Be/B star fraction may not strongly
  depend on spectral subtype.}
% conclusions heading (optional)
{}
\maketitle
%
%%%%%%%%%%%%%%%%%%%%%%%%%%%%%%%%%%%%%%%%%%%%%%%%%%%%%%%%%%%%%%%%%%%%%%%%%%
%%%%%%%%%%%%%%%%%%%%%%%%%%%%%%%%%%%%%%%%%%%%%%%%%%%%%%%%%%%%%%%%%%%%%%%%%%

\section{Introduction}\label{sec:intro}

Be stars are non-supergiant B stars that show or have shown H$\alpha$
emission, as defined by \citet{1981BeSN....4....9J}. Emission does not only
occur in the first members of the Balmer line series, but can affect the
continuum and line profiles of other species as well, most often singly
ionized metals, such as \ion{Fe}{ii}.  It is generally agreed that, in
classical Be stars, this emission is due to the presence of a gaseous
Keplerian disk, concentrated in the equatorial plane. This disk is a decretion
disk, i.e., the source of the disk material is the central star, generated by
the equatorial flow of stellar material. One of the key factors in creating
the disk is supposed to be the very high rotational velocity. In fact, Be
stars are known to have higher rotational velocities than normal B-type stars
\citep{2013A&A...550A..79C}.
For a complete review on the topic, see \citet{2003PASP..115.1153P} and
\citet{2013A&ARv..21...69R}.

Classical Be stars are known to vary both in brightness and spectral line
appearance, with a large range of time scales from years to minutes
\citep{1997A&A...318..548O,2002A&A...394..137F,2007ASPC..362..260K}.  While
long-term variations are associated with formation and dissipation of the disk
\citep{1997A&A...318..548O}, the origin of short-term variability is usually
attributed to pulsations within the B star photosphere
\citep{2000ASPC..214..178B,2009A&A...506...95H}.  Photometric studies show
that earlier type Be stars are more likely to be variable
\citep[e.g.][]{1998A&A...335..565H}.

\citet{2015AJ....149....7C} have demonstrated that Be stars can often be found
among stars observed for the purpose of removing telluric absorption in the
near-infrared domain, because main sequence B stars are among the preferred
objects for this task. Inspired by this example, we decided to search for Be
stars in a similarly extensive database of telluric standard star
observations, namely the one taken at the VLT with the X-shooter instrument.

%%%%%%%%%%%%%%%%%%%%%%%%%%%%%%%%%%%%%%%%%%%%%%%%%%%%%%%%%%%%%%%%%%%%%%%%%%
%%%%%%%%%%%%%%%%%%%%%%%%%%%%%%%%%%%%%%%%%%%%%%%%%%%%%%%%%%%%%%%%%%%%%%%%%%
\section{Observations and Data reduction}\label{sec:obs}

%%%%%%%%%%%%%%%%%%%%%%%%%%%%%%%%%%%%%%%%%%%%%%%%%%%%%%%%%%%%%%%%%%%%%%%%%%
\begin{figure*}[t]
\begin{center}
\includegraphics[angle=0,width=14cm,clip]{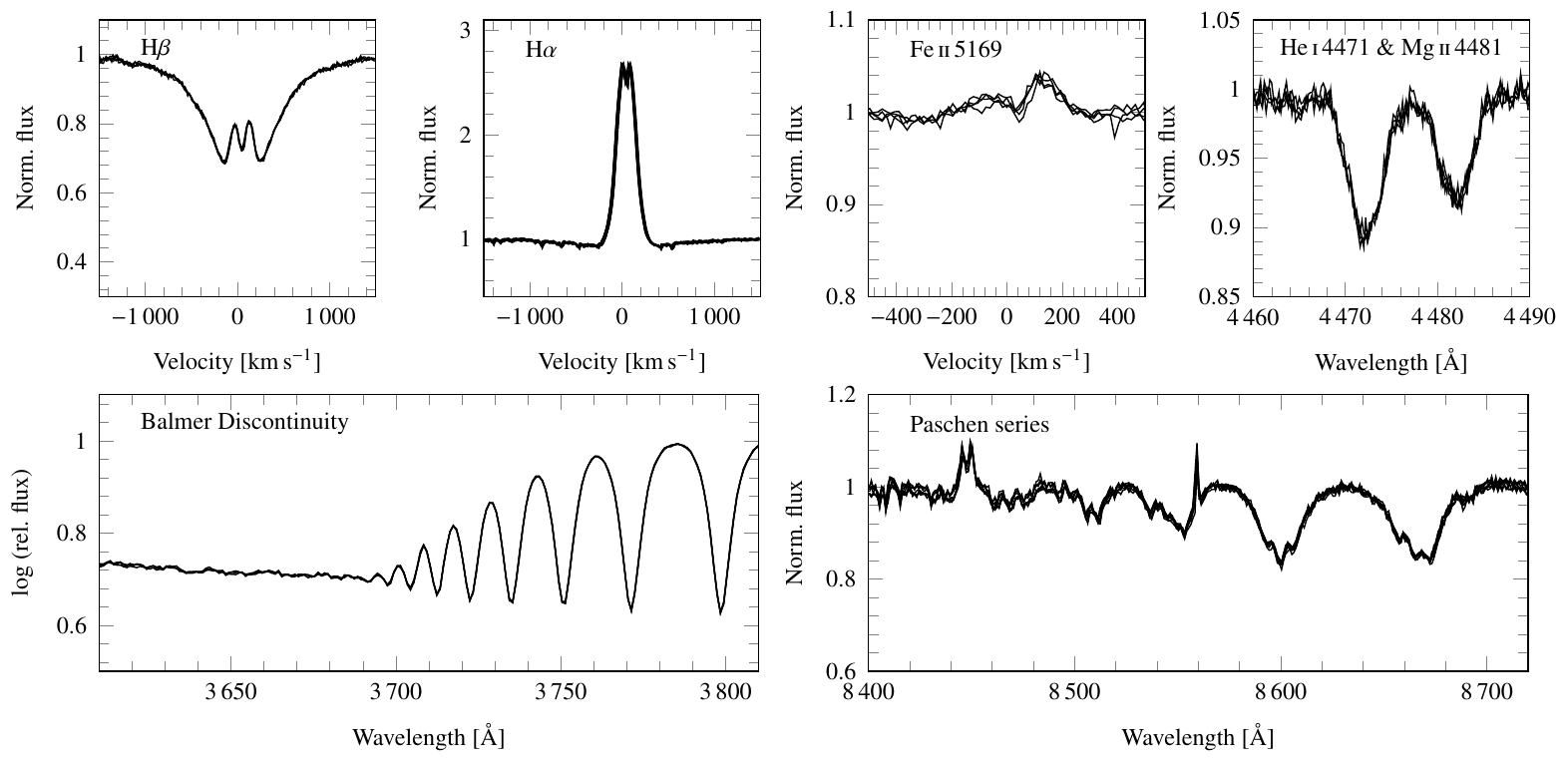}
\end{center}
\caption{\label{fig:spec}Spectrum overview plot for Hip\,11116. All available
  spectra are overplotted, giving an indication of the spectral
  variability. For the remaining 77 active Be stars, see
  Appendix~\ref{app:spec}.}
\end{figure*}
%%%%%%%%%%%%%%%%%%%%%%%%%%%%%%%%%%%%%%%%%%%%%%%%%%%%%%%%%%%%%%%%%%%%%%%%%%

All spectra have been acquired with the VLT/X-shooter instruments. Most data
were taken as telluric standard stars for other observations, from the
commissioning of X-shooter in 2007 until 2015. Once the value of this database
for science on its own was realized, additional spectra of stars of interest
were acquired in a dedicated observing program in ESO period 93. A few
datasets for stars of interest were also taken under a number of different
program IDs, and were downloaded from the archive as well. The main program
IDs under which data was observed for this work, however, are 60.A-9022,
60.A-9024, and 093.D-0415. The meteorological observing conditions, dates, and
nightlog excerpts are available from the ESO data archive together with the
raw data.

X-shooter is a multi-wavelength medium-resolution spectrograph mounted at the
Cassegrain focus of UT2 of the Very Large Telescope (VLT) at ESO Paranal that
has a mirror diameter of 8.2\,m.  X-shooter's three arms are named UVB,
covering 300--550\,nm, VIS, covering 550--1010\,nm, and NIR, covering
1000--2500\,nm. The resolution depends on the chosen slit-width and ranges
from $R = 1890$ to $9760$ in the UVB, 3180 to 18110 in the VIS, and 3900 to
11490 in NIR arm, respectively \citep{2011A&A...536A.105V}.

To select the Be stars and other possibly interesting objects from this huge
archival sample, first the raw  frames in the VIS arm were obtained. A window,
containing only the spectral order with the H$\alpha$ line, was cut from the
entire raw frame. From the inter-order space in that window the local
background, effectively the sum of bias\,$+$\,dark\,$+$\,scattered light, was
estimated and subtracted. The result was integrated into a 1D-spectrum,
divided by a generic blaze shape obtained from flat field frames, and then
divided by the counts in the continuum. This procedure yields an approximately
normalized H$\alpha$ line profile in the units of pixel vs.\ flux (i.e.,
without any wavelength calibration) that is sufficient to judge upon the line
shape for the presence of emission and other features of interest.

The resulting more than 10\,000 profiles of 1334 stars were inspected visually
for emission or any other curious appearance, such as binarity or strong
profile shape distortions. 1093 stars were found to be spectroscopically
normal BA main sequence objects, and 16 of later spectral type. Of the
remaining objects, 89 were spectroscopic binaries of type SB2 (as seen in
H$\alpha$) and 48 were found to be supergiants with winds, or otherwise not
quite as expected for a single main-sequence star. The procedure yielded a
number of emission line stars for further inspection, of which two are known
as Herbig Be stars (\object{Hip\,56379} and \object{Hip\,85755}), and four are
mass-transferring binaries of various types (\object{Hip\,33237},
\object{Hip\,45311}, \object{Hip\,88615}, and \object{Hip\,93502}). This
leaves 78 Be stars identified from the X-shooter data. In addition, among the
un-suspicious stars four could be identified using the SIMBAD database as
currently inactive Be stars and hence were added to the sample
(\object{Hip\,15188}, \object{Hip\,25950}, \object{Hip\,108022}, and
\object{Hip\,108975}), bringing the number to a total of 82 Be stars. We note
that not all stars could be used for every analysis below, so the number of
Be stars used for some results might be lower.

The spectra of the Be star sample, identified in the above way, were reduced
with the REFLEX workflow for X-shooter \citep{2013A&A...559A..96F}. The ESO
Recipe Flexible Execution Workbench is a workflow environment to run ESO VLT
pipelines. It provides an easy and interactive way to reduce VLT science
data.  The steps executed by the ESO X-shooter pipeline (v.2.6.0) include bias
subtraction, flat-fielding, wavelength and flux calibration, and order
merging.

For the flux calibration the master response calibration provided by the ESO
archive was used, except for UVB data obtained between 2009 and beginning of
2012. Using the master calibration on these spectra produces an obviously
spurious dip in the region immediately bluewards of the Balmer
discontinuity. This is most likely due to the use of two flatfield lamps in
the very blue, where their flux ratio was not entirely stable. To avoid the
problem, spectra taken in these years were flux calibrated with a flux
standard taken in the same night or not more than a few nights before or
after, that was reduced with the same flatfield exposures.

Spectra that had obvious faults, like bad flux calibration, too little flux,
or were overexposed in the relevant parts of the continuum, were discarded
from the following analysis. This left ten Be stars without suitable data to
determine stellar parameters.

For each spectrum, a flux calibrated version and a
normalized version, using a global spline fit to continuum regions, were
produced for analysis.

In Fig.~\ref{fig:spec} an overview of the data of the Be star Hip\,11116 is
shown. Similar plots for the remaining 77 identified Be stars are presented in
Appendix~\ref{app:spec}.

%%%%%%%%%%%%%%%%%%%%%%%%%%%%%%%%%%%%%%%%%%%%%%%%%%%%%%%%%%%%%%%%%%%%%%%%%%
%%%%%%%%%%%%%%%%%%%%%%%%%%%%%%%%%%%%%%%%%%%%%%%%%%%%%%%%%%%%%%%%%%%%%%%%%%
\section{Analysis methods}\label{sec:methods}

%%%%%%%%%%%%%%%%%%%%%%%%%%%%%%%%%%%%%%%%%%%%%%%%%%%%%%%%%%%%%%%%%%%%%%%%%%
\subsection{Fundamental parameters via the BCD method}\label{ssec:num3}

The primary goal of this work was to obtain fundamental parameters of Be
stars. In addition the line broadening $v \sin i$ and Balmer emission
equivalent widths were obtained.

%%%%%%%%%%%%%%%%%%%%%%%%%%%%%%%%%%%%%%%%%%%%%%%%%%%%%%%%%%%%%%%%%%%%%%%%%%
\subsubsection{Measuring the Balmer discontinuity}

Stellar parameters were determined with a procedure akin to the BCD method
\citep[Named after the main contributors to the method, Barbier, Chalonge, and
  Divan. See App.~A of][for a description, as well as the references in that
  work for the history of the method]{2009A&A...501..297Z}. The BCD method
uses the height ($D_\star$) and position ($\lambda_1$) of the Balmer
discontinuity for spectral classification. Some adaptations were necessary,
though, since the original BCD method is designed to work on low-resolution
photographic spectra with $\Delta\lambda=8$\,\AA\ at the position of the BD, or
about $R=460$.

\begin{figure}
\includegraphics[angle=0,width=8.8cm,clip]{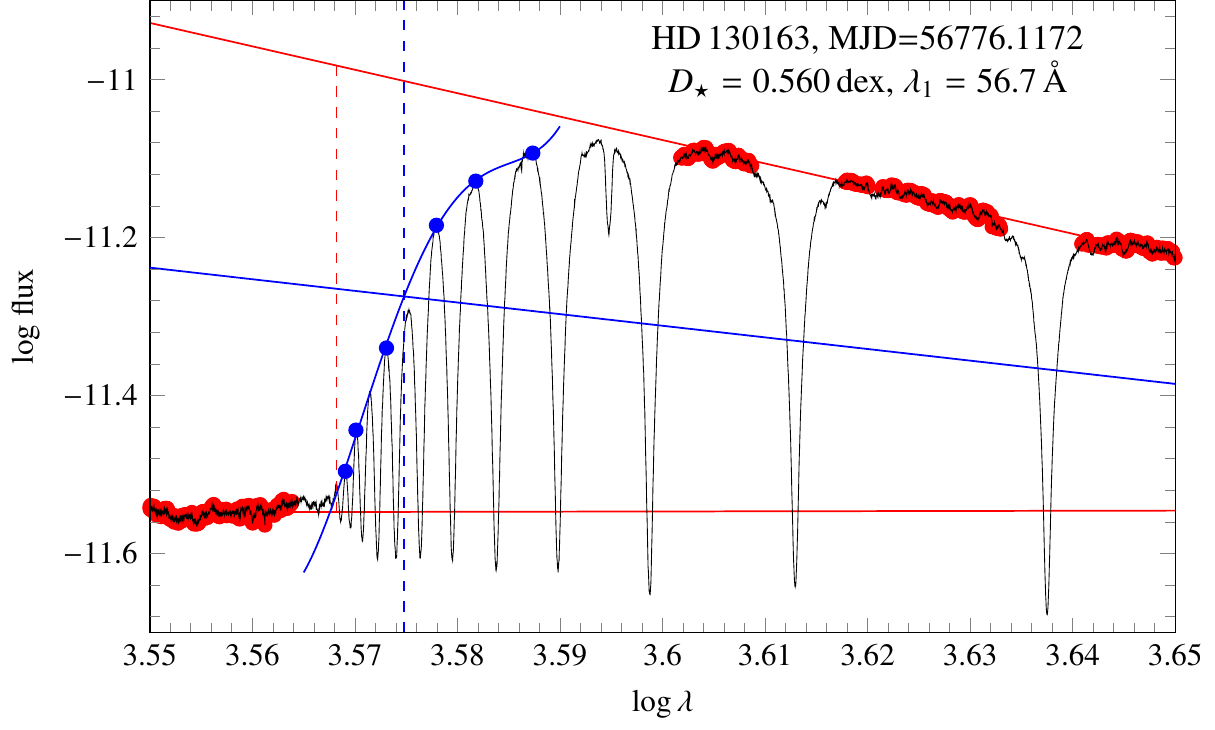}%

\includegraphics[angle=0,width=8.8cm,clip]{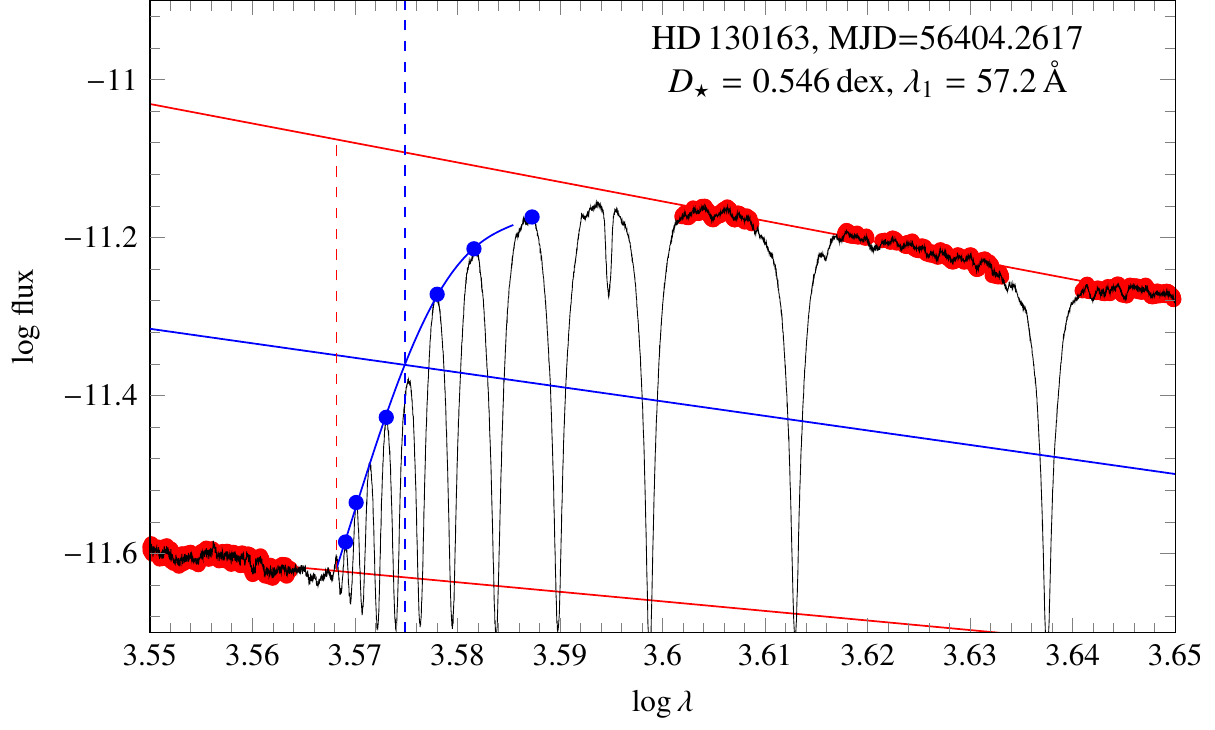}%

\includegraphics[angle=0,width=8.8cm,clip]{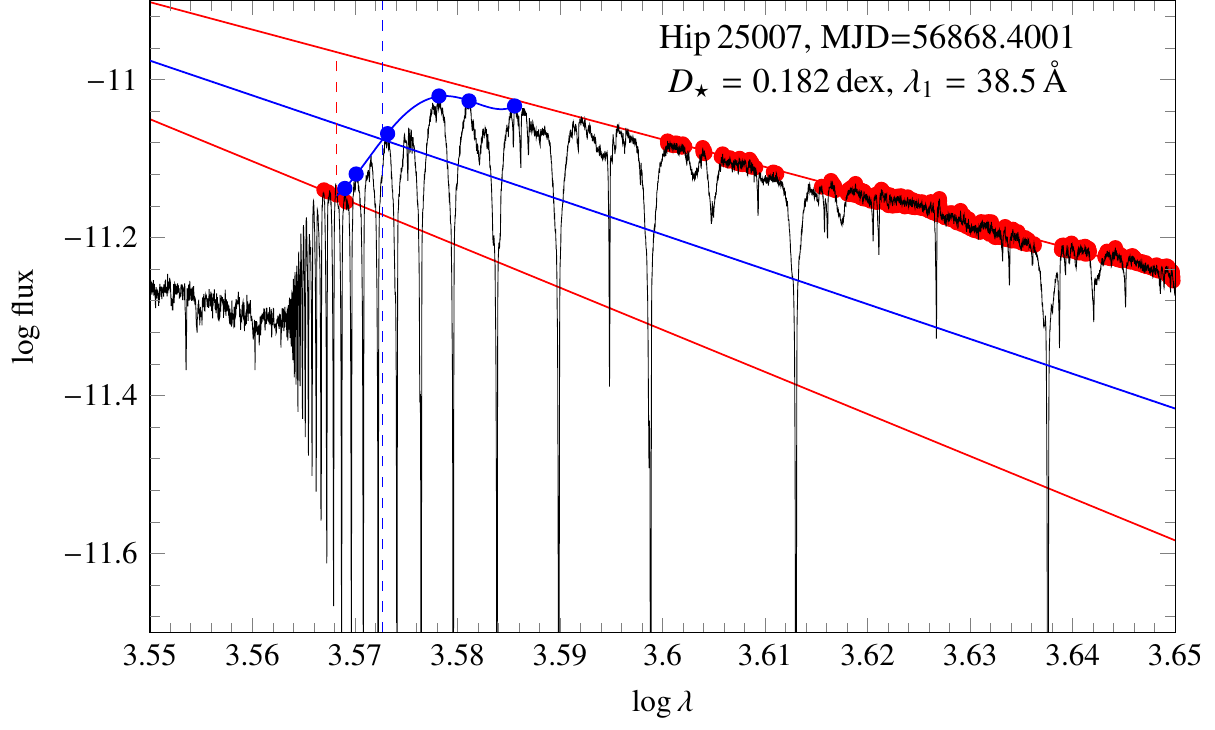}%

\caption{For HD\,130163 in the two upper panels the spectra with the most
  discrepant calibration of the UV flux slope. The steps to determine
  $D_\star$ are indicated in red: selection of continuum points to fit the
  slopes and the difference between the two fits at 3700\,\AA\ as the vlaue of
  $D_\star$.  In blue for $\lambda_1$: the midpoints between the continuum
  fits in red, the upper envelope of the BD. The intersection of this envelope
  with the midpoints to determine $\lambda_1$.  For Hip25007, the lower panel,
  a clear double BD is present, and the Balmer continuum must be determined by
  choosing {\it bona fide} continuum points manually. }
\label{fig:BCD}
\end{figure}

$D_\star$ is measured by extrapolating the Balmer and Paschen continua to a
wavelength of 370\,nm and taking the flux difference in dex, see the solid and
dashed red lines in Fig.~\ref{fig:BCD}. For this, both the spectra with
re;ative flux calibration and the traditionally normalized spectra were
used. In the normalized spectrum, a selection was made to chose the points on
which the continuum fit would be performed in the flux-calibrated
spectrum: All points with a normalized intensity between 0.95 and and 1.05
were selected for this purpose, see the red marked points in
Fig.~\ref{fig:BCD}.
\begin{itemize}
\item To determine the {\bf Paschen continuum}, a selection of these points
  between $\lambda= 398$ and 450\,nm was made. A linear regression was
  computed to the flux-calibrated spectrum in log-log space.
\item For the {\bf Balmer continuum}, and only in cases in which no double
  Balmer discontinuity (see below) was apparent, the wavelength selection was
  made between 350 and 368\,nm, and then again a linear regression was
  computed in log-log space. 
\end{itemize}
The difference between the two regressions at 370\,nm, which corresponds to
3.568 in log scale, is then taken as $D_\star$.

A special problem for Be stars is that the Balmer discontinuity can appear
twice, once due to the stellar photosphere, and a second time due to the disk,
bluewards from the photospheric one. A clearly visible second BD mostly occurs
in strong shell and pole-on Be stars. This is due to the lower temperature and
pressure broadening in the circumstellar environment
\citep{1987A&AS...67..203K,2012A&A...544A..64A}.  {\bf The term shell star
  here denotes a Be star in which the photosphere is (partly) obscured by the
  disk. A pole-on star, to the contrary, is a Be star seen at very low
  inclination. In the former the additonal line opacity is the reason for the
  second BD, in the latter it is the line emission, that in pole-on stars is
  typically much higher, in terms of units of the local continuum, than in Be
  stars of intermediate inclination. While this does not exclude a secondary
  BD in intermediate inclination Be stars, it makes it much harder to detect,
  as it is typically weaker. Indeed, all stars showing a strong secondary BD
  are either shell stars, or stars with very narrow photospheric and emission
  lines.}  The above method of determining the Balmer continuum would fail
here, as it would inevitably focus on the second, circumstellar Balmer
discontinuity. For stars which have such a clearly double Balmer discontinuity
in their spectrum, the position of points for the linear regression for the
Balmer continuum around 370\,nm was defined by hand by selection the points
with the highest flux between Balmer line cores, as demonstrated in the
lowermost panel of Fig.~\ref{fig:BCD}.

The second parameter measured in the BCD system, $\lambda_1$ is the mean
spectral position of the Balmer discontinuity, usually given as differential
from 370\,nm, i.e., $\lambda_1 = \lambda_\mathrm{mean BD}-370$. Following the
original BCD convention, $\lambda_1$ is given in \AA ngstr\"om in this
work. To measure $\lambda_1$, the two linear regressions to Paschen and Balmer
continuum are averaged to a third linear function between the two. Next, the
midpoints between the Balmer lines, which are local flux maxima, are used for
a 4$^\mathrm{th}$ order polynomial fit to estimate an upper envelope of the
flux curve. The wavelength at which this upper envelope intersects the
midpoint between Balmer and Paschen continuum is taken to measure
$\lambda_1$. We note that this position is not independent of the spectral
resolution, in particular for lower resolutions as used in the original BCD
system.

The principle of measurement is illustrated in Fig.~\ref{fig:BCD}. For the
non-Be star HD\,130163, the results and fit for the two most discrepant
observations, in terms of slopes of the Balmer and Paschen continua, are
compared. For the strong Be-shell star Hip\,25007 the procedure in case of a
clear double Balmer discontinuity is illustrated.

%%%%%%%%%%%%%%%%%%%%%%%%%%%%%%%%%%%%%%%%%%%%%%%%%%%%%%%%%%%%%%%%%%%%%%%%%%
%%%%%%%%%%%%%%%%%%%%%%%%%%%%%%%%%%%%%%%%%%%%%%%%%%%%%%%%%%%%%%%%%%%%%%%%%%

%%%%%%%%%%%%%%%%%%%%%%%%%%%%%%%%%%%%%%%%%%%%%%%%%%%%%%%%%%%%%%%%%%%%%%%%%%
\subsubsection{Model grid}\label{ssec:num4}

To determine stellar parameters the measurements will be compared against a
model grid, rather than an existing calibration of the BCD method, also since
calibrations exist only for low resolution (much lower than the lowest
available for X-shooter data). We use the B4 model, {which in terms of
  physics is identical to the {\sc Bruce3} model described in Sect.~3 of
  \citet{2013MNRAS.429..177R}, but has been updated computationally to make
  use of GPU parallel computing. The model creates a surface grid of a
  rotating star based on the Roche model, then assigns each point a local
  value of $\log g$ and $T\mathrm{eff}$ using the Roche model and von Zeipel
  relations, here using a traditional value of $\beta=0.25$. The
  computationally new code then interpolates these pre-computed intensity
  spectra, which take into account the aspect of the line of sight. Finally,
  the numercially updated part integrates the surface grid over the visible
  surface, a task GPUs were explicitly designed for, into the observed
  spectrum. This is done first for input spectra including all spectral lines,
  and then for an input grid of spectra with continuum emission only. The
  resulting spectra can then either be normalized traditionally, i.e., in the
  same way as the observations, or ``perfectly'', with the computed continuum
  spectrum.

}

The set of stellar parameters for the input grid is based on Table 6 of
\citet{2009A&A...501..297Z}, i.e., computing models based on their parameters
for B0 to A1, and for LC’s V to III. Note that any sufficiently dense and even
sampling of the $D_\star$--$\lambda_1$-plane would have produced the same
numerical results for the determined parameters, except it would not have been
possible to give an estimate of spectral types. This grid of models was
produced for five X-shooter instrumental resolutions in the UVB, which depend
on slit-width, as mentioned in section~\ref{sec:obs}.

For the spectrophotometric slit of $5\arcsec$ the resolution is effectively
governed by the seeing, for which a somewhat worse than typical value for
Paranal observatory of $1.0\arcsec$ is assumed. The BCD parameters were
measured in the model spectra in the same way as in the observed data. 

To obtain the stellar parameters, the three models forming a triangle that
encloses the observed value in the $D_\star$--$\lambda_1$-plane were chosen,
and the barycentric coordinates of the observational ($D_\star, \lambda_1$)
pair in this triangle were computed (see Fig.~\ref{fig:exampleBCD}). With
these coordinates, the physical parameters were interpolated. The spectral
type is assigned as the one of the most nearby grid point. We note that this
is not a spectral classification scheme, which would have to rely on actual
standard stars, but is rather meant as an indication only. In a few cases
where stars are outside the grid limits, but still close (see
Fig.~\ref{fig:sampleBCD}), extrapolation was used instead of the barycentric
interpolation.

As Be stars are typically fast rotators, the grid was computed for two values
of $\omega = \Omega/\Omega_\mathrm{crit}$, a slow rotation grid with $\omega =
0.25$ and a fast one with $\omega = 0.85$. The former was computed for only
one value of $v \sin i =20\mathrm\,{kms^{-1}}$, a total of $5\times51$ models,
the latter for inclinations from a pole-on value of $v \sin i
=10\,\mathrm{kms^{-1}}$ to the equator on case in steps of
$10\,\mathrm{kms^{-1}}$, a total of $5\times651$ models. For the fast rotation
grid, at low temperatures and low surface gravities, the equatorial parameters
were outside the grid of input spectra, and no model spectra were produced for
these parameters. {The full grids will be published in CDS as Tables 6 to 15.}

Most program stars are of late type and some are at the very edge of the
$D_\star$--$\lambda_1$-range covered by the models at $\omega = 0.85$.  We
investigate the influence of rapid rotation on the parameter determination
first for a number of test stars only.

%%%%%%%%%%%%%%%%%%%%%%%%%%%%%%%%%%%%%%%%%%%%%%%%%%%%%%%%%%%%%%%%%%%%%%%%%%
\subsubsection{Test cases}\label{ssec:num5}
Three non-Be stars with a large number of observations with the different slit
widths were selected from the archive for the purpose of assessing stability
and reliability of the method. These are Hip\,72362 (HD\,130163, A0\,V, $V =
6.9$\,mag), Hip\,98926 (HD\,190285, A0\,V, $V = 7.2$\,mag), and Hip\,01115
(HD\,955, B3/5\,V, $V = 7.4$\,mag). 

In addition, program stars with ten or more observations were analyzed
in a thorough way similar to the three test stars, before applying the method
to the bulk of objects with fewer observations. All stars used for tests are
given in Table~\ref{tab:teststars}.

An absolute flux calibration is not very reliable across the different slit
widths, so only the relative flux calibration is assessed. For this, all
spectra were normalized to the mean flux in the interval from 404 to
406\,nm. The spectra have a large individual scatter, but the averages for the
different slit widths are indistinguishable. This indicates that, at the
resolutions offered by X-shooter, the differences in resolution does not have
much influence for the method employed.

\begin{figure}
\centering
\includegraphics[width=8.8cm]{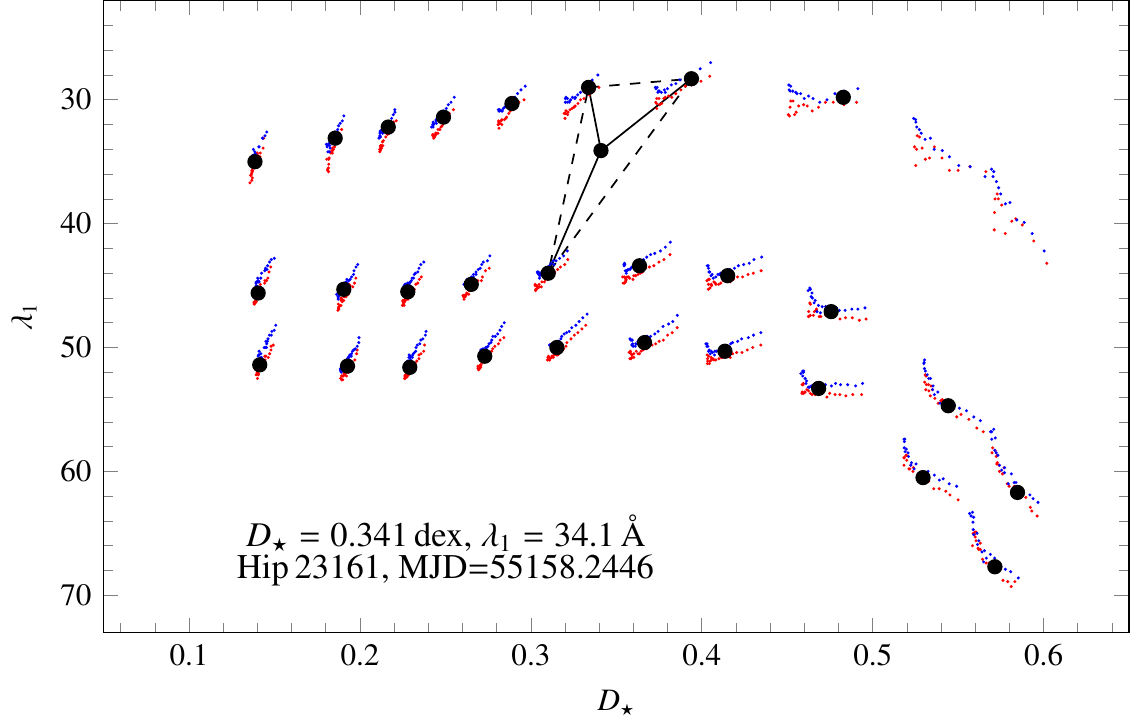}
\caption{The BCD plane and interpolation for one star. Shown are all
  $(D_\star,\lambda_1,)$ points for all computed $v \sin i$ in the models for
  $R=3300$ (red) and $R=9100$ (blue). The chosen sub-grid to analyze this
  observation ($R=5100$ for 1\arcsec\ slit width, $v \sin
  i=150\,\mathrm{kms^{-1}}$), is shown in black, and the surrounding triangle
  is indicated by dashed lines, together with the barycentric coordinates to
  obtain the weighted parameter values by solid lines. The nearest grid point,
  used to associate a spectral type, is for a B7\,III star.
}\label{fig:exampleBCD}
\end{figure}

In particular, there is no systematic difference between the observation with
the $5\arcsec$ slit, which does not suffer from any slit-loss, vs.\ the ones
with smaller slit-widths. This is confirmed by measurements in the model grid,
in which the differences between the resolutions turned out well to be below
the scatter of the measurements and other systematic errors discussed here
(see also Fig.~\ref{fig:exampleBCD}). The last two columns of
Table~\ref{tab:teststars} give the mean and standard deviation of the measured
BCD parameters. Both height and position of the Balmer discontinuity $D_\star$
and $\lambda_1$ can be very well measured in X-shooter data. An imperfect flux
calibration turns out not to be a problem, since, even if not perfect, it is
reasonably stable across the region of interest.  In a sense the fitting
procedure can be regarded as self-calibrating, and only strong slopes,
curvatures, or discontinuities in the flux calibration around 370\,nm would
have a strongly detrimental effect on the derived parameters.

In the next step, and for the non-Be stars only, we investigate how well the
$D_\star$ and $\lambda_1$ values translate into physical parameters.  As
Table~\ref{tab:teststarresults} shows, there are systematic effects between
the slow rotation and the high rotation model grids. The same BCD parameters
analyzed with the slow rotation grid will give systematically higher effective
temperatures, well outside the statistical scatter. For the effective gravity,
the effect is less severe, giving lower $\log g$ for slow rotation, but still
within the limit of the statistical error, {by which the $3\,\sigma$ limit is
  meant, traditionally employed in astronomy vs.\ the more conservative
  $5\,\sigma$ limit often found in other fields of physics}.

In turn, the differences of the BCD measurements with high rotation, but
different inclinations, is negligible. In this context, it does not matter
that we do not know the actual rotation of the non-Be test stars. It only
matters that we do know it for the Be stars, namely that they are rapid
rotators, much closer to 85\% than to 25\%. It follows that the Be stars must
be analyzed with the 85\% grid to avoid the identified systematic
errors. However, we do not need to know the inclination, or even $v \sin i$ of
the Be star with high precision, since this choice does not have a strong
effect on the determined BCD parameters.

Combining the errors listed in Tables~\ref{tab:teststars} and
\ref{tab:teststarresults}, one can estimate a typical error of about 50\,K and
0.03 in $\log g$ for a late type B stars, for which the BCD method has the
highest power of distinction (since the BD parameters change steeply at this
spectral type), and about 400\,K and 0.05 in $\log g$ for a mid type B
star. Since for most targets only one or two spectra are available, we use
these numbers as the typical accuracy.

%%%%%%%%%%%%%%%%%%%%%%%%%%%%%%%%%%%%%%%%%%%%%%%%%%%%%%%%%%%%%%%%%%%%%%%%%%
%%%%%%%%%%%%%%%%%%%%%%%%%%%%%%%%%%%%%%%%%%%%%%%%%%%%%%%%%%%%%%%%%%%%%%%%%%
\begin{table*}
%\footnotesize
\caption{Objects, their derived spectral types, and the number of observations
  at each slit width to test the method and obtain the statistical scatter and
  the mean measured $D_\star$ and $\lambda_1$ values.}
\label{tab:teststars}
\begin{center}
\renewcommand{\arraystretch}{1.1}% Tighter
%\footnotesize
\begin{tabular}{l|lllllllll}

  Star      & Sp Type & \multicolumn{6}{c} {N of observations/slit}  &  $D_\star$                & $\lambda_1$-3700       \\
            &       & {$0.5\arcsec$} &{$0.8\arcsec$} &{$1.0\arcsec$}& {$1.3\arcsec$}& {$1.6\arcsec$}&  {$5.0\arcsec$}    &  dex              & \AA             \\[2mm]
  HD\,130163  & A1V   & ---  & ---  &  38  &  5   &   6  &  1  & 0.545 $\pm$ 0.006 & 63.65 $\pm$ 0.97 \\ 
  HD\,190285 &  A1V  &  14  &   8  &  15  &  1   &   5  & --- & 0.565 $\pm$ 0.013 & 68.16 $\pm$ 1.1  \\
  Hip\,1115 & B3IV  & ---  &   2  &   2  & ---  &   5  &  2  & 0.253  $\pm$ 0.006 & 45.65 $\pm$ 2.0 \\[1mm]
  Hip\,32474 & A0IIIe &  1   &  --- &   7  & ---  &   2  & --- & 0.522  $\pm$ 0.01 & 44.525 $\pm$ 3.5 \\
  Hip\,39483 & B4IIIe & ---  &   3  &   5  &  1   &   1  & --- & 0.281 $\pm$ 0.008 & 34.44 $\pm$ 1.0  \\
  Hip\,52977 & B4IIIe &  7   &  --- &   5  &  8   &   1  & --- & 0.263 $\pm$ 0.017 & 34.44 $\pm$ 3.04 \\
  Hip\,71974 & A0IIIe &  5   &  --- &   5  &  1   &   2  & --- & 0.522 $\pm$ 0.007 & 47.89 $\pm$ 1.1  \\ 
  Hip\,85138 & B7IIIe &  6   &  --- &   4  & ---  &   3  & --- & 0.393 $\pm$ 0.011 & 34.44 $\pm$ 2.4  \\
  Hip\,85195 & B9IIIe &  5   & ---  &  16  &  3   &   6  & --- & 0.479 $\pm$ 0.007 & 38.92 $\pm$ 1.1  \\
  Hip\,88374 & B9IVe  &  1   &   2  &  10  &  2   &   2  &  2  & 0.443 $\pm$ 0.006 & 44.52 $\pm$ 2.07 \\
  Hip\,89486 & A0IIIe &  4   & ---  &  6   & ---  &   4  & --- & 0.512 $\pm$ 0.01 & 41.16 $\pm$ 0.88  \\
  Hip\,94986 & B4IIIe &   8  &   5  &  7   &  1   &   5  & --- & 0.278 $\pm$ 0.02  & 36.68 $\pm$ 3.52 \\
  Hip104508 & B7IIIe & ---  &   6  &  14  & ---  &   1  & --- & 0.431 $\pm$ 0.014 & 36.68 $\pm$ 2.95 \\
 % \noalign{\smallskip}
      \end{tabular}
      \end{center}
      \end{table*}

\begin{table*}
%\footnotesize
\caption{Stellar parameter for the non-Be stars from Table 1, derived under
  different assumptions for the stellar rotation.}\quad
\label{tab:teststarresults}
\begin{center}
\renewcommand{\arraystretch}{1.1}% Tighter
%\footnotesize
\begin{tabular}{l|cc|cc|ccc}
Star     &  \multicolumn{2}{c|}{$\omega=0.25$, $v\sin i=20\mathrm{km\,s}^{-1}$ } 
    &  \multicolumn{2}{c|}{$\omega=0.85$, $v\sin i=20\mathrm{km\,s}^{-1}$ } 
    &  \multicolumn{3}{c}{$\omega=0.85$, max converging or measured $v \sin i$ } \\
         &   $T_\mathrm{eff}$ & $\log g_\mathrm{pole}$   &   $T_\mathrm{eff}$ & $\log g_\mathrm{pole}$   &  $v \sin i$ & $T_\mathrm{eff}$ & $\log g_\mathrm{pole}$  \\
         & [K]  &    & [K]  &      & [$\mathrm{km\,s}^{-1}$]  & [K]    &               \\
 HD\,130163 & $9763\pm64$ & $3.76\pm0.03$ & $9481\pm63$ & $3.89\pm0.03$ &120 & $9592\pm54$ & $3.86\pm0.03$ \\
HD\,190285 & \multicolumn{2}{c|}{outside of grid} &  \multicolumn{2}{c|}{outside of grid} &130 & $9358\pm32$ & $3.94\pm0.02$ \\
Hip\,1115  & $15296\pm324$ & $3.78\pm0.05$ & $14970\pm349$ & $3.91\pm0.05$ &150& $15119\pm358$ & $3.92\pm0.05$ \\
\end{tabular}

     \end{center}
        \end{table*}
%%%%%%%%%%%%%%%%%%%%%%%%%%%%%%%%%%%%%%%%%%%%%%%%%%%%%%%%%%%%%%%%%%%%%%%%%%
%%%%%%%%%%%%%%%%%%%%%%%%%%%%%%%%%%%%%%%%%%%%%%%%%%%%%%%%%%%%%%%%%%%%%%%%%%

%%%%%%%%%%%%%%%%%%%%%%%%%%%%%%%%%%%%%%%%%%%%%%%%%%%%%%%%%%%%%%%%%%%%%%%%%%
\subsection{Projected rotational velocity}\label{sssec:num9}

When available, $v \sin i$ was taken from the literature; otherwise for each
observed spectrum the rotational parameter $v \sin i$ was fitted using
synthetic spectra to the \ion{Mg}{ii}\,4481\,\AA\ line, which is reasonably
strong across the entire range of spectral types investigated in this work,
mostly B5 to B9 with a few earlier ones only.

For the fit the line profiles were computed with two different sets of
underlying model atmospheres. For effective temperatures below 15\,000\,K
ATLAS9 LTE model atmospheres \citep{1969tons.conf..375K} were used. Above that
temperature TLUSTY NLTE atmospheres \citep{2007ApJS..169...83L} were used. As
values for $T_\mathrm{eff}$ and $\log g$ the ones obtained by the BCD method
were used.

The results of $v \sin i$ for each star are given in Table~\ref{tab:params},
and Appendix~\ref{app:vsini} shows the observed and fitted line profiles for
each star.

%%%%%%%%%%%%%%%%%%%%%%%%%%%%%%%%%%%%%%%%%%%%%%%%%%%%%%%%%%%%%%%%%%%%%%%%%%
\subsection{Disk variability and other observations of interest}\label{ssec:num6}

Often Be stars show variability in their emission equivalent width (EW) and in
the line profiles \citep{2013A&A...550A..79C}.  This was checked visually in
the spectra, as shown in Appendix~\ref{app:spec}, and flagged in
Table~\ref{tab:params}.

In addition, equivalent widths were measured for the Balmer lines H$\alpha$
and H$\beta$, together with the equivalent widths in the models for these
stars, see Table~\ref{table:balmdec}. The goal to measure the Balmer
decrement, i.e., the ratio of emission strength, however, was not achieved
with acceptable accuracy. This is because the emission in general is often
weak, and the Balmer decrement is steep for late-type Be stars. This is seen
in the total H$\beta$ EWs in Table~\ref{table:balmdec}, none of which is
negative, i.e. all are still dominated by the photospheric absorption.  The
values for H$\beta$ after subtracting the model photospheric EW are dominated
by the systematic errors arising from the stellar parameters and the resulting
D34 Balmer decrement does not allow to draw any reliable conclusion.

Some stars show the infrared \ion{Ca}{ii} triplet in emission. It has been
speculated that this is connected to binarity, in particular accretion onto a
secondary \citep{1976IAUS...70..401P}. \citet{2012arXiv1205.2259K} reject
this, but accede that the presence of this line must be due to some not
further specified peculiarity in the circumstellar environment. The stars for
which a clear or possible \ion{Ca}{ii} triplet emission is observed are also
flagged in Table~\ref{tab:params}.

%%%%%%%%%%%%%%%%%%%%%%%%%%%%%%%%%%%%%%%%%%%%%%%%%%%%%%%%%%%%%%%%%%%%%%%%%%
%%%%%%%%%%%%%%%%%%%%%%%%%%%%%%%%%%%%%%%%%%%%%%%%%%%%%%%%%%%%%%%%%%%%%%%%%%

%======================================================================================================
\section{Results}\label{sec:num4} 
In this section the overall results concenring the sample are reported, for
notes and observations on individual stars see Appendix~\ref{sec:notes}.

%%%%%%%%%%%%%%%%%%%%%%%%%%%%%%%%%%%%%%%%%%%%%%%%%%%%%%%%%%%%%%%%%%%%%%%%%%
%%%%%%%%%%%%%%%%%%%%%%%%%%%%%%%%%%%%%%%%%%%%%%%%%%%%%%%%%%%%%%%%%%%%%%%%%%
\begin{table*}																			
\caption{Program stars with the number of valid observations, whether
  H$\alpha$ variability, a single or double BD, and the \ion{Ca}{ii} IR
  triplet is observed, the measured $D_\star$and $\lambda_1$ (mean values in
  case of more than one observation) and stellar parameters. For the double BD
  and \ion{Ca}{ii} flags ``Y'' and ``N'' are clear statements, ``:'' means
  uncertain, and ``---'' mean no suitable spectra were available. Newly
  identified Be stars are marked in bold face. In case of H$\alpha$
  variability, ``---'' means that all spectra, even if more than one, were
  taken in the same night. }
\label{tab:params}	
\begin{center}																			
\begin{tabular}{llllllllllll}																		
Star	& \# of obs. &  Sp   & H$\alpha$ & Double &IR \ion{Ca}{ii} &  $D_\star$ &$\lambda_1$ -3700  & $T_\mathrm{eff}$	& $\log g_\mathrm{{pol}}$&    $v\sin i$ 	\\
        &  U/V/N    &  Type & var?      & BD?   &    emiss.?         &[dex]    &          [\AA]   & [K]           &	[dex]	       & [$\mathrm{km\,s^{-1}}$]	\\								
\hline																				
	\object{Hip\,11116}	&4/4/4	&B8III		& N	& N	& N	&0.35	&29.97	&11994	&3.47	&190	\\
	\object{Hip\,15188} 	&1/1/1	&B3: 		& ---	& ---	& N	&---	&---	&---	&---	&---	\\
{\bf \object{Hip\,23161}}	&2/2/2	&B7III		& ---	& N	& N	&0.335	&29.97	&12272	&3.45	&150	\\
	\object{Hip\,24475}	&3/4/3	&B9II 		& Y	& N	& N	&0.44	&25.50	&10643	&3.29	&220	\\
	\object{Hip\,25007}	&6/6/6	&B3IV   	& Y	& Y	& N	&0.183	&38.30	&18089	&3.63	&---	\\
{\bf \object{Hip\,25690}}	&4/4/4	&A0IV   	& ---	& N	& N	&0.525	&50.14	&9691	&3.70	&90	\\
	\object{Hip\,25950}	&1/2/0	&B7:    	& N     & ---   & N     &---	&---	&---	&---	&---	\\
	\object{Hip\,26368}	&2/2/2	&A1III  	& N	& N	& Y	&0.535	&45.65	&9550	&3.57	&190	\\
	\object{Hip\,26964}	&2/2/1	&B4V   		& ---	& Y	& N	&0.262	&50.2	&15220	&4.00	&200	\\
	\object{Hip\,28561}	&3/3/3	&B8IV 		& Y	& Y	& Y	&0.399	&40.4 	&11554	&3.73	&70	\\
{\bf \object{Hip\,29635}}	&8/8/9	&B8IV  		& N	& N	& Y	&0.395	&41.16	&11409	&3.75	&230	\\
{\bf \object{Hip\,31362}}	&1/1/1	&B8IV  		& ---	& N	& N	&0.443	&38.92	&10769	&3.64	&270	\\
{\bf \object{Hip\,32474}}	&10/14/16&A0III		& Y	& N	& N	&0.524	&44.52	&9733	&3.60	&120	\\
	\object{Hip\,33509}	&1/1/1	&B4V   		& ---	& Y	& N	&0.226	&45.65	&15747	&3.92	&160	\\
{\bf \object{Hip\,34144}}	&1/1/1	&B8III 		& ---	& N	& N	&0.4	&27.73	&11204	&3.38	&200	\\
{\bf \object{Hip\,36009}}	&1/1/1	&B4V   		& ---	& N	& N	&0.23	&50.14	&16435	&4.00	&40	\\
{\bf \object{Hip\,37007}}	&6/6/6	&B7IV  		& N	& N	& Y	&0.34	&36.68	&12245	&3.70	&230	\\
{\bf \object{Hip\,39183}}	&2/2/2	&A0III 		& Y	& N 	& :	&0.481	&25.50	&10167	&3.22	&190	\\
{\bf \object{Hip\,39483}}	&10/11/11&B6IV 		& Y	& N	& N	&0.28	&34.44	&13583	&3.60	&130	\\
{\bf \object{Hip\,39595}}	&1/1/1	&B9IV  		& ---	& N	& Y	&0.455	&38.92	&10592	&3.63	&240	\\
{\bf \object{Hip\,41085}}	&1/1/1	&B5IV  		& ---	& Y	& N	&0.245	&41.16	&14860	&3.84	&250	\\
{\bf \object{Hip\,41268}}	&2/2/2	&B8III 		& ---	& Y	& N	&0.384	&29.97	&11460	&3.45	&240	\\
{\bf \object{Hip\,42060}}	&1/1/1	&B7V   		& ---	& N	& N	&0.35	&45.65	&12235	&3.95	&250	\\
	\object{Hip\,43073}	&1/1/1	&B9IV  		& ---	& N	& N	&0.466	&38.92	&10413	&3.62	&250	\\
{\bf \object{Hip\,43114}}	&1/1/0	&B5IV:  	& --- 	& N 	& N	&---	&---	&---	&---	&---	\\
{\bf \object{Hip\,44423}}	&2/3/3	&B7IV  		& N	& N	& N	&0.362	&38.92	&11909	&3.74	&160	\\
	\object{Hip\,46329}	&2/2/2	&B5IV  		& Y	& N	& N	&0.25	&45.65	&14763	&3.95	&160	\\
	\object{Hip\,47868}	&1/1/1	&B0III:		& ---	& N	& N	&0.067	&23.70	&---	&---	&---	\\
{\bf \object{Hip\,47962}}	&1/3/3	&A0IV  		& N	& N	& N	&0.517	&45.65	&9786	&3.65	&220	\\
{\bf \object{Hip\,48582}}	&4/6/6	&B5IV  		& Y	& Y	& Y	&0.26	&51.26	&15396	&4.01	&---	\\
	\object{Hip\,48943}	&2/5/2	&B5IV  		& N	& Y	& Y	&0.284	&43.95	&14077	&3.81	&190	\\
	\object{Hip\,51444}	&1/2/2	&B5III 		& Y	& N	& N	&0.23	&25.50	&14795	&3.30	&250	\\
	\object{Hip\,51491}	&3/3/4	&B9III 		& Y	& N	& Y	&0.475	&38.92	&10341	&3.60	&230	\\
{\bf \object{Hip\,51546}}	&2/2/2	&A0V:  		& ---	& N	& N	&---	&---  	&---	&---	&210	\\
{\bf \object{Hip\,52977}}	&21/21/21&B6III 	& Y	& N	& N	&0.26	&34.44	&14076	&3.57	&200	\\
{\bf \object{Hip\,56393}}	&1/1/1	&A1IV  		& ---	& N	& :	&0.56	&52.39	&9200	&3.62	&270	\\
	\object{Hip\,57861}	&4/6/6	&B6IV  		& Y 	& N 	& N	&0.29	&38.92	&13390	&3.76	&270	\\
{\bf \object{Hip\,59970}}	&2/2/2	&A1IV-III	& ---	& N	& :	&0.574	&48.45	&---	&---	&---	\\
{\bf \object{Hip\,64501}}	&2/2/2	&B8IV  		& ---  	& N 	& N	&0.39	&38.92	&11512	&3.77	&---	\\
{\bf \object{Hip\,64867}}	&5/6/6	&A0III 		& N	& N	& N	&0.524	&36.68	&9708	&3.43	&170	\\
	\object{Hip\,66339}	&1/1/1	&B3V  		& ---	& Y	& N	&0.222	&52.7	&17493	&4.00	&210	\\
	\object{Hip\,66351}	&2/2/2	&B9III 		& ---	& N	& N	&0.469	&37.80	&10404	&3.57	&140	\\
	\end{tabular}
	\end{center}	
\end{table*}	                	

\setcounter{table}{2}
\begin{table*}																			
\caption{Continued}
\label{tab:params2}	
\begin{center}																			
\begin{tabular}{llllllllllll}																		
Star	& \# of obs. &  Sp   & H$\alpha$ & Double &IR \ion{Ca}{ii} &  $D_\star$ &$\lambda_1$ -3700  & $T_\mathrm{eff}$	& $\log g_\mathrm{{pol}}$&    $v\sin i$ 	\\
        &  U/V/N   &  Type & var?      & BD?   &    emiss.?         &[dex]    &          [\AA]   & [K]           &	[dex]	       & [$\mathrm{km\,s^{-1}}$]	\\								
\hline

{\bf \object{Hip\,68100}}	&2/2/2	&B6V   		& ---	& N	& N	&0.326	&46.77	&12804	&3.96	&100	\\
{\bf \object{Hip\,69429}}	&4/6/6	&A0IV  		& N	& N	& N	&0.521	&45.65	&9749	&3.63	&190	\\
	\object{Hip\,71668} 	&3/3/3	&B2V: 		& N	& N	& N	&---	&---	&---	&---	&---	\\
{\bf \object{Hip\,71974}}	&12/141/4&A0IV  	& N	& N	& N	&0.522	&47.89	&9709	&3.64	&160	\\
{\bf \object{Hip\,78375}}	&1/1/1	&B9: 		& N	& N	& :	&---	&---	&---	&---	&---	\\
{\bf \object{Hip\,80577}}	&3/3/3	&B8IV  		& Y	& N	& N	&0.39	&47.89	&11472	&3.97	&100	\\
{\bf \object{Hip\,80820}}	&3/3/3	&B9III 		& Y	& N	& N	&0.44	&25.50	&10576	&3.28	&150	\\
{\bf \object{Hip\,81321}}	&0/2/2	&A0V: 		& --- 	&---	& N	&---	&---	&---	&---	&---	\\
	\object{Hip\,82874}	&1/1/0	&B7III 		& ---	& N	& N	&0.311	&32.20	&12755	&3.50	&240	\\
{\bf \object{Hip\,83278}}	&3/3/3	&B9IV  		& N	& N	& N	&0.437	&36.68	&10816	&3.59	&230	\\
{\bf \object{Hip\,84184}}	&3/3/3	&B8IV  		& Y	& Y	& N	&0.37	&38.92	&11808	&3.77	&270	\\
{\bf \object{Hip\,85138}}	&14/15/15&B8IV 		& Y	& N	& N	&0.39	&34.44	&11365	&3.61	&180	\\
{\bf \object{Hip\,85195}}	&28/29/29&B9III		& Y	& N	& N	&0.48	&38.92	&10253	&3.58	&250	\\
{\bf \object{Hip\,85566}}	&3/3/3	&A1III 		& N	& N	& :	&0.53	&38.92	&9572	&3.45	&240	\\
	\object{Hip\,87032}	&3/3/3	&B8V:   	& N     & N     & Y     &---	&---	&---	&---	&300	\\
{\bf \object{Hip\,87698}}	&1/2/2	&A0III 		& N	& N	& N	&0.523	&41.16	&9725	&3.54	&210	\\
	\object{Hip\,88172}	&2/5/3	&B8IV  		& N	& N	& N	&0.39	&35.56	&11436	&3.65	&180	\\
{\bf \object{Hip\,88374}}	&21/25/25&B9IV 		& Y	& N	& N 	&0.445	&45.65	&10730	&3.84	&140	\\
{\bf \object{Hip\,89486}}	&14/14/15&A0III		& Y	& N	& N	&0.51	&41.16	&9874	&3.57	&180	\\
{\bf \object{Hip\,89500}}	&5/5/5	&B9III 		& N	& N	& N	&0.46	&29.97	&10443	&3.39	&240	\\
	\object{Hip\,90096}	&4/5/5	&B8IV  		& N	& N	& N	&0.41	&36.68	&11170	&3.65	&200	\\
{\bf \object{Hip\,90509}}	&2/2/2	&B8IV  		& ---	& N	& :	&0.42	&38.92	&11062	&3.71	&250	\\
{\bf \object{Hip\,91460}}	&3/3/3	&A1III 		& N	& N	& N	&0.54	&32.20	&9503	&3.25	&150	\\
	\object{Hip\,91975}	&2/3/3	&B9III 		& Y	& N	& :	&0.455	&35.56	&10600	&3.55	&250	\\
{\bf \object{Hip\,92038}}	&3/3/3	&B7III 		& Y	& Y	& Y	&0.303	&32.20	&13002	&3.54	&170	\\
	\object{Hip\,93993}	&3/3/3	&A0III 		& N	& N	& N	&0.524	&43.40	&9694	&3.58	&{10}\\
	\object{Hip\,94770}	&2/2/2	&B8IV  		& ---	& N	& Y	&0.41	&35.56	&11145	&3.60	&260	\\
{\bf \object{Hip\,94859}}	&1/3/3	&B7III 		& Y	& N	& :	&0.33	&35.56	&12673	&3.65	&240	\\
{\bf \object{Hip\,94986}}	&24/36/36&B6III 	& Y	& N	& N	&0.28	&36.68	&13706	&3.68	&120	\\
{\bf \object{Hip\,95109}}	&1/2/2	&B8V   		& N	& N	& N	&0.395	&49.01	&11518	&3.98	&260	\\
{\bf \object{Hip\,96453}}	&1/3/3	&B6III 		& Y	& N	& N	&0.26	&29.97	&14172	&3.44	&230	\\
	\object{Hip\,99457}	&7/9/3	&B1IV  		& Y	& N	& N	&0.101	&41.16	&22907	&3.71	&---	\\
	\object{Hip\,100664}	&2/3/3	&A0III 		& N	& N	& N	&0.506	&36.68	&9908	&3.47	&190	\\
{\bf \object{Hip\,104508}}	&18/24/24&B9IV 		& Y	& N	& N	&0.43	&36.68	&10915	&3.61	&270	\\
	\object{Hip\,108022}	&2/5/5	&B6III 		& N	& N	& N	&0.29	&27.73	&13100	&3.40	&110	\\
	\object{Hip\,108402}	&5/5/5	&B7V   		& N	& Y	& Y	&0.391	&61.39	&12682	&4.03	&250	\\
	\object{Hip\,108597}	&3/3/3	&B5IV  		& N	& Y	& Y	&0.256	&38.92	&14473	&3.76	&---	\\
	\object{Hip\,108975}	&2/2/2	&B4IV  		& N	& N	& N	&0.264	&38.65	&14648	&3.66	&---	\\
						
	\end{tabular}	
	\end{center}	
	\end{table*}

\begin{table*}[]
%\begin{center}	
\caption{Equivalent widths of H$\alpha$ and H$\beta$ for the program stars
  with clearly measurable emission.}\bigskip
 \label{table:balmdec}
 \begin{center}

\begin{tabular}{ccccccccccccc}

 Star         & $T_\mathrm{eff}$  & \multicolumn{3}{c}{H$\alpha$}  &  \multicolumn{3}{c}{H$\beta$}    \\
              & [K]  & (tot)      &(phot)       &  (emi)        &(tot)     &(phot)     & (emi)            \\[2mm]

Hip\,11116	& 12240  & 	-7.94 	& 4.20	& -12.15 & 5.23	& 7.24 & -2.00	\\ % &  2.32	
Hip\,26368	& 9874	 &       5.39 	& 6.83	& -1.435 & 10.8	& 12.1 & -1.28	\\ % &  0.43	
Hip\,29635	& 11822  & 	-1.69 	& 5.39	& -7.085 & 7.76	& 9.55 & -1.78	\\ % &  1.52	
Hip\,33509	& 17064  & 	-16.4 	& 3.57	& -20.06 & 3.31	& 6.40 & -3.08	\\ % &  2.49	
Hip\,37007	& 15093  & 	-5.95 	& 4.53	& -10.48 & 6.06	& 7.57 & -1.50	\\ % &  2.68	
Hip\,39183	& 11231  & 	 0.65 	& 5.89	& -5.235 & 7.68	& 8.51 & -0.82	\\ % &  2.44	
Hip\,39595	& 12234  & 	-1.65 	& 5.68	& -7.337 & 9.12	& 10.1 & -1.03	\\ % &  2.71	
Hip\,46329	& 14599  & 	-8.91 	& 4.27	& -13.18 & 5.28	& 6.85 & -1.56	\\ % &  3.24	
Hip\,51491	& 11453  & 	 4.44 	& 5.94	& -1.493 & 8.39 & 10.5 & -2.12	\\ % &  0.26	
Hip\,80577	& 11697  & 	 0.43 	& 5.94	& -5.503 & 9.07	& 10.1 & -1.09	\\ % &  1.93	
Hip\,82874	& 14099  & 	-6.90 	& 3.93	& -10.83 & 5.13	& 6.75 & -1.61	\\ % &  2.57	
Hip\,87032	& 11936  & 	-6.37 	& 6.70	& -13.07 & 7.89	& 11.1 & -3.22	\\ % &  1.55	
Hip\,90096	& 13441  & 	 3.05 	& 5.20	& -2.146 & 8.56	& 9.02 & -0.45	\\ % &  1.82	
Hip\,90509	& 13123  & 	 0.44 	& 5.37	& -4.925 & 8.76	& 9.51 & -0.74	\\ % &  2.53	
Hip\,91975	& 11951  & 	 1.64 	& 5.48	& -3.830 & 7.70	& 9.45 & -1.74	\\ % &  0.84	
Hip\,92038	& 15139  & 	-14.4 	& 3.80	& -18.23 & 3.80	& 5.98 & -2.18	\\ % &  3.21	
Hip\,94770	& 12721  & 	-2.43 	& 5.12	& -7.561 & 7.70	& 8.84 & -1.13	\\ % &  2.55	
Hip\,94859	& 15311  & 	-2.25 	& 4.26	& -6.523 & 6.80	& 8.12 & -1.31	\\ % &  1.91	
Hip\,95109	& 11847  & 	-2.64 	& 5.96	& -8.601 & 8.56	& 10.2 & -1.68	\\ % &  1.96	
Hip\,96453	& 14338  & 	 2.81 	& 4.52	& -1.706 & 5.51	& 5.77 & -0.25	\\ % &  2.55	
Hip\,99457	& 23061  & 	-1.37 	& 2.51	& -3.889 & 2.73	& 3.71 & -0.97	\\ % &  1.53	
\end{tabular}
\end{center}
\end{table*}

%%%%%%%%%%%%%%%%%%%%%%%%%%%%%%%%%%%%%%%%%%%%%%%%%%%%%%%%%%%%%%%%%%%%%%%%%%
%%%%%%%%%%%%%%%%%%%%%%%%%%%%%%%%%%%%%%%%%%%%%%%%%%%%%%%%%%%%%%%%%%%%%%%%%%

\begin{table*}[]
\begin{center}	
\caption{Statistics of Be stars observed with X-shooter}
 \label{tab:stat}

\begin{tabular}{l|cccc|ccccc|cccccc}
Bin name    & \multicolumn{3}{c}{early}  & $\Sigma$& \multicolumn{4}{|c}{mid} &$\Sigma$& \multicolumn{5}{|c}{late} &$\Sigma$\\
Sp.\ type          & B0 & B1 & B2 & &B3 & B4 & B5 & B6 &                & B7 & B8 & B9 & A0 & A1  \\[2mm]
\# of all stars    & 4  & 10 & 85 & \it 99  &129& 38 & 112& 34    &  \it 313  & 57 & 211& 347& 52 & 14   &\it 681   \\
\# of Be stars     & 1  & 1  &  1 & \it 3  &3 &  4 &  7 &  7      &  \it  21  & 9  & 16 &  13& 13 & 5    &\it  56  \\
\# of new Be stars & 0  & 0  &  0 & \it 0   &0  &  1 &  3 & 5     &  \it   9  & 6  &  10&  8 & 11 & 4    &\it  39  \\[1mm]
   LC V            & 0  & 0  & 1  & \it 1  &1 &  3 &  0 & 1       &  \it   5  & 2  &  2 & 0  &  1 & 0    &\it   5 \\
   LC IV           &  0 & 1  & 0  &\it 1   &1  &  1 &  5 & 2      &  \it   9  & 2  &  11& 5  &  5 & 2    &\it  25  \\
   LC III          & 1  &  0 & 0  & \it 1  &0 &  0 &  1 & 4       &  \it   5  & 4  &  3 & 6  &  7 & 3    &\it  23  \\[1mm]
\ion{Ca}{ii} emi/abs&  0 & 0  & 0  &\it 0 &  0 &  0 &  3 &  0     &  \it  3   & 4  & 5  & 4  &  1 &  4   &\it 18   \\[1mm]
Variability detectable?& 0  & 1  & 1  & \it 2 &1  & 1  &  5 & 6   &  \it  13  & 6  & 11 &  10&  10& 3    &\it  40  \\
H$\alpha$ variable &  0 & 1  & 0  & \it 1  &1  &  0 &  3 &  5      &  \it   9  & 2  &  4 &  6 &  2 &  0   &\it  14  \\
%\# of stars    &    &    &    &    &    &    &    &    &    &    &    &     \\
%\# of stars    &    &    &    &    &    &    &    &    &    &    &    &     \\

\end{tabular}
\end{center}
\end{table*}

%%%%%%%%%%%%%%%%%%%%%%%%%%%%%%%%%%%%%%%%%%%%%%%%%%%%%%%%%%%%%%%%%%%%%%%%%%
%%%%%%%%%%%%%%%%%%%%%%%%%%%%%%%%%%%%%%%%%%%%%%%%%%%%%%%%%%%%%%%%%%%%%%%%%%
%\section{Discussion}

%%%%%%%%%%%%%%%%%%%%%%%%%%%%%%%%%%%%%%%%%%%%%%%%%%%%%%%%%%%%%%%%%%%%%%%%%%
\begin{figure}
\centering
\includegraphics[width=8.8cm]{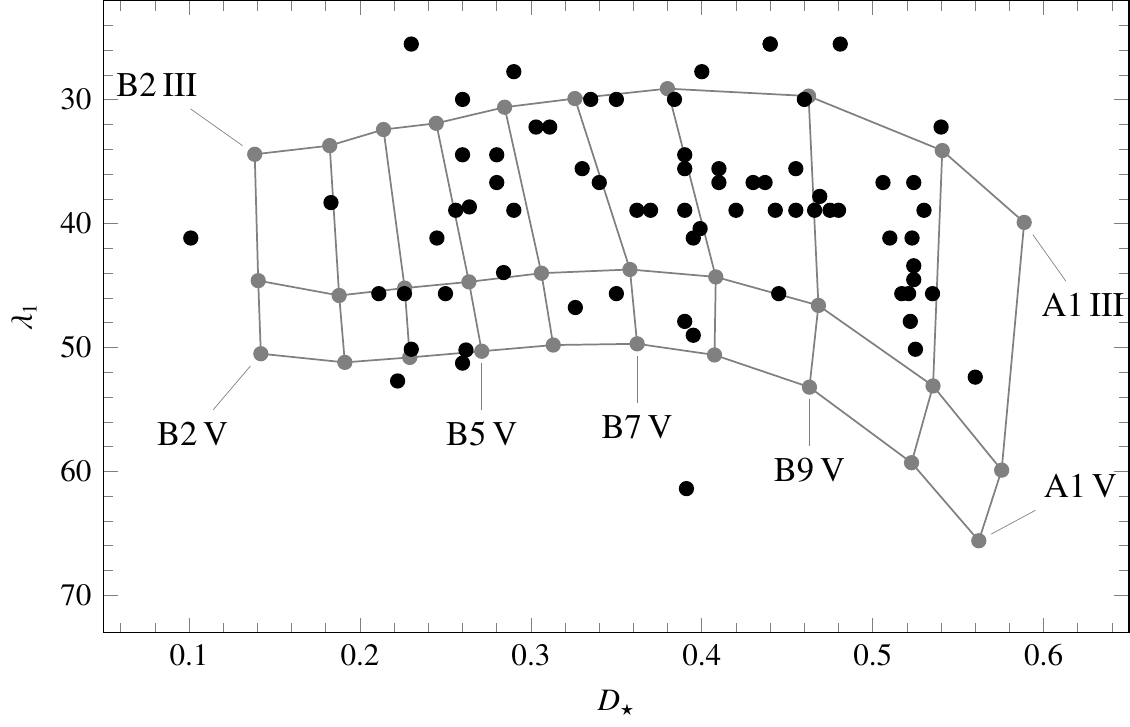}

\includegraphics[width=8.8cm]{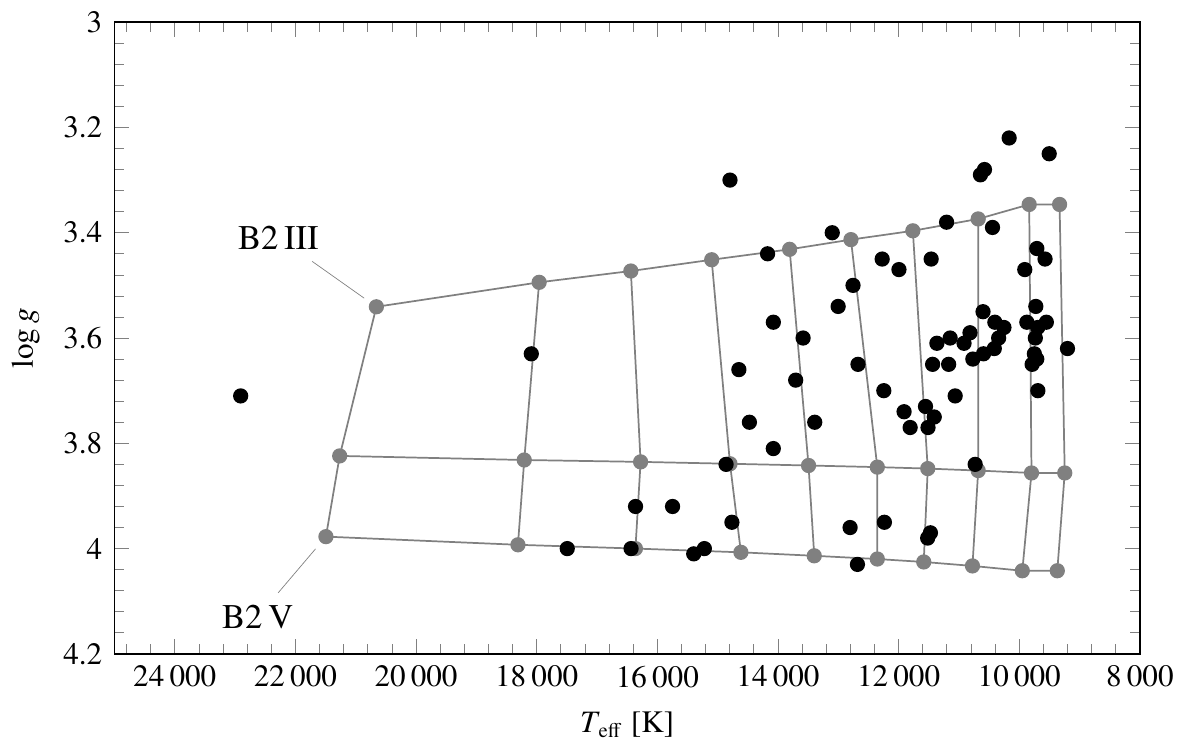}
\caption{Distribution of the Be stars in the $D_\star$-$\lambda_1$ plane (top)
  and $T_\mathrm{eff}$-$\log g$ plane (bottom). The model
  grid points symbolizing the center of the respective spectral types are shown
  in grey.
  The bias towards later type Be stars is clearly apparent from the lower
  panel, with only one star hooter than about 20\,000\,K. }
\label{fig:sampleBCD}
\end{figure}
%%%%%%%%%%%%%%%%%%%%%%%%%%%%%%%%%%%%%%%%%%%%%%%%%%%%%%%%%%%%%%%%%%%%%%%%%%

%%%%%%%%%%%%%%%%%%%%%%%%%%%%%%%%%%%%%%%%%%%%%%%%%%%%%%%%%%%%%%%%%%%%%%%%%%
\subsection{Incidence of Be stars}

To assess the impact of this work on the statistics of Be stars, the selection
biases of the sample need to be known and discussed. In fact, the target list
is indeed heavily biased towards later type B stars, which is due to the
selection policy for telluric standard stars at the VLT: Unless explicitely
specified otherwise by the PI of the observations, late type B and early type
A main sequence stars are preferred.  Consequently, the distribution among the
luminosity classes V to III is less biased, even though a bias favouring
non-giant stars still exists, as not all observers are familiar with the
broader definition of main sequence among B stars vs.\ solar type telluric
stars.  Another bias is that Be stars are known to be unsuited as telluric
standards, hence there is a bias against known Be stars. Two of those biases
can be well seen from Table~\ref{tab:stat}, in that there are generally fewer
early type stars, and almost no Be stars among them, even if the observed Be
star incidence is highest among the early type B stars
\citep{1997A&A...318..443Z}.

Some care is needed to interprete Table~\ref{tab:stat}, however. The spectral
types for the non-Be stars were taken from SIMBAD, i.e., they are collected
from a large number of quite inhomogeneous sources. For instance, the original
definition of the MK spectral classification scheme did not include standard
stars for all subtypes, in particular B4, B6, and B7 were missing and not
classified in that scheme at all, while already quite early the intermediate
type of B0.5 was included \citep[see, e.g., Table 1
  in][]{1979RA......9..389A}. The system has evolved since, and more
intermediate subtypes as well as standard stars for the missing integer
subtypes have been proposed. Nevertheless, the first line of our
Table~\ref{tab:stat} shows a lack of B4, B6, and B7 for exactly this
reason. Also, for the purpose of the Table, all intermediate classifications
have been rounded to the next earlier integer sub-type.

Photometric or spectrophotometric classification systems, like BCD, on the
other hand, define B0 to B9 much more linearly, and also without intermediate
types. Therefore, in order to avoid systematic effects arising due to these
inconsistencies as much as possible, the sample is grouped into early, mid,
and late type stars, which we define as B0--B2, B3--B6, and B7--A1. Then there
are 3 Be stars out of 99 early B stars, 21 out of 313 mid B stars, and 56 out
of 681 late ones.

This lack of early type Be stars in the sample is due to known Be stars being
avoided. The detection probability for an earlier type Be star is much higher,
as discussed by, e.g., \citet{2007ASPC..361..539Z}. The reason for this bias
can be seen from Appendix~\ref{app:spec}: Almost all newly discovered Be stars
have quite weak H$\alpha$ emission, and often no H$\beta$ at all. Since the
traditional spectral classification wavelength range does not include either,
but only H$\gamma$ and bluer Balmer lines, the Be nature is easily missed for
stars with steep Balmer decrement.

Fig.~3 of \citet{2007ASPC..361..539Z} suggests that the statistics of Be stars
is sufficiently complete for spectral types as late as about B6, but becomes
increasingly incomplete for B8 and later sub-types. Updating the numbers of
\citet{2007ASPC..361..539Z} with our findings does not entirely restore the
suggested trend in their Fig.~3, but given that our search was not designed to
achieve completeness, it certainly strengthens the suggestion of
\citet{2007ASPC..361..539Z}, that the probability of a B star to become a Be
during its life star does not (strongly) depend on spectral subtype.

\subsection{Stellar parameters}\label{ssec:num7}

The values of $D_\star$ and $\lambda_1$ measured by the method described
above, as well as the stellar parameters $T_\mathrm{eff}$ and $\log g$
obtained from them, are given in Table~\ref{tab:params} and plotted in the
upper panel of Fig.~\ref{fig:sampleBCD}. Comparing this to Fig.~1 of
\citet{2005A&A...441..235Z} the values of $\lambda_1$ show a systematic
offset. This is because of the much higher resolution of X-shooter data
vs.\ nominal BCD method data. At the nominal resolution for the BCD method,
the value of $\lambda_1$ is strongly affected by the convolution of the
stellar spectrum with the instrumental resolution. This is why the original
BCD method puts strong emphasis of using spectra at a given resolution of
$\Delta\lambda=8$\,\AA\ at the BD, and the published BCD calibrations cannot
be used fort X-shooter data. On the other hand, the X-shooter instrumental
resolution is so high that its contribution to the BCD parameters is
negligible, and our mapping of $T_\mathrm{eff}$ and $\log g$ onto $D_\star$
and $\lambda_1$ is valid for all medium to high resolution data.

Table~\ref{tab:stat} also confirms the finding of \citet{2005A&A...441..235Z}
that among the later type Be stars the higher luminosity classes are more
common, i.e., later-type Be stars are more likely found in the second half of
their main sequence life. This is in agreement with the idea of the Be phase
being a consequence of rotational evolution during the main sequence
\citep[e.g.,][]{2013A&A...553A..25G}. In that hypothesis a Be star with
moderate rotation at the ZAMS will, due to the internal evolution and angular
momentum transport from core to surface, at some point approach critical
rotation at the surface as the star ages. To prevent the surface rotation to
go above the critical threshhold, angular momentum must be transported
away. The means of this transport then is the circumstellar decretion disk
\citep{2011A&A...527A..84K}.

%%%%%%%%%%%%%%%%%%%%%%%%%%%%%%%%%%%%%%%%%%%%%%%%%%%%%%%%%%%%%%%%%%%%%%%%%%
\subsection{Disk properties as a function of spectral type}

Although it was not possible to determine the Balmer decrement for our sample
stars (see Table~\ref{table:balmdec} for measurements of the equivalent
widths), already this provides information: Because the emission is weak and
the Balmer decrement too steep, the disks found by X-shooter are too tenuous
to allow a reliable measurement of the Balmer decrement. This is in agreement
with \citet{2017MNRAS.464.3071V}, who found that late type Be stars have less
dense disks than early type ones. As mentioned above, there is also a general
agreement that late type Be stars show less variability than early-type
ones. This is again confirmed by the numbers shown in Table~\ref{tab:stat},
where in the early and mid-type Be stars 2/3 are found to be variable, but
only about 1/3 among the late subtypes.

%%%%%%%%%%%%%%%%%%%%%%%%%%%%%%%%%%%%%%%%%%%%%%%%%%%%%%%%%%%%%%%%%%%%%%%%%%
\subsection{The IR \ion{Ca}{ii} triplet}
We clearly detect the IR \ion{Ca}{ii} triplet in emission in 13 stars. This is
in agreement with the reported about 20\% of Be stars showing this feature
\citep[e.g.,][]{2012arXiv1205.2259K}.  In absorption and possibly emission the
IR \ion{Ca}{ii} triplet is seen in another 8 stars.

There is no obvious correlation of the presence of the IR \ion{Ca}{ii}
triplet, in either emission or absorption, with spectral type of the Be
star. Although our sample (heavily biased towards later type Be stars) does
not include stars with \ion{Ca}{ii} triplet emission earlier than B5, a
literature search does reveal such stars
\citep{1976IAUS...70...59P,1976IAUS...70..401P,1981A&A...103....1B,2012arXiv1205.2259K}.

The emission morphology closely resembles that of the \ion{O}{i}\,8446
line, which is supposed to trace the hydrogen Ly$\beta$ formation region,
since its upper level is excited by fluorescence from this transition
\citep{2012ApJ...753...13M}. However, while the \ion{O}{i}\,8446 can be
clearly linked to processes originating in the Be star, this is not the case
for the IR \ion{Ca}{ii} triplet. 

The emission strength of the \ion{Ca}{ii} triplet can vary without similar
changes taking place in \ion{O}{i}\,8446 or the Balmer lines. It can even be
transient without a major change in the Balmer line emission properties, e.g.,
the \ion{Ca}{ii} triplet was not detected in $\gamma$\,Cas by
\citet{1981A&A...103....1B}, but is reported to be present by
\citet{2012arXiv1205.2259K} to be anti-correlated with the Balmer emission for
the same star.

This suggests that, while the formation region is the same as for the other
lines, i.e., the disk around the Be star, the excitation process is not
originating in the saem source as for the other spectral lines formed in the
disk.  \citet{2012arXiv1205.2259K} investigate the correlation of the
\ion{Ca}{ii} triplet with binarity and conclude that binarity is not the
responsible mechanism, but suggest some other, not further specified
peculiarity of the circumstellar disk.

For Young Stellar Objects the common presence of the \ion{Ca}{ii} triplet in
emission is suspected to be linked to either magnetic processes or accretion
\citep[e.g.,][]{2011MNRAS.411.2383K,2013RAA....13.1189M}, or a combination of
both. In cataclysmic variables, the formation of the triplet is more
specifically traced to external UV irradiation of an optically thin gas
\citep{2004ARep...48..476I}. If we combine that with the current understanding
of Be stars, which do not show any trace of large scale magnetic fields, this
leaves UV photons formed in accretion shocks as the most promising mechanism
to power the \ion{Ca}{ii} triplet. The self re-accretion from the viscous disk
is probably not sufficient, as otherwise almost all Be stars should have
\ion{Ca}{ii} emission.

This leaves binarity. As \citet{2012arXiv1205.2259K} point out, several known
binaries do not show \ion{Ca}{ii} in their data. However, plain binarity is
not sufficient, the companion must also accrete to form the UV flux to excite
\ion{Ca}{ii}. Hence binarity remains a possible hypothesis to explain the
infrared \ion{Ca}{ii} triplet emission.

\section{Conclusions}
Searching the X-shooter database of telluric standards, 78 Be stars were
detected in emission, of which 48 had not been reported before. The sample is
strongly biased towards later-type Be stars. In some sense, this is an
advantage, because later type Be stars, owing to their lack of variability and
often less dense disks, are less well studied than earlier type ones. In
particular, we could confirm, or at least strengthen, a number of findings and
hypotheses:
\begin{itemize}
\item The Galactic Be star fraction drops less steep towards the later
  spectral (sub-)types than previously known numbers suggest. It may even be
  constant, as proposed by \citet{2005A&A...441..235Z}.
\item Late type Be stars show less variablility of their disks than early type
  ones.
\item Late type Be stars have less dense disks than early type ones.
\item Be stars are more likely to be closer to the TAMS than to the ZAMS.
\item The presence of the IR \ion{Ca}{ii} in emission may be linked to
  accretion onto a companion, but the emission itself originates from the Be
  disk proper.
\end{itemize}
Some of these points either clearly are, or may well be, linked to stellar
evolution and its timescales. For instance the lower density of disks around
later subtype Be stars could be a natural consequence of their slower
evolution, if indeed the disk is the means by which the star stays below
critical rotation: The amount of angular momentum to lose over a given time is
simply less. The same might explain the lower variability of the later
subtypes.

In summary, while late type Be stars are less well investigated than the earlier
ones, it might actually be this lack of ``interesting'' behavior in them that
will enable new insights on the origin and evolution of Be stars.

%------------------------------------------------------------------
\begin{acknowledgements}
We dedicate this work to the memory of Prof.~M.~Hamdy.
 
A. Shokry would like to acknowledge the Egyptian Ministry of Higher Education
(MoHE) for providing the financial support for his Joint scholarship, as well
as the ESO observatory for offering facilities and tools needed through his
stay in ESO (Chile), and the Kottamia Center of Scientific Excellence for
support. A. Shokry is deeply indebted to Prof.\ S. Saad for her continued
support.

R. Townsend acknowledges support from NASA grant NNX12AC72G.

DMF acknowledges support from FAPESP grant 2016/16844-1.

This research has made use of NASA’s Astrophysics Data System Service, as well
as of the SIMBAD database, operated at CDS, Strasbourg, France.
\end{acknowledgements}

%-------------------------------------------------------------------

\bibliographystyle{bibtex/aa}
\bibliography{ref}

%------------------------------------------------------------------------------------

\clearpage
\begin{appendix}

%%%%%%%%%%%%%%%%%%%%%%%%%%%%%%%%%%%%%%%%%%%%%%%%%%%%%%%%%%%%%%%%%%%%%%%%%%
%%%%%%%%%%%%%%%%%%%%%%%%%%%%%%%%%%%%%%%%%%%%%%%%%%%%%%%%%%%%%%%%%%%%%%%%%%
\section{Notes on individual Be stars}\label{sec:notes} 
For each of the identified Be stars, in the following observations of interest
are noted, together with literature values of parameters and stellar rotation,
where available, and whether the star is a known or newly identified Be star.

\begin{itemize}

\item 
Hip\,11116 (HD\,14850). \citet{1970MNRAS.148...79B} report emission in
H$\beta$.  The analysis by \citet{2006MNRAS.371..252L} gives spectral type
B8\,Ve, and $T_\mathrm{eff} = 13500 \pm 550$\,K, $\log g = 3.60 \pm 0.10$, $v
\sin i = 150 \pm 20\mathrm{kms^{-1}}$. The parameters determined by the BCD
method in this work are within {their $3\,\sigma$ errors}, but not vice
versa.

\item Hip\,15188 is listed as a B3\,Ve star with $v\sin i \approx
  130\,\mathrm{kms^{-1}} $ in the SIMBAD database. The spectral appearance is in
  agreement with the early type, but the Be star seems inactive at the moment,
  i.e., there is no trace of circumstellar line emission.
 
\item 
 Hip\,23161 (HD\,31764, HR\,1600). \citet{1964ApJ...139.1105T} classified the
 star as a variable in a binary system. They identified the components as as
 B8-7\,III and B6-7\,IV, suggesting a period of $P = 230$\,yr. In our spectra
 no companion is obvious, nor would any detectable radial velocity change be
 expected for such a long period, but it is a newly identified weak Be star.

\item 
Hip\,25007 (AN\,Col, HD\,35165, HR\,1772), has been listed as B5\,IVnp by
\citet{1969ApJ...157..313H}, and as B5\,IVnpe by \citet{1988A&AS...74..497M}
This star is a strong shell star showing a clear double Balmer discontinuity,
which, if not taken into account, leads to a later spectral type
classification. Consequently, we derive B3\,IVe-sh as spectral type, in
agreement with \citet{2006MNRAS.371..252L} who gave $T_\mathrm{eff} =
21500\pm500$\,K and $\log g =3.77\pm0.10$\,dex, with a high $v \sin i
=350\pm23\,\mathrm{kms^{-1}}$. Its spectral variability seems to be due to a
long-term $V/R$ cycle type behavior.

\item
 Hip\,25690 (HD\,37027) is a newly identified weak late-type Be star.

\item Hip\,25950 (HD\,36408, HR\,1847) is a very narrow lined star that shows
  no trace of emission in the X-shooter data. The data is unsuitable for BCD
  parameter determination. It is a well separated double star, and
  \citet{2009ApJS..180..138H} classified one component as B7\,IIIe. The star
  observed by X-shooter was possibly the other, non-Be component, judging from
  the spectra shown by \citet{2009ApJS..180..138H}. As this is not certain,
  however, it is kept in this list.

\item  	
Hip\,26368 (HD\,37935, HR\,1960), spectral type B9.5\,Ve according
to \citet{2002A&A...385..488C}.

\item  	 
HIP 26964, (V731 Tau, HD\,37967, HR\,1961), a spectral type of B2.5\,Ve as
reported in \citet{1984A&AS...58..685B}, \citet{1988AJ.....96..346G}
classified the star as B3\,V, $T_\mathrm{eff} = 21000$\,K, $\log g = 4.0$,
while \citet{2005A&A...440..305F} determine $T_\mathrm{eff} = 16543\pm 264$\,K
and $\log g = 3.850\pm 0.041$ and $v\sin i = 210\pm
10\,\mathrm{kms^{-1}}$. The discrepancy is probably due to its double BD,
filled in by emission, as we conclude the same values as
\citet{2005A&A...440..305F}, i.e., a stronger BD when taking into account the
doubling. {The Balmer emission lines have a very narrow, single peaked
appearance, supporting a pole-on star designation, even if the $v\sin i$ seems
quite high for that.}

\item 
Hip\,28561 (HD\,40724, HR\,2116) is the brightest newly identified Be star
among the ABE sample of \citet{2015AJ....149....7C}. It shows a double BD in
emission and strong IR \ion{Ca}{ii} triplet emission, and the UVB range shows a
clear composite spectrum, i.e., this star is a binary. Whether it is a
classical Be star or an interacting binary cannot be decided with the data at
hand.

\item 
Hip\,29635 (HD\,44533) is a newly discovered late type Be star.

\item
Hip\,31362 (HD\,46936) is a newly discovered late type Be star.

\item  	 
Hip\,32474 (HD\,49147, HR\,2502) is a newly discovered late type Be star. It has
been used as calibrator for interferometry and comparison star for polarimetry,
a role for which a Be star is not suitable. In particular, the emission is
variable; there is a clear signature of forming a weak disk in the X-shooter
spectra.

\item
Hip\,33509 (HD\,51506) is a known early to mid type Be star. It is usually
classified earlier than in our study, and has a double BD in emission.

\item  
Hip\,34144 (HD\,53296) is a newly discovered late type Be shell star.

\item  
Hip\,36009 (HD\,58630) is a newly discovered mid type Be shell star.

\item 
Hip 37007 (HD\,61950, HR\,1243) is a newly discovered late type Be star. It
shows IR \ion{Ca}{ii} triplet emission, and curiously, while the Balmer and
other emission lines do not vary in strength, the IR \ion{Ca}{ii} triplet
emission does.

\item  	 
Hip 39183 (HD\,65804) is a newly discovered late type Be star. There is a trace
signature of the IR \ion{Ca}{ii} triplet in the Paschen lines it is blended
with, but it is impossible to tell whether this is in absorption or emission.

\item  	 
Hip 39483 (HD\,66594, HR\,1243) is a newly discovered mid type Be star. It
showed transient activity forming a disk that subsequently decayed. At the
limit of the X-shooter resolution one may see a pulsational signature of a
{low inclination} Be star in the \ion{Mg}{ii}\,4481 line.

\item 
Hip 39595 (HD\,66956) is a newly discovered late type Be star showing the IR
\ion{Ca}{ii} triplet in emission.

\item
Hip\,41085 (HD\,70948) is a newly discovered mid type Be shell star with a
double BD.

\item  	
Hip 41268 (HD\,71255) is a newly discovered late type Be shell star with a
double BD.

\item 
Hip 42060 (HD\,72973) is a newly discovered late type Be  star. 

\item  	 
Hip 43073 (HD\,208213, HR\,3488) is a known late type Be star with weak emission
that has not been investigated in detail.

\item  
Hip 43114 (AI Pyx, HD\,75112) is a newly discovered mid type Be star.  The
X-shooter spectrum is overexposed, making a reliable determination of the BCD
parameters impossible. We note that in the data acquired by
\citet{2007A&A...470.1201D}, available from the ESO archive, the same
H$\alpha$ signature is present, but the star was not flagged as a Be star by
them. They classified the star as B5\,V. The \ion{Mg}{ii}\,4481 profile is
curious in both X-shooter and HARPS spectra, the star might be a binary.

\item  	
Hip 44423 (HD\,77907, HR\,3611) is a newly discovered mid type Be star. 

\item  	
Hip 46329 (HD\,81753, HR\,3745) is a B6\,Ve star according to
\citet{1988A&AS...74..497M} with  $v\sin i = 300 \mathrm{kms^{-1}}$. 
\citet{1967IBSH...11...34K} give the spectral type as B5\,Ve. In the X-shooter
spectra the star shows clear disk growth and may show pulsational signature in
the \ion{Mg}{ii}\,4481 profiles.

\item  	
Hip 47868 (HD\,84567, HR\,38780) is a known early type Be star, discovered by
\citet{1999A&AS..134..359G}. The BeSS database give the spectral type as
B0.5IIIne. Its BCD parameters are outside our grid, both in $D_\star$ and in
$\lambda_1$, which may indicate a higher luminosity class. We note that
luminosity classification for very early B-type stars with rotationally
broadened lines is tricky, as the differences between luminosity classes Ib to
about III or even IV are very small. The star may therefore not be a classical
Be star, but a supergiant with a rotationally modified wind, such as, for
instance, $\gamma$\,Ara.

\item  
Hip 47962 (HD\,84929) is a newly discovered late type Be star. 

\item  
Hip 48582 (HD\,85834) is flagged as an emission line star in SIMBAD. Since this
was communicated by C.\ Martayan on the basis of the same X-shooter data used
here, we consider it as a new discovery for the purpose of assessing the
statistical impact on the Be star frequency. The star shows strong $V/R$
variability in H$\alpha$. The observed change takes place in 23\,days, making
it very likely that the star is either a binary, in which case the secondary
affects the disk, or a hierarchical triple, in which case the variability is
due to the linear superposition of the RV curve of the components Ba+Bb. We
note that the star has strong IR \ion{Ca}{ii} triplet in emission. The
spectrum shows also a small double BD.

\item 
Hip 48943 (OY Hya, HD\,86612, HR\,3946) is a well known Be star of spectral
type B4\,Ve \citep{1969ApJ...157..313H}.  $v\sin i = 229$\,km\,s$^{-1}$ is
given by \citet{1983A&A...126..192Z}, also \citet{1986A&A...163...67B}
determine $v\sin i=230$\,km\,s$^{-1}$. The star has a pronounced double BD
character. Some of the H$\alpha$ spectra are overexposed, cross-checking with
H$\beta$ reveals that the apparent variability is purely due to that. However,
in the Paschen regime the variability of the Paschen lines blended with the IR
\ion{Ca}{ii} triplet is far stronger than that of the non-blended lines. Most
likely the IR \ion{Ca}{ii} triplet is in emission, and variably so.

\item 
Hip 51444 (LX Vel, HD\,91188) is a known emission line star. It is a
photometrically variable, for which \citet{1990MNRAS.245...92B} report as the
most likely frequency $f = 0.684$\,d$^{-1}$. In the X-shooter spectra the
circumstellar emission is clearly variable, and the strength of the central
absorption in H$\alpha$ and its variability in the Paschen lines suggests a
shell nature of the star.

\item 
Hip 51491 (HD\,91120, HR\,4123) is a well known Be shell star. It shows the IR
\ion{Ca}{ii} triplet is in emission, but it is interesting to note that while
H$\alpha$ does vary in the three available VIS spectra, the \ion{Ca}{ii}
triplet remains constant.

\item Hip 51546 (HD\,91373) is a newly discovered late type Be star.  No
  reliable stellar parameters can be given due to over exposure of the UVB
  continuum.

\item  	
Hip 52977 (HD\,94097) is a newly discovered mid to late type Be star. It shows
some low level of variability in H$\alpha$ and possibly \ion{Mg}{ii}\,4481.

\item  	
Hip 56393 (HD\,100528) is a newly discovered late type Be star. The
Paschen line profiles blended with the IR \ion{Ca}{ii} triplet look
suspiciously different from the other Paschen lines, but not enough so for a
conclusive statement.

\item  	 
Hip 57861 (HD\,103077)  is a newly discovered mid to late type Be star. It
showed clear emission line variability in the X-shooter spectra, and as well
\ion{He}{i}\,4471 and \ion{Mg}{ii}\,4481 are variable in their line profiles.

\item
Hip\,59970 (HD\,106965) is a newly discovered late type Be star. Physical
parameters cannot be given since the BCD values are outside our grid. This
could be because the star might have a weak double BD, or BDs merging into
each other without being clearly separated, which we could not unambiguously
identify. Indeed, it seems to be a shell star, and it should also be noted
that the \ion{He}{i}\,4471 line is surprisingly strong for its spectral type
(i.e., height of BD). The IR \ion{Ca}{ii} triplet is certainly present in
absorption, whether there is also an emission component is uncertain, but
possible.

\item  	 
Hip 64501 (HD\,114531 is a newly discovered late type Be star. 

\item  	
Hip 64867 (HD\,115415)  is a newly discovered late type Be star. 

\item  	
Hip 66339 (HD\,118246, GP Vir) is a known mid type Be shell star. Its spectral
type was given as B3e, with $v\sin i$ of 270\,$\mathrm{kms^{-1}}$ by
\citet{1996PASP..108..833H}, while \citet{1997PASP..109....1S} classified it
as B5\,IVe, and determine a $v\sin i$ of above 350 $\mathrm{kms^{-1}}$. It
shows a clear double BD, favoring the earlier spectral type, i.e., B3\,V.

\item  
Hip 66351 (HD\,117872)  is a newly discovered late type Be star. 

\item  
Hip 68100 (HD\,120845) is a newly discovered mid to late type {low
  inclination, near to pole-on Be star, with $v \sin i =100
  \mathrm{kms^{-1}}$.}

\item  
Hip 69429 (HD\,124176)  is a newly discovered late type Be star.

\item 
Hip 71668 (CK Cir HD\,128293) is a known Be star.  The spectral types
published range from B5\,Ve, with $v\sin i= 216 \mathrm{kms^{-1}}$
\citep{1975MmRAS..78...51B} to B2\,IVe \citep{1992A&AS...95..535J}. The
X-shooter spectrum favors the earlier type when looking at the He/Mg balance.

\item  	
Hip 71974 (HD\,129433, HR\,5484, 4Lib) is a newly discovered late type Be
star.

\item  	
Hip 78375 (HD\,143513) is a newly discovered late type Be shell star. The
situation is very similar to Hip\,59970: Physical parameters cannot be given
since the BCD values are outside our grid. This could be because the star has
a double BD which we failed to identify.  The IR \ion{Ca}{ii} triplet is
certainly present in absorption, whether there is also an emission component
is uncertain, but possible.

\item  	
Hip 80577 (HD\,147747) is a newly discovered late type Be star, seen at {low inclination, with $v \sin i =100 \mathrm{kms^{-1}}$.}

\item  	
Hip 80820 (HD\,148382) is a newly discovered late type Be star. The H$\alpha$
emission is slightly variable.

\item  	
Hip 81321 (HD\,149595)  is a newly discovered, very weak late type Be star. No
UVB spectrum is available, but the  H$\alpha$ profile is clearly indicating a
Be star.

\item  
Hip 82874 (HD\,152541)	is a known but little studied late type Be star.

\item  
Hip 83278 (HD\,153608) is a newly discovered late type Be star. 

\item  
Hip 84184, is a newly discovered late type strong Be shell star. The strength
of the shell changed considerably in the X-shooter spectra over about two
years, showing a double BD only when the stronger shell is present.

\item  
Hip 85138 (HD\,156709) is a newly discovered late type Be star. The H$\alpha$
emission is slightly variable.

\item 
Hip 85195 (HD\,157546, HR\,6473) is a newly discovered late type Be star. The
H$\alpha$ emission is slightly variable.

\item 
Hip 85566 (HD\,158419) is a newly discovered late type Be star. However, it was
reported by \citet{2016ApJ...830...84K} as a disk candidate based on its
infrared excess, indicating cold dust. It might, therefore, rather be a Herbig
Ae star or a $\beta$\,Pictoris type object. The IR \ion{Ca}{ii}
triplet is in absorption, but might have an emission component as well.

\item 
Hip 87032 (HD\,161734) was classified as an emission line star by
\citet{2002AJ....124..989G}. As the flux calibration of the Balmer continuum
is obviously wrong, we do not give BCD parameters. The IR \ion{Ca}{ii}
triplet is in emission.

\item  	
Hip 87698 (HD\,162888) is a newly discovered weak, but obvious late type Be
star.

\item 
Hip 88172  (V974 Her, HD\,164447, HR\,6720) is a known Be star.  
 
\item  	 
Hip 88374 (HD\,164716, HR\,6732) is a newly discovered late type Be
star. The H$\alpha$ emission is clearly variable.

\item  
Hip 89486 (HD\,167230)  is a newly discovered late type Be
star. The H$\alpha$ emission is clearly variable.	

\item  
Hip 89500 (HD\,167095) is a newly discovered very weak late type Be star. The
only indication for a Be nature is a slight filled in absorption flank of
H$\alpha$. Since this is however not present in another spectrum, this is a
good indication for a variable amount of circumstellar material, i.e., a Be
star. Computing the difference spectra reveals the usual double peak emission
signature.

\item  	
Hip 90096 (HD\,169033, HR\,6881) \citet{1943ApJ....98..153M} classified as
B8\,Ve, as did \citet{1980A&AS...42..103J}, who also gave $v\sin i =
220\mathrm{kms^{-1}}$

\item  	
Hip 90509 (HD\,165338) is a newly discovered late type Be star. The IR
\ion{Ca}{ii} triplet is in absorption, but might have an emission component as
well.

\item  	
Hip 91460 (HD\,172054) is a newly discovered late type Be star. 

\item 
Hip 91975 (4 Aql, HD\,173370,  HR\,7040) is a well known Be star.
\citet{1975ApJ...196..773I} observed weak emission in H$\alpha$, with a strong
central reversal. The IR \ion{Ca}{ii} triplet is in absorption, but might have
an emission component as well.

\item 
Hip 92038 (HD\,173375) is a newly discovered late type Be star.  The emission
strongly increased during the observations with X-shooter over about three
years, to the point at which a double BD became apparent. The IR \ion{Ca}{ii}
triplet is in emission as well, but its strength decreased, i.e., behaved in
opposite to the Balmer emission.

\item  	
Hip 93993 (HD\,178075, HR\,7246) is a newly discovered late type Be star. It
shows an extreme pole-on appearance {$v \sin i =10 \mathrm{kms^{-1}}$},
with only H$\alpha$ in emission. The IR \ion{Ca}{ii} triplet is very weakly
present in absorption, which is probably photospheric and typical for the late
spectral type.

\item  	
Hip 94770 (HD\,179419) is a known (but largely ignored) late type Be star,
given as B8\,Ve by \citet{1983A&AS...52..471A}. The change in the H$\alpha$ is
due to slight over exposure in one of the two spectra and not real. The IR
\ion{Ca}{ii} triplet is in emission.

\item  	
Hip 94859 (HD\,180699) is a newly discovered late type Be star. The IR
\ion{Ca}{ii} triplet is in absorption, emission might possibly be present as
well.

\item  	
Hip 94986 (HD\,180885, HR\,7316) is a newly discovered {low
  inclination} early type Be star. It is variable in H$\alpha$, and
the emission is not always present. Pulsational variability is clearly seen in
\ion{Mg}{ii}\,4481.

\item  	
Hip 95109 (HD\,181751) is a newly discovered late type Be star.
The spectral type was given as B8 by \citet{1983IBVS.2376....1S}.

\item  	
Hip 96453 (HD\,184597) is a newly discovered mid type Be star. It shows clear
signs of an outburst in one of the two spectra acquired in the VIS arm of
X-shooter.

\item  	 
Hip 99457 (BE\,Cap, HD\,191639, HR\,7709) is a known early type Be star in
SIMBAD, even though it is not entirely clear where this was first reported. It
shows a clear and strong outburst event in the X-shooter data.

\item  	 
Hip 100664 (HD\,194244, HR\,7803) is a known late type Be star, first reported
by \citet{2005ApJS..156..237N}.

\item  	 
Hip 104508 (HD\,201317)  is a newly discovered late type Be star. The H$\alpha$
emission shows clear variability.

\item  	 
Hip 108022 (16 Peg, HD\,208057, HR\,8356) is a known mid type Be star. It was
announced by \citet{1943ApJ....98..153M} and listed as MWC 644. No other study
since reported Balmer emission in this star, and also in the X-shooter spectra
there is no evidence for a Be nature.

\item  	
Hip 108402 (HD\,208612)  is a known late type Be shell star. It shows both a
double BD and the IR \ion{Ca}{ii} triplet is in emission.

\item 
Hip 108597 (VV PsA HD\,208886) is a known late type Be shell star
discovered by \citet{1976ApJS...30..491H}, who found H$\alpha$ to be a very
sharp, moderate to weak emission line and H$\beta $ to be in absorption, from
1949-1952 objective prism plates.  It shows both a double BD and the IR
\ion{Ca}{ii} triplet is in emission.

\item  	 
Hip 108975 (UU PsA, HD\,209522, HR\,8408) has been reported as early type Be
star by \citet{1943ApJ....98..153M} as MWC 650. No recent study has found
Balmer emission in this star, and also in the X-shooter spectra there is no
evidence for emission.

\end{itemize}

\clearpage

\onecolumn
\section{Projected rotational velocity fits}
\label{app:vsini}

\begin{figure*}[h]
\centering
\subfloat{
    \includegraphics[width=4cm,height=4cm] {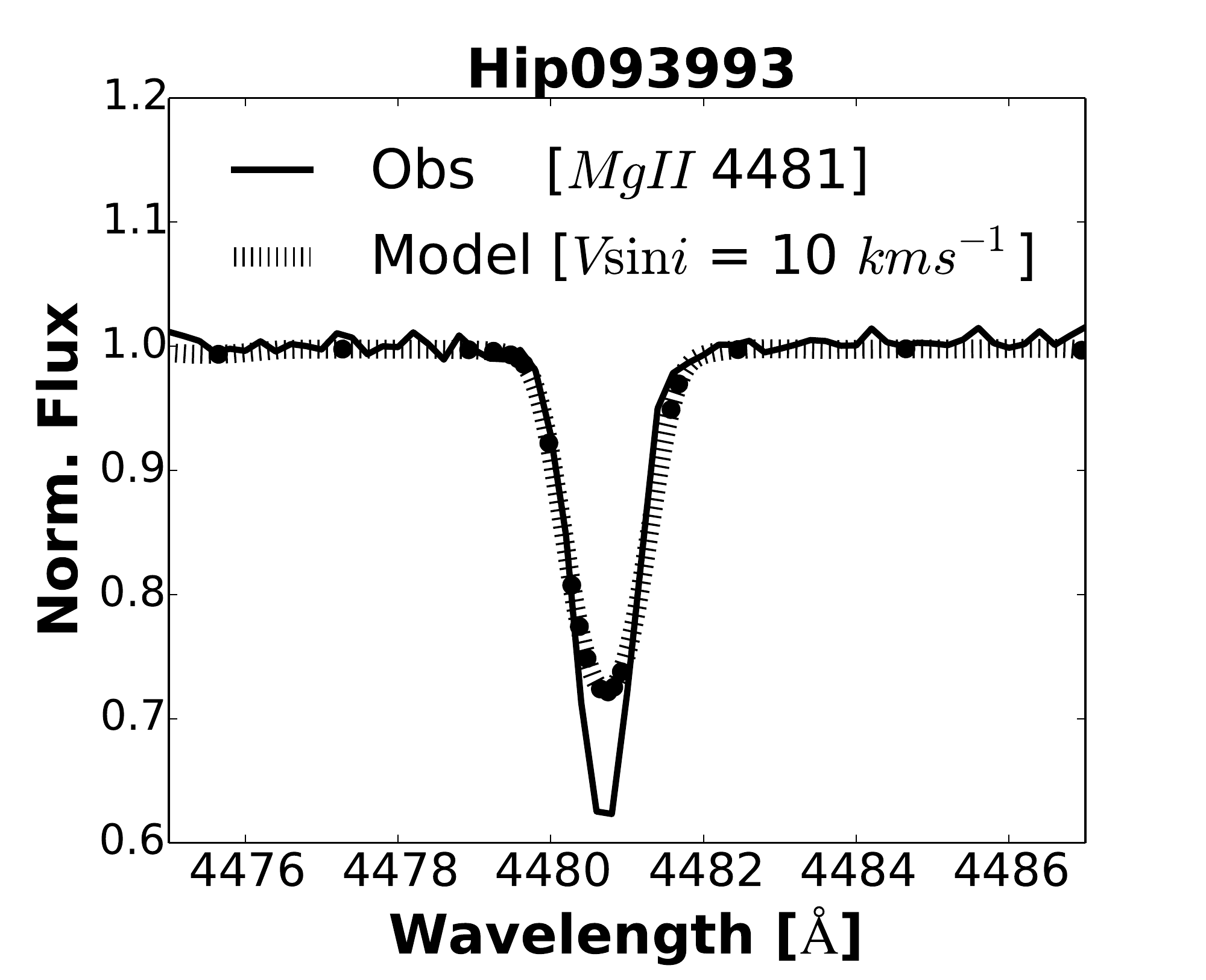} 
    \includegraphics[width=4cm,height=4cm] {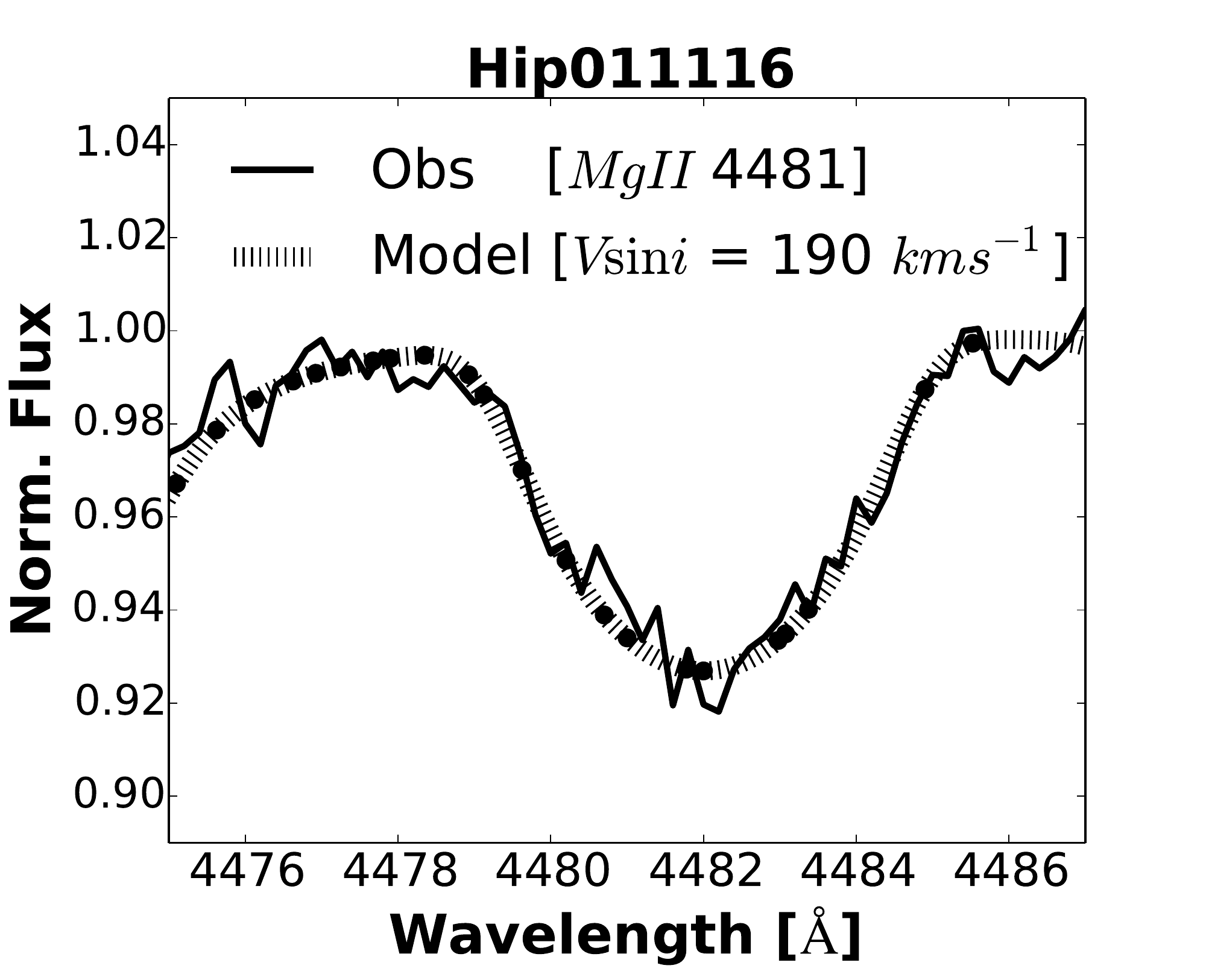}  
    \includegraphics[width=4cm,height=4cm] {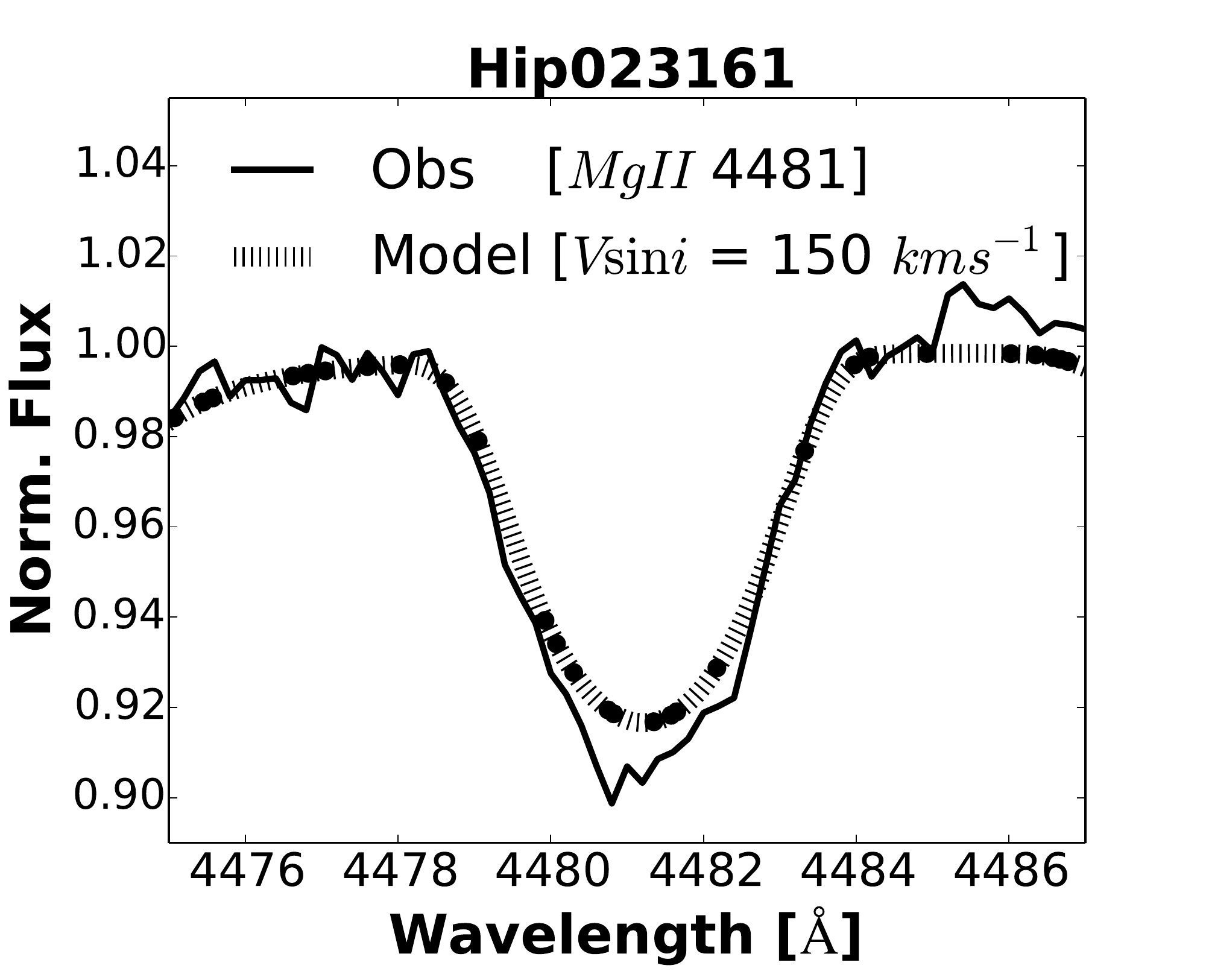} 
    \includegraphics[width=4cm,height=4cm] {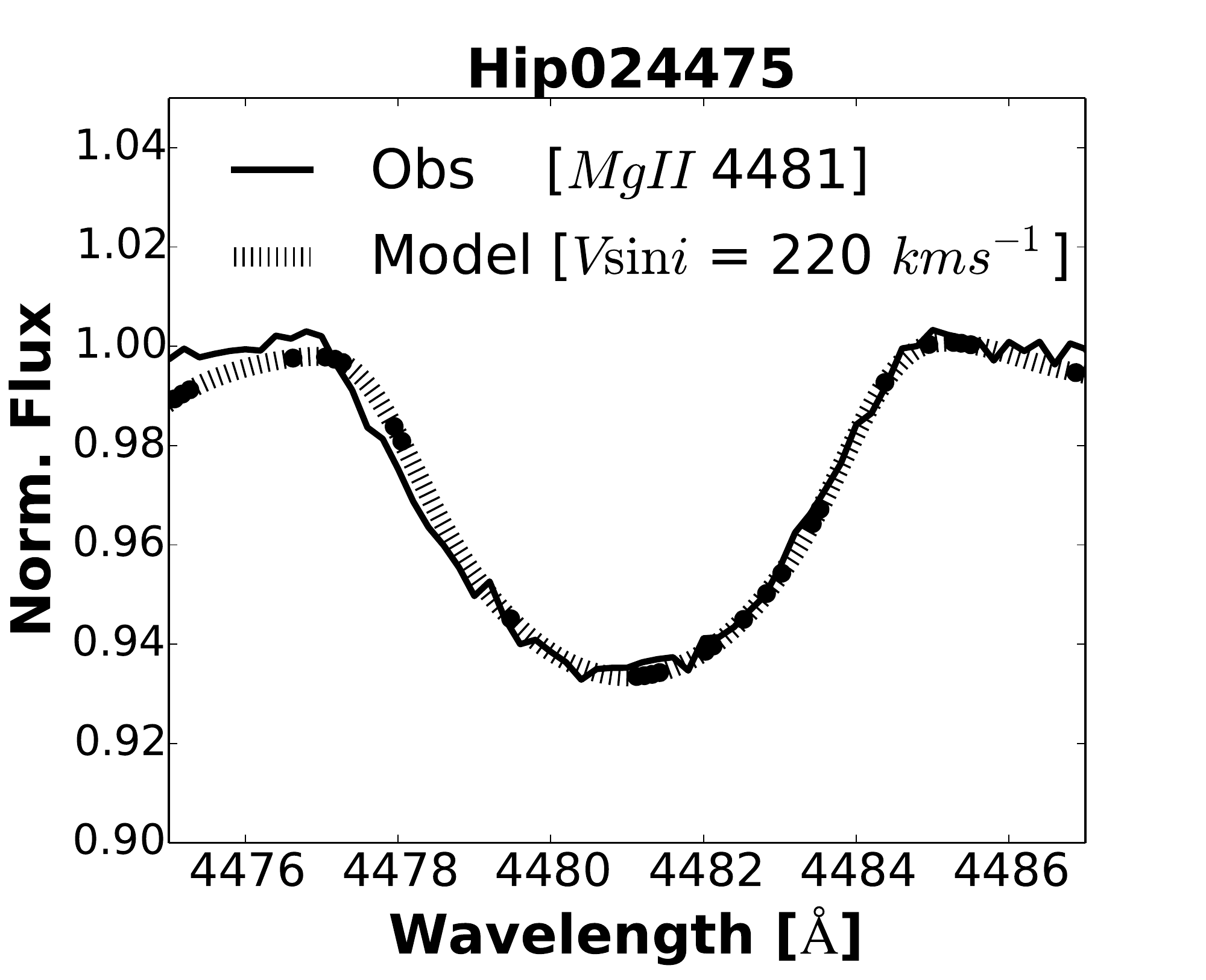} 
 }
 
 \subfloat{
  \includegraphics[width=4cm,height=4cm] {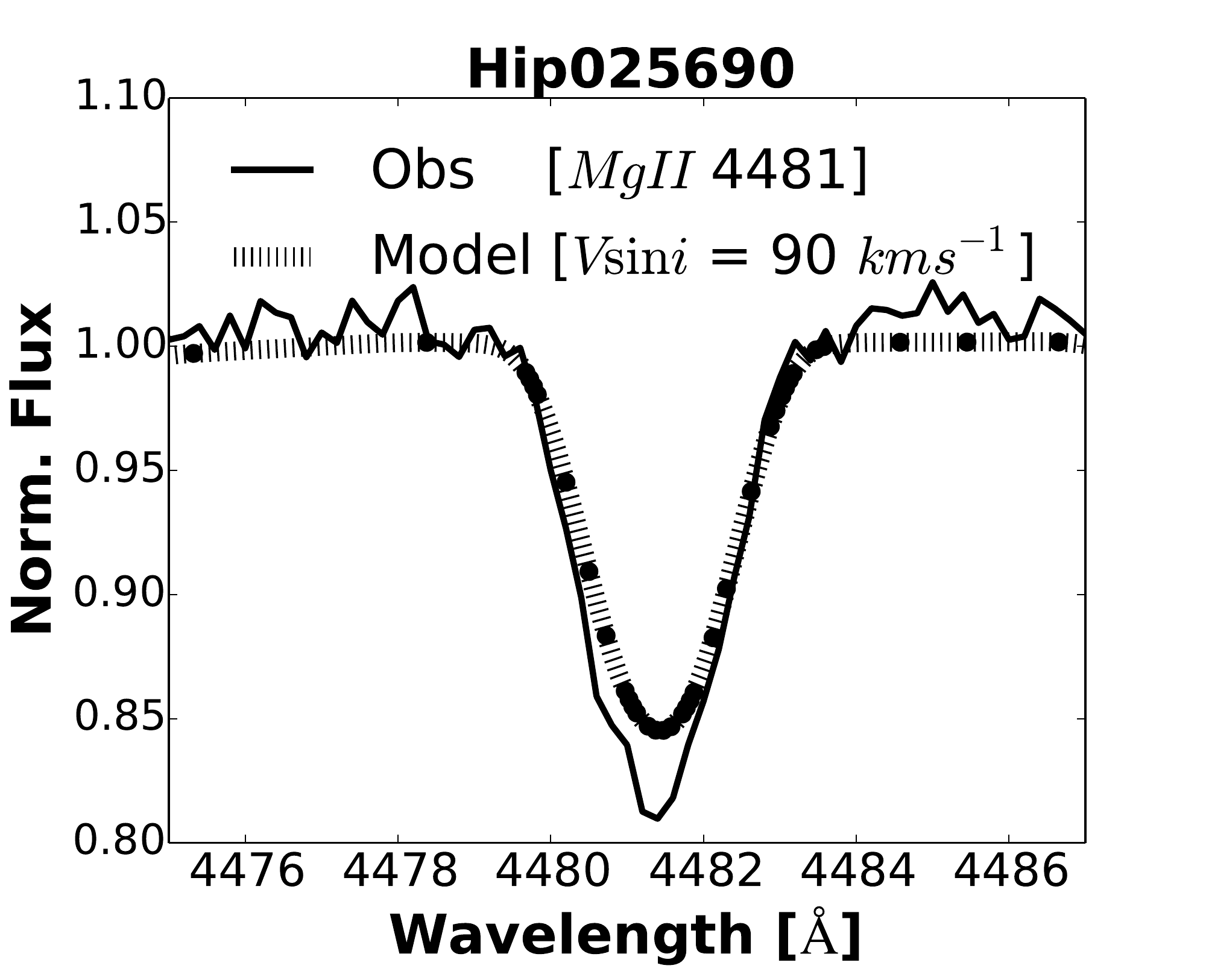}  
  \includegraphics[width=4cm,height=4cm] {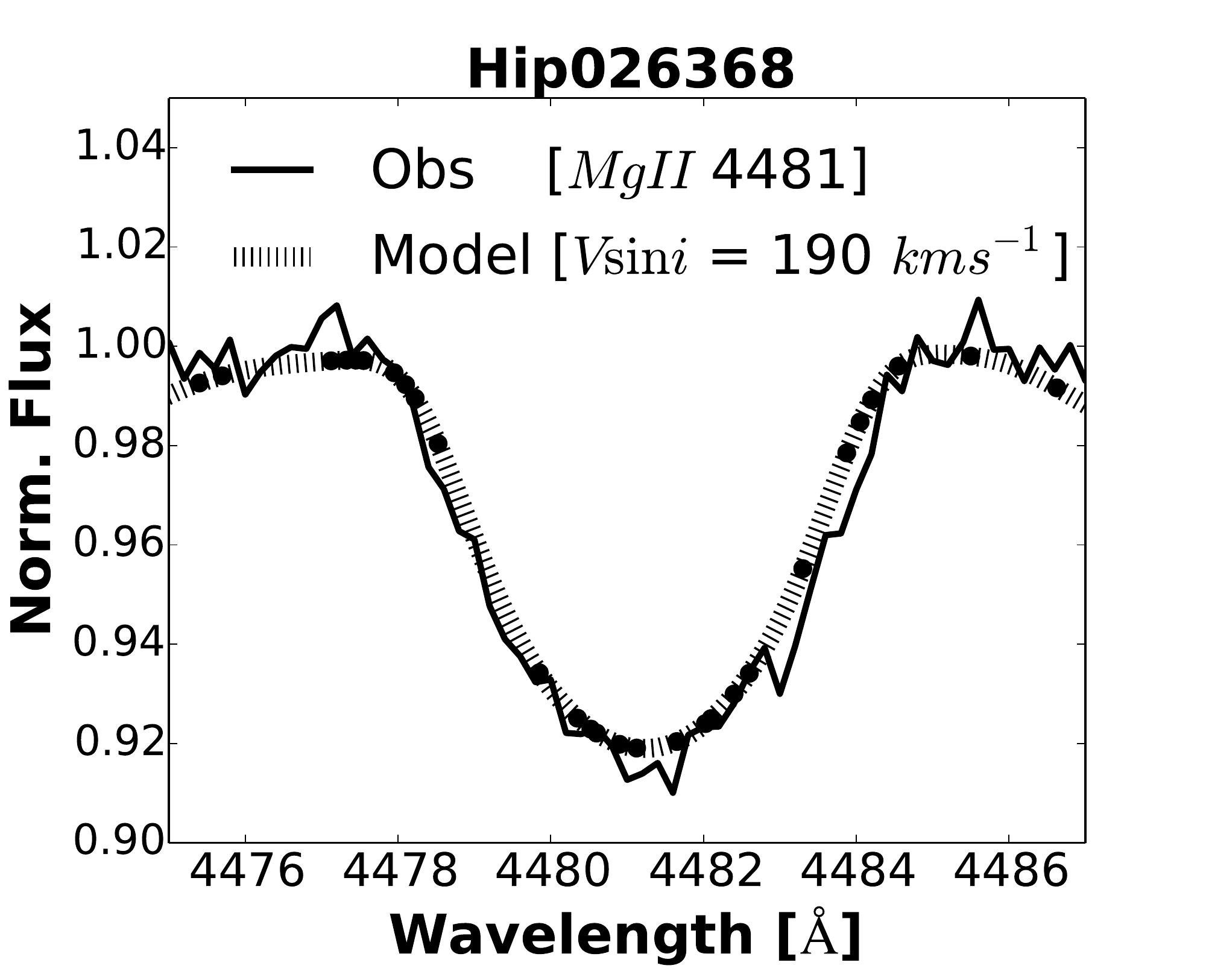} 
  \includegraphics[width=4cm,height=4cm] {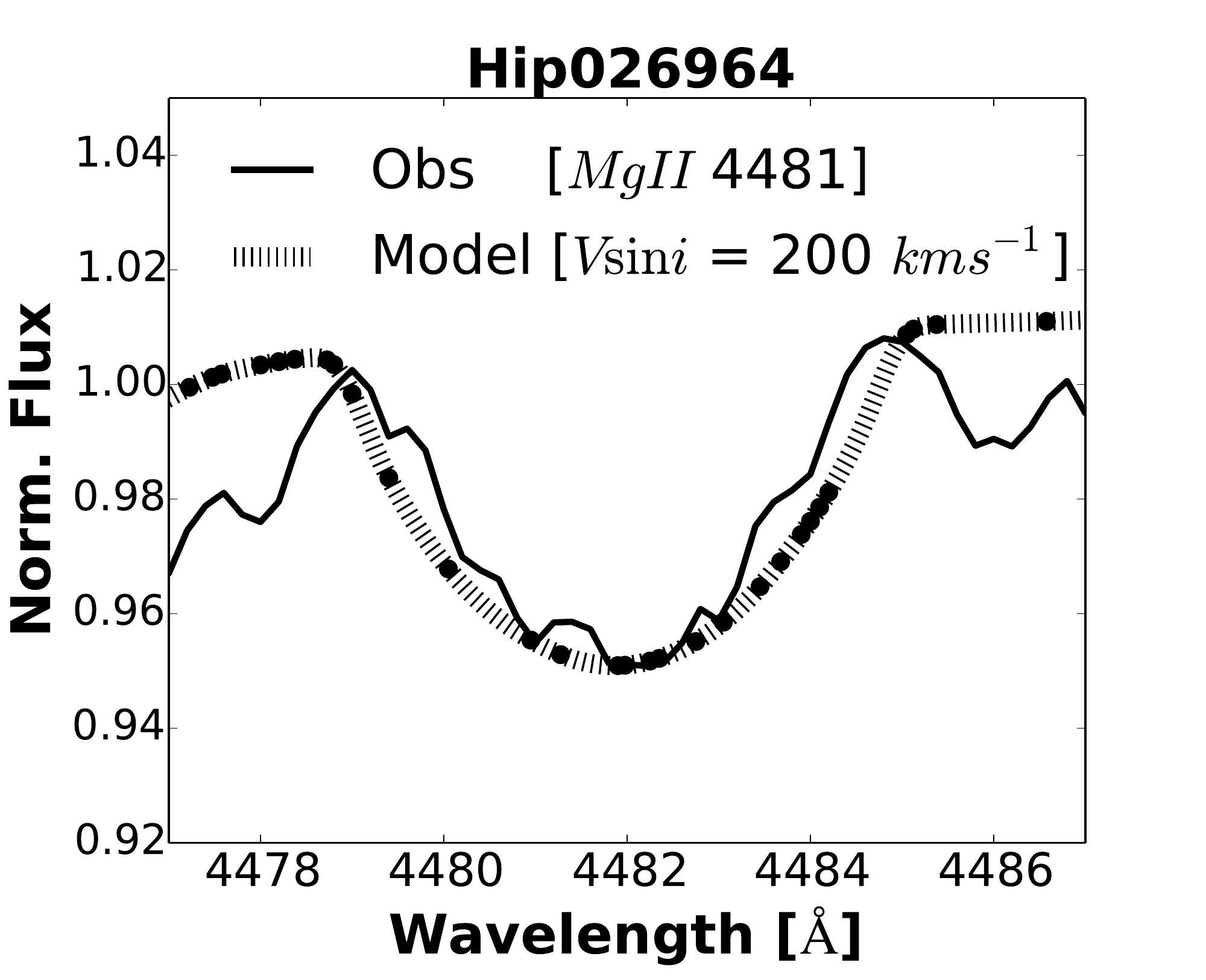} 
  \includegraphics[width=4cm,height=4cm] {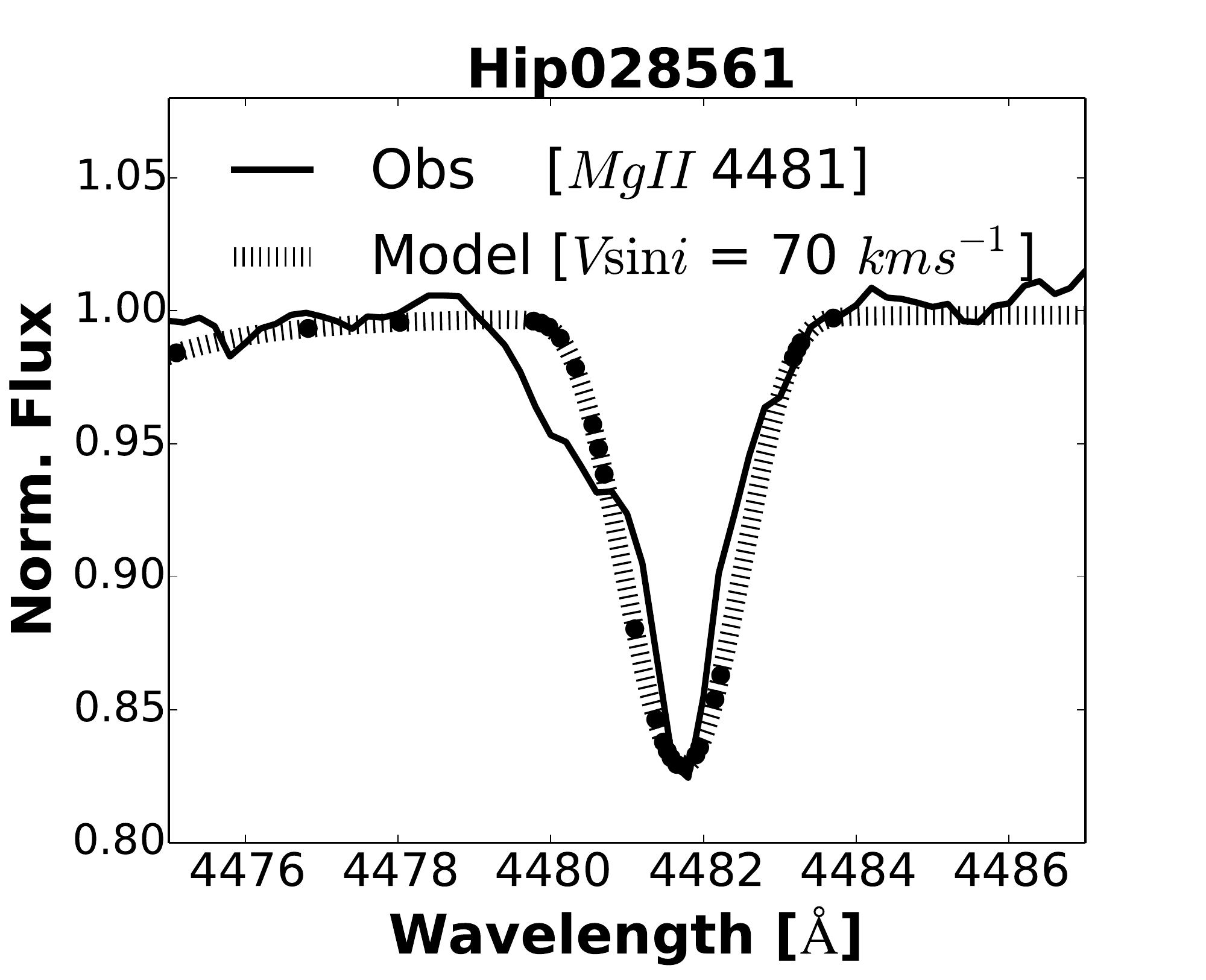}
 } 
 
   \subfloat{
       \includegraphics[width=4cm,height=4cm] {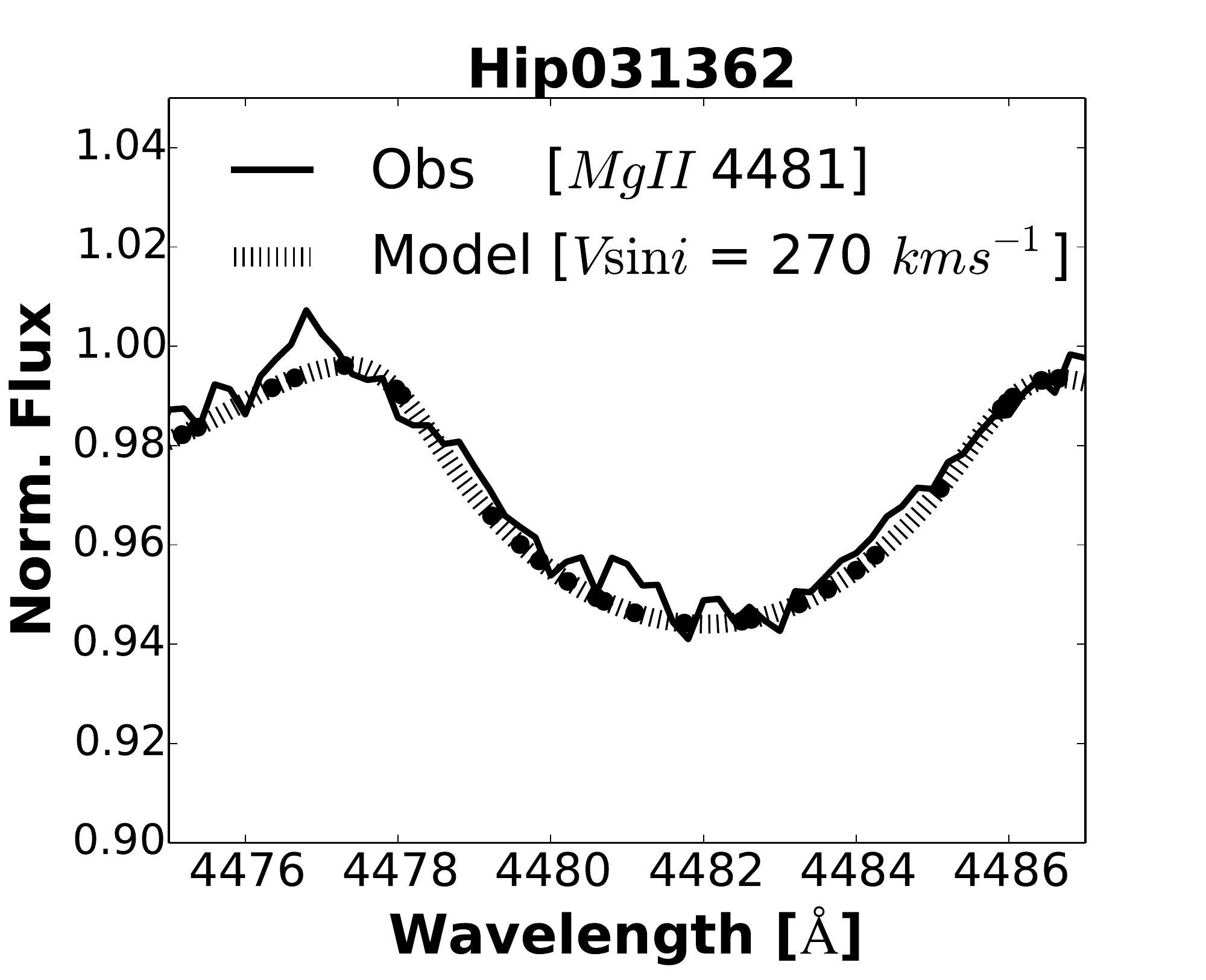}  
        \includegraphics[width=4cm,height=4cm] {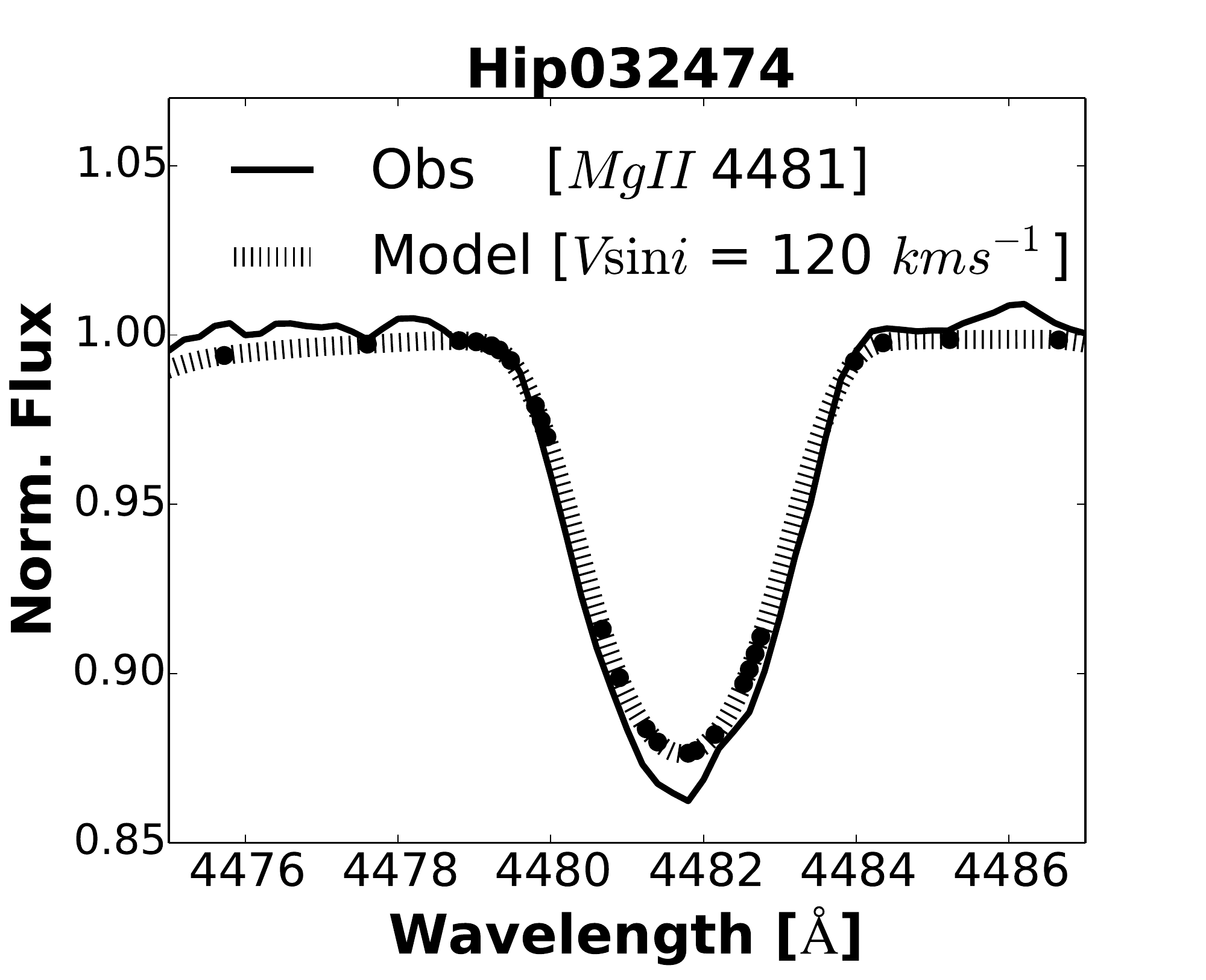} 
            \includegraphics[width=4cm,height=4cm] {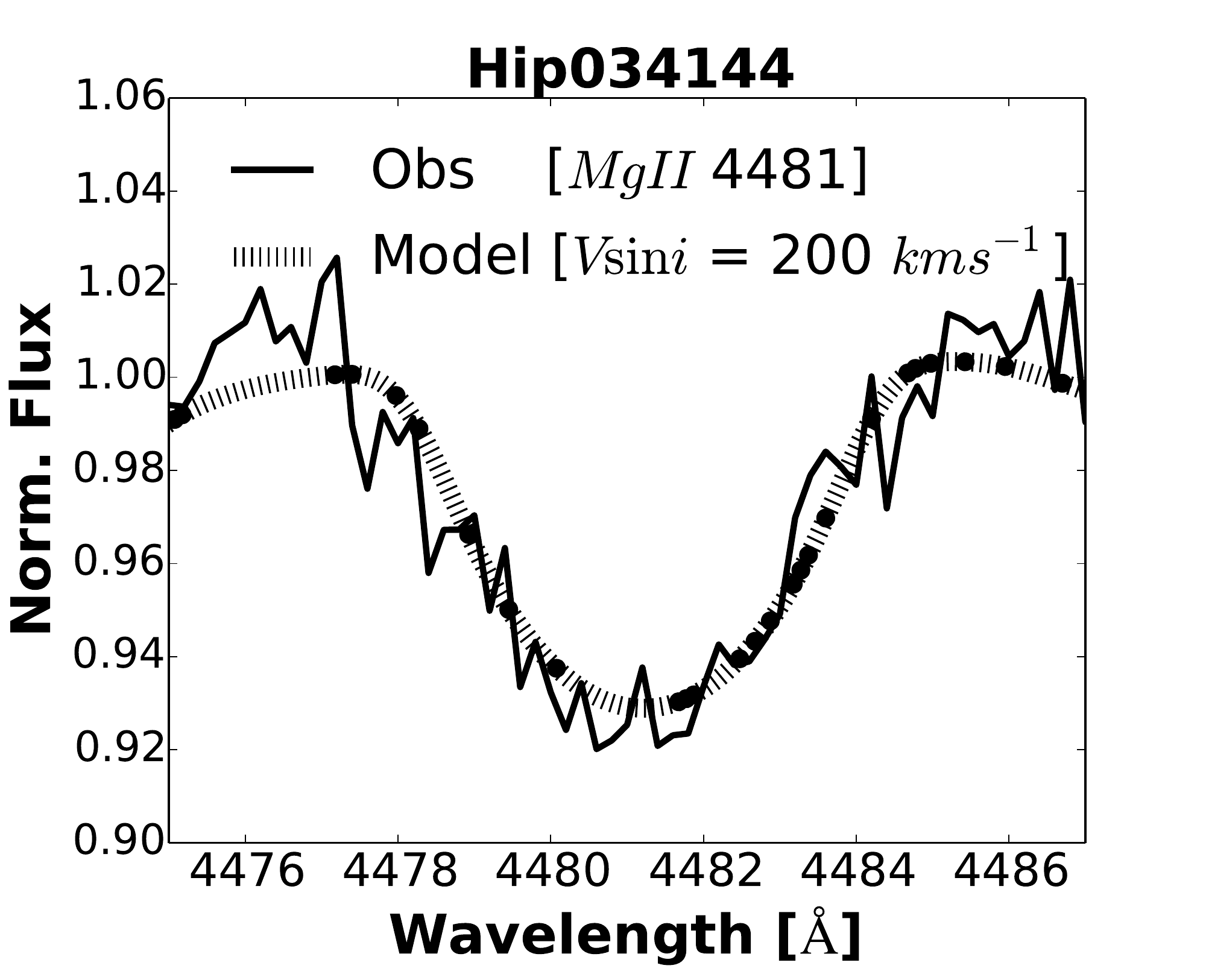}     
    \includegraphics[width=4cm,height=4cm] {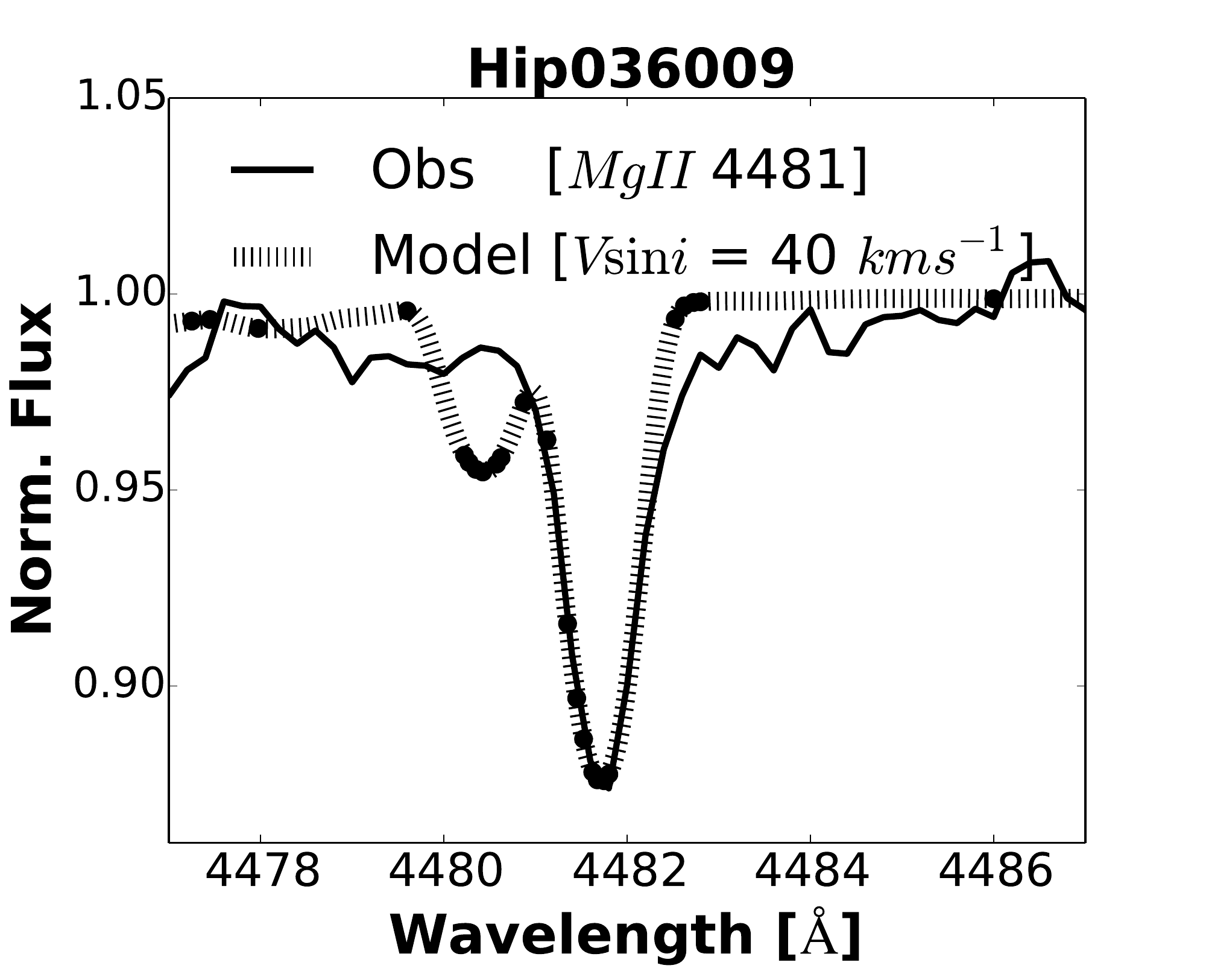} 
   }
  
\subfloat{
  \includegraphics[width=4cm,height=4cm] {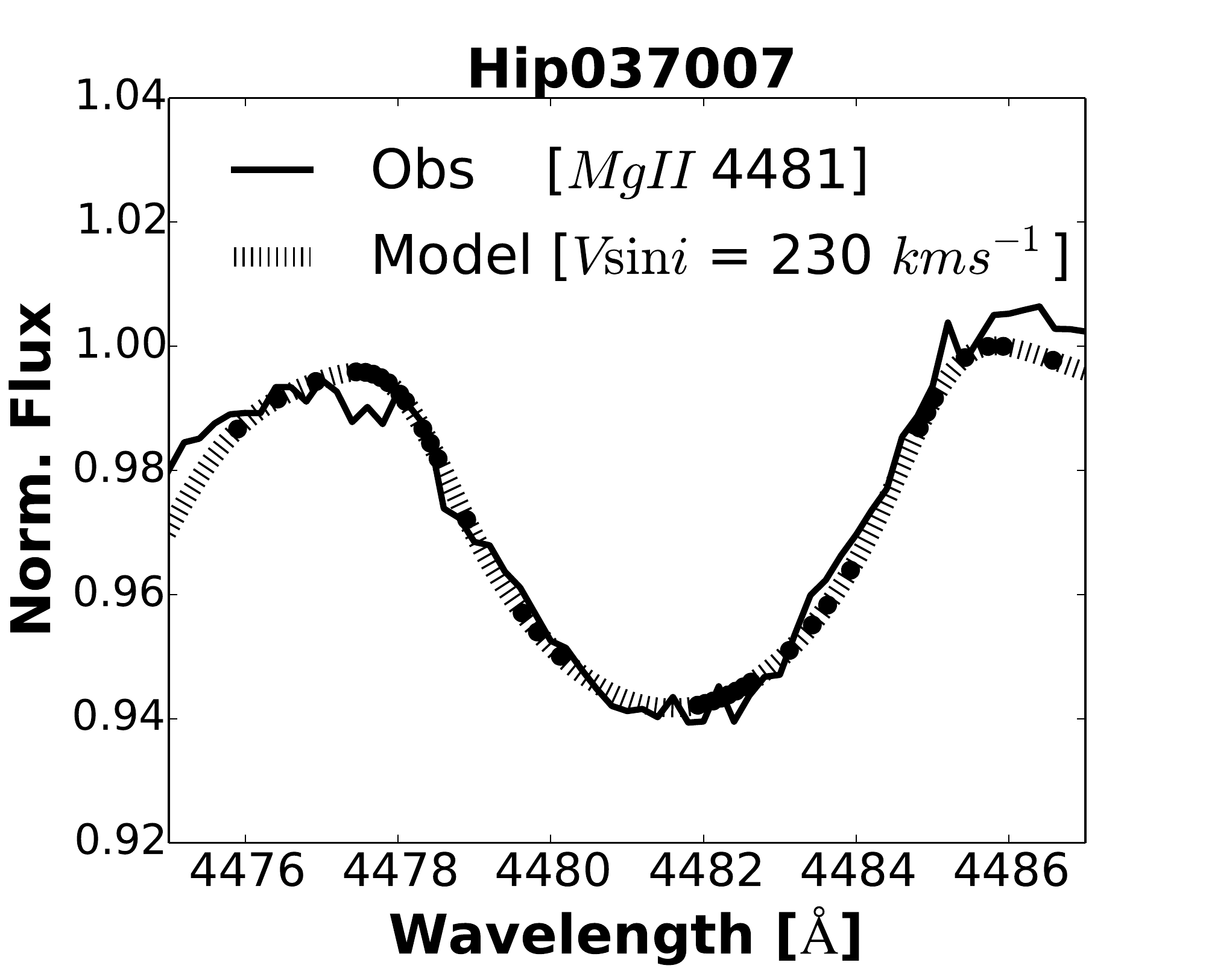}  
    \includegraphics[width=4cm,height=4cm] {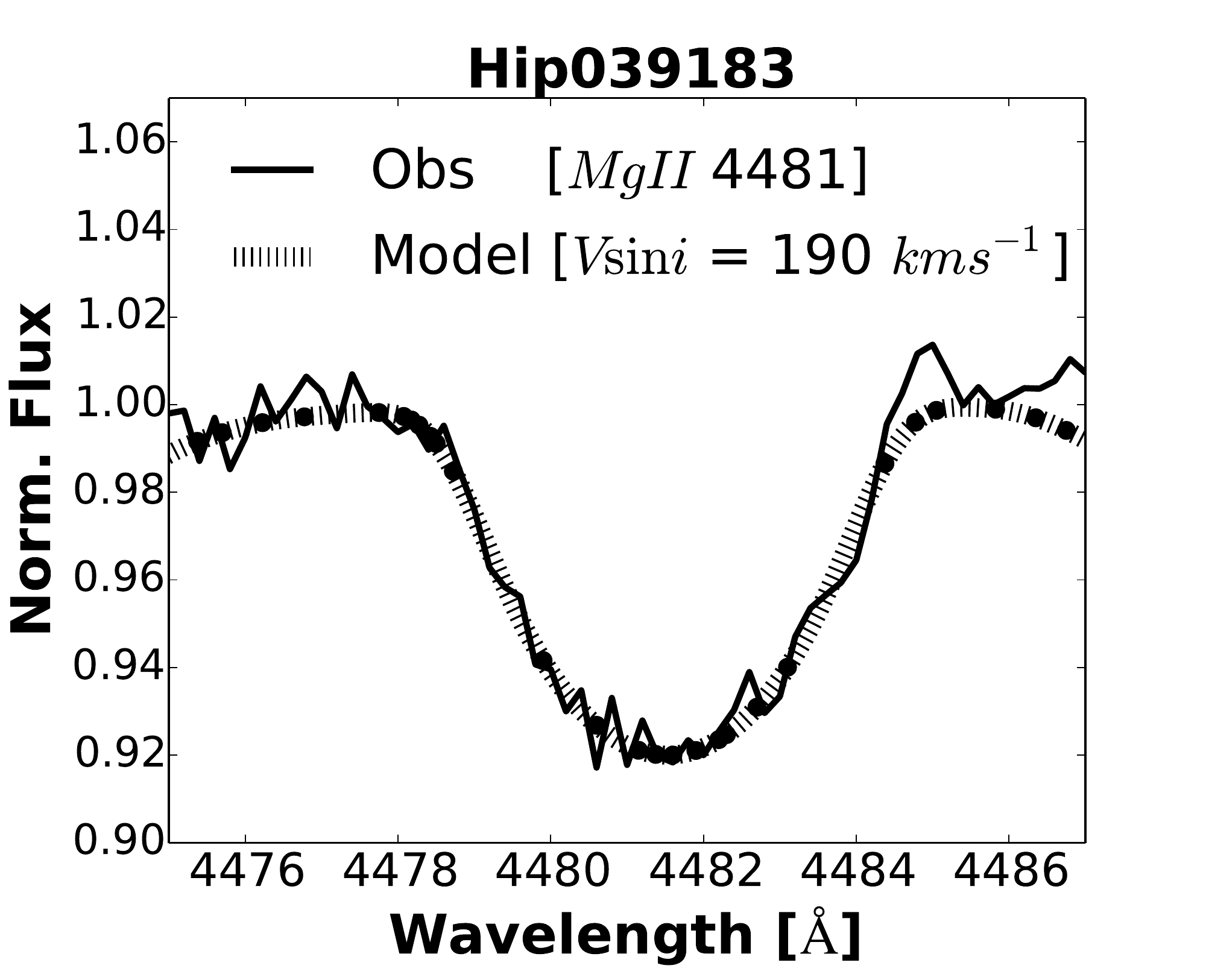} 
      \includegraphics[width=4cm,height=4cm] {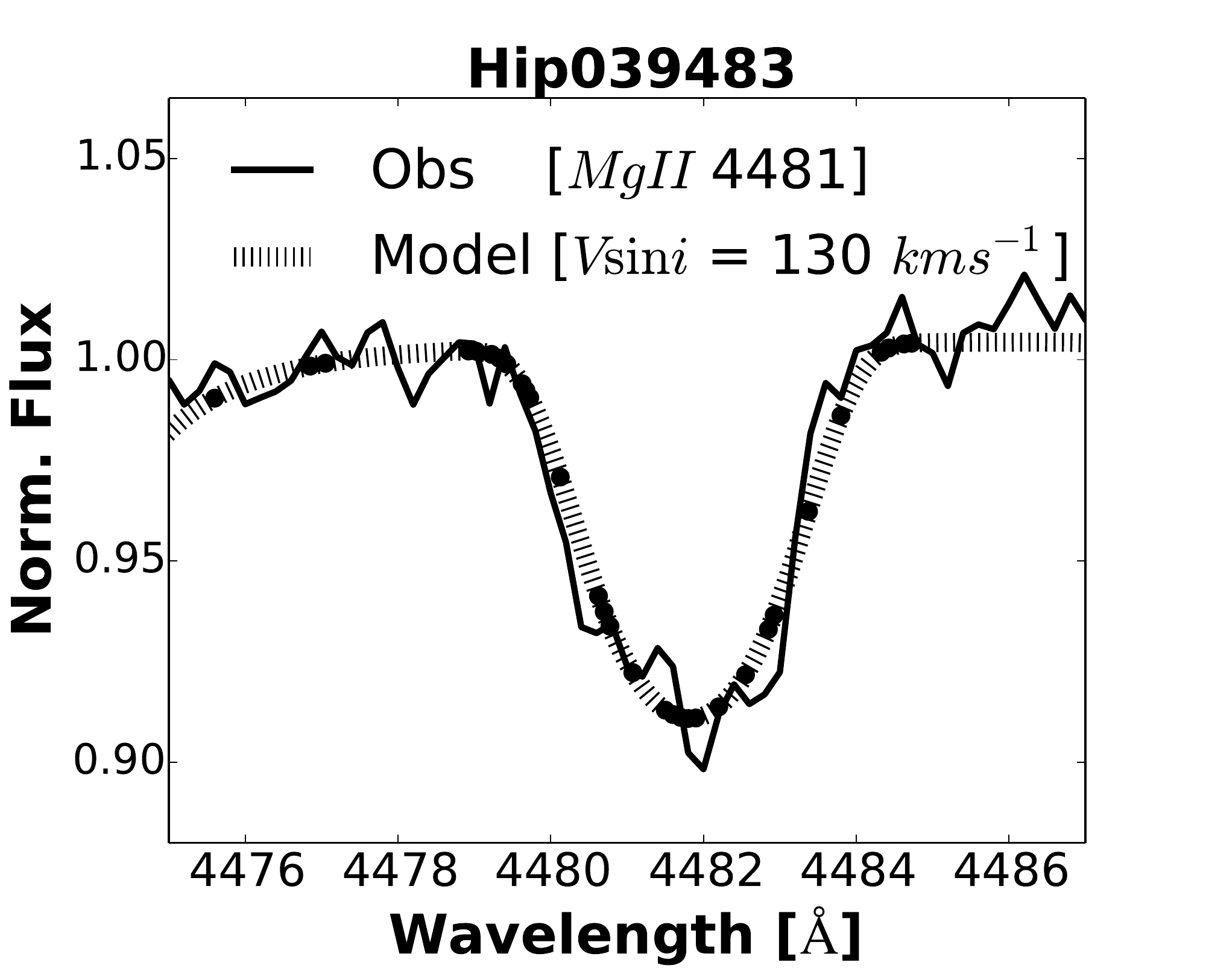} 
      \includegraphics[width=4cm,height=4cm] {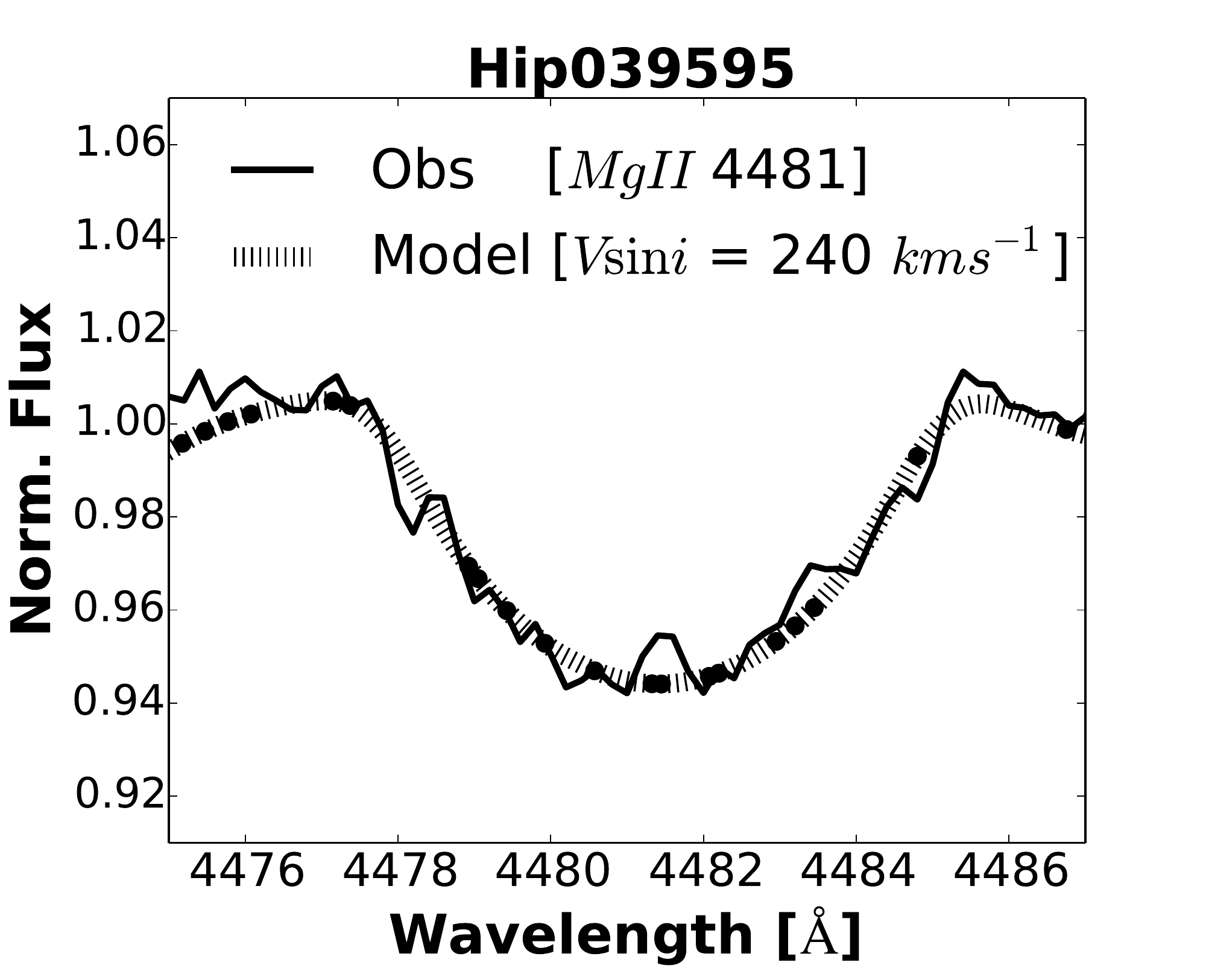}  
} 
\caption{Best fit models for projected rotational velocities for the program Be
  stars, for \ion{Mg}{ii}\,4481\,\AA . Solid lines mark observed and dotted ones
  theoretical profiles.}
\label{fig:vrot}
\end{figure*}

\begin{figure*}
\ContinuedFloat
\centering

\subfloat{
      \includegraphics[width=4cm,height=4cm] {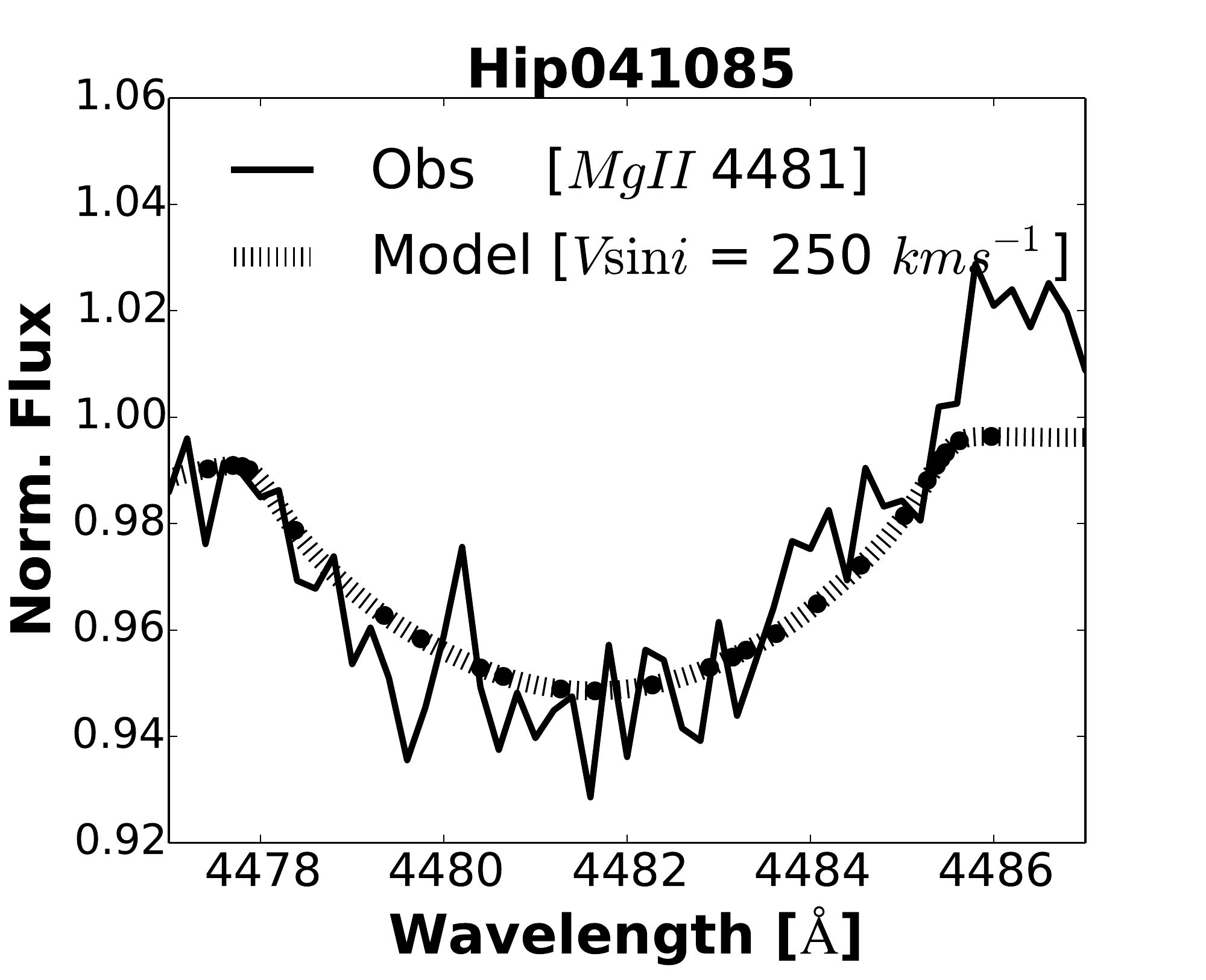}  
     \includegraphics[width=4cm,height=4cm] {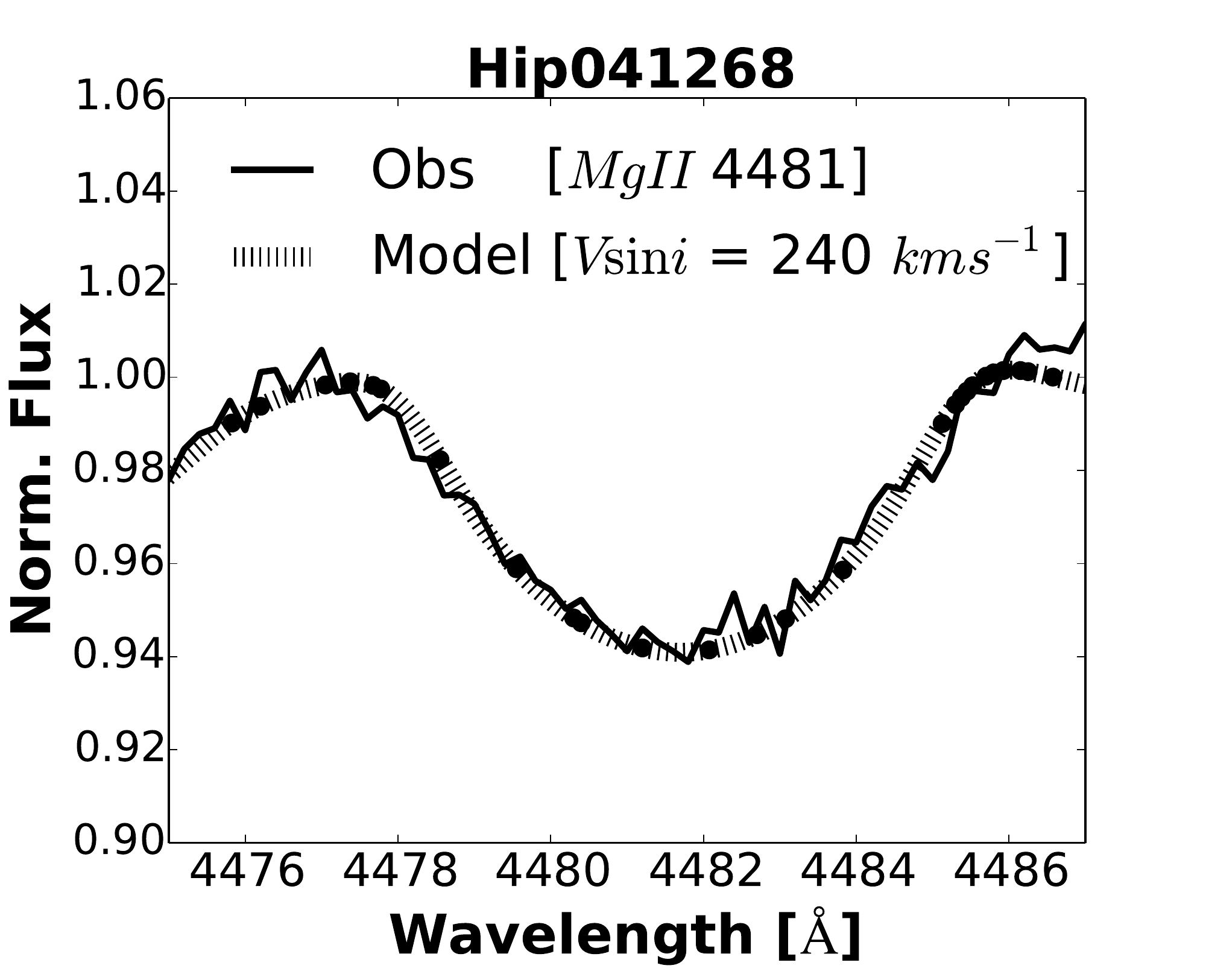} 
  \includegraphics[width=4cm,height=4cm] {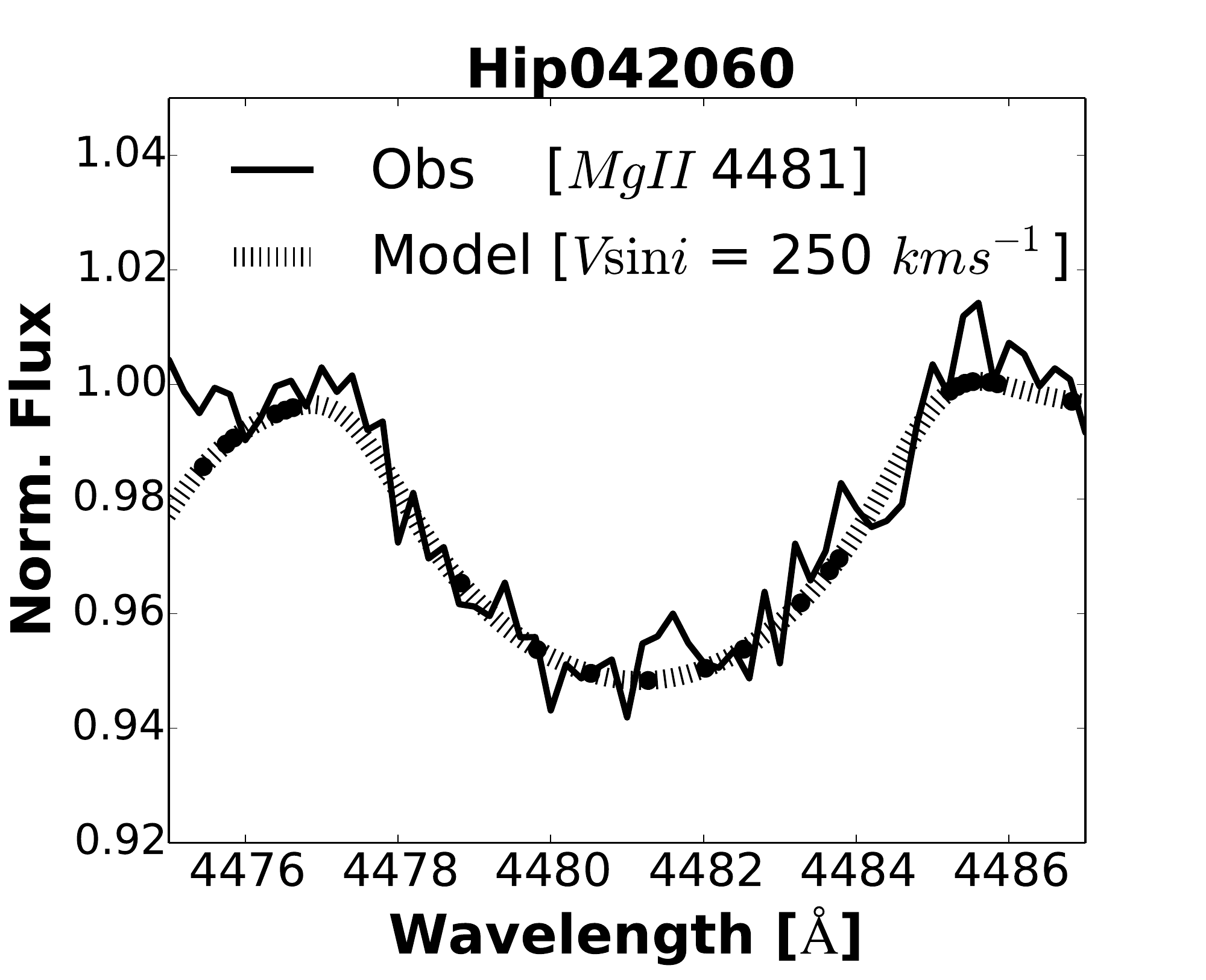} 
     \includegraphics[width=4cm,height=4cm] {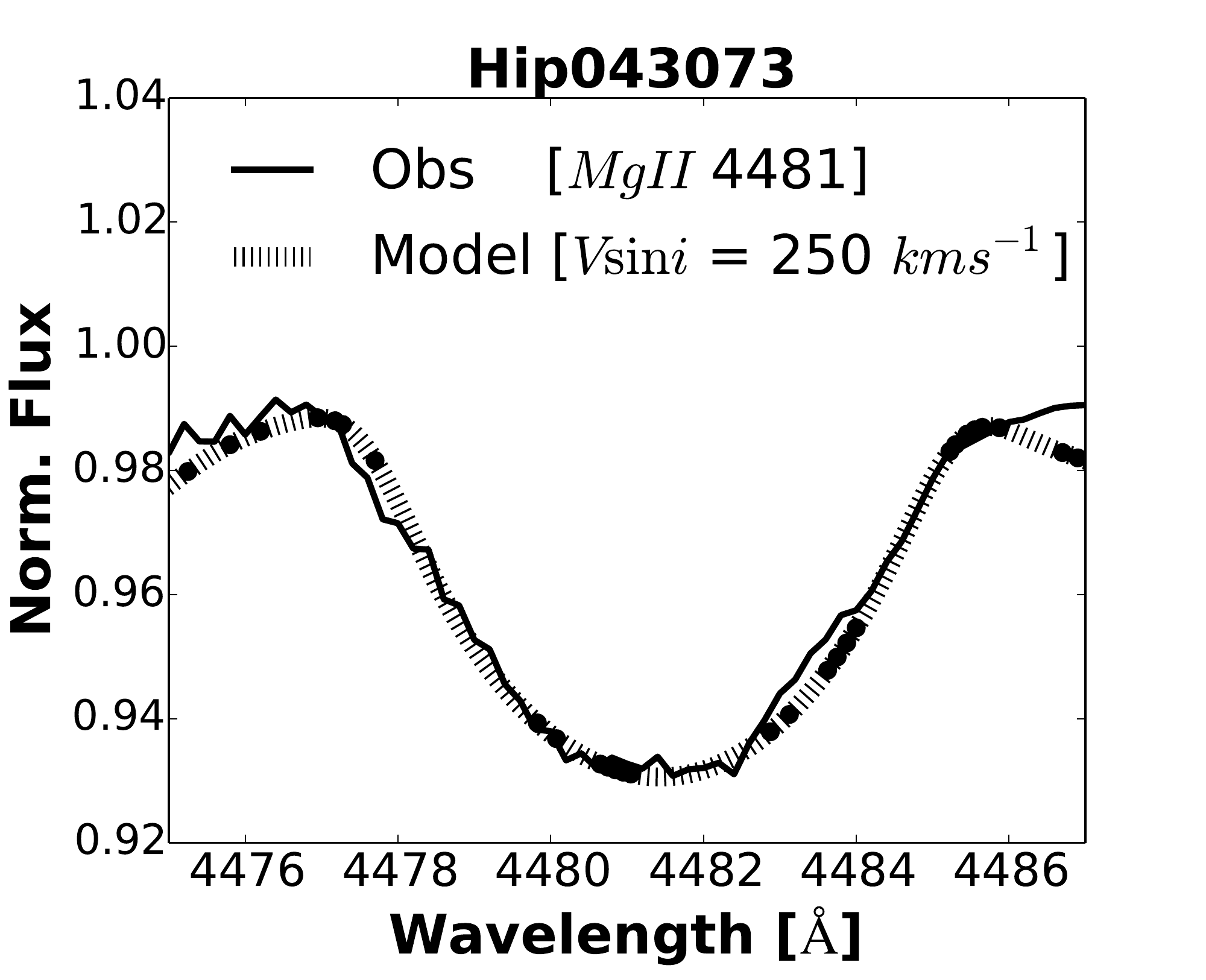}  
   }

\subfloat{
   \includegraphics[width=4cm,height=4cm] {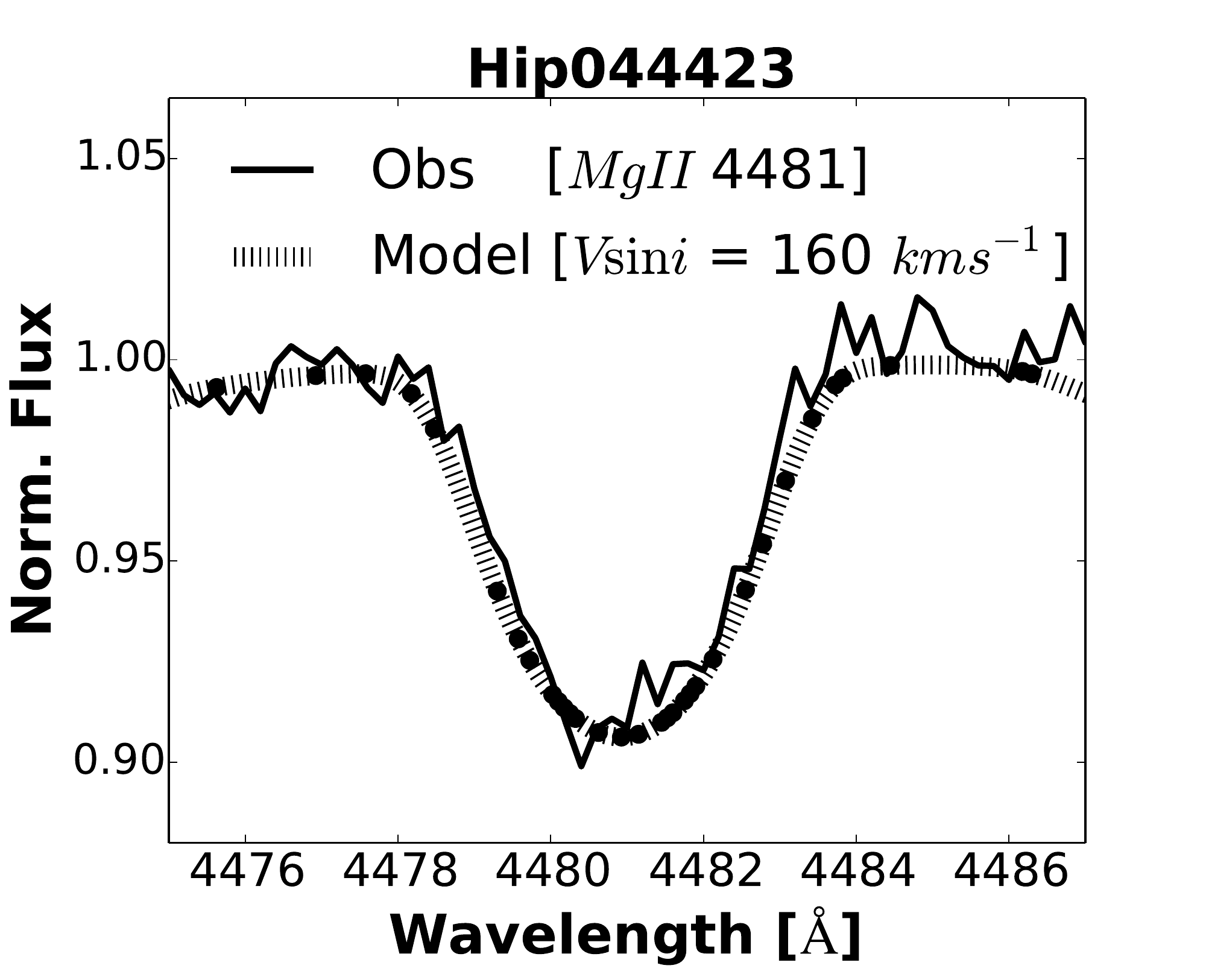} 
     \includegraphics[width=4cm,height=4cm] {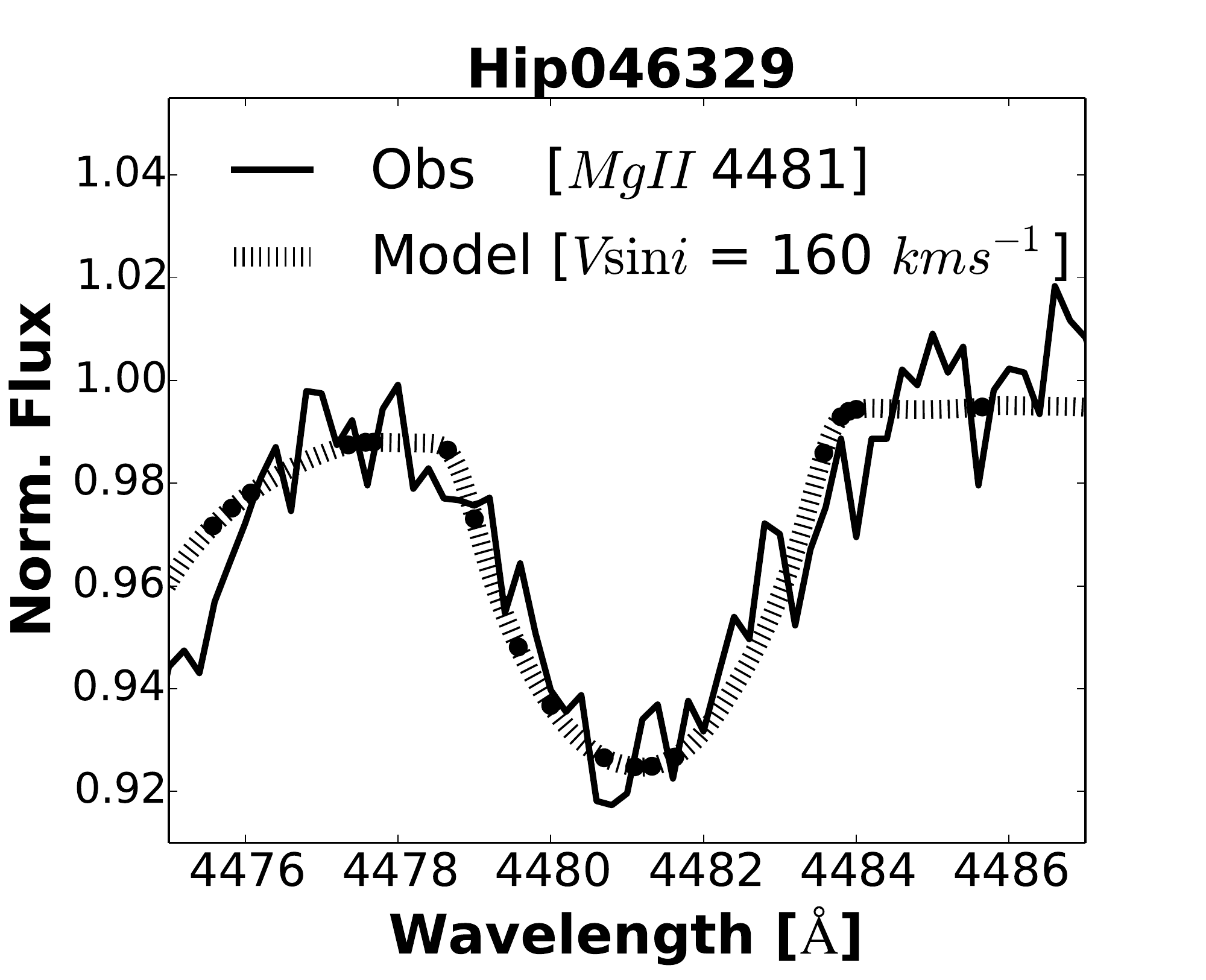} 
     \includegraphics[width=4cm,height=4cm] {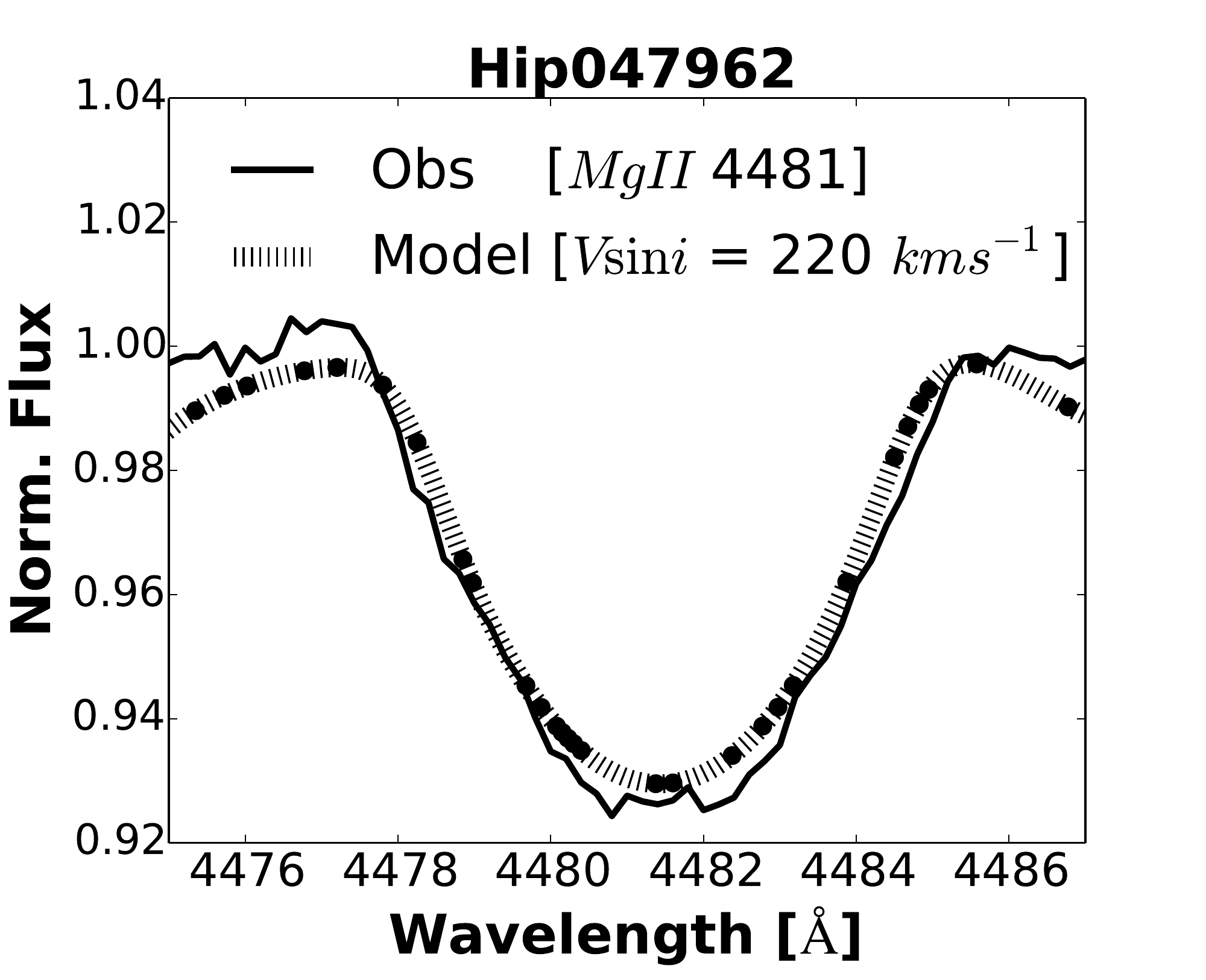} 
\includegraphics[width=4cm,height=4cm] {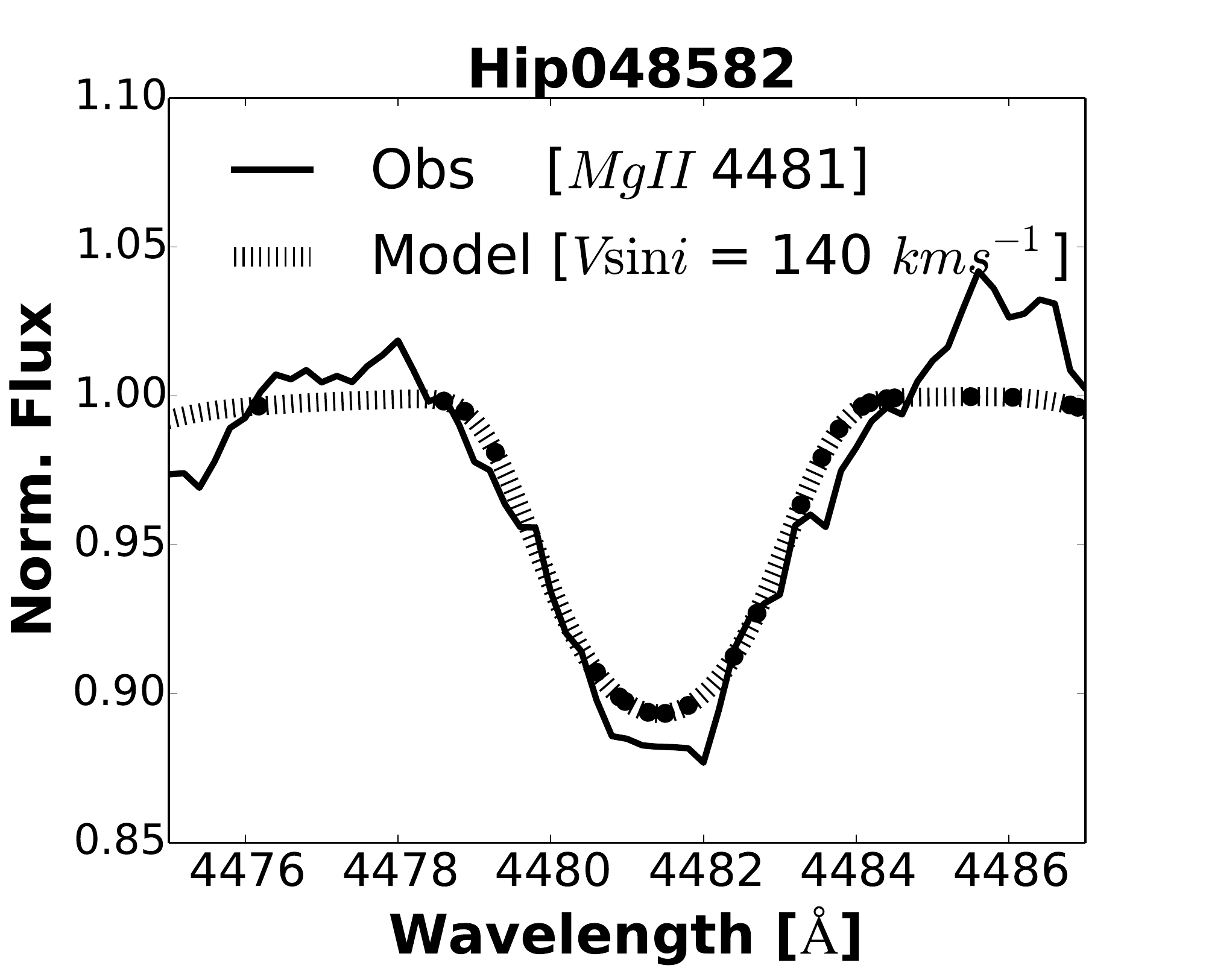}            
}

\subfloat{
    \includegraphics[width=4cm,height=4cm] {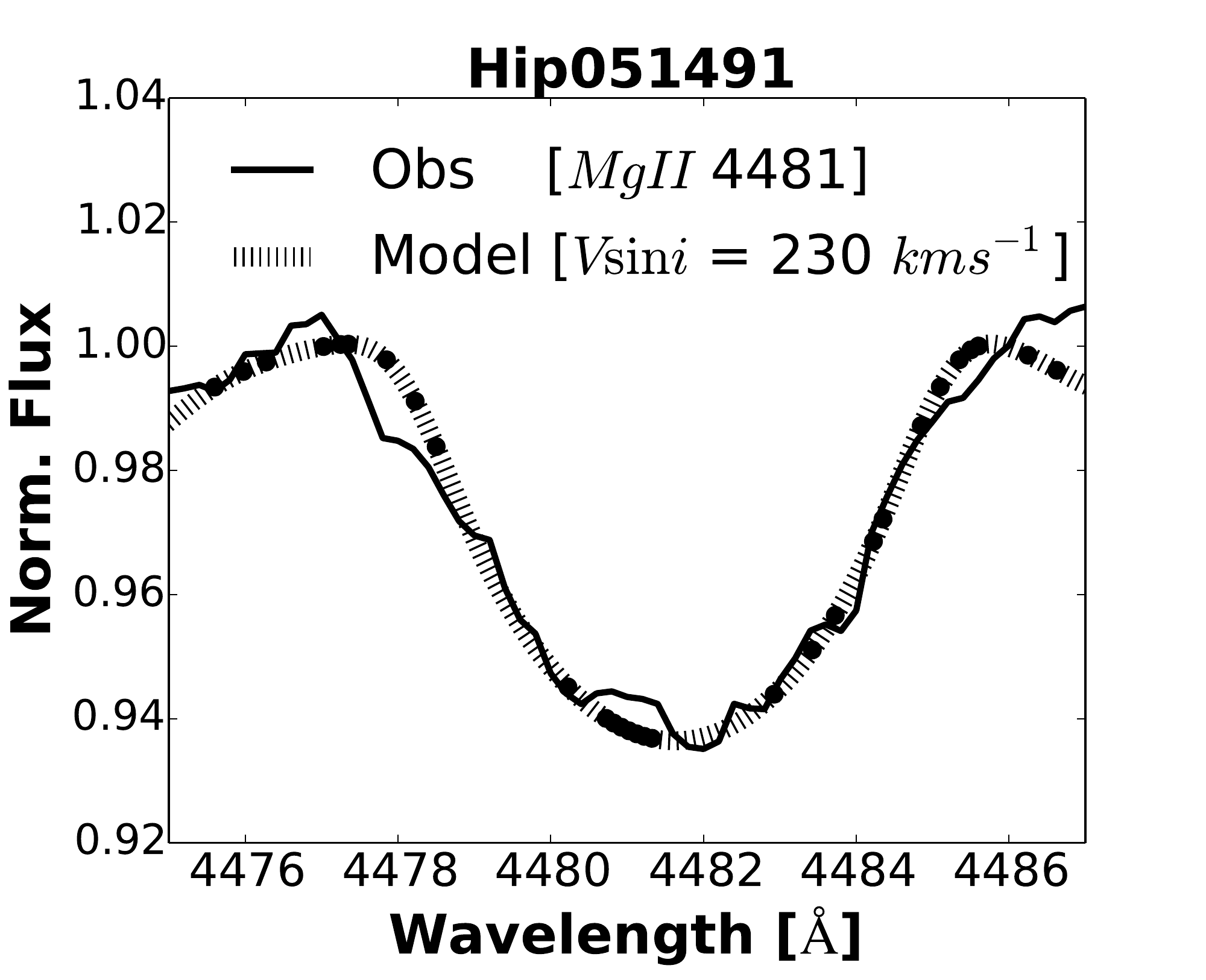} 
  \includegraphics[width=4cm,height=4cm] {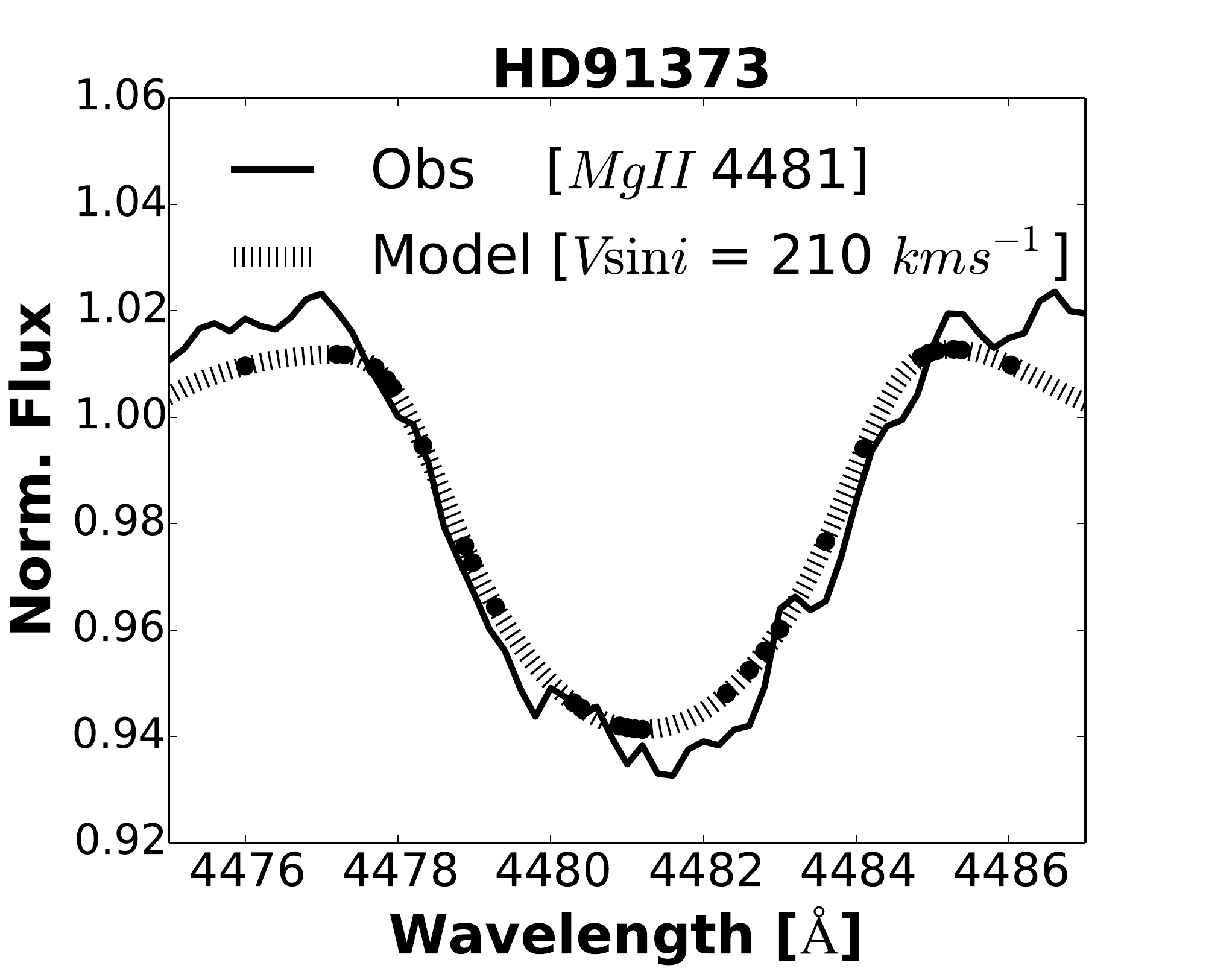} %Hip51546
    \includegraphics[width=4cm,height=4cm] {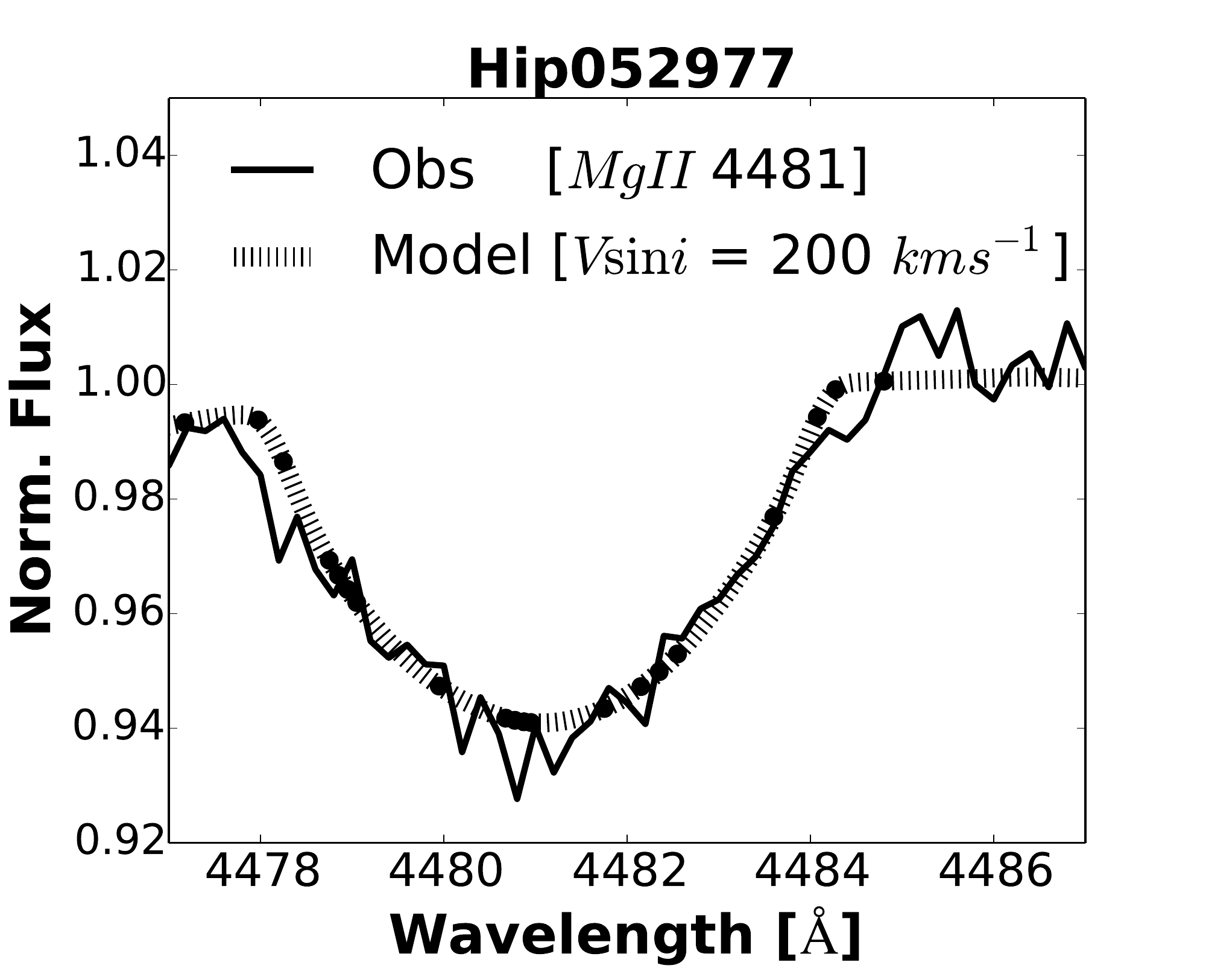} 
  \includegraphics[width=4cm,height=4cm] {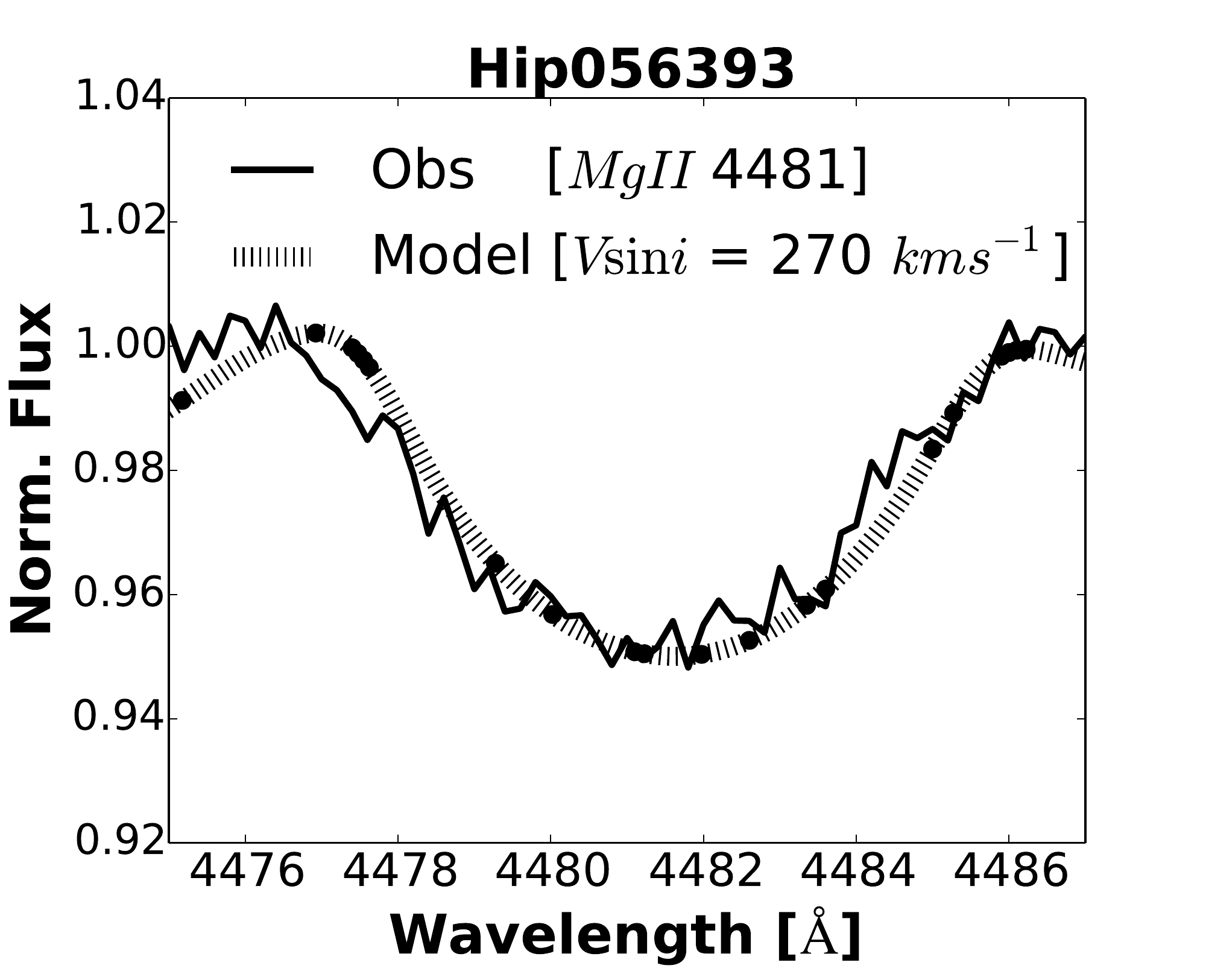} 
}

\subfloat{
  \includegraphics[width=4cm,height=4cm] {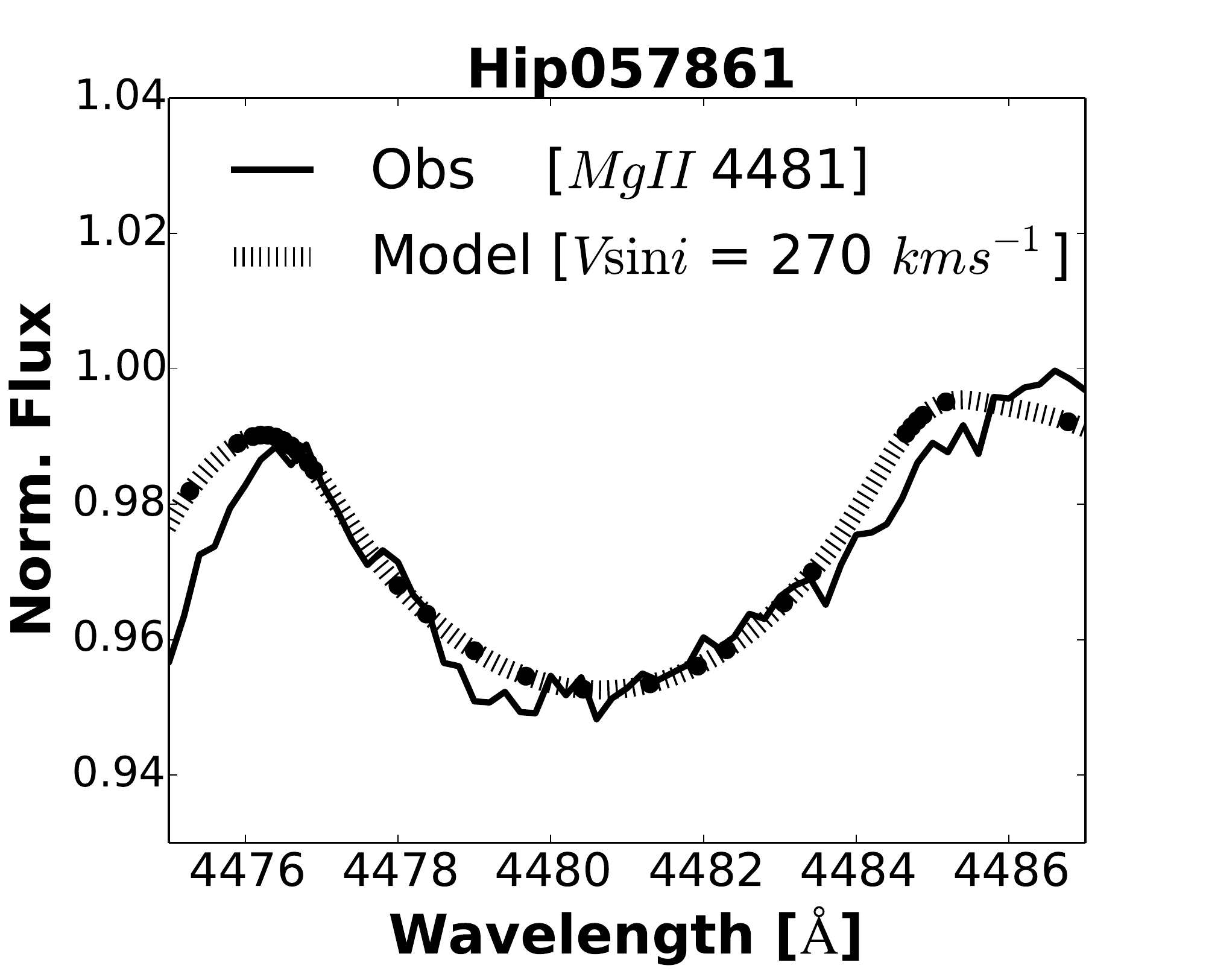} 
  \includegraphics[width=4cm,height=4cm] {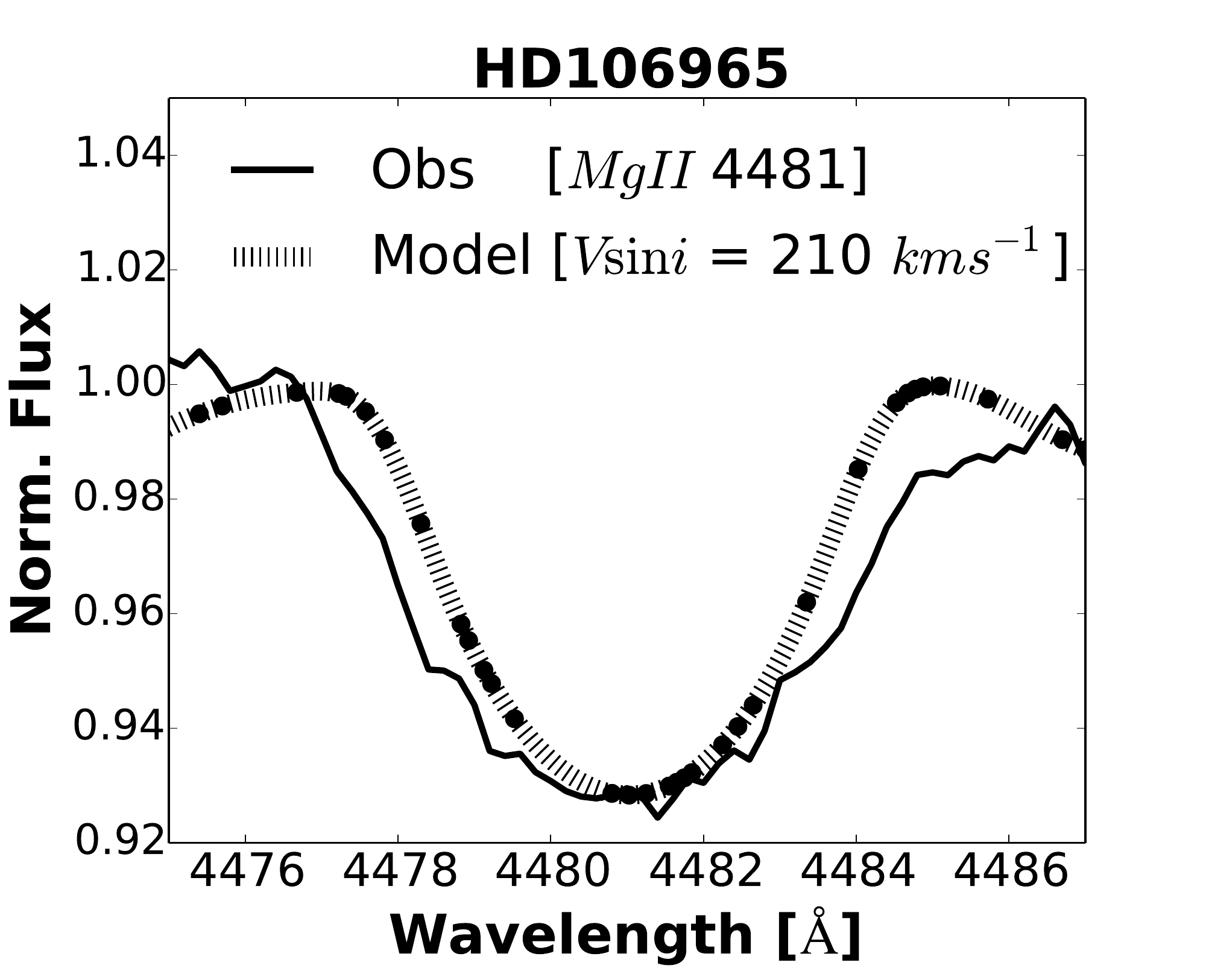}%Hip59970 
  \includegraphics[width=4cm,height=4cm] {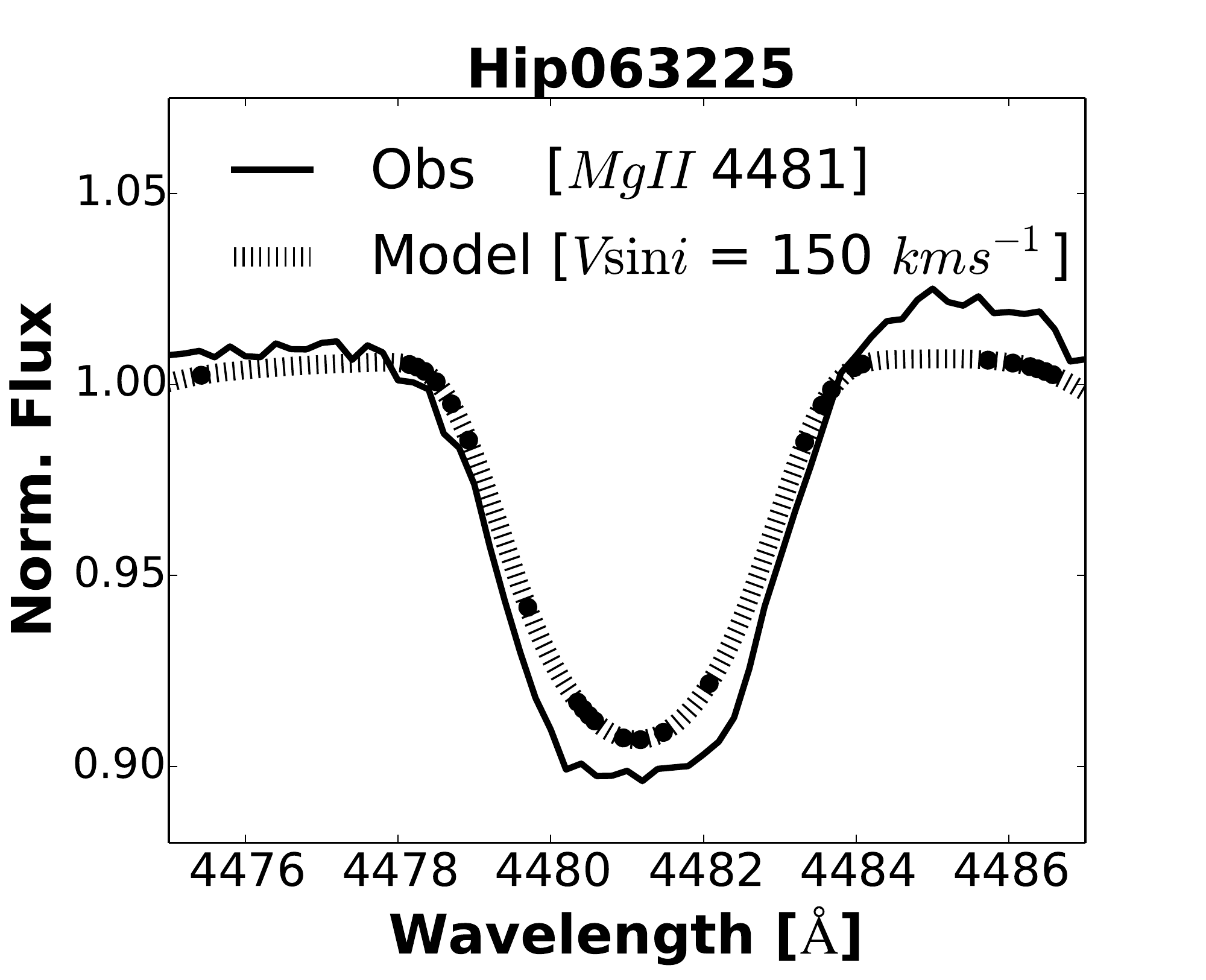}  
   \includegraphics[width=4cm,height=4cm] {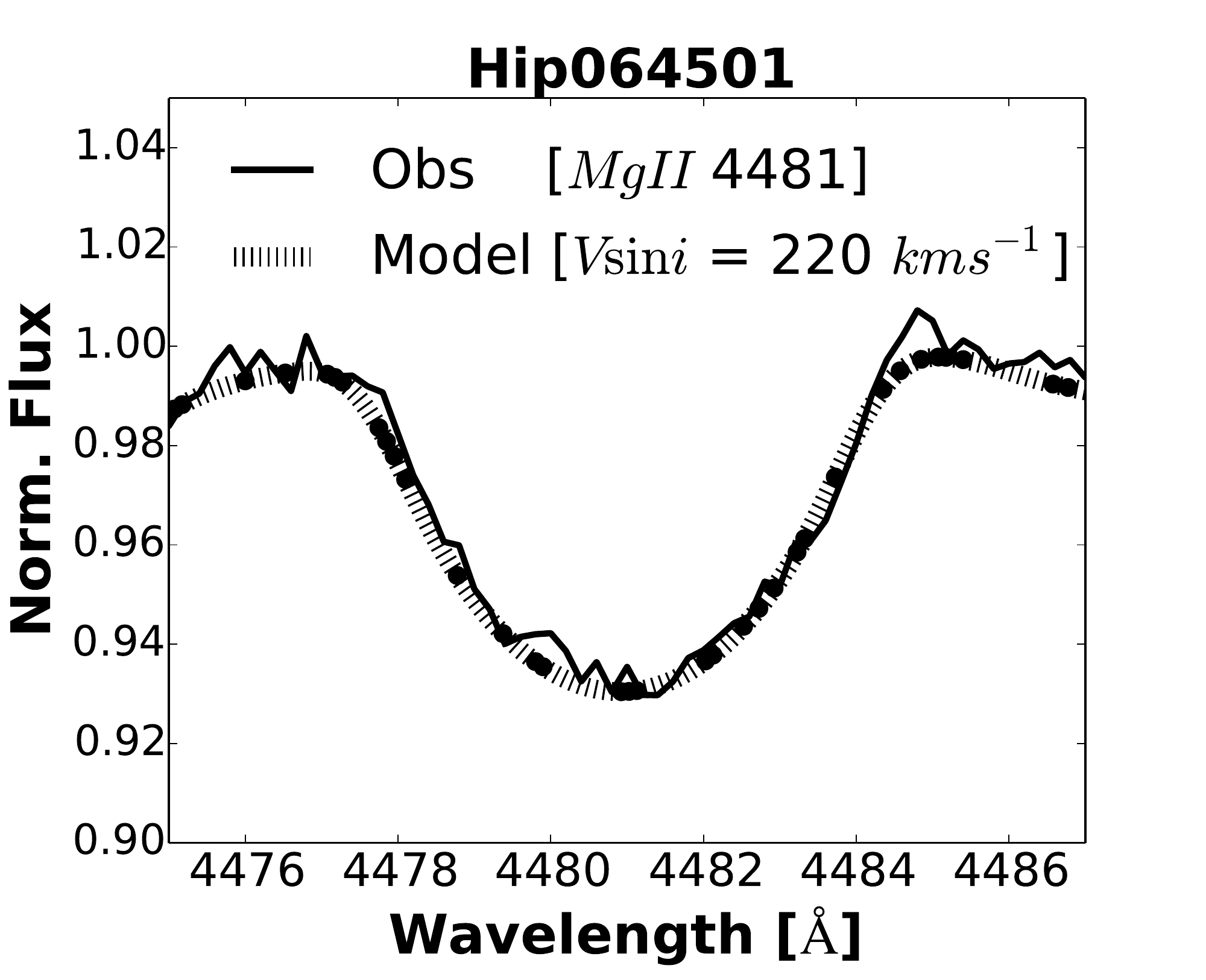}          
}

\subfloat{
   \includegraphics[width=4cm,height=4cm] {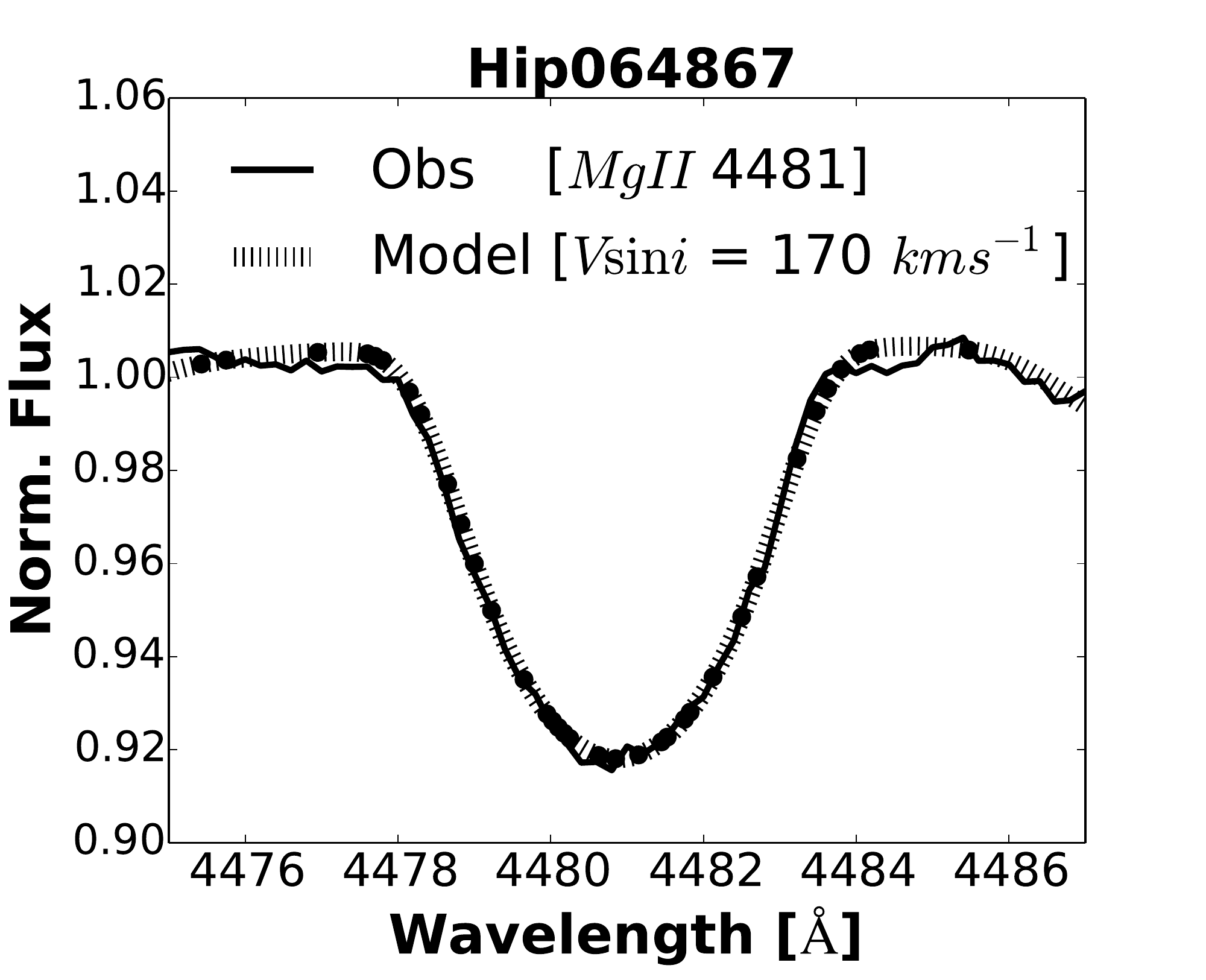} 
    \includegraphics[width=4cm,height=4cm] {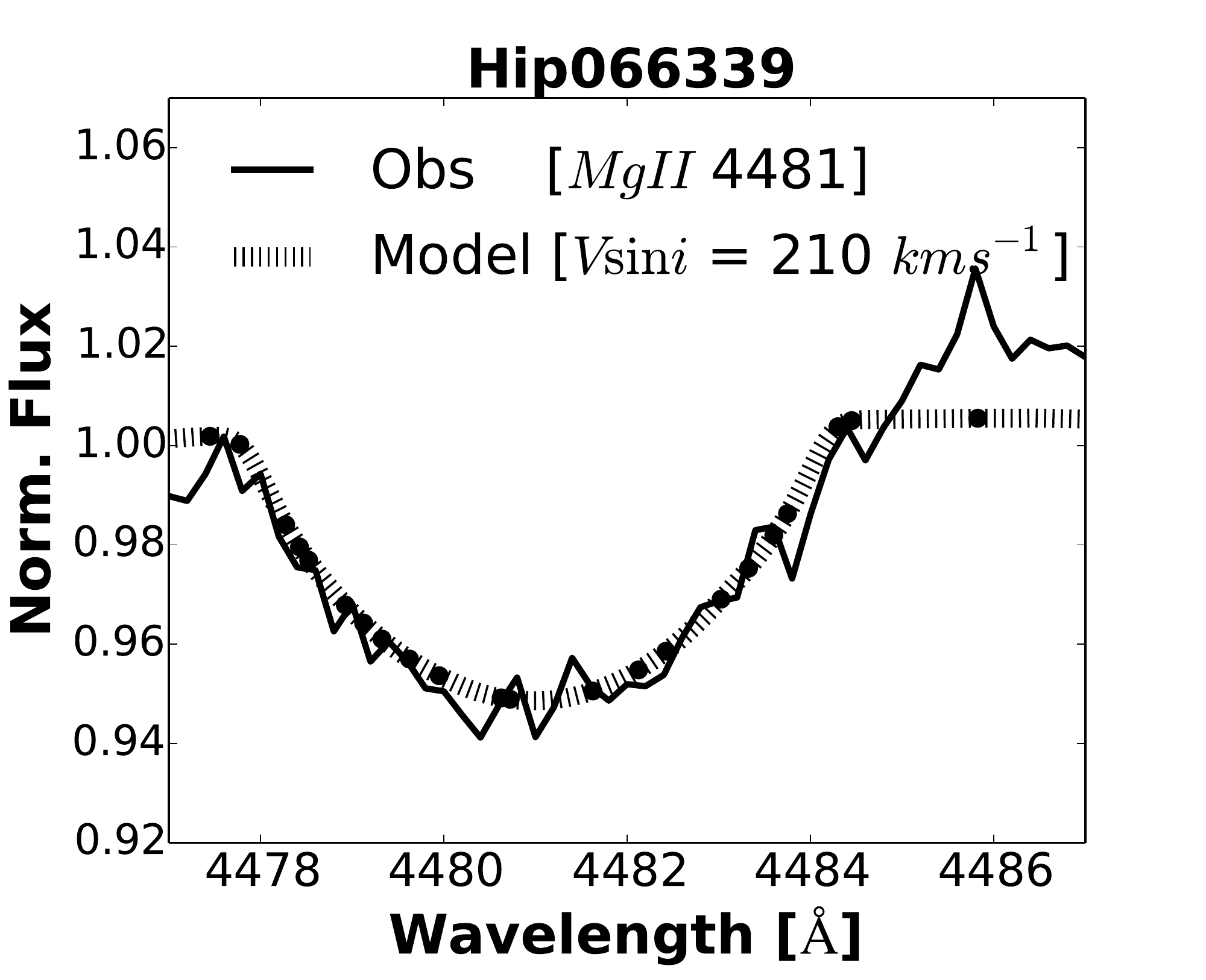} 
    \includegraphics[width=4cm,height=4cm] {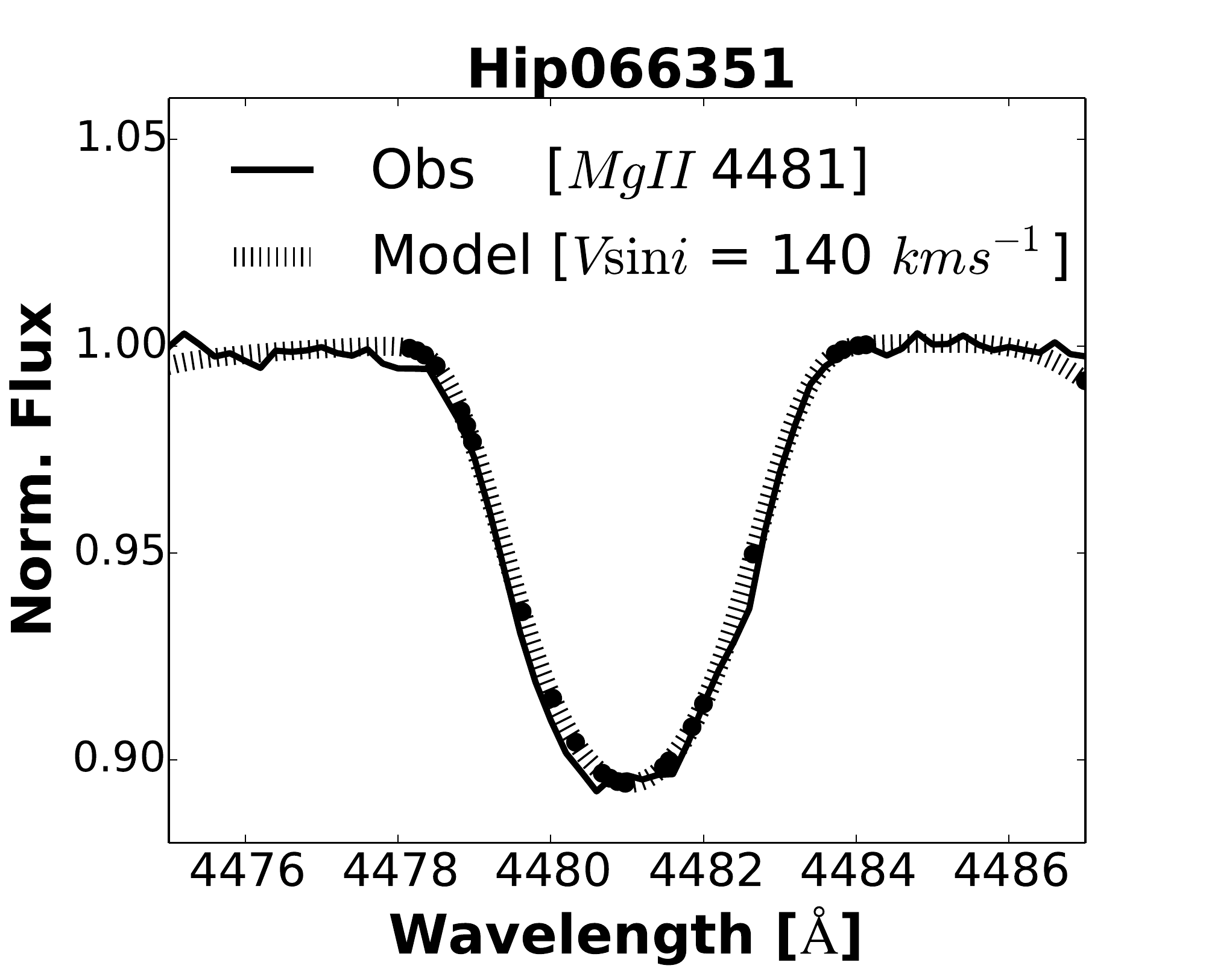} 
     \includegraphics[width=4cm,height=4cm] {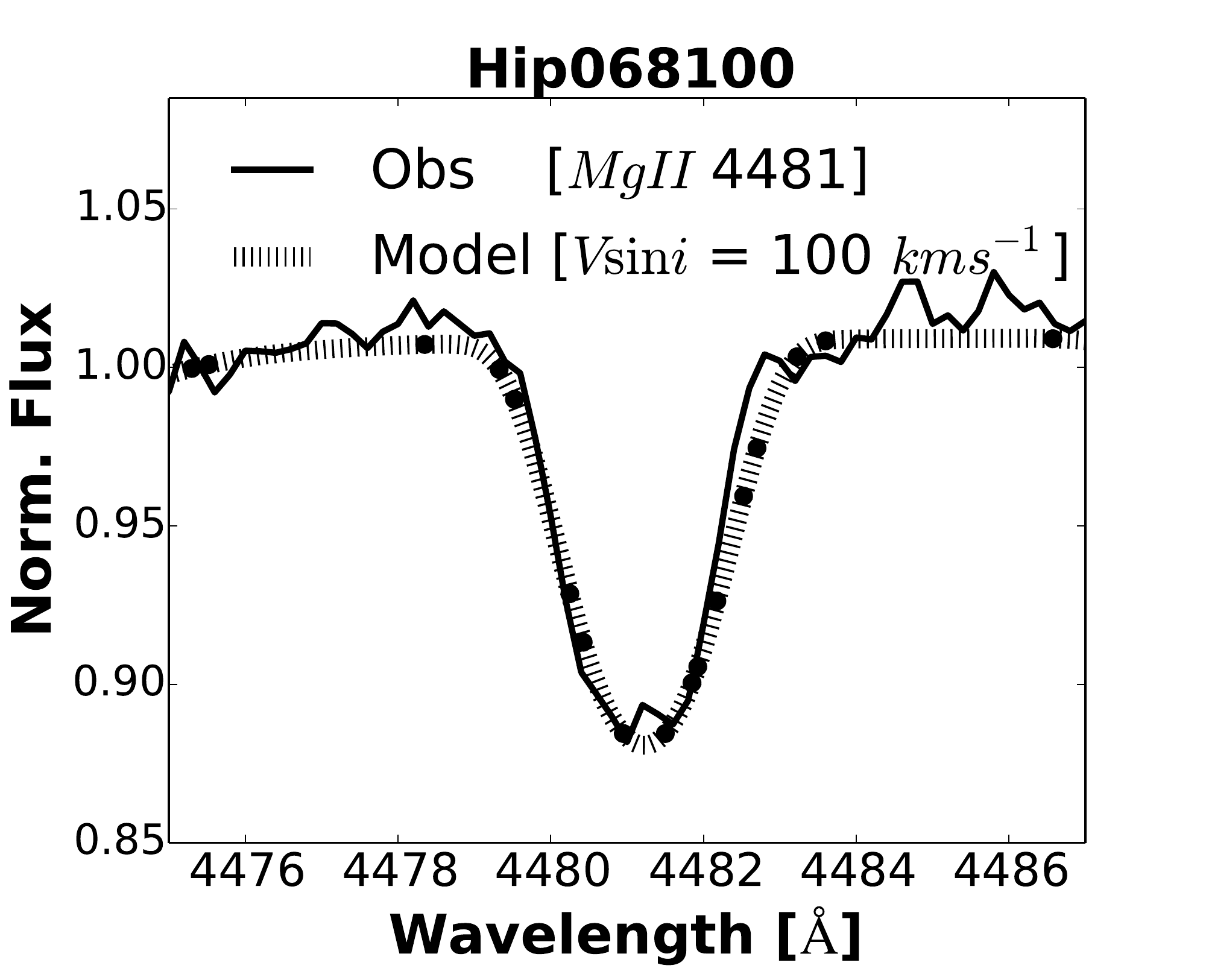} 
}

\caption{Continued}
\end{figure*}

\begin{figure*}
\ContinuedFloat
\centering

\subfloat{
        \includegraphics[width=4cm,height=4cm] {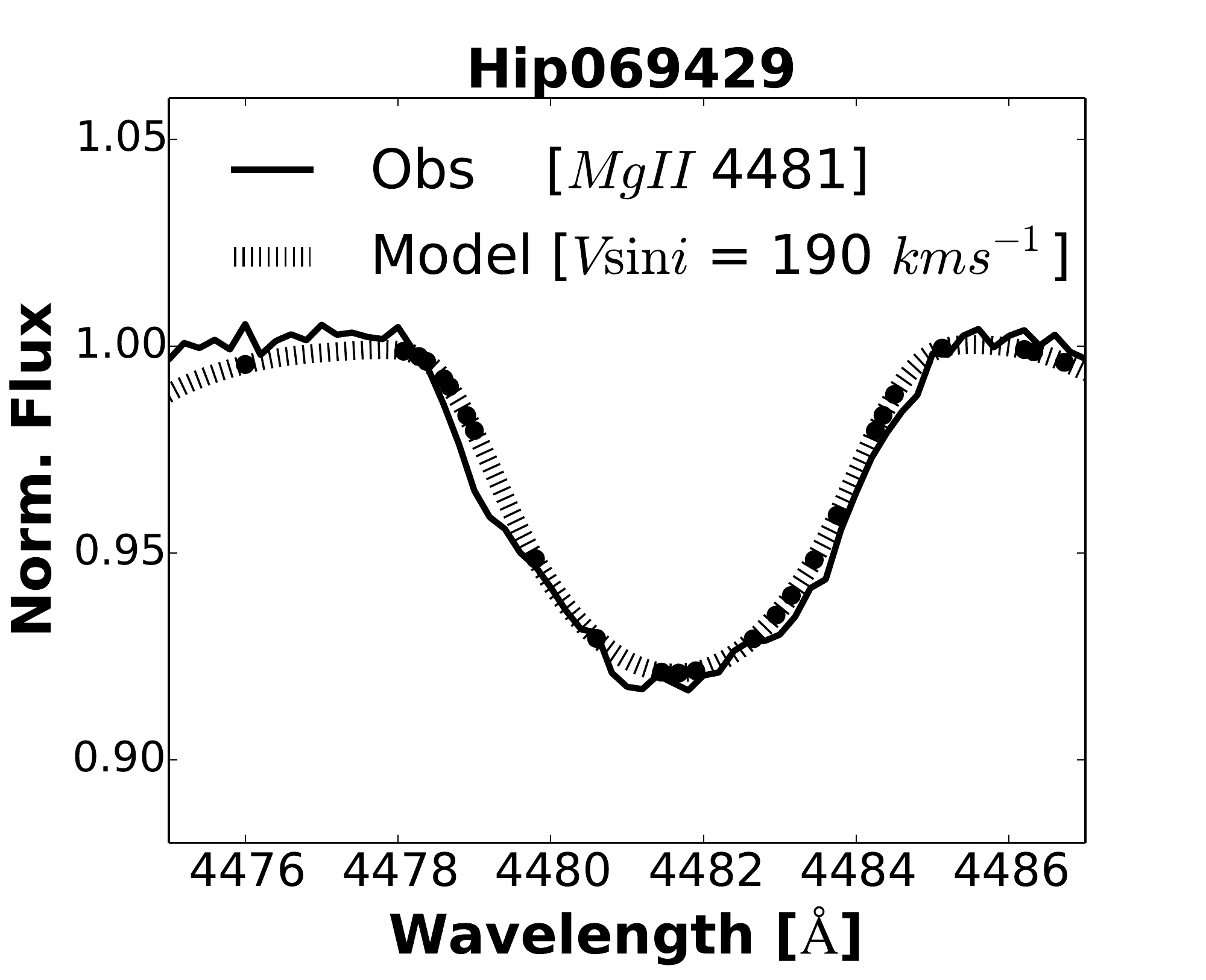}     
   \includegraphics[width=4cm,height=4cm] {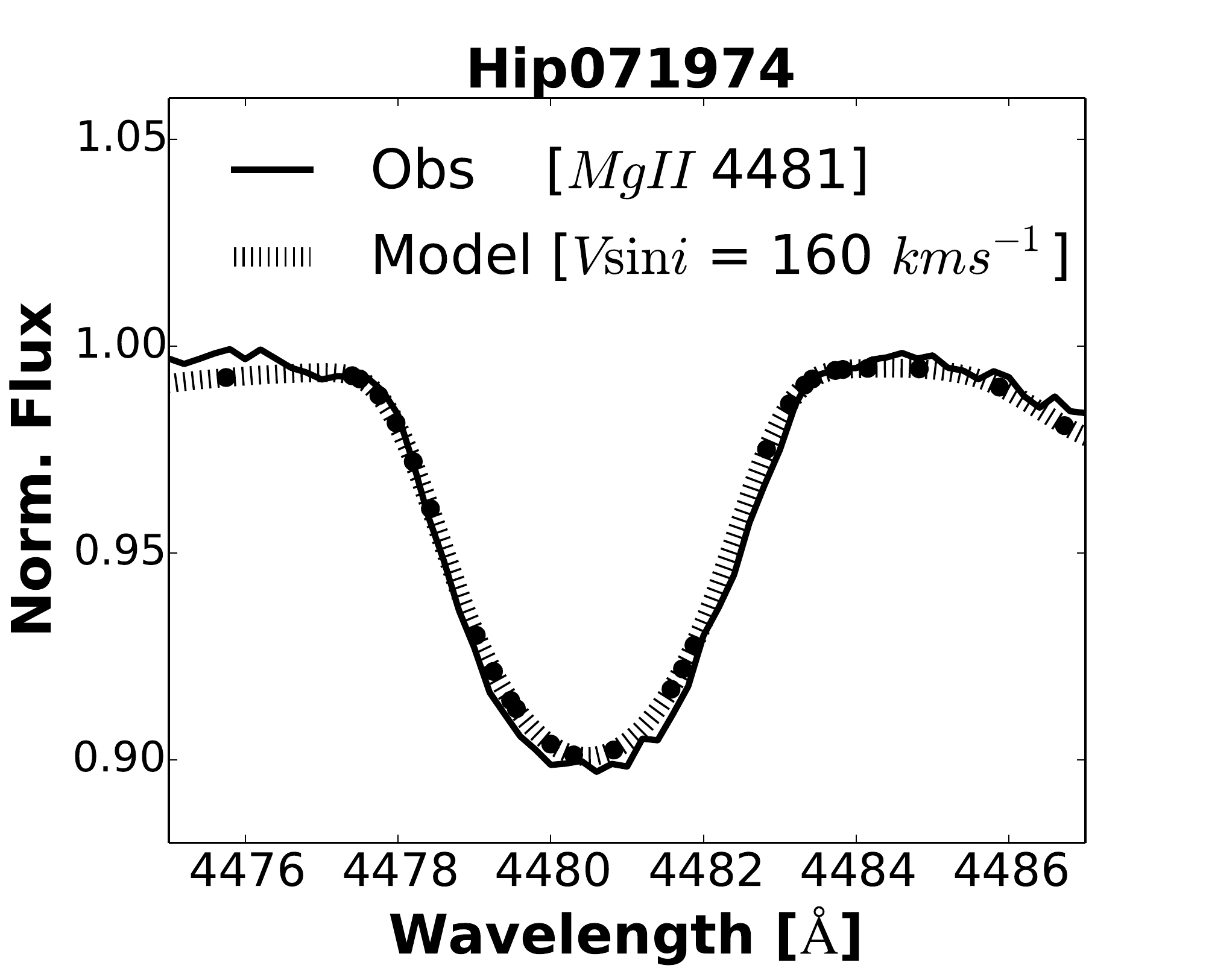} 
    \includegraphics[width=4cm,height=4cm] {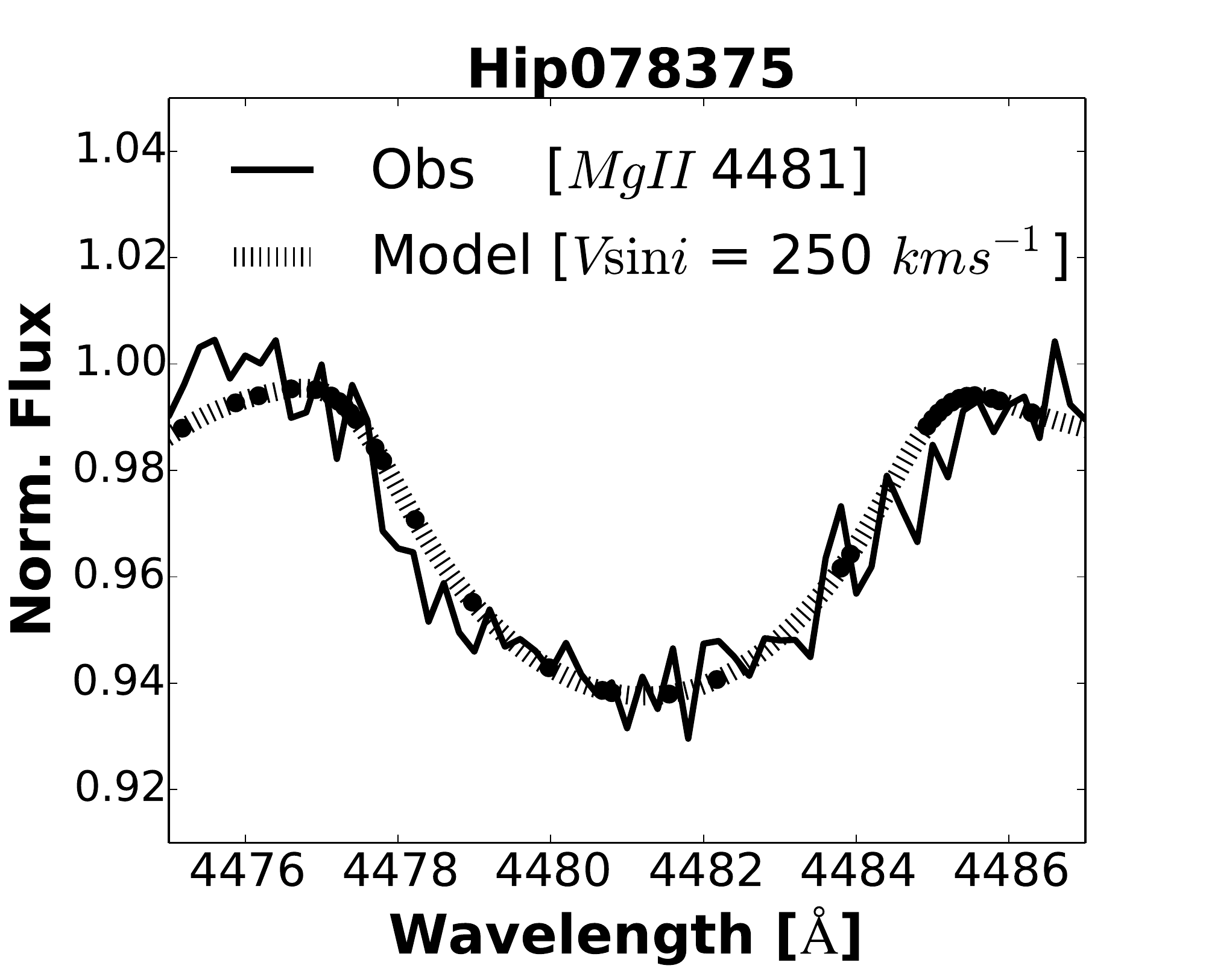} 
    \includegraphics[width=4cm,height=4cm] {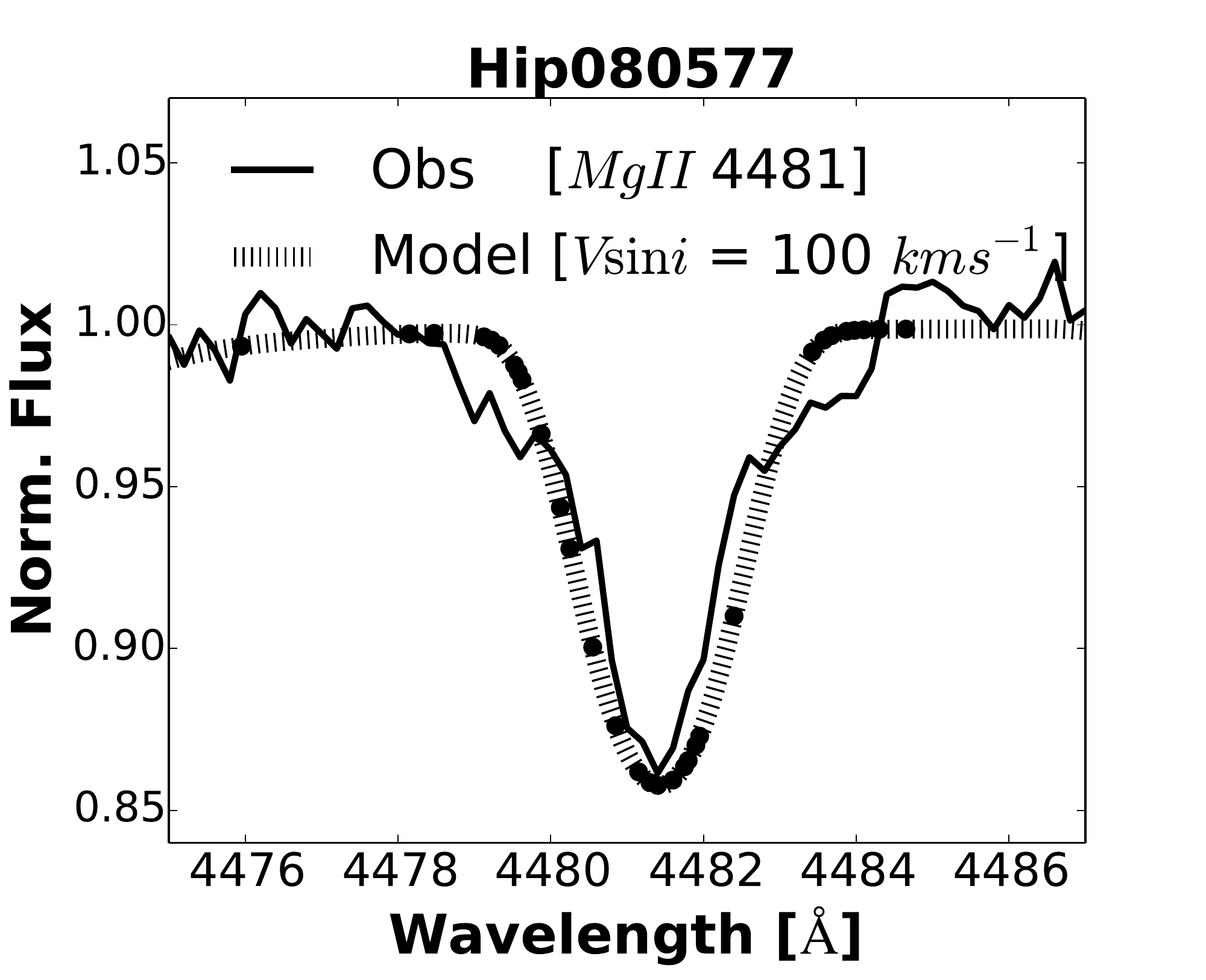}  
}

\subfloat{
  \includegraphics[width=4cm,height=4cm] {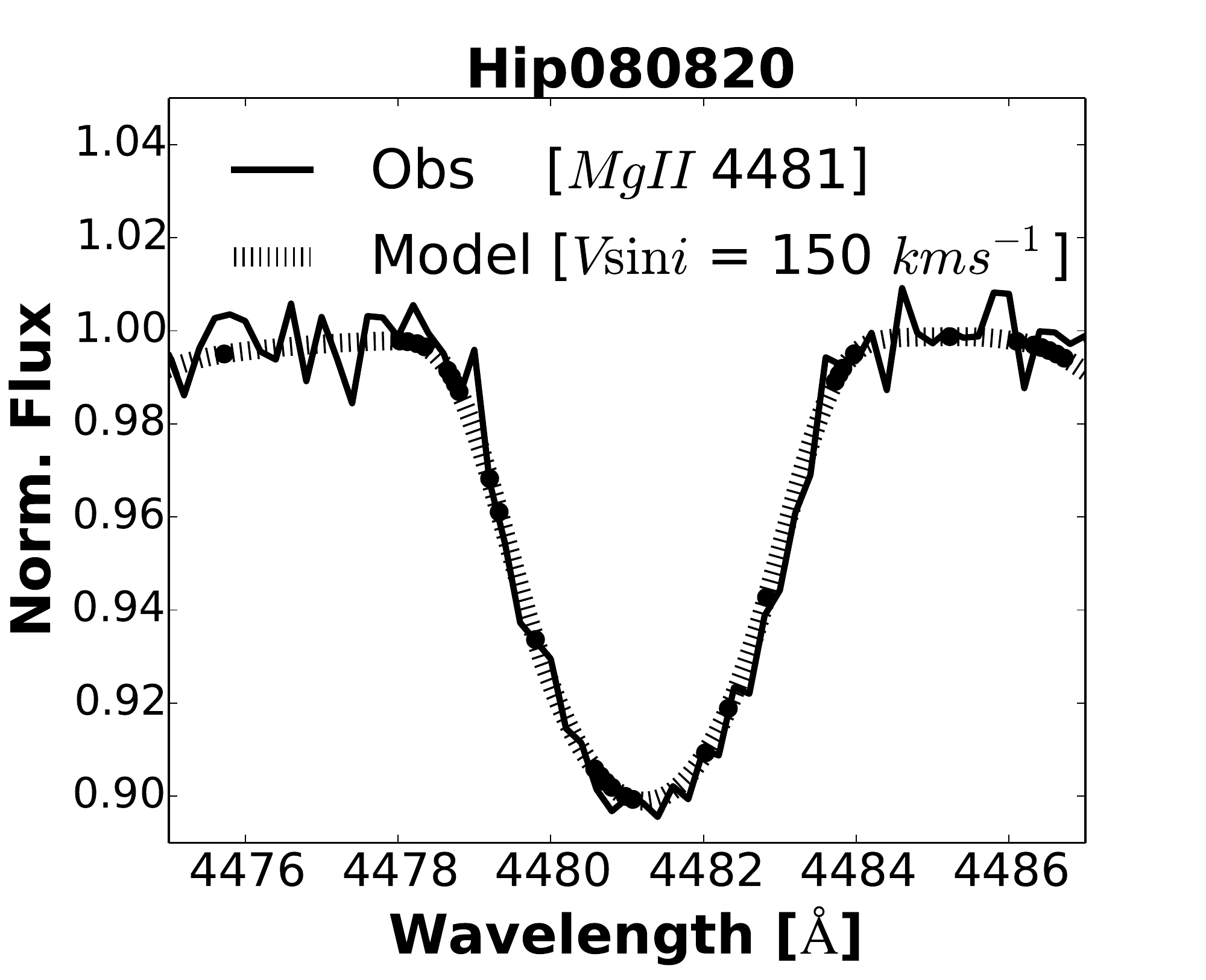} 
  \includegraphics[width=4cm,height=4cm] {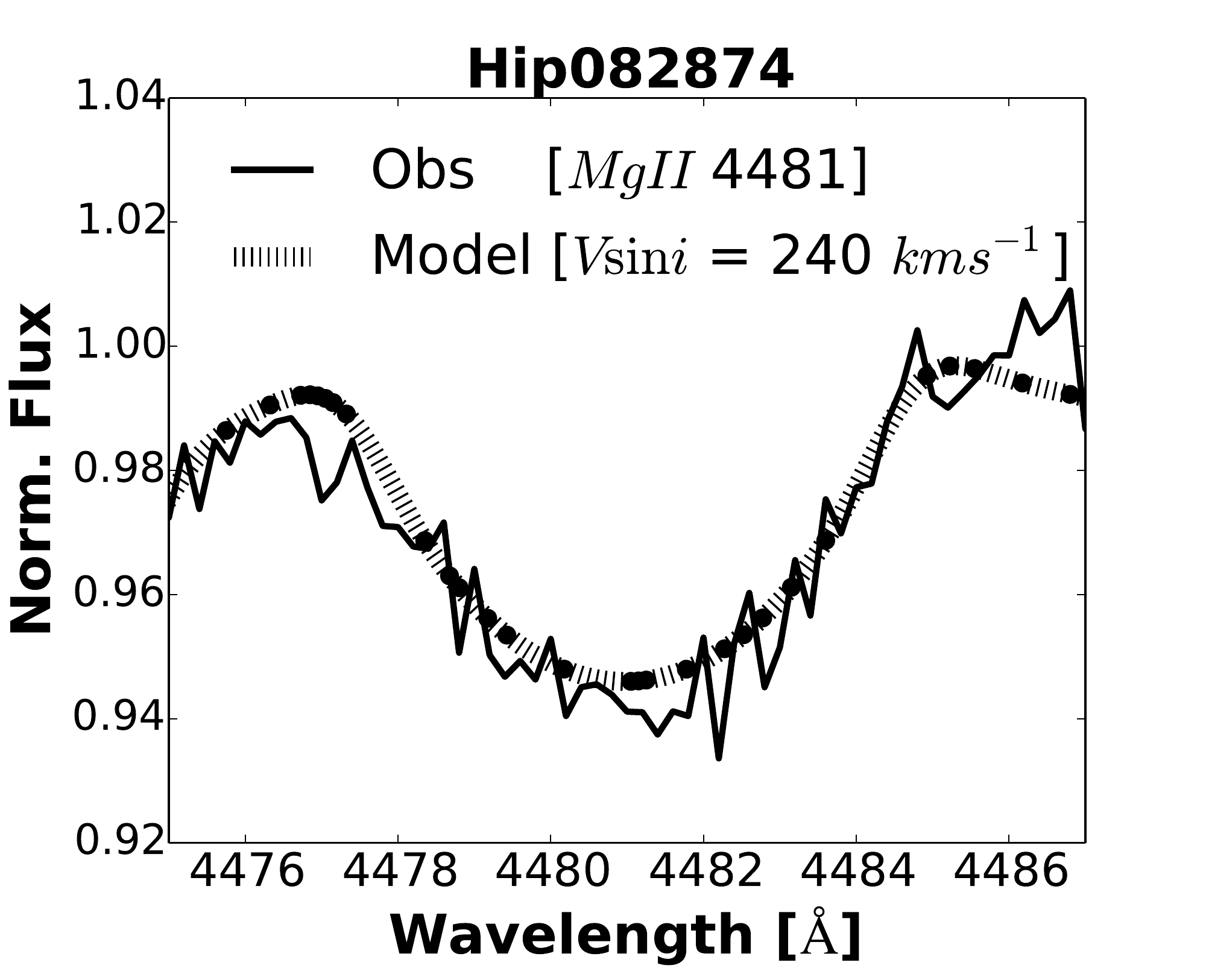} 
   \includegraphics[width=4cm,height=4cm] {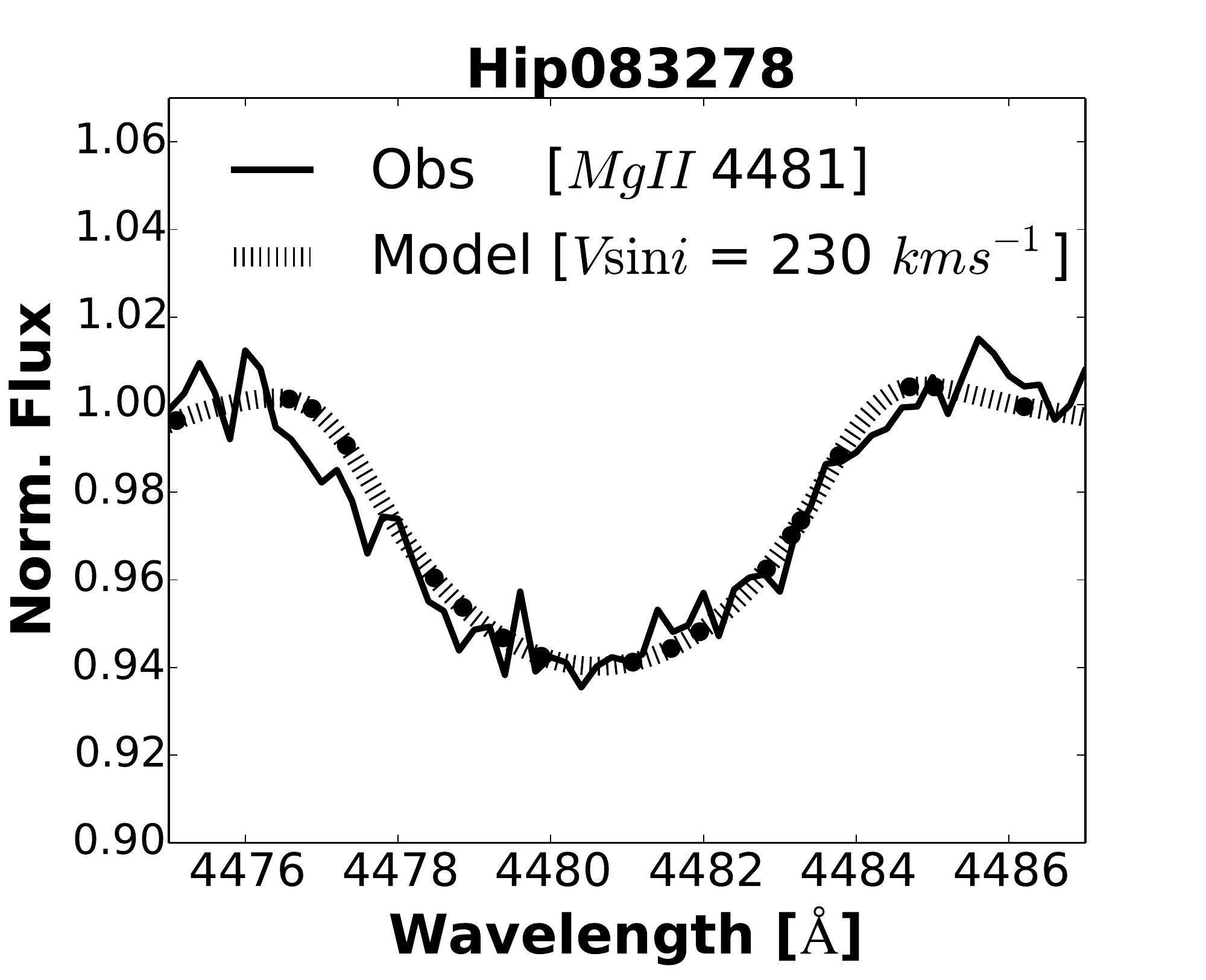} 
  \includegraphics[width=4cm,height=4cm] {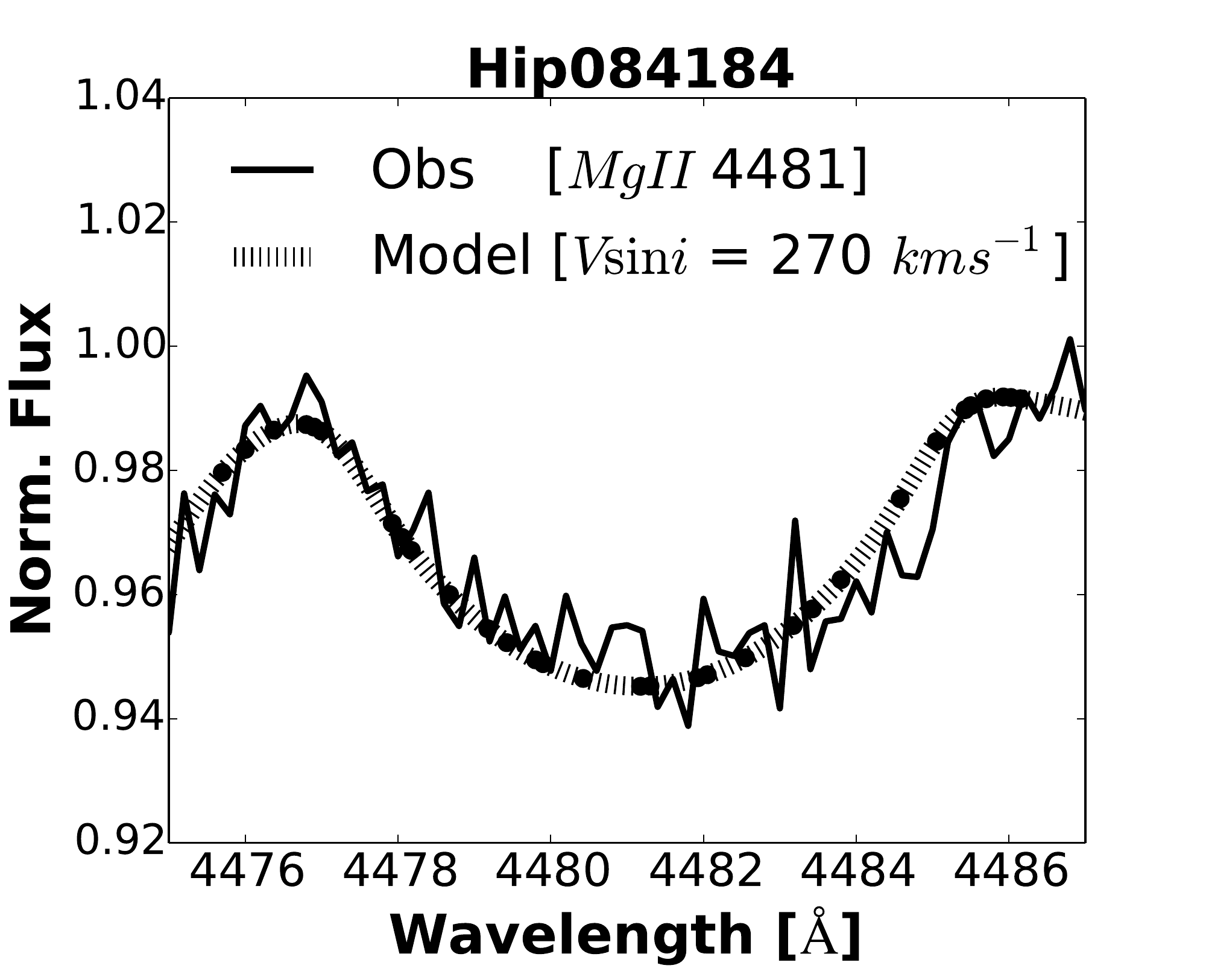} 
}

\subfloat{
 \includegraphics[width=4cm,height=4cm] {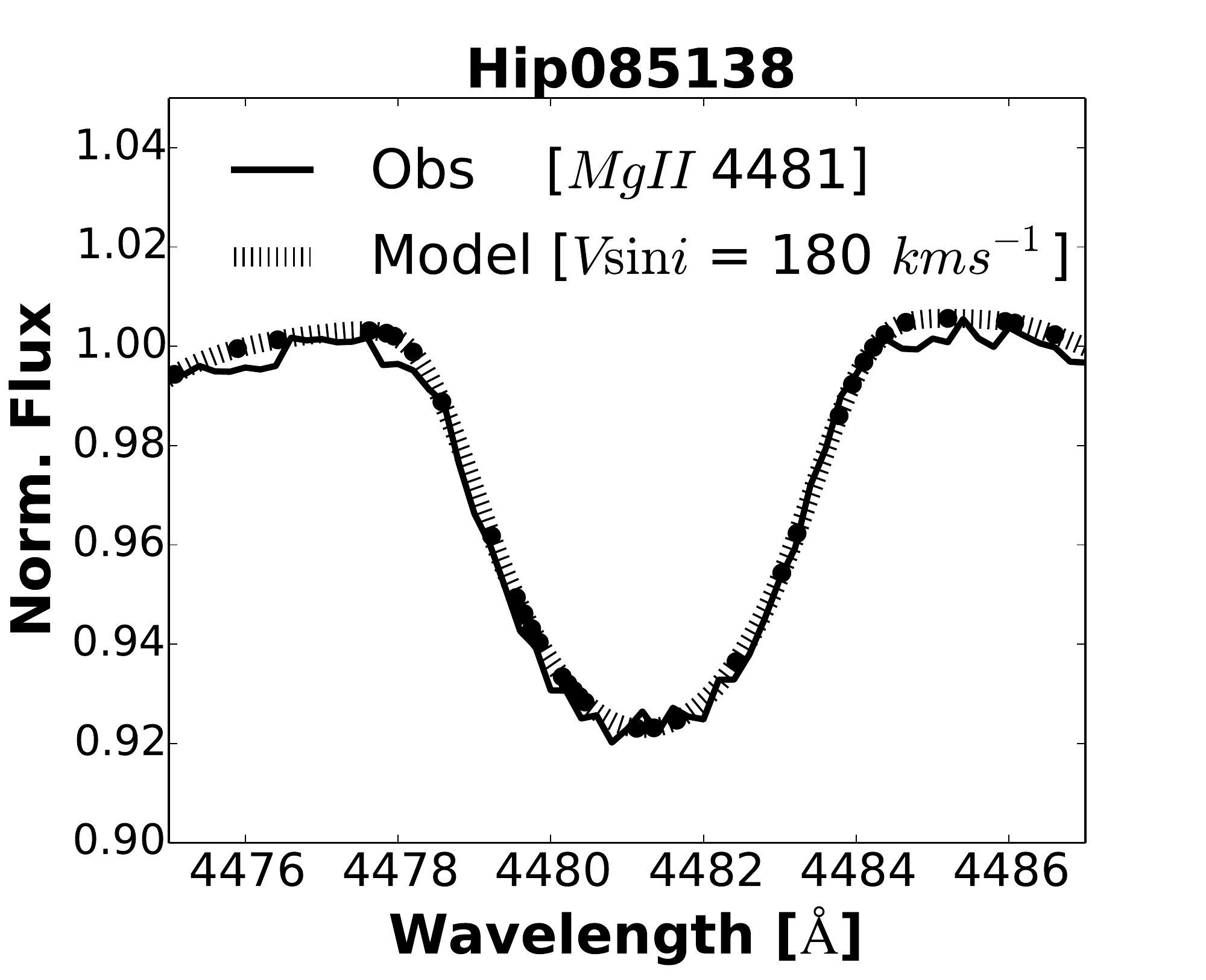} 
  \includegraphics[width=4cm,height=4cm] {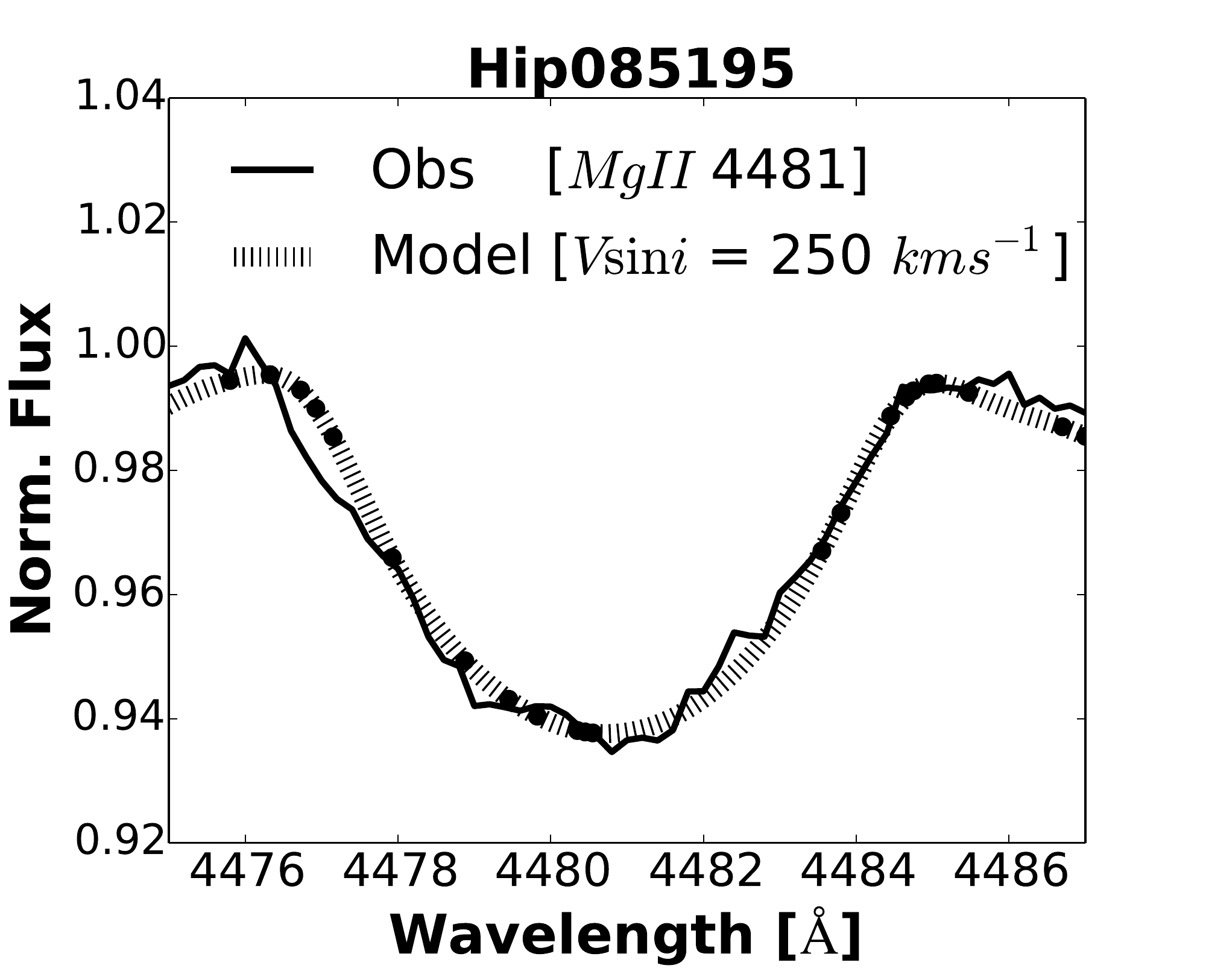} 
    \includegraphics[width=4cm,height=4cm] {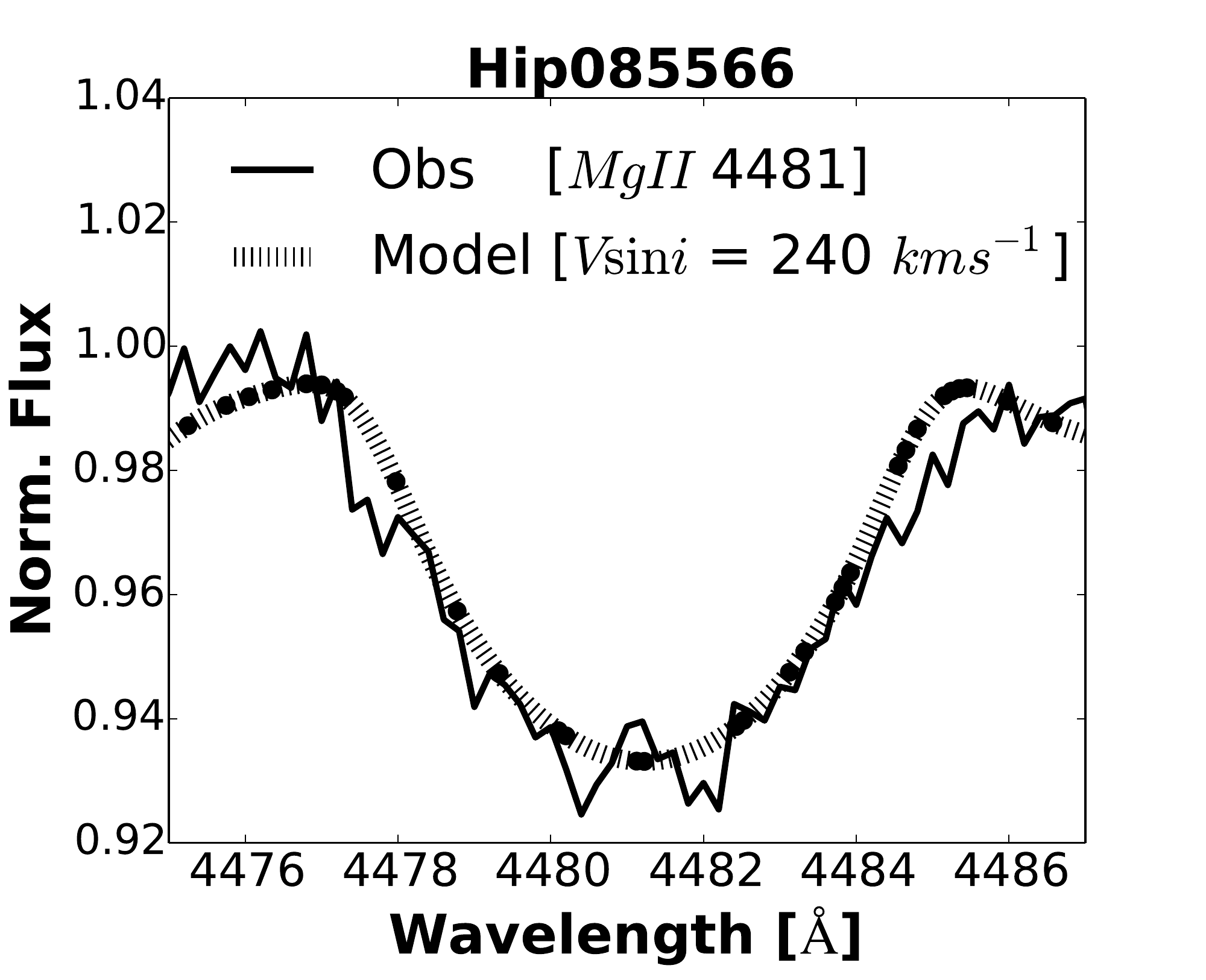} 
 \includegraphics[width=4cm,height=4cm] {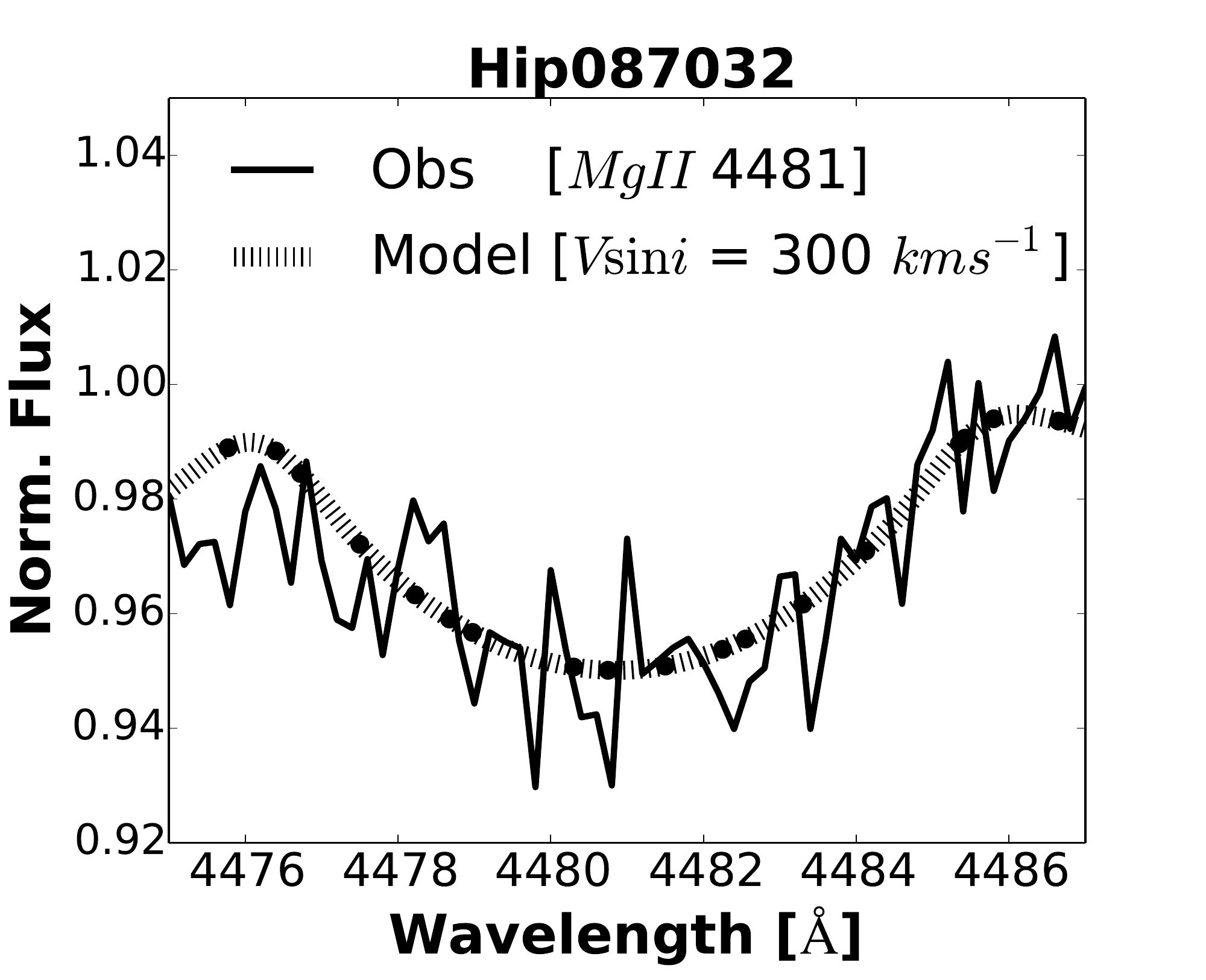} 
 
}

\subfloat{
     \includegraphics[width=4cm,height=4cm] {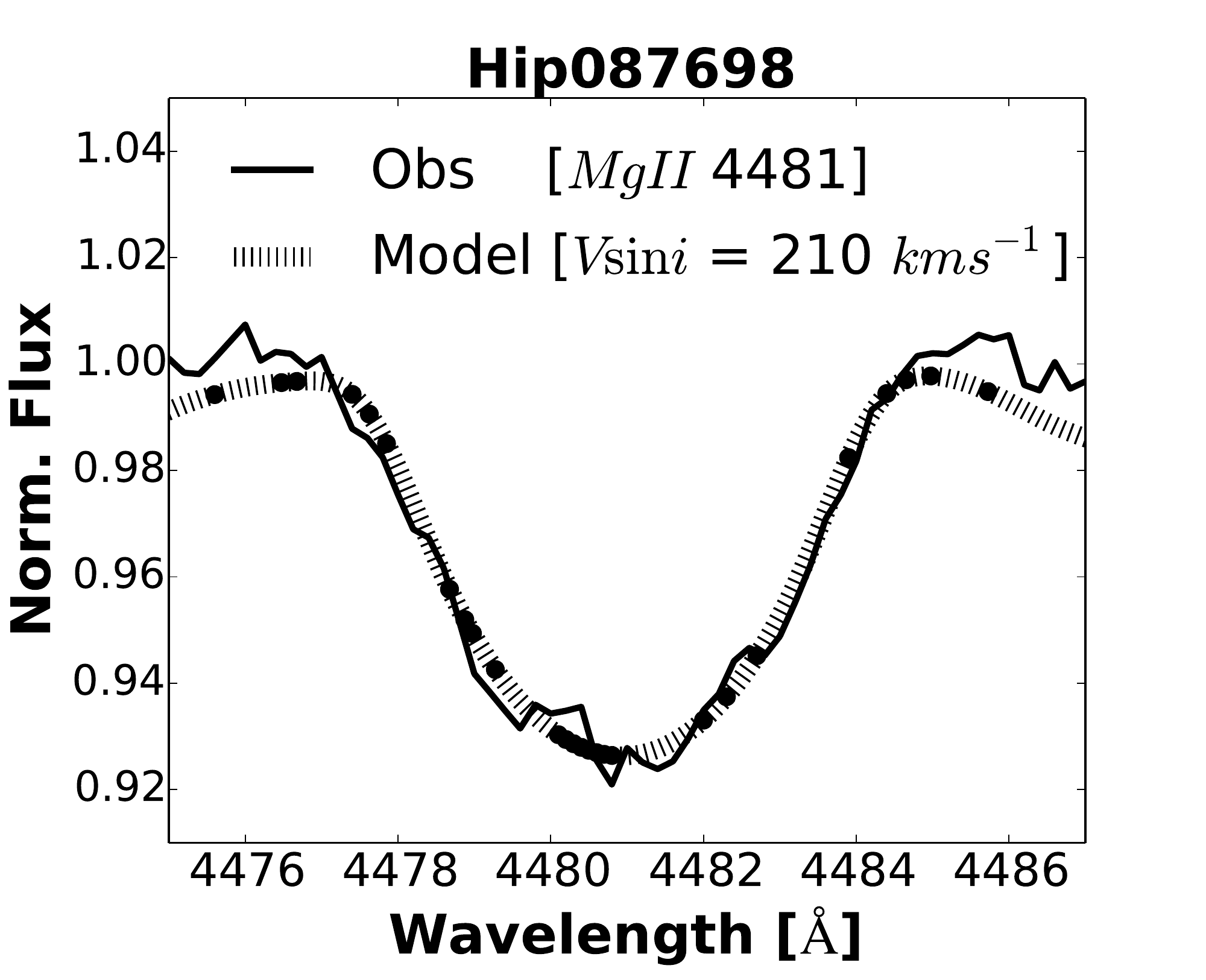} 
   \includegraphics[width=4cm,height=4cm] {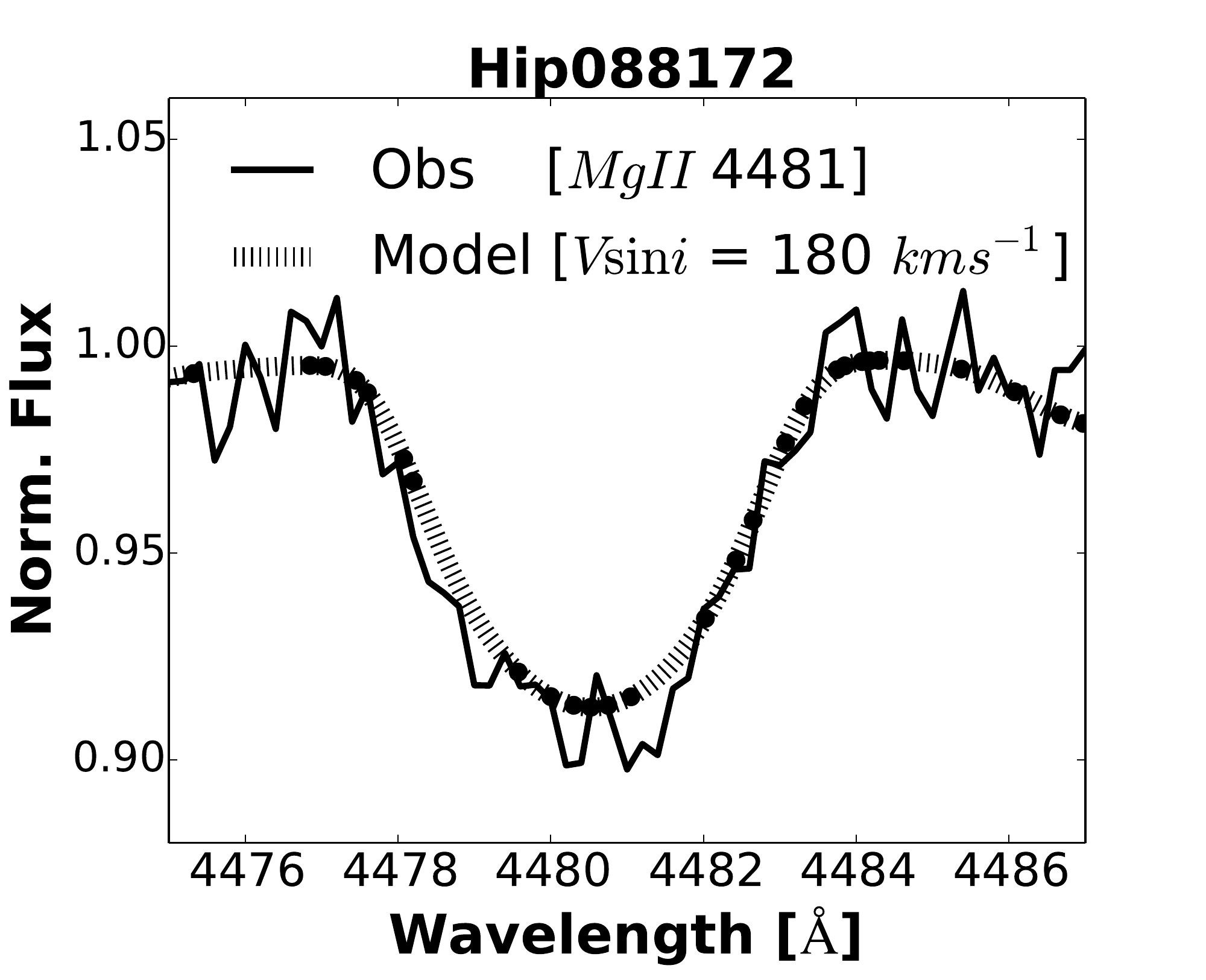} 
 \includegraphics[width=4cm,height=4cm] {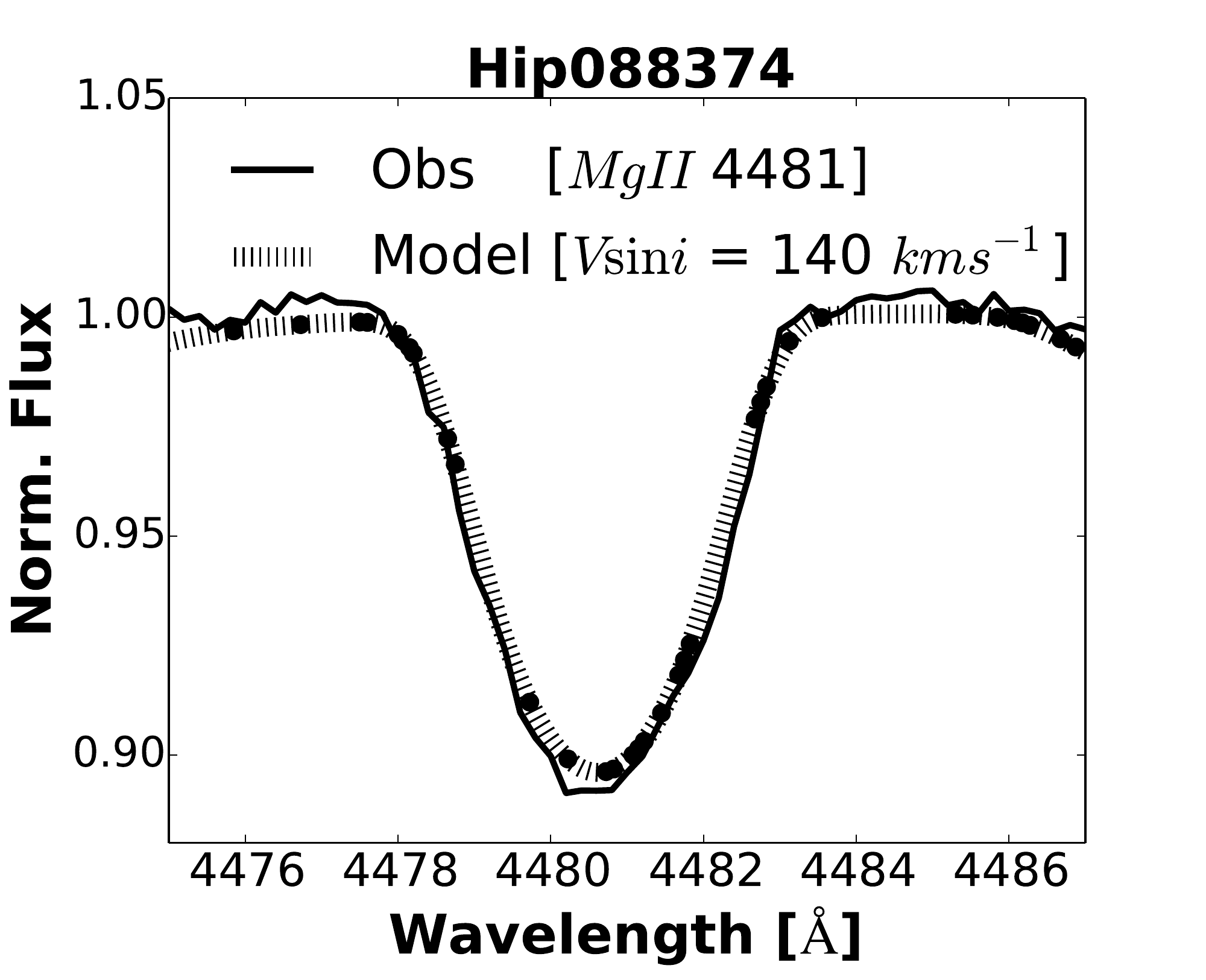} 
   \includegraphics[width=4cm,height=4cm] {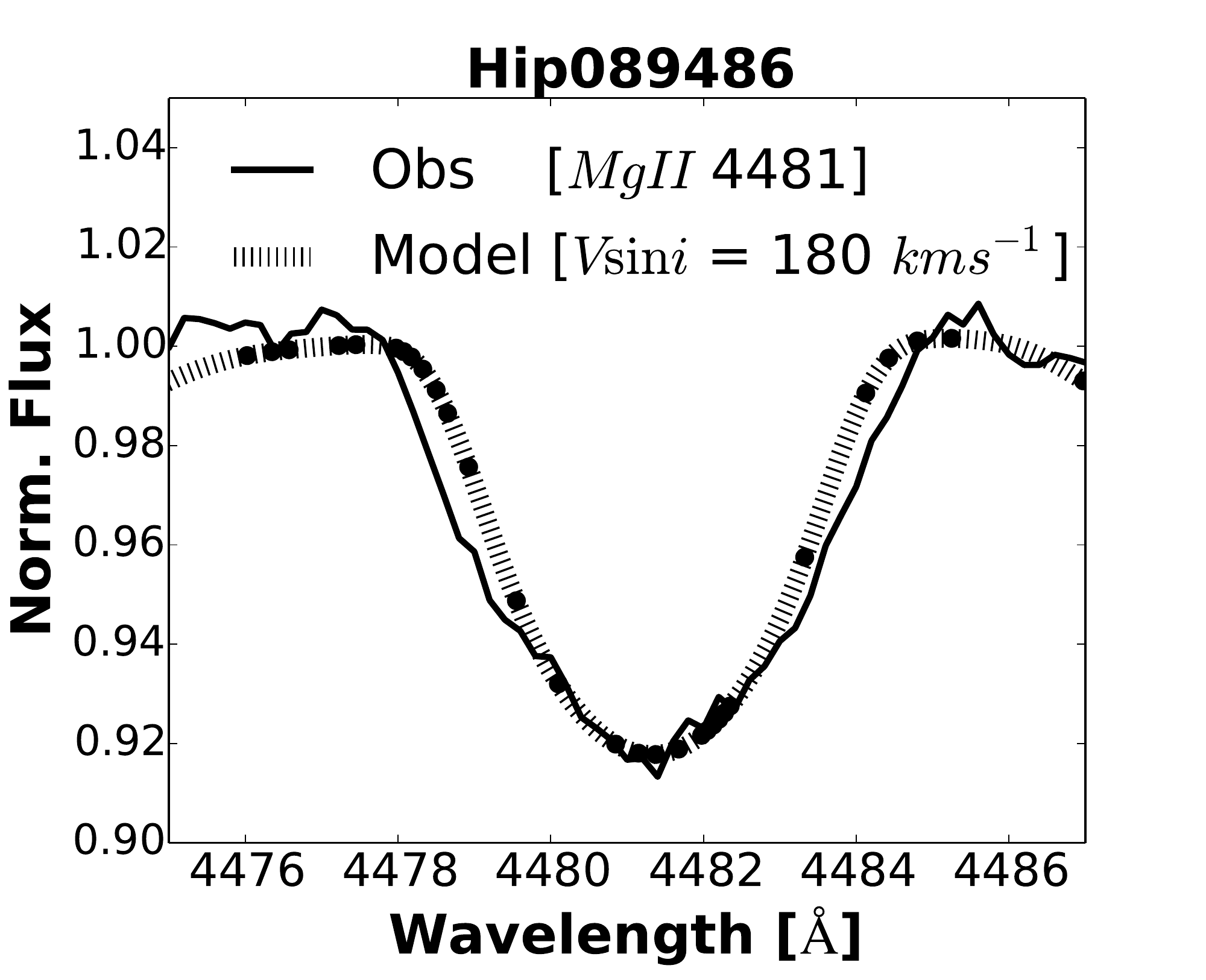} 

}

\subfloat{
 \includegraphics[width=4cm,height=4cm] {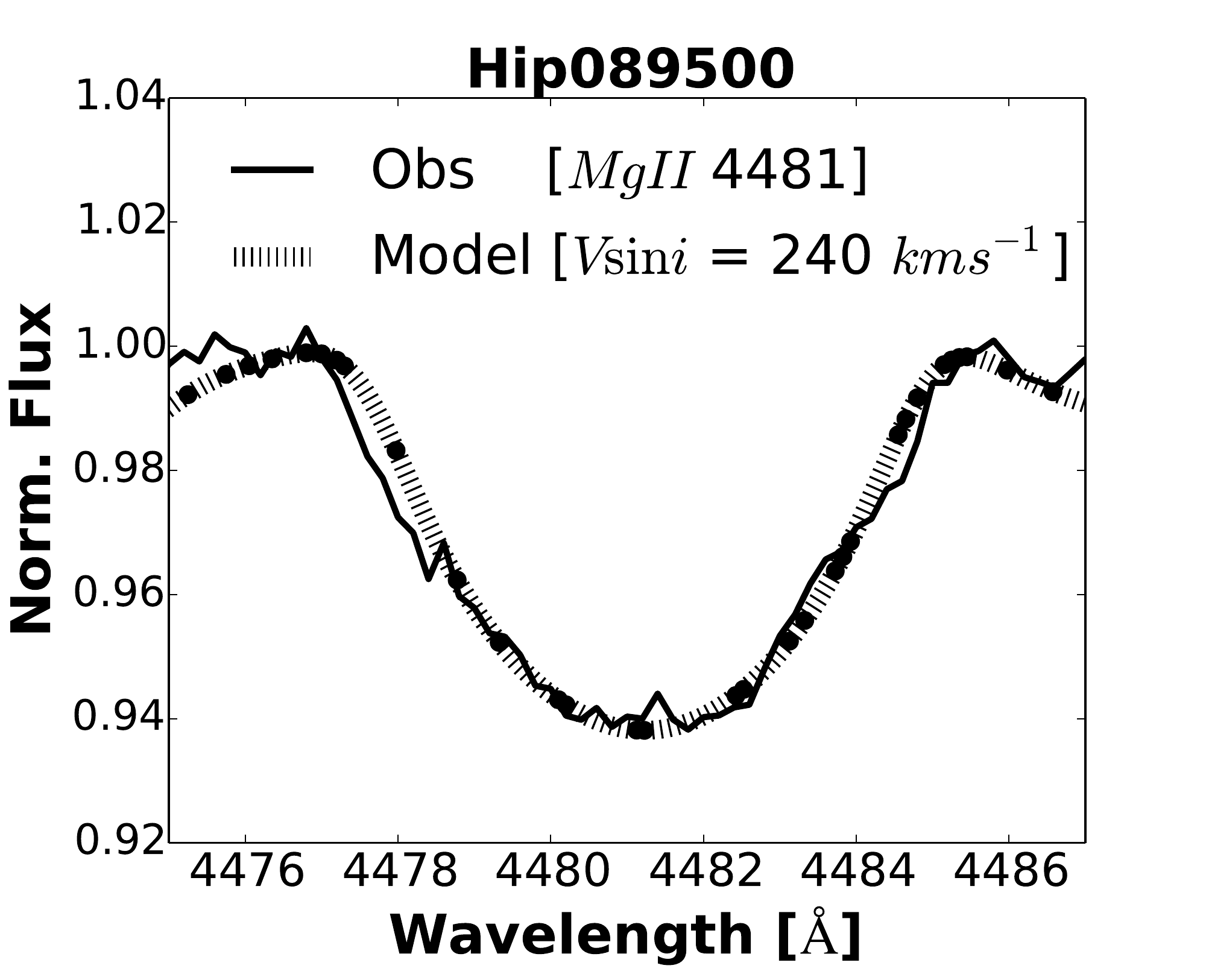} 
  \includegraphics[width=4cm,height=4cm] {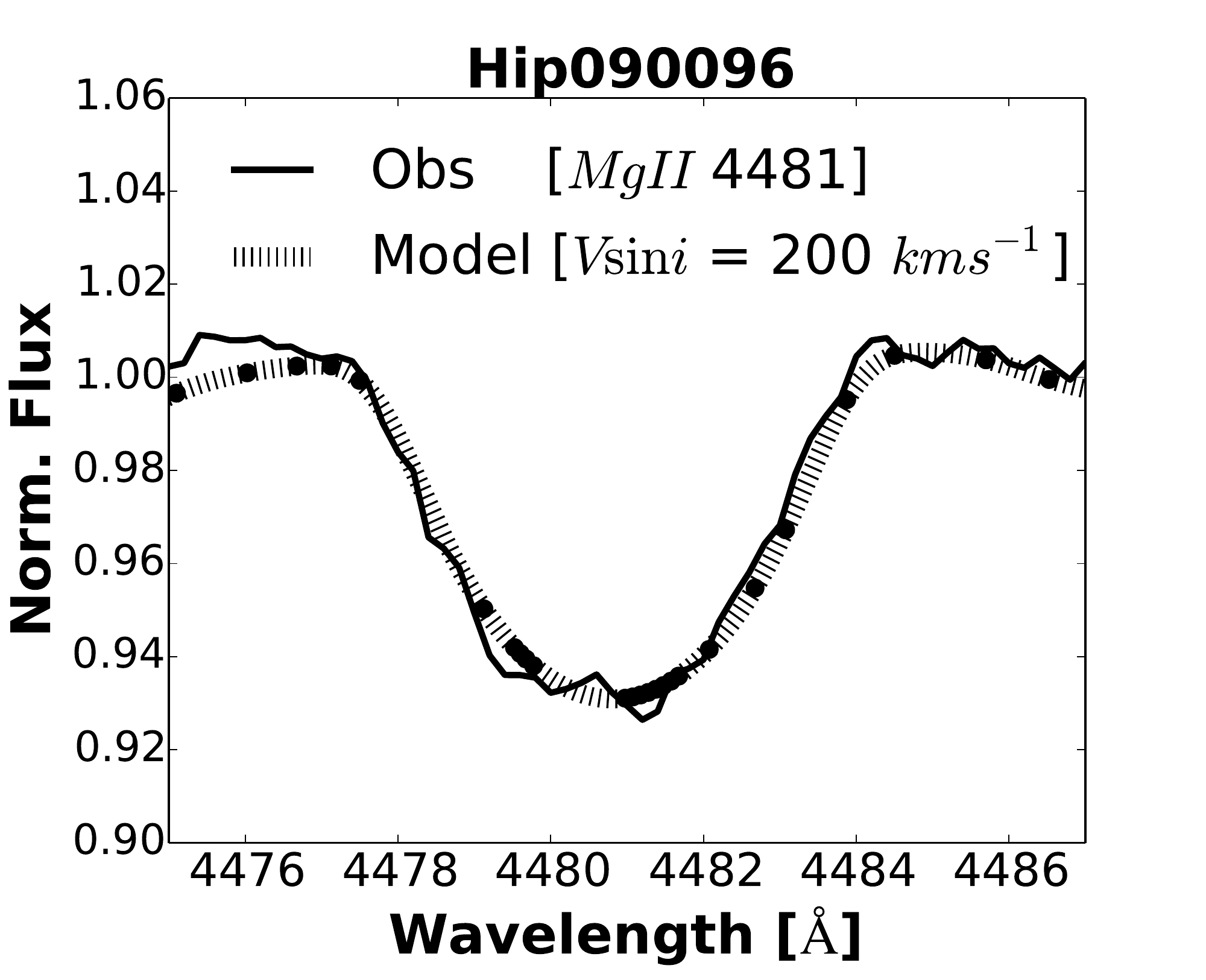} 
     \includegraphics[width=4cm,height=4cm] {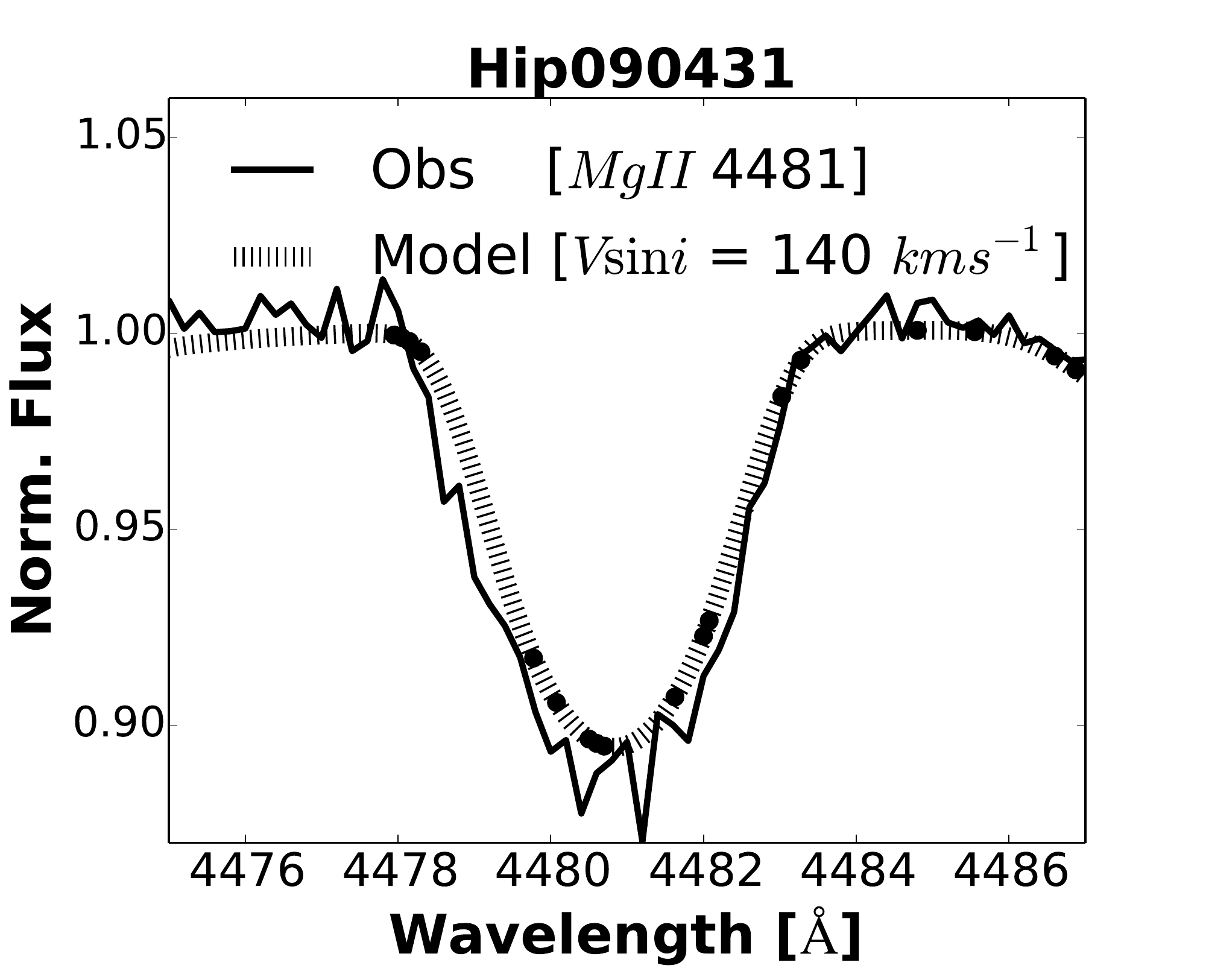}         
  \includegraphics[width=4cm,height=4cm] {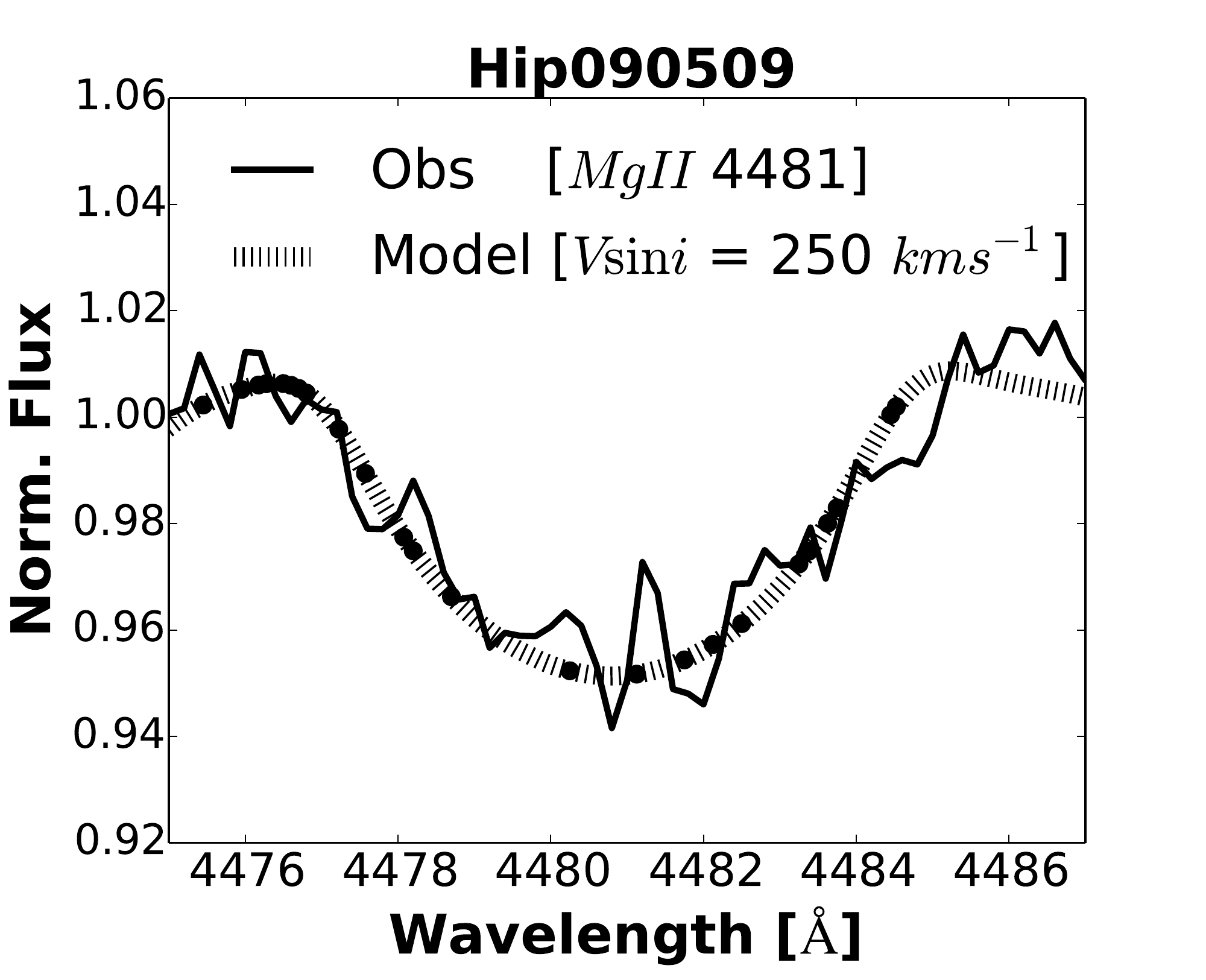}  
}

\caption{Continued}
\end{figure*}

\begin{figure*}
\ContinuedFloat
\centering

\subfloat{
    \includegraphics[width=4cm,height=4cm] {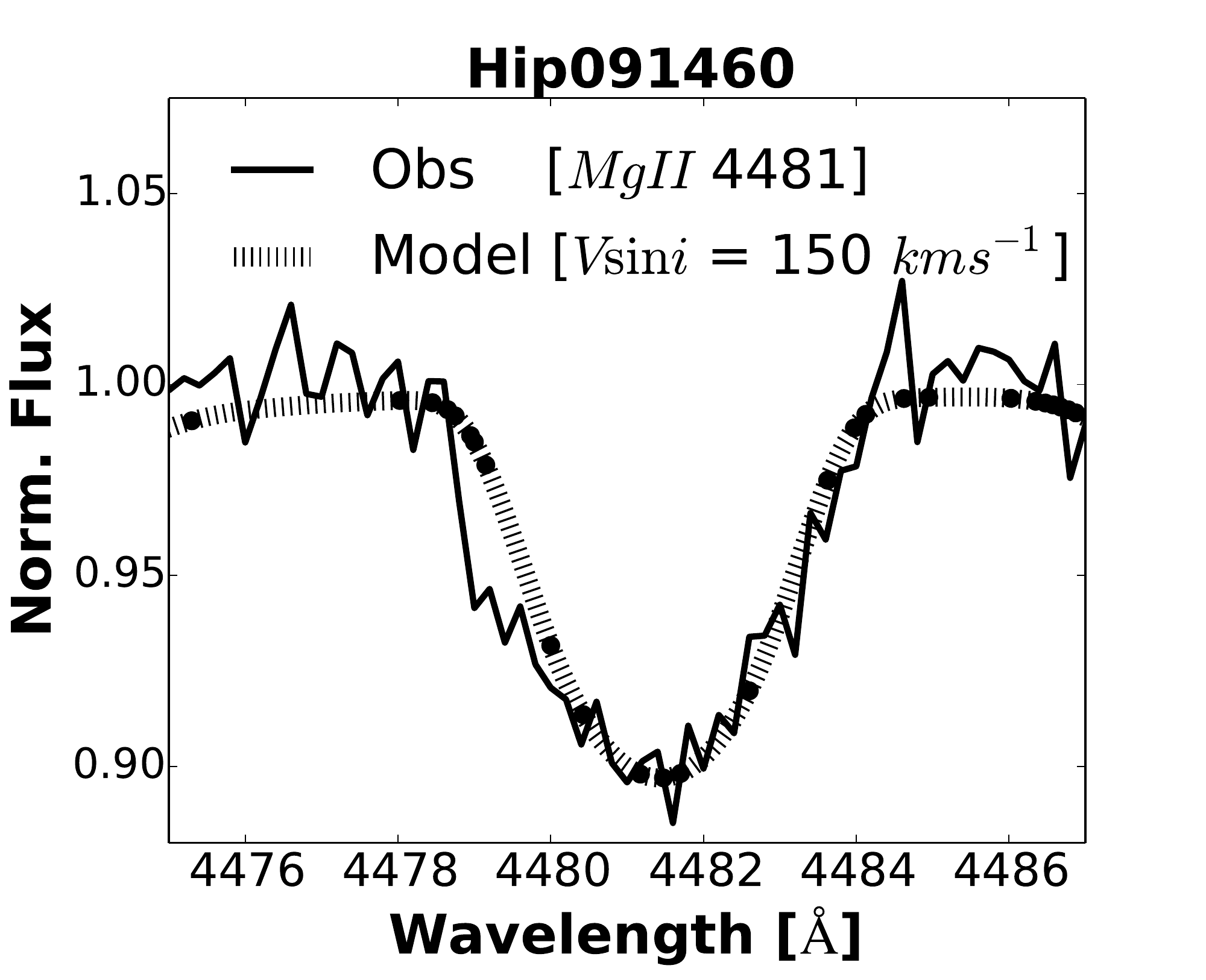} 
   \includegraphics[width=4cm,height=4cm] {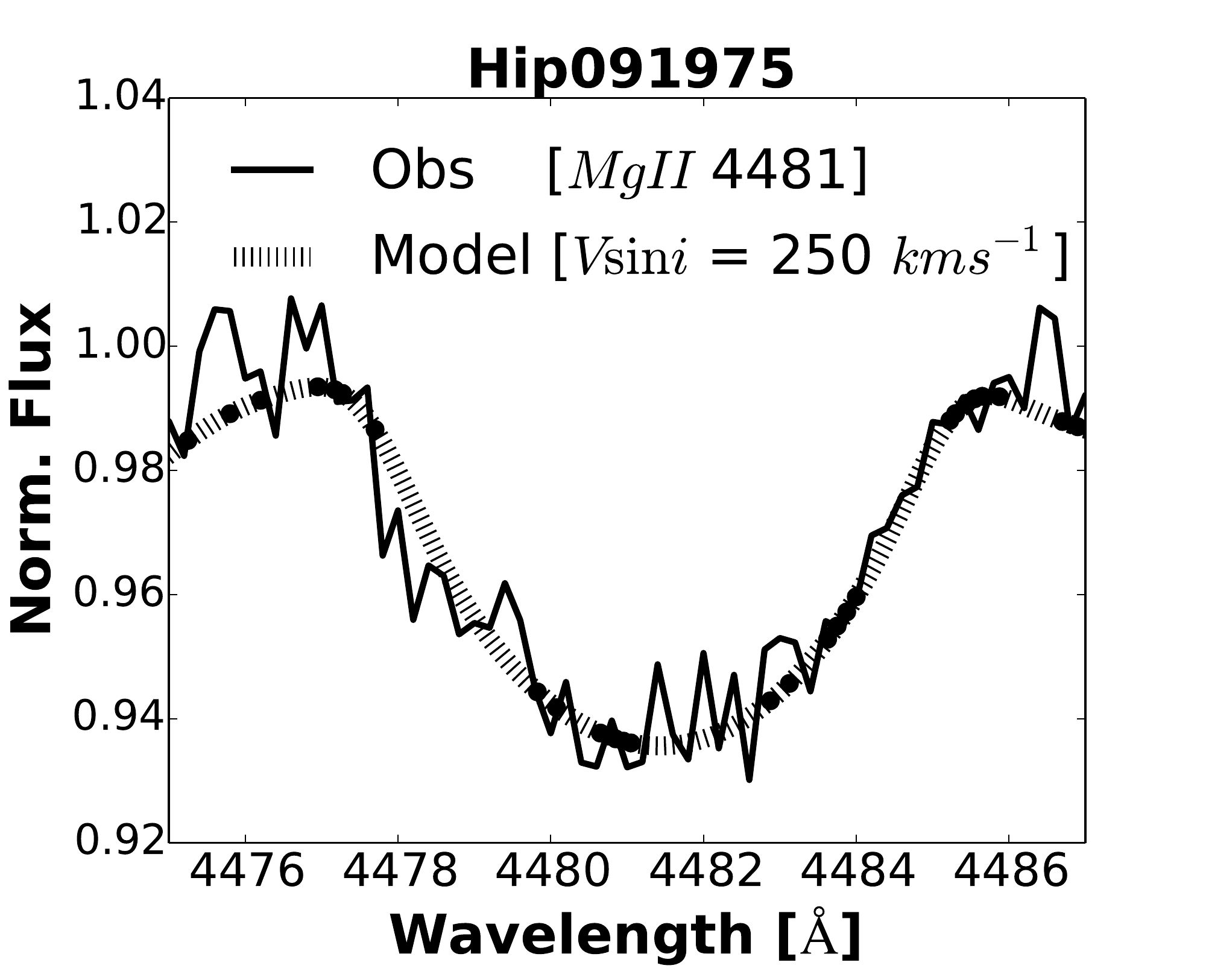} 
    \includegraphics[width=4cm,height=4cm] {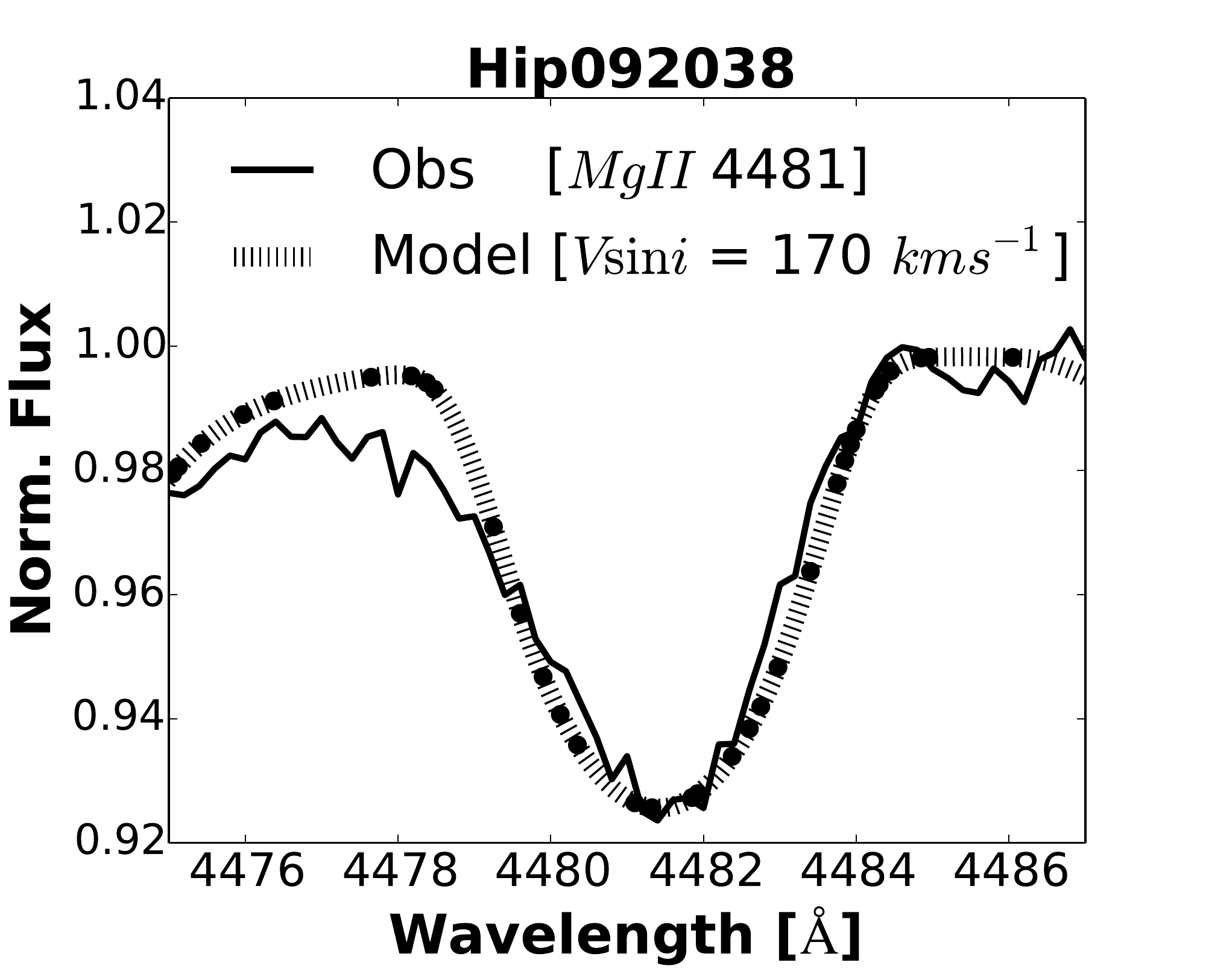} 
    \includegraphics[width=4cm,height=4cm] {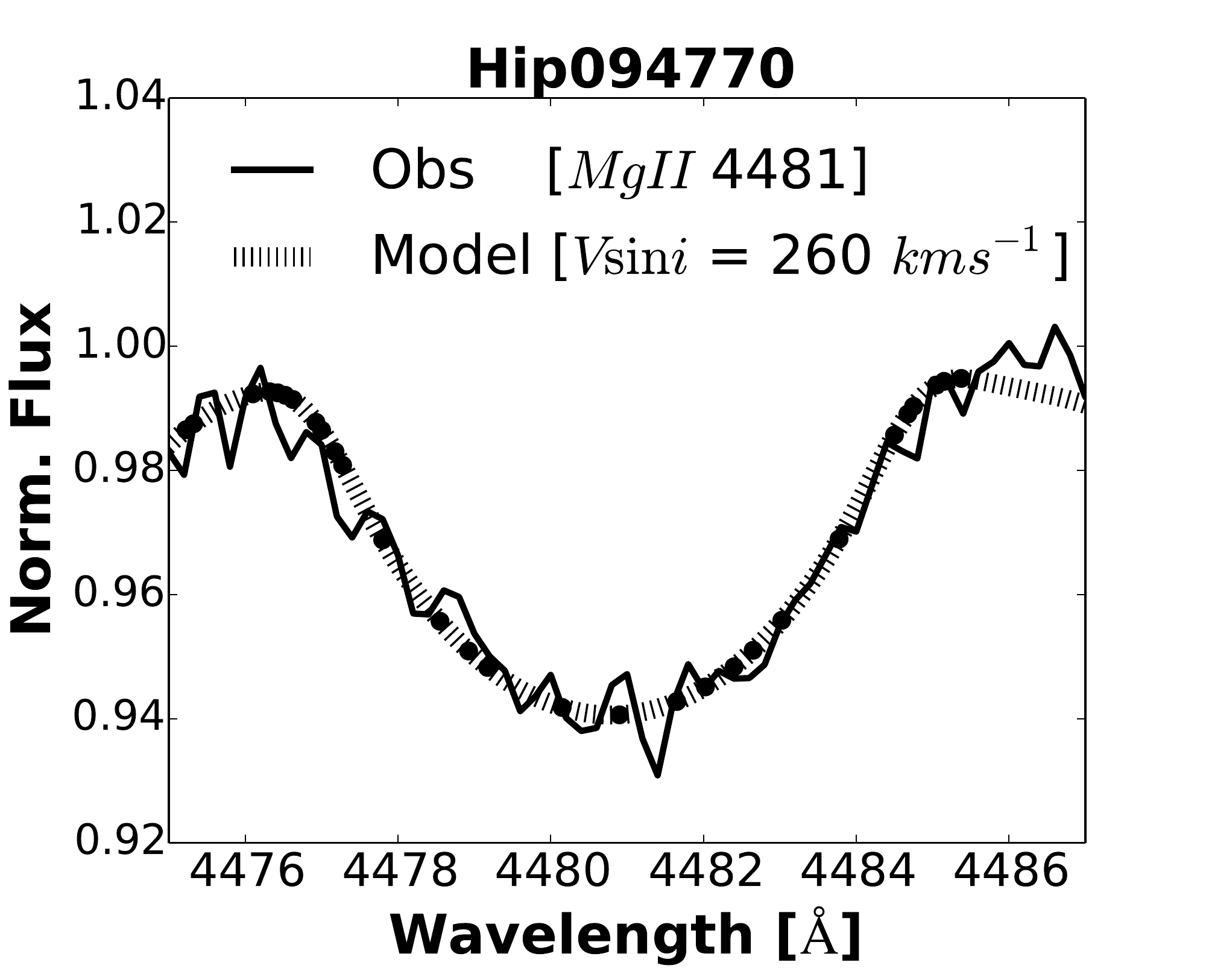}        

}

\subfloat{
  \includegraphics[width=4cm,height=4cm] {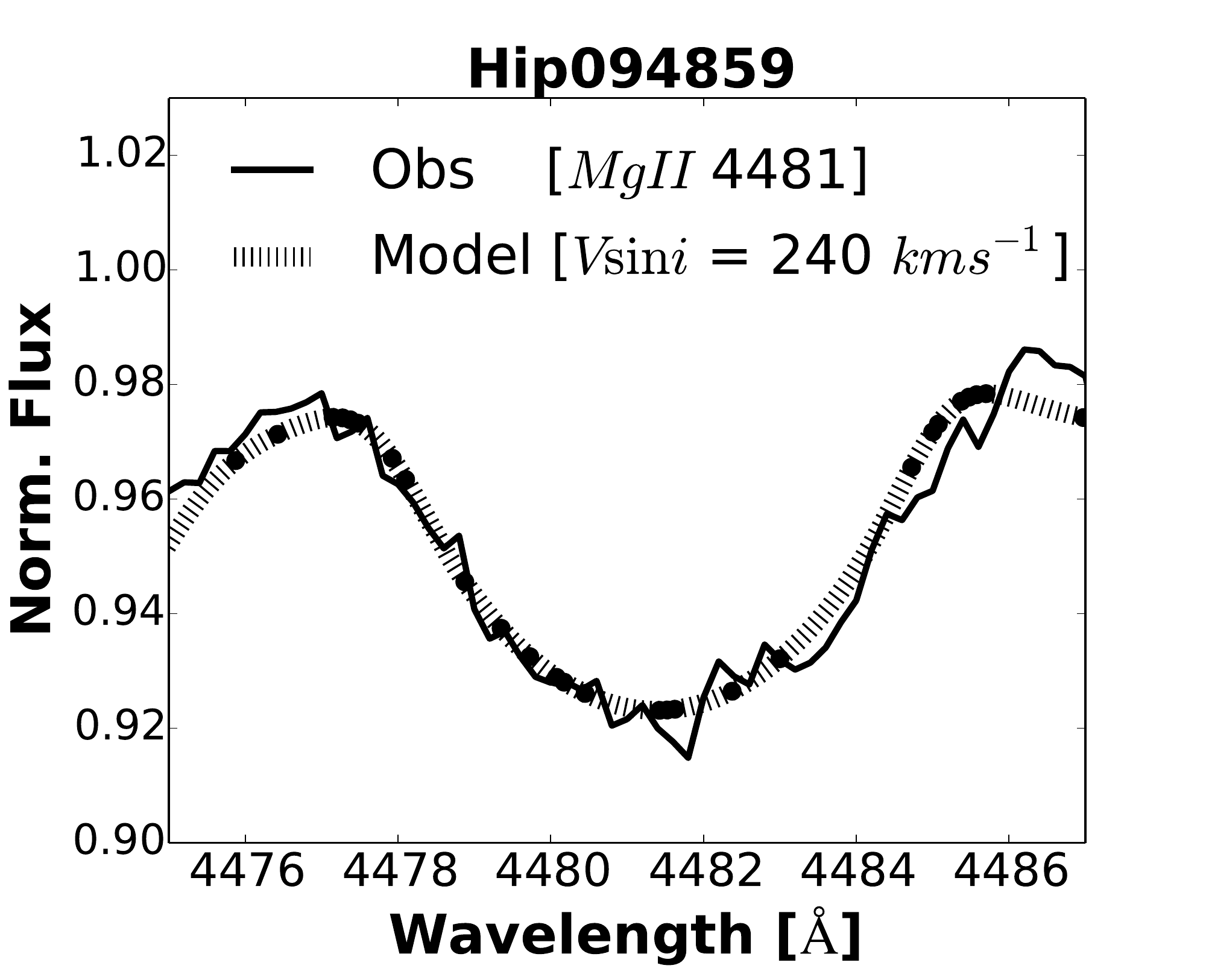} 
   \includegraphics[width=4cm,height=4cm] {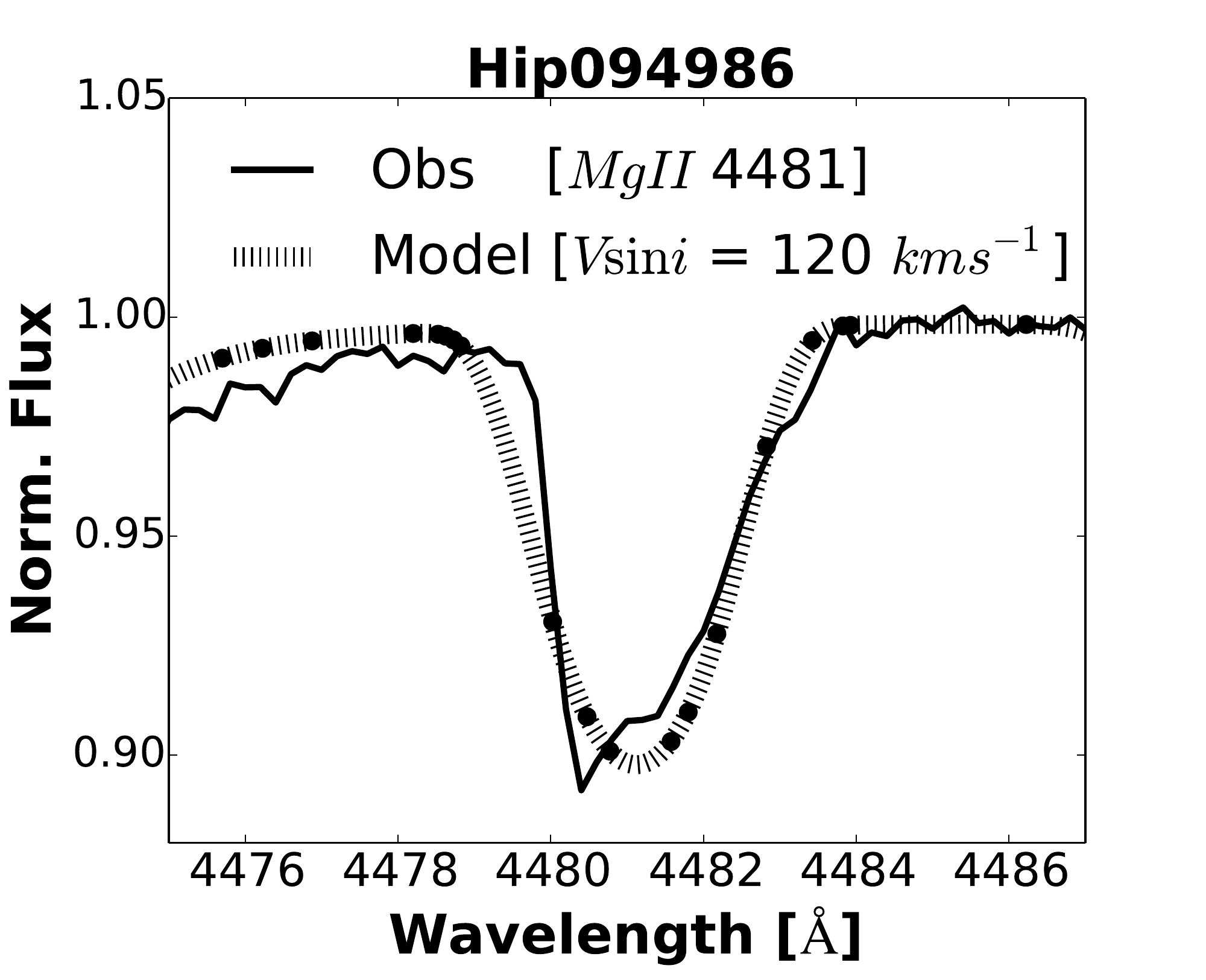} 
    \includegraphics[width=4cm,height=4cm] {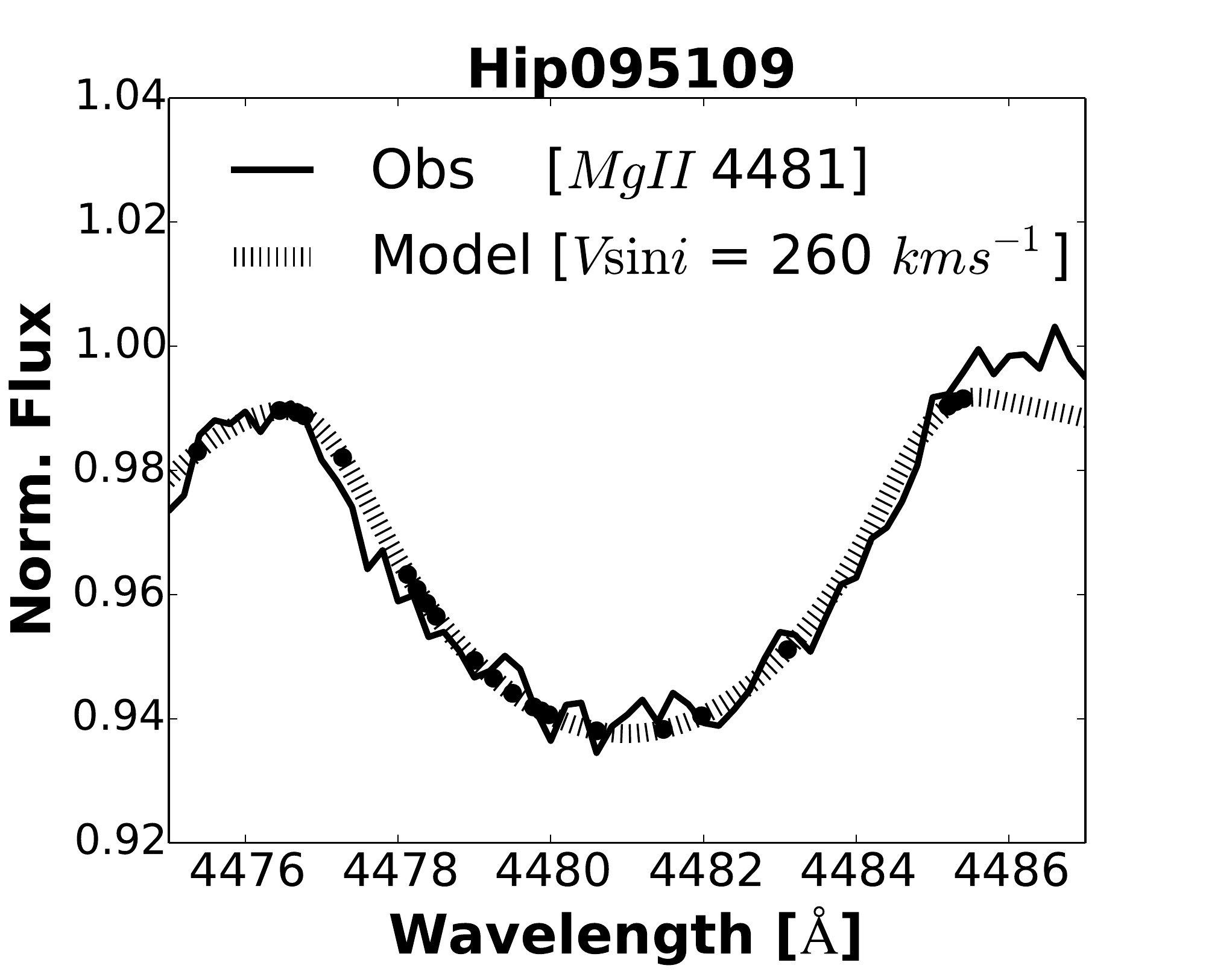} 
     \includegraphics[width=4cm,height=4cm] {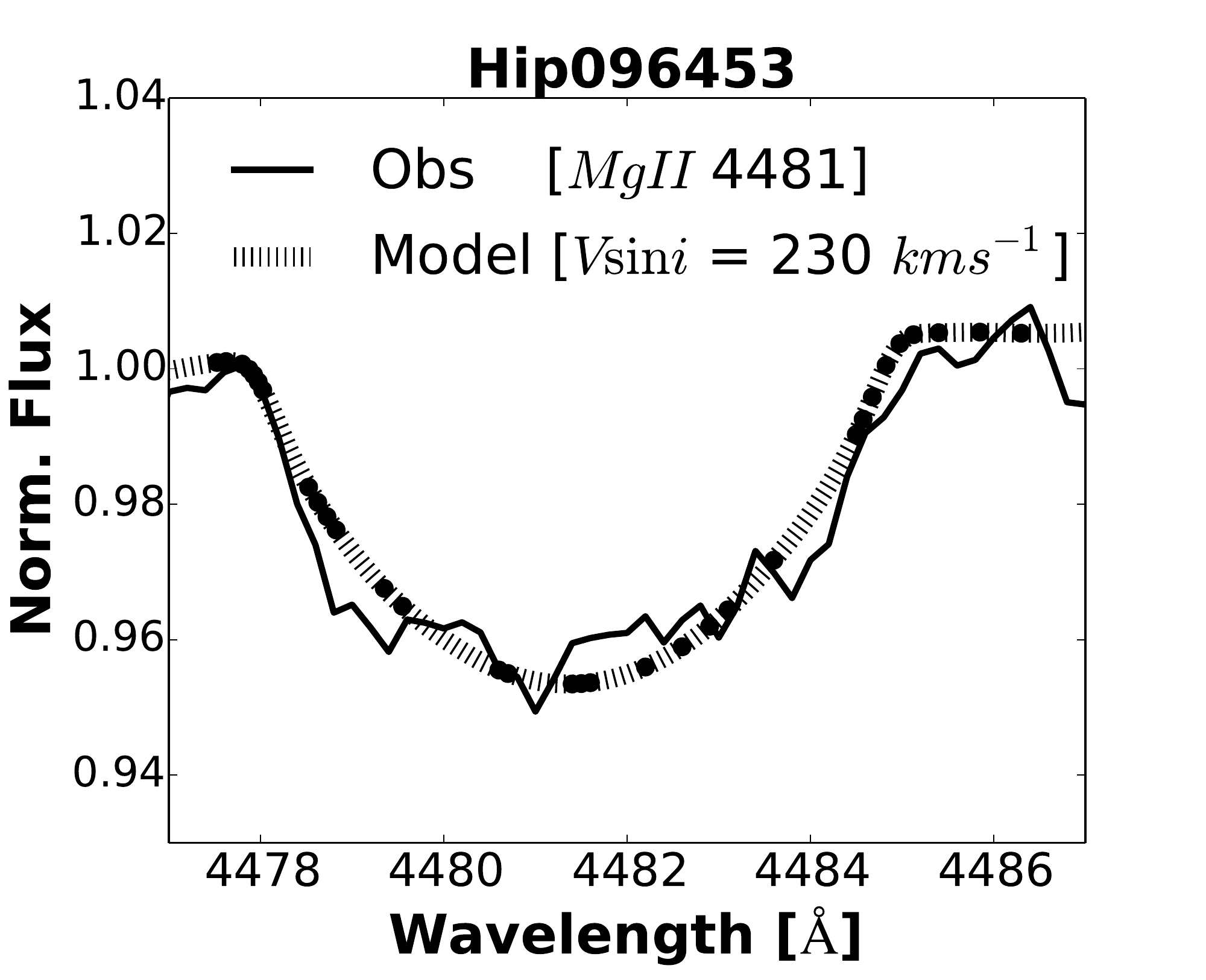} 

}

\subfloat{
 \includegraphics[width=4cm,height=4cm] {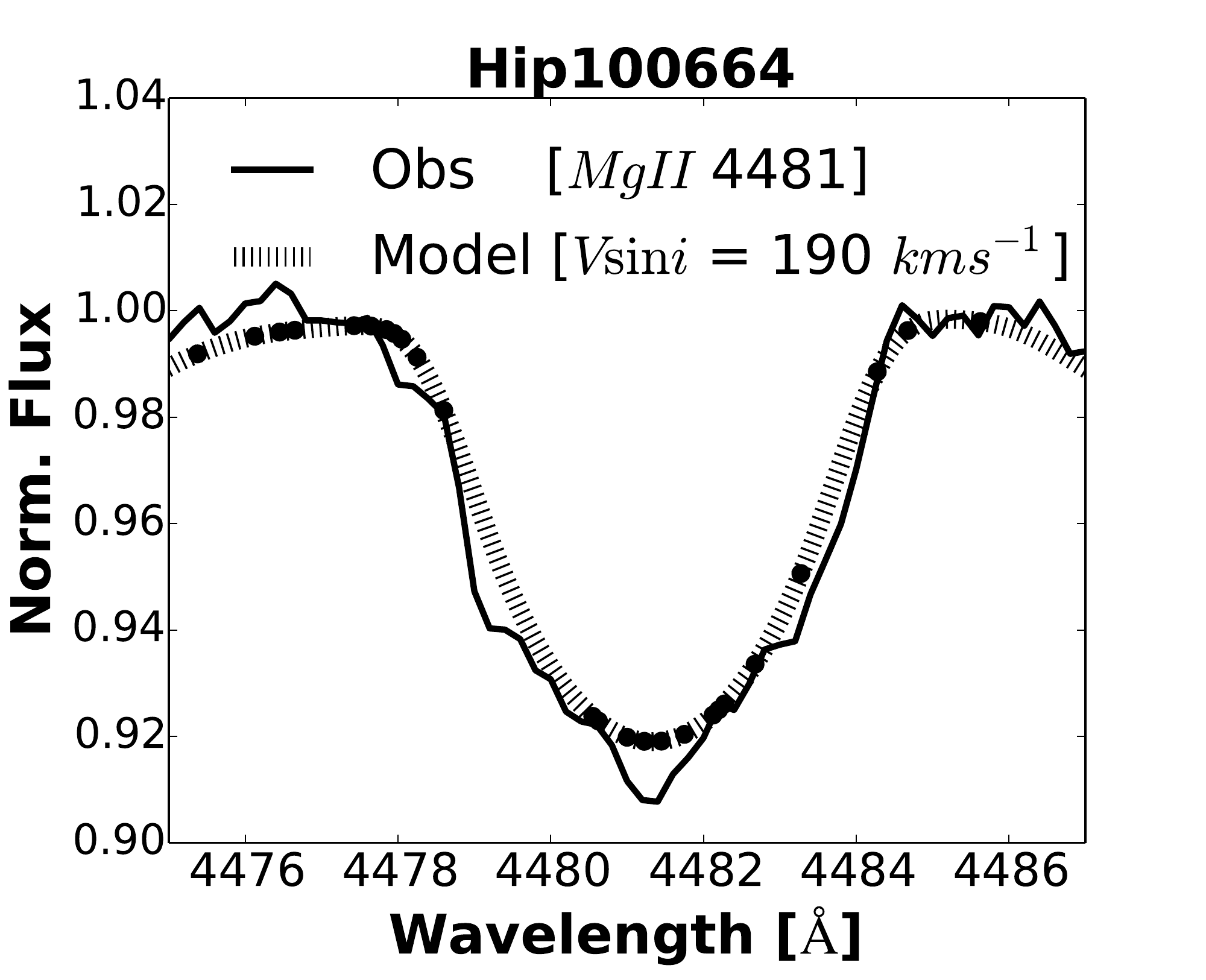} 
    \includegraphics[width=4cm,height=4cm] {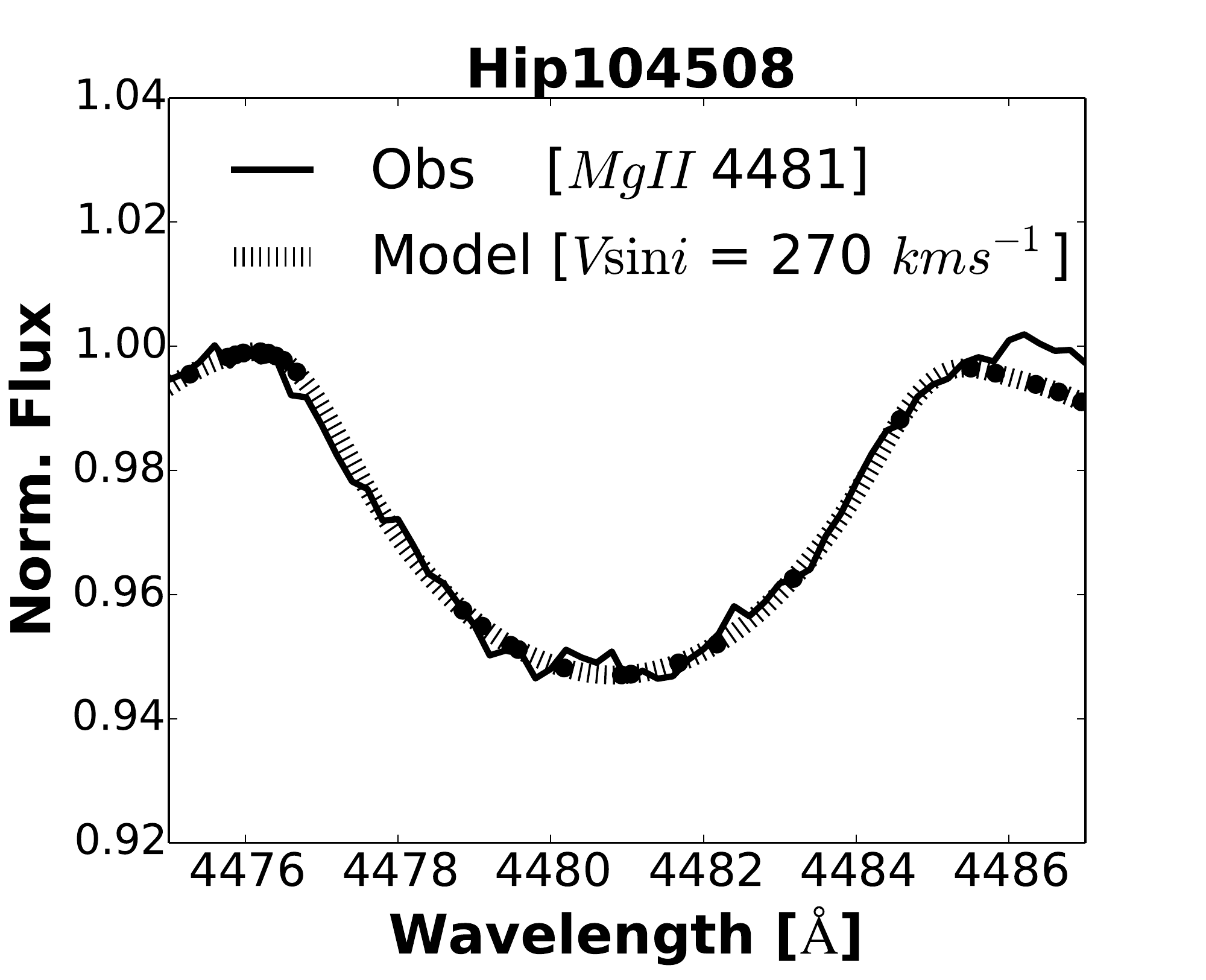} 
  \includegraphics[width=4cm,height=4cm] {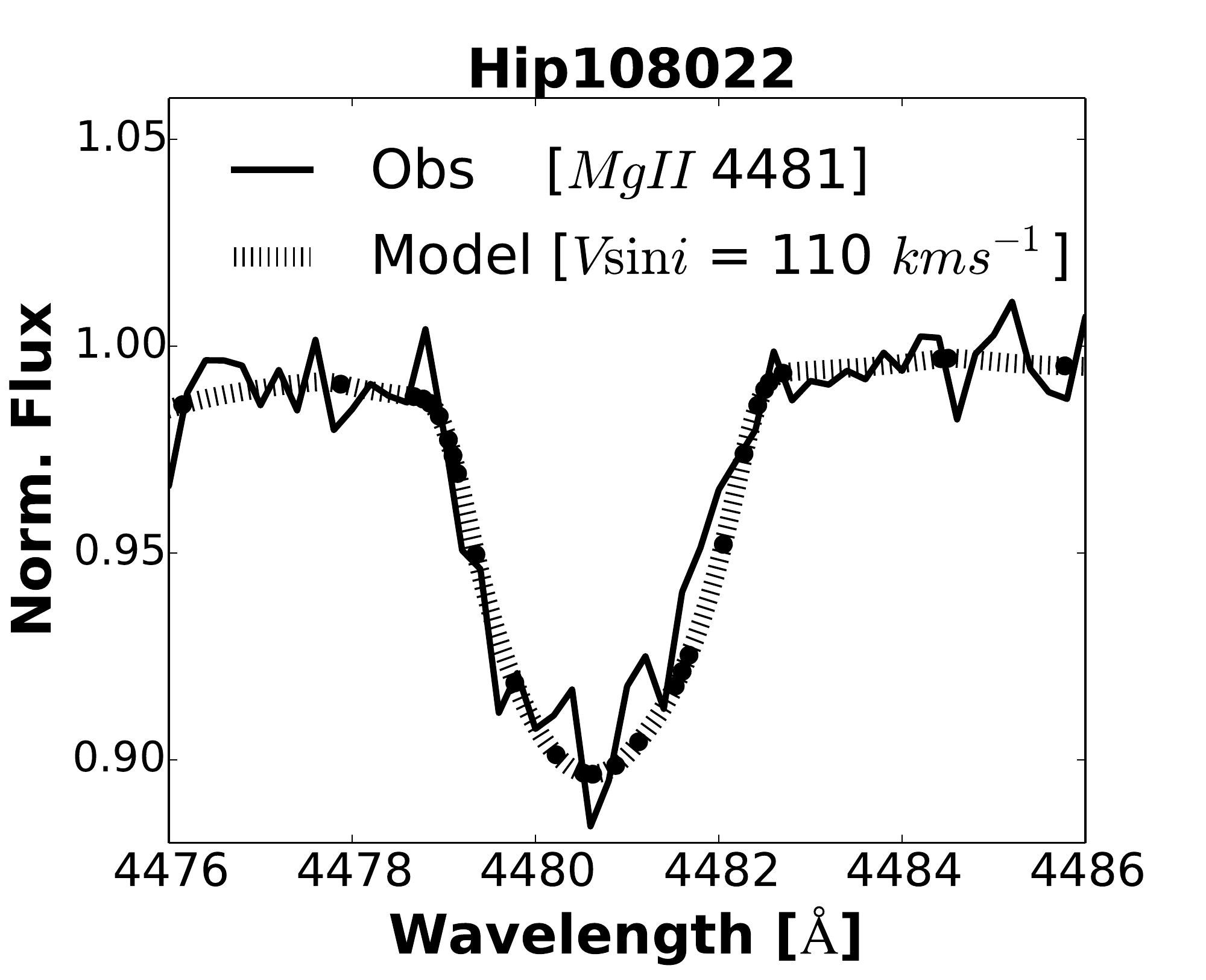} 
  \includegraphics[width=4cm,height=4cm] {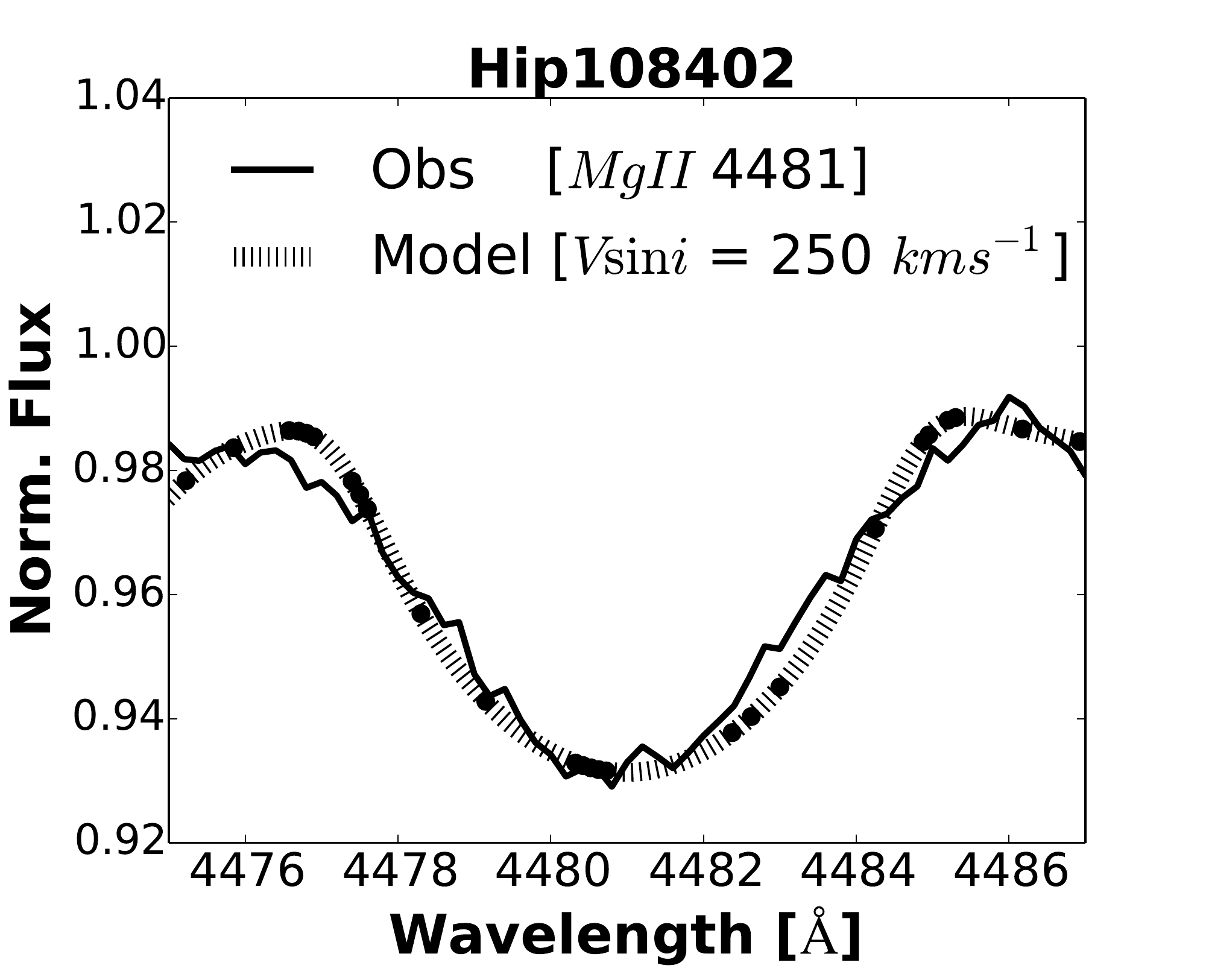}  
   
}

\caption{Continued}
\end{figure*}

\clearpage
\onecolumn

\begin{minipage}[t][7cm][t]{\textwidth}
\twocolumn
\section{Spectral appearance of the identified Be stars}
\label{app:spec}
In this Appendix excerpts of the spectra of the identified Be stars are shown,
except for Hip\,11116, which is shown in Fig.~\ref{fig:spec}. In some cases
noisy or otherwise unsuitable spectra were excluded from the plots.

In the upper frow, from left to right, profiles of H$\beta$, H$\alpha$, the
\ion{Fe}{ii}\,5169, and the \ion{He}{i}\,4471 and \ion{Mg}{ii}\,4481 lines are
shown. These illustrated the presence and variablity of Balmer emission as
well as the Balmer decrement, the presence of circumstellar emission or shell
absorption in \ion{Fe}{ii}, and the balance of He vs.\ Mg may serve as a
sanity check on the obtained effective temperatures and spectral types.

In the lower row, the Balmer discontinuity is shown and the higher lines of the
Paschen series, that include the \ion{O}{i}\,8446 and \ion{Ca}{ii} triplet
lines.  

\onecolumn

\end{minipage}

\begin{figure*}[h]
\begin{center}
\includegraphics[angle=0,width=14cm,clip]{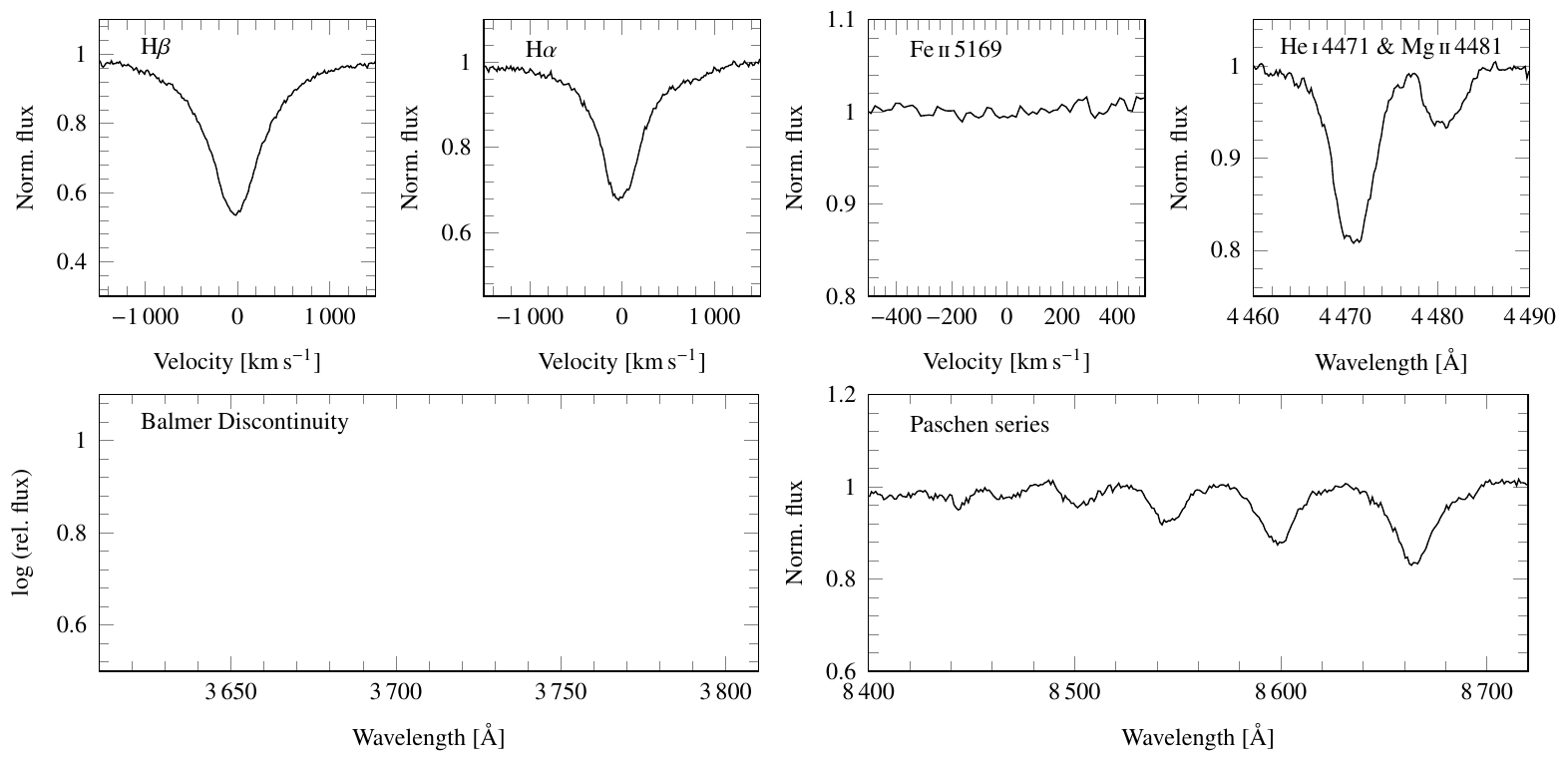}
\end{center}
\caption{Spectrum overview plot for Hip\,15188. No usable UVB spectrum is
available for this star.}
\end{figure*}

\begin{figure*}[h]
\begin{center}
\includegraphics[angle=0,width=14cm,clip]{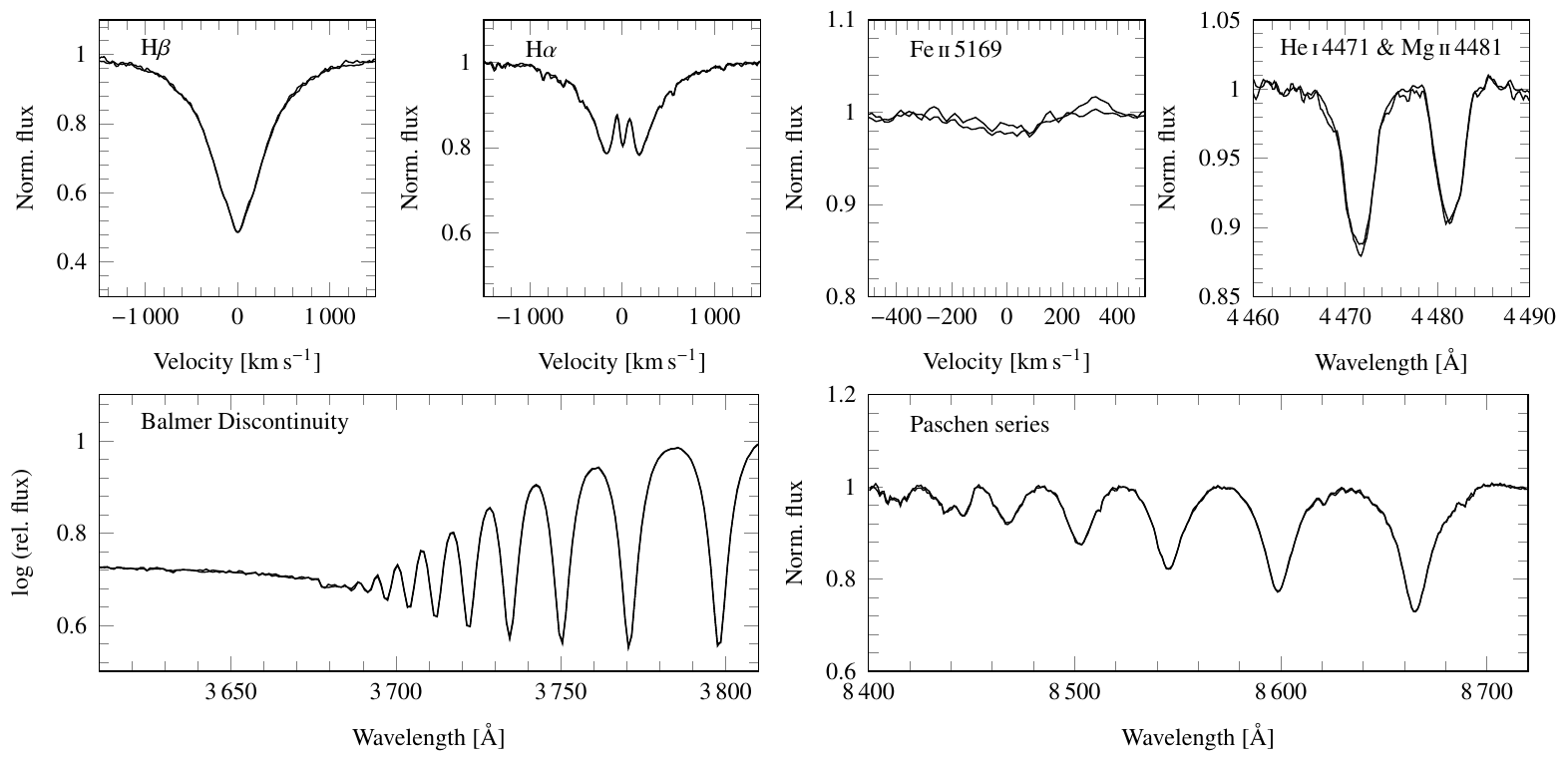}
\end{center}
\caption{Spectrum overview plot for Hip\,23161}
\end{figure*}
\clearpage
\begin{figure*}
\begin{center}
\includegraphics[angle=0,width=14cm,clip]{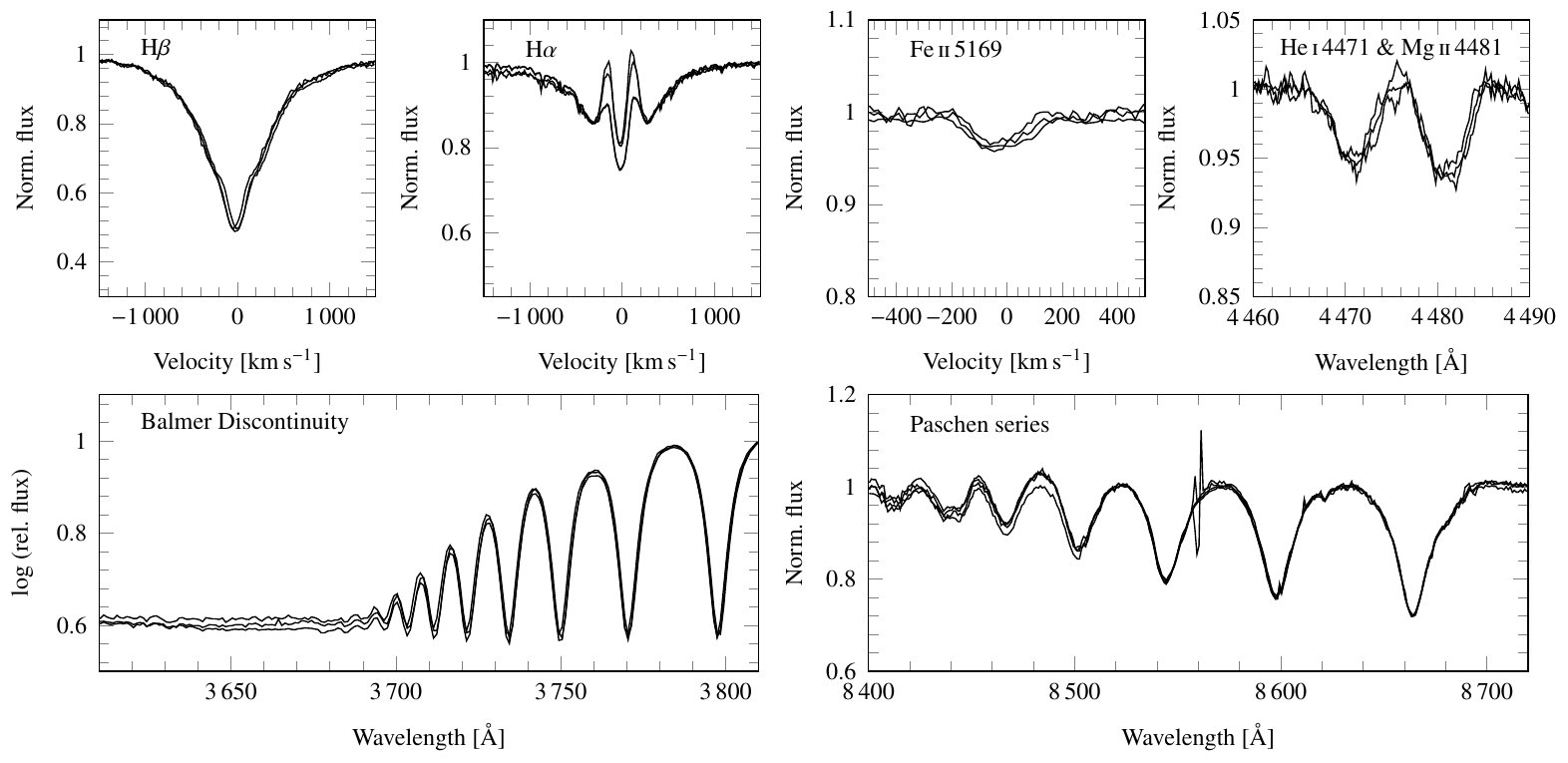}
\end{center}
\caption{Spectrum overview plot for Hip\,24475}
\end{figure*}

\begin{figure*}
\begin{center}
\includegraphics[angle=0,width=14cm,clip]{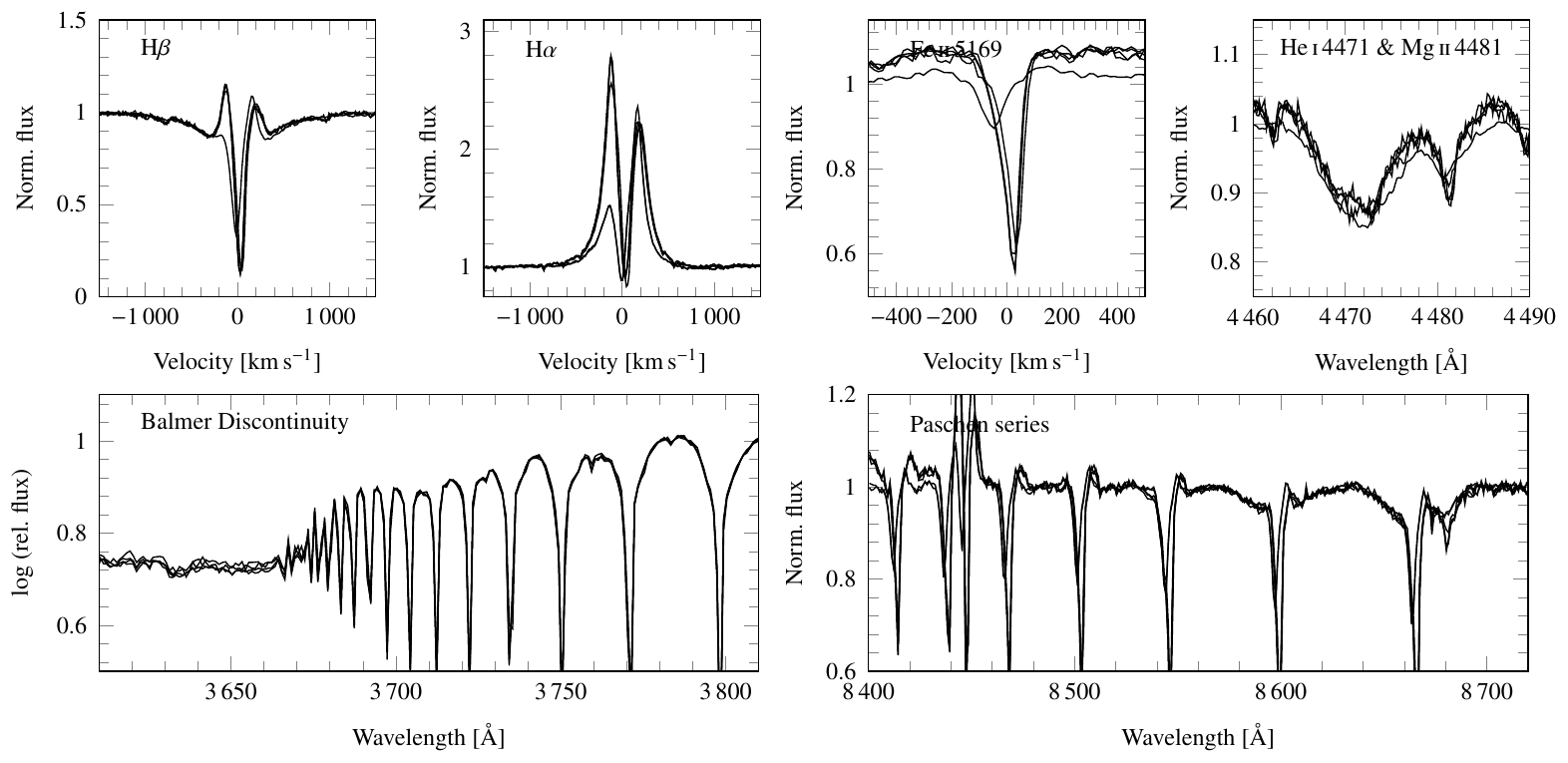}
\end{center}
\caption{Spectrum overview plot for Hip\,25007}
\end{figure*}

\begin{figure*}
\begin{center}
\includegraphics[angle=0,width=14cm,clip]{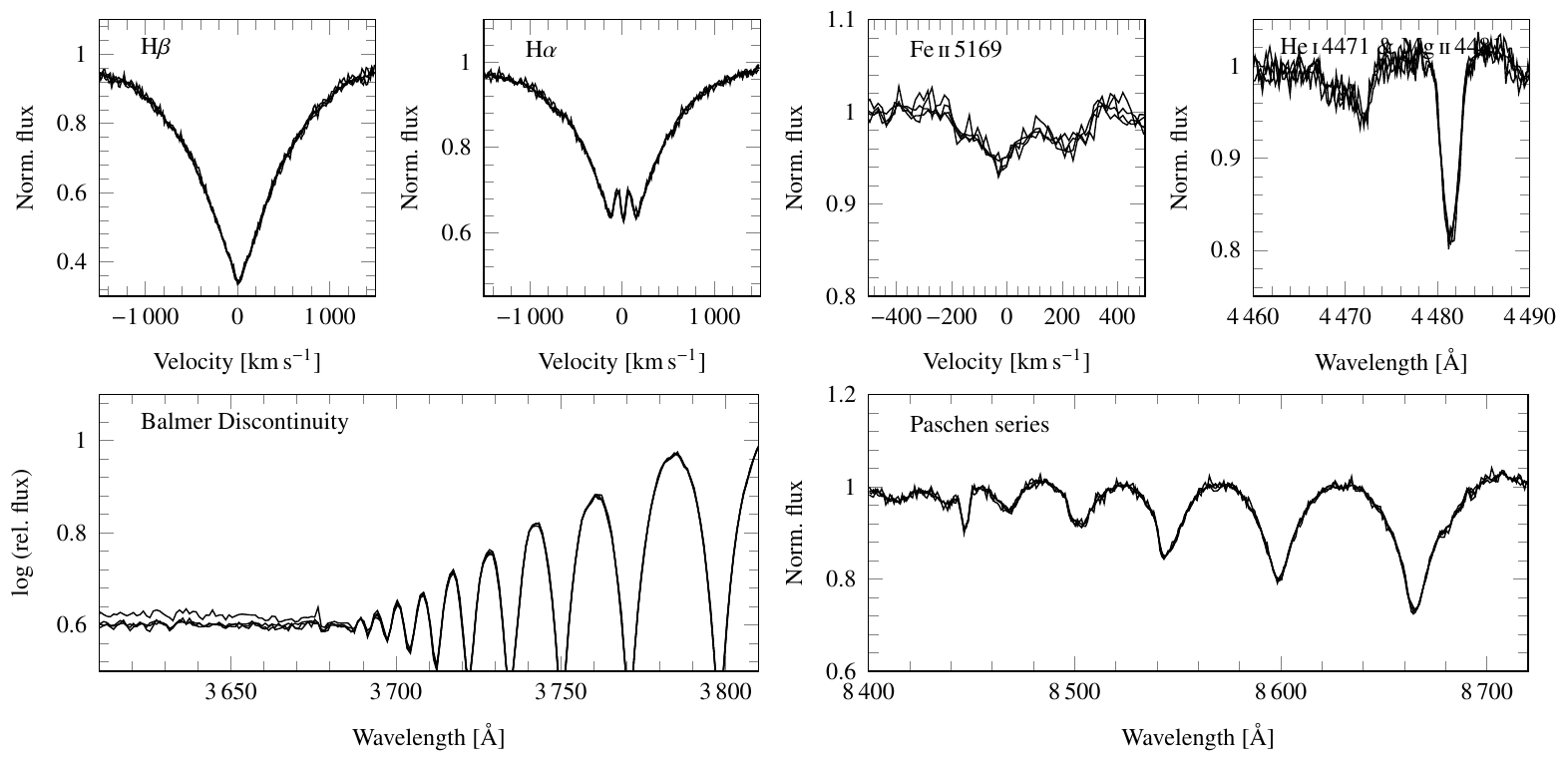}
\end{center}
\caption{Spectrum overview plot for Hip\,25690}
\end{figure*}
\clearpage
\begin{figure*}
\begin{center}
\includegraphics[angle=0,width=14cm,clip]{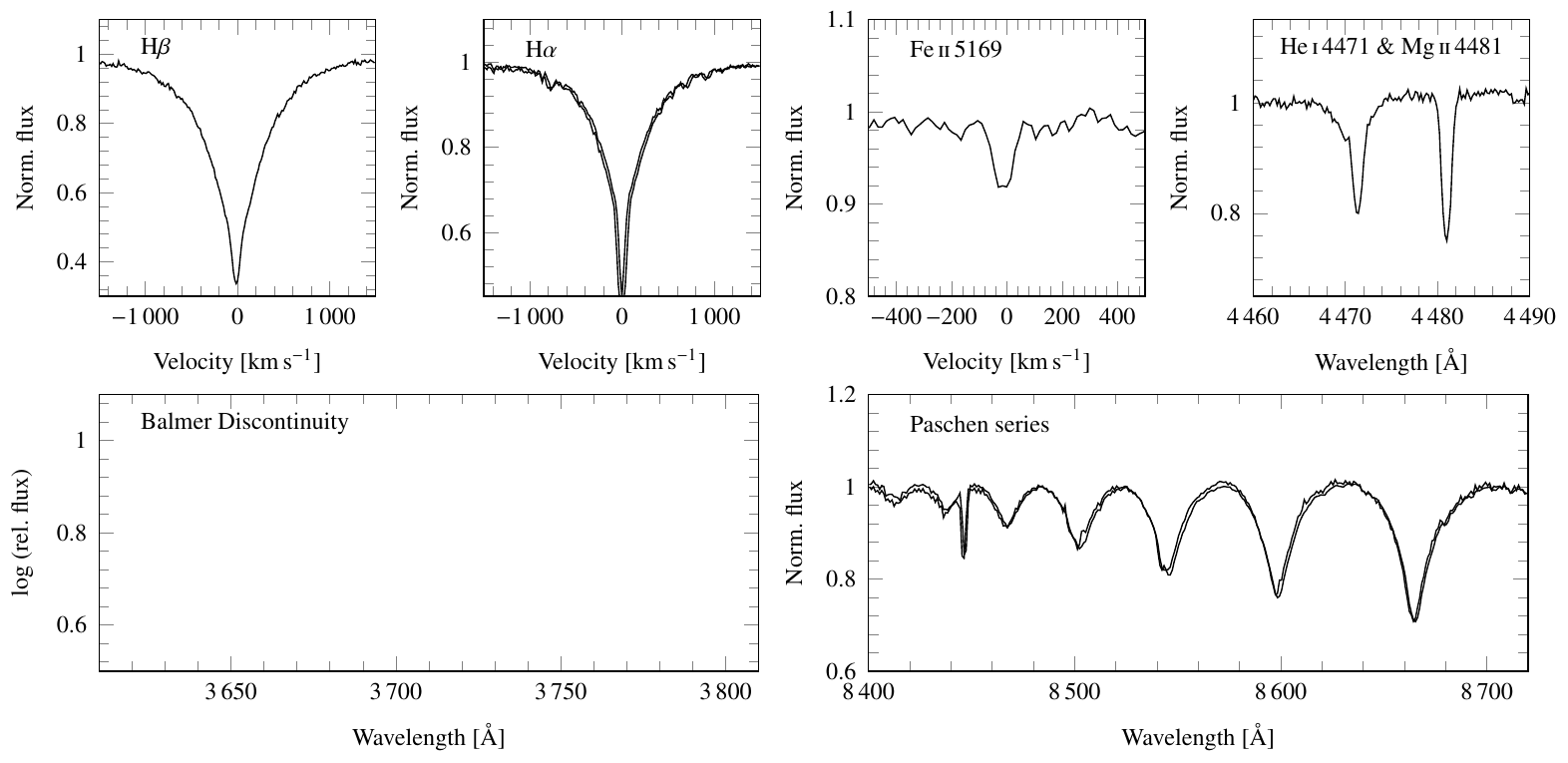}
\end{center}
\caption{Spectrum overview plot for Hip\,25950}
\end{figure*}

\begin{figure*}
\begin{center}
\includegraphics[angle=0,width=14cm,clip]{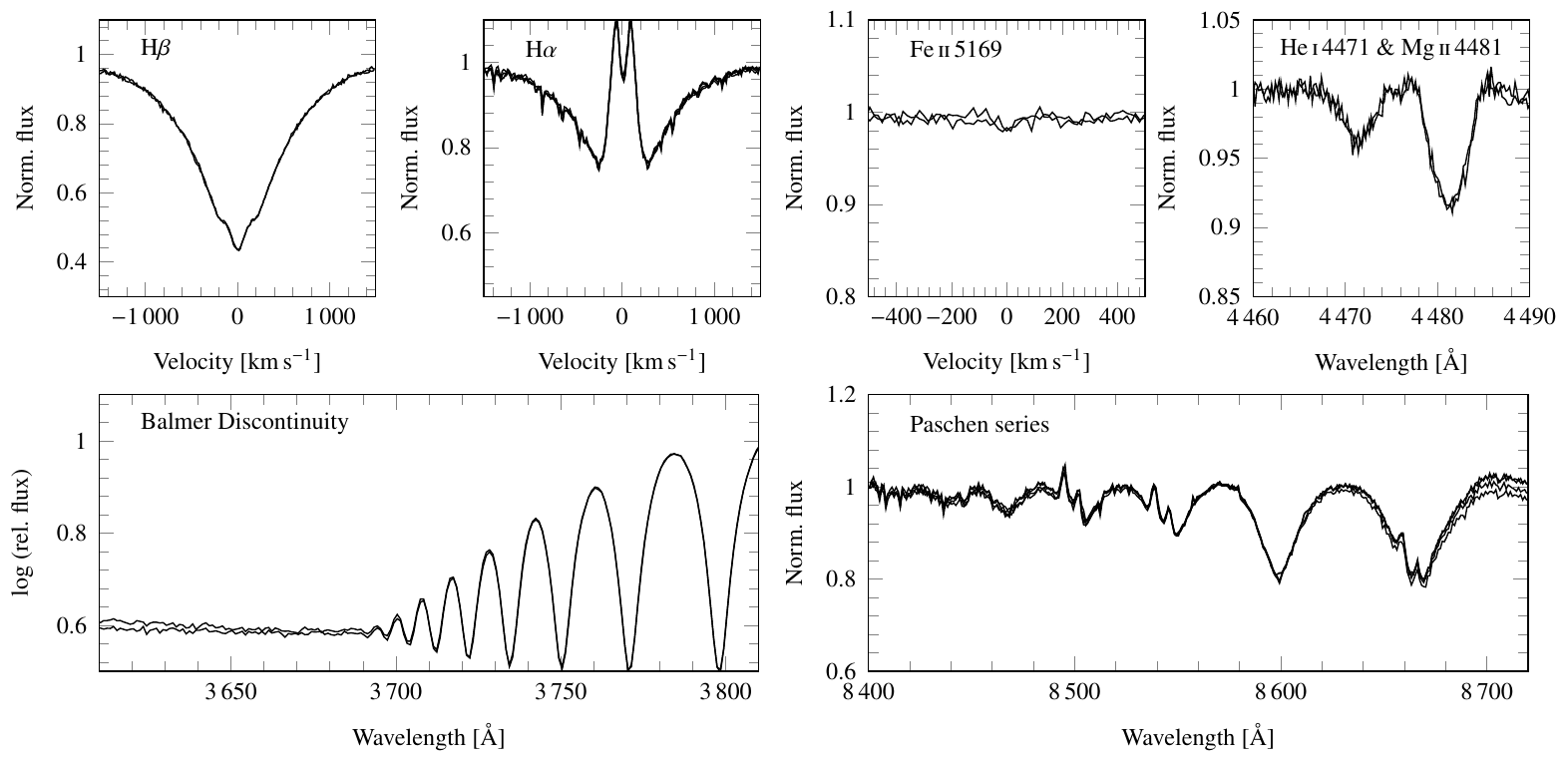}
\end{center}
\caption{Spectrum overview plot for Hip\,26368}
\end{figure*}

\begin{figure*}
\begin{center}
\includegraphics[angle=0,width=14cm,clip]{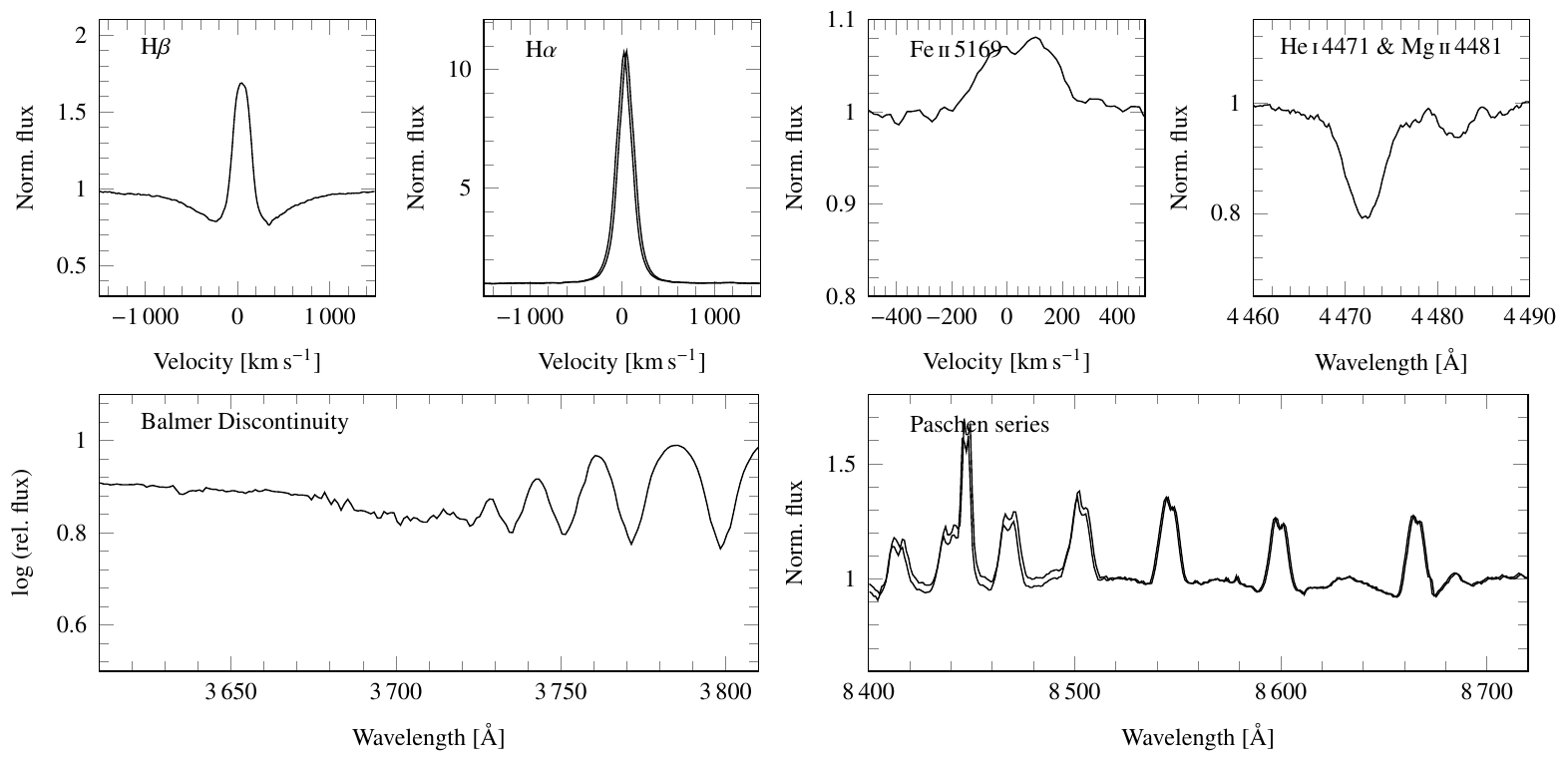}
\end{center}
\caption{Spectrum overview plot for Hip\,26964}
\end{figure*}
\clearpage

\begin{figure*}
\begin{center}
\includegraphics[angle=0,width=14cm,clip]{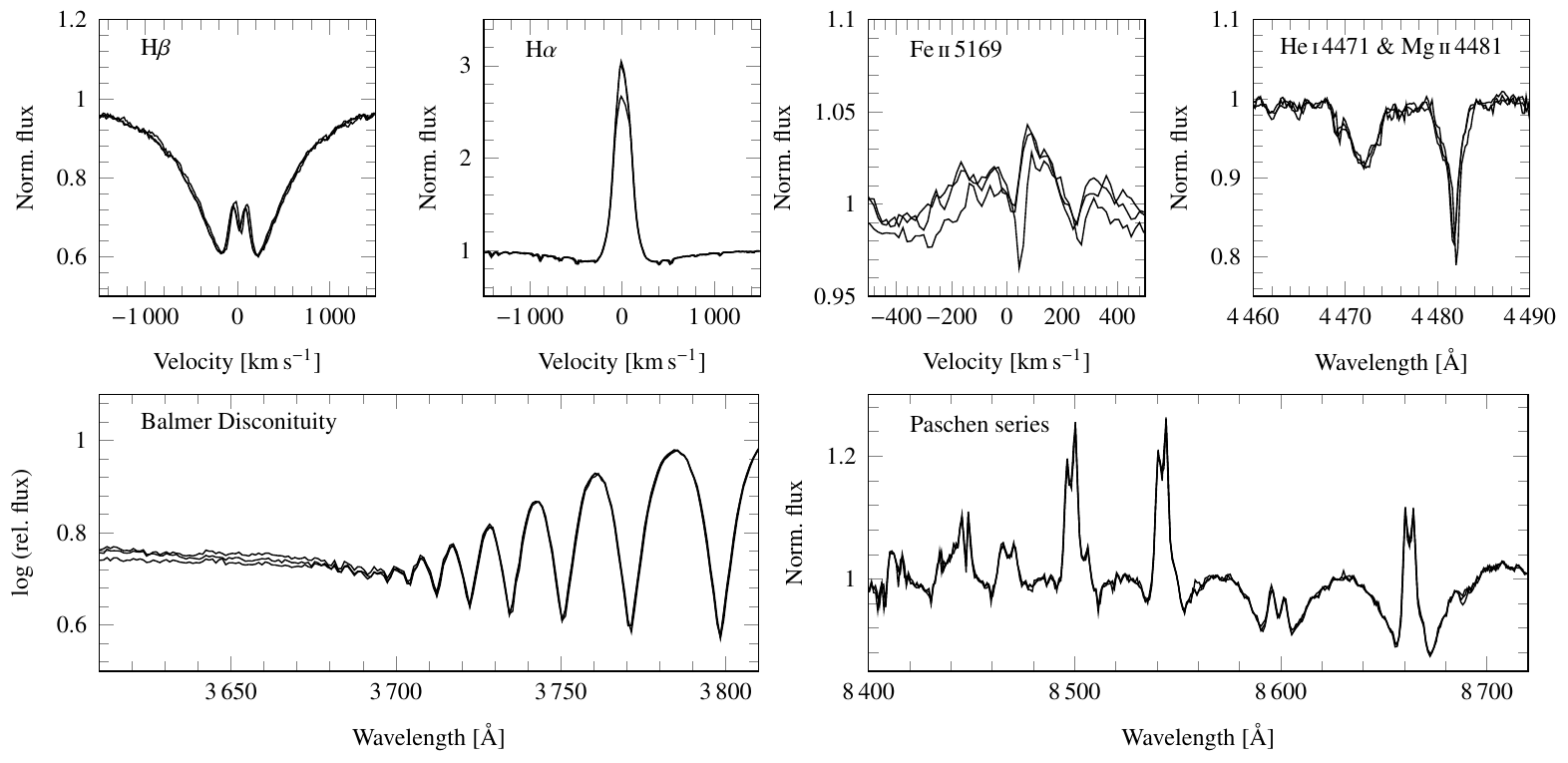}
\end{center}
\caption{Spectrum overview plot for Hip\,28561}
\end{figure*}

\begin{figure*}
\begin{center}
\includegraphics[angle=0,width=14cm,clip]{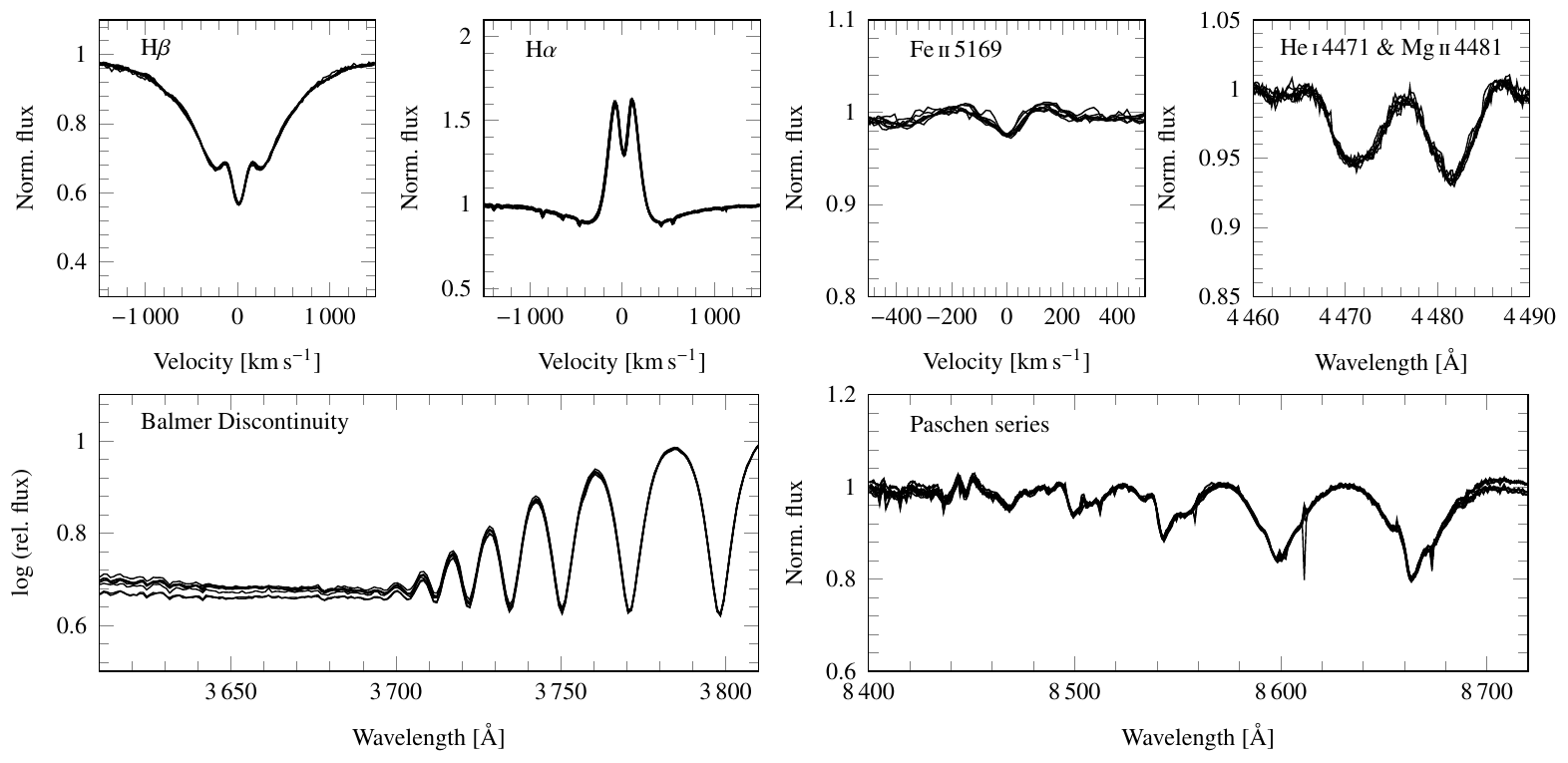}
\end{center}
\caption{Spectrum overview plot for Hip\,29635}
\end{figure*}

\begin{figure*}
\begin{center}
\includegraphics[angle=0,width=14cm,clip]{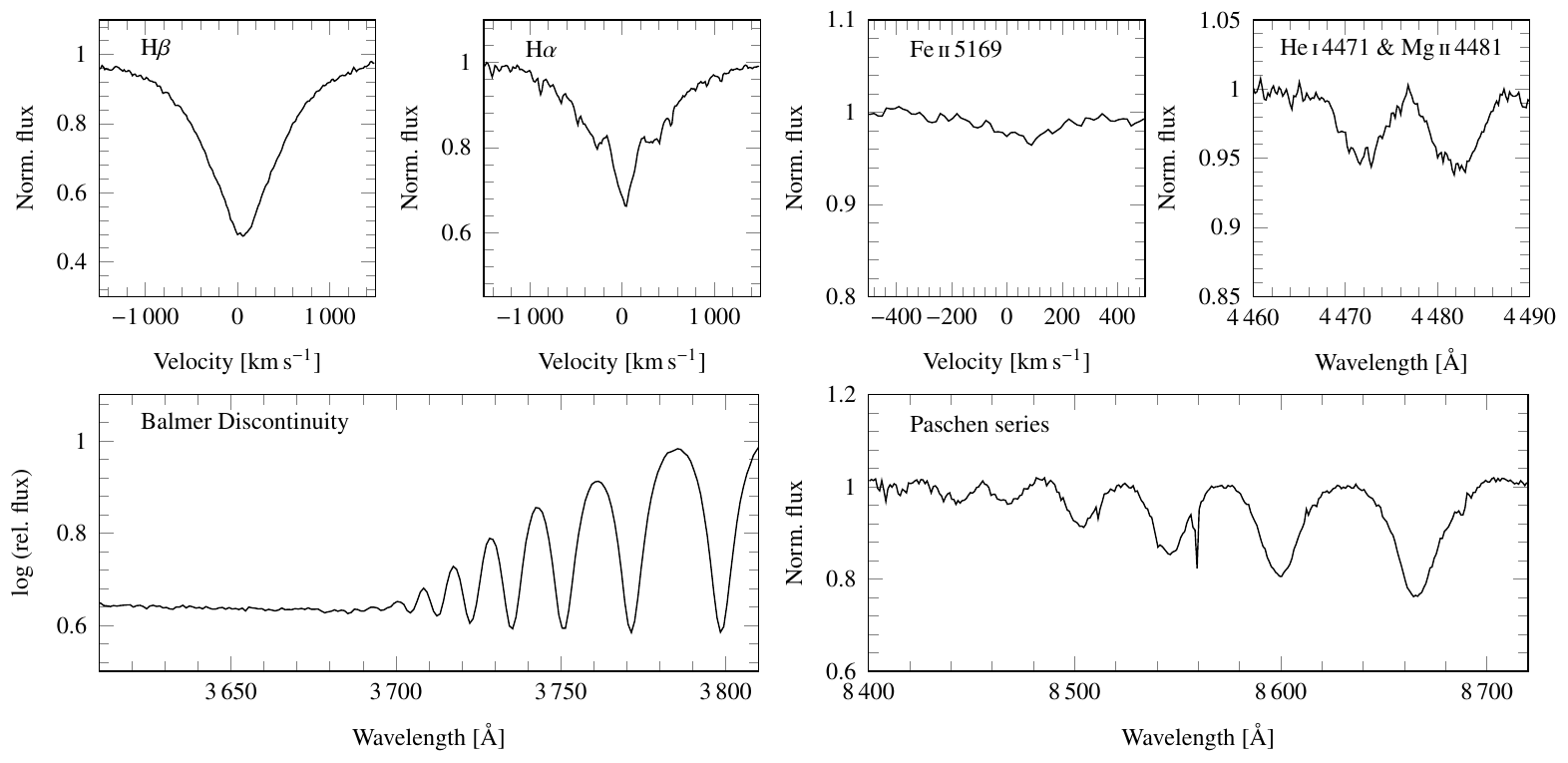}
\end{center}
\caption{Spectrum overview plot for Hip\,31362}
\end{figure*}
\clearpage

\begin{figure*}
\begin{center}
\includegraphics[angle=0,width=14cm,clip]{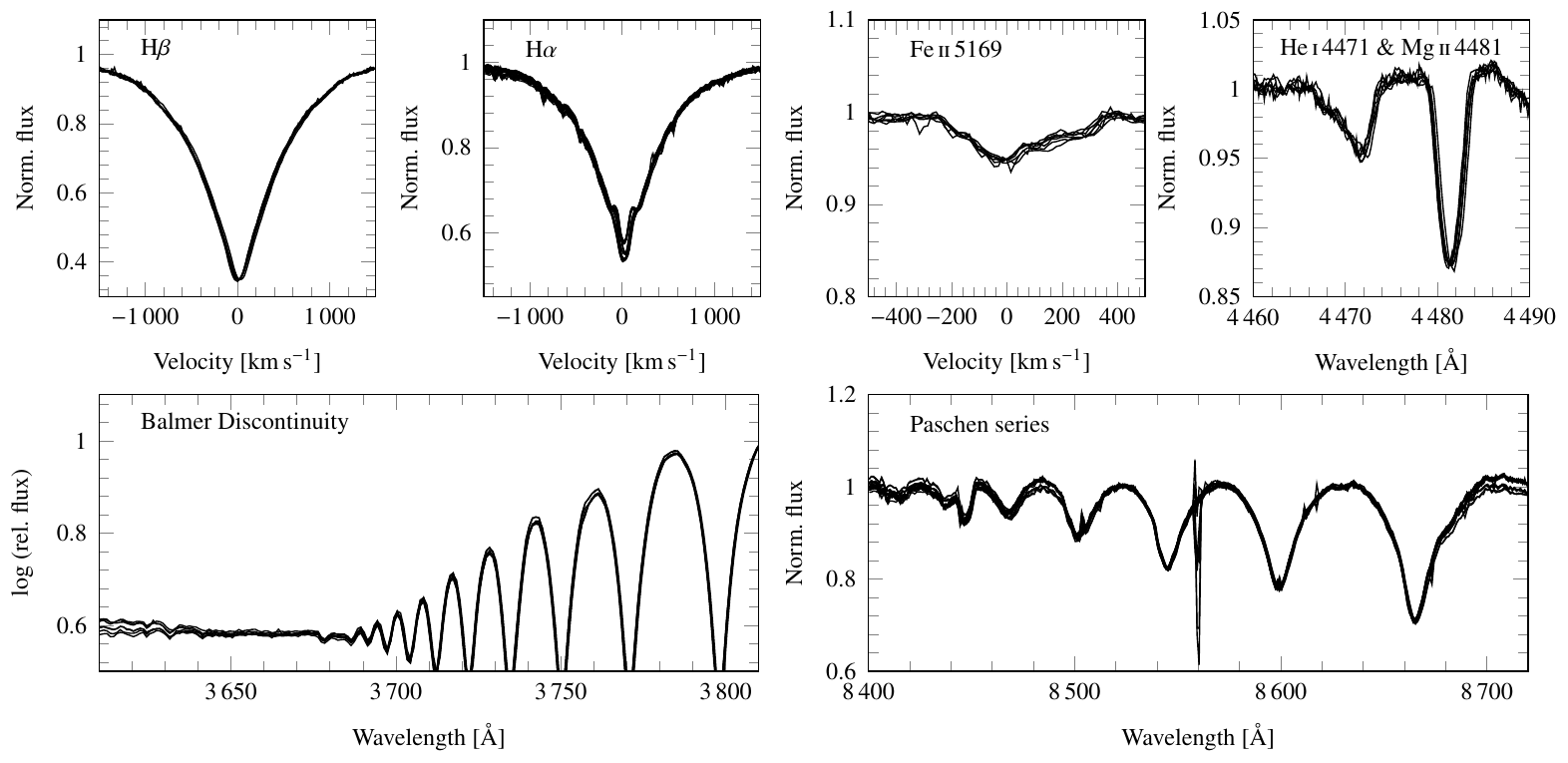}
\end{center}
\caption{Spectrum overview plot for Hip\,32474}
\end{figure*}

\begin{figure*}
\begin{center}
\includegraphics[angle=0,width=14cm,clip]{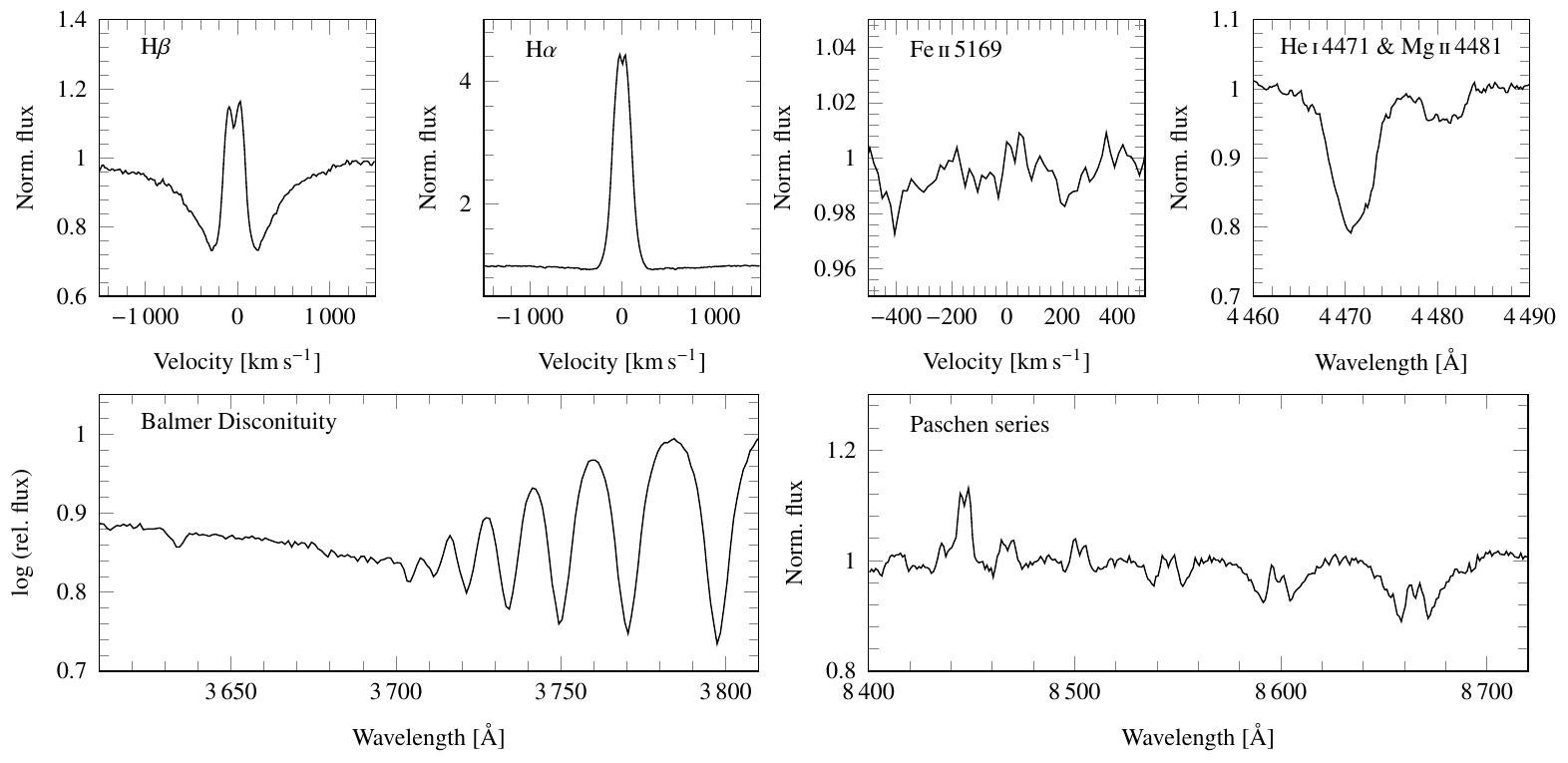}
\end{center}
\caption{Spectrum overview plot for Hip\,33509}
\end{figure*}

\begin{figure*}
\begin{center}
\includegraphics[angle=0,width=14cm,clip]{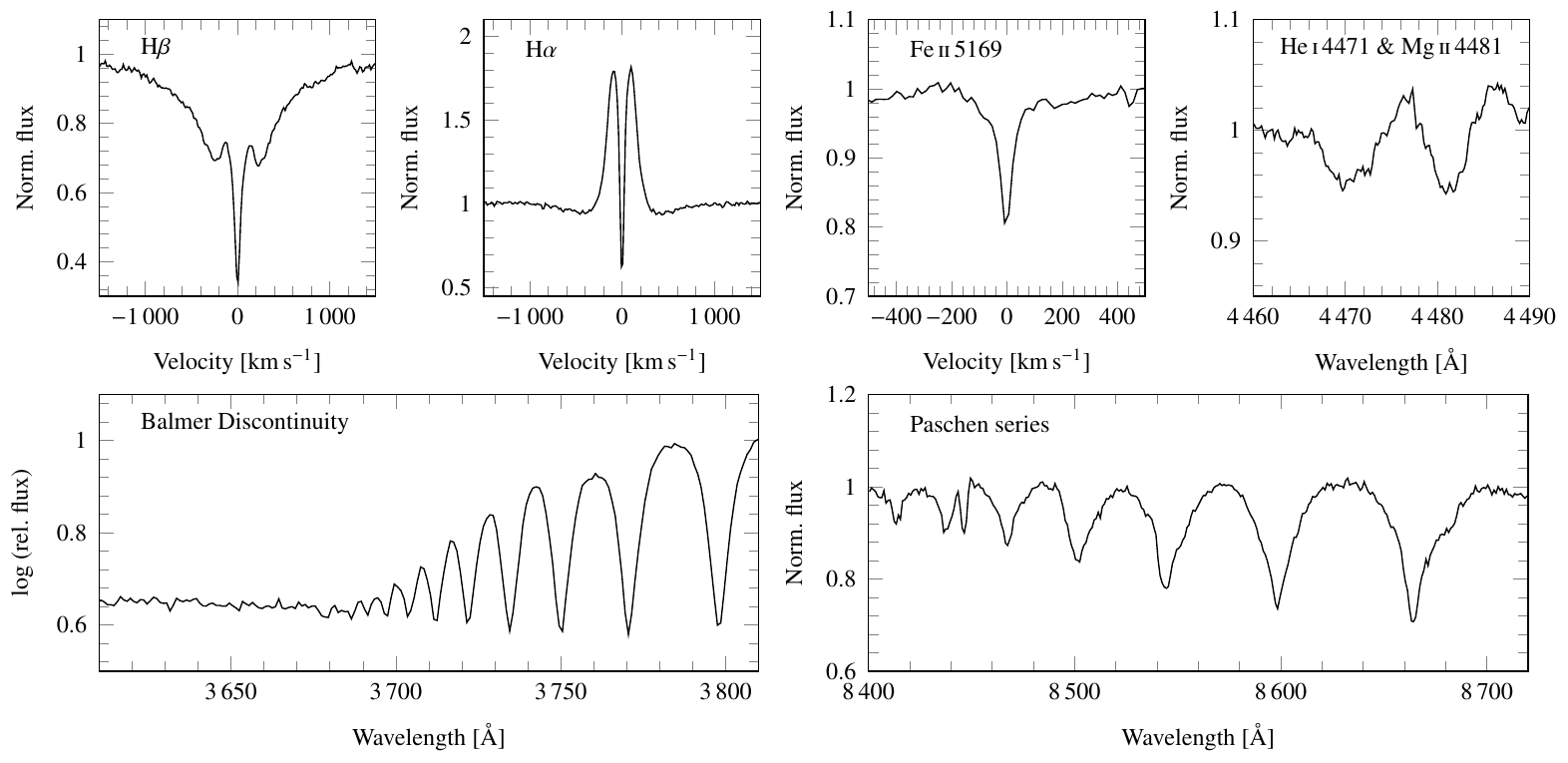}
\end{center}
\caption{Spectrum overview plot for Hip\,34144}
\end{figure*}
\clearpage

\begin{figure*}
\begin{center}
\includegraphics[angle=0,width=14cm,clip]{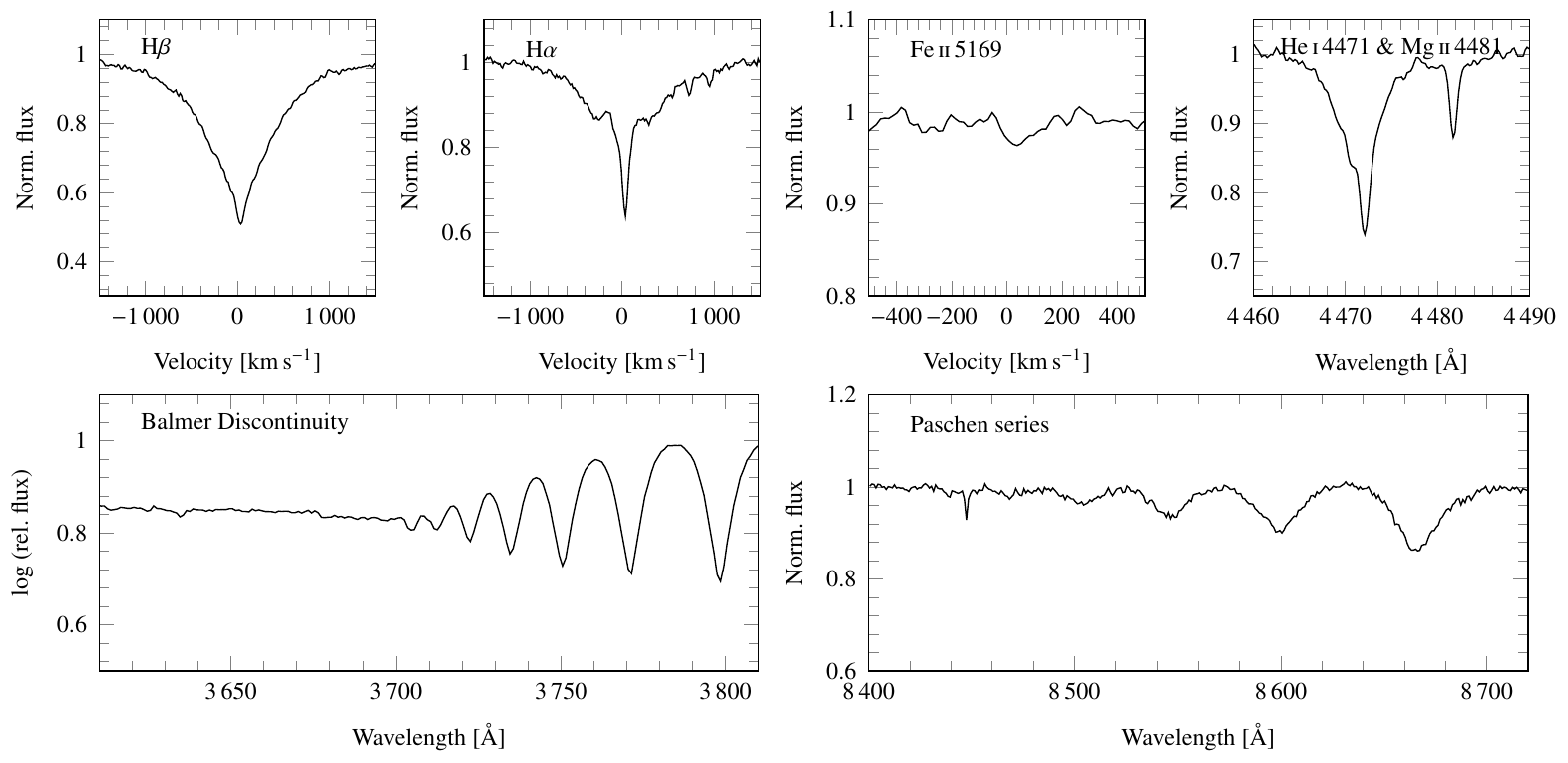}
\end{center}
\caption{Spectrum overview plot for Hip\,36009}
\end{figure*}

\begin{figure*}
\begin{center}
\includegraphics[angle=0,width=14cm,clip]{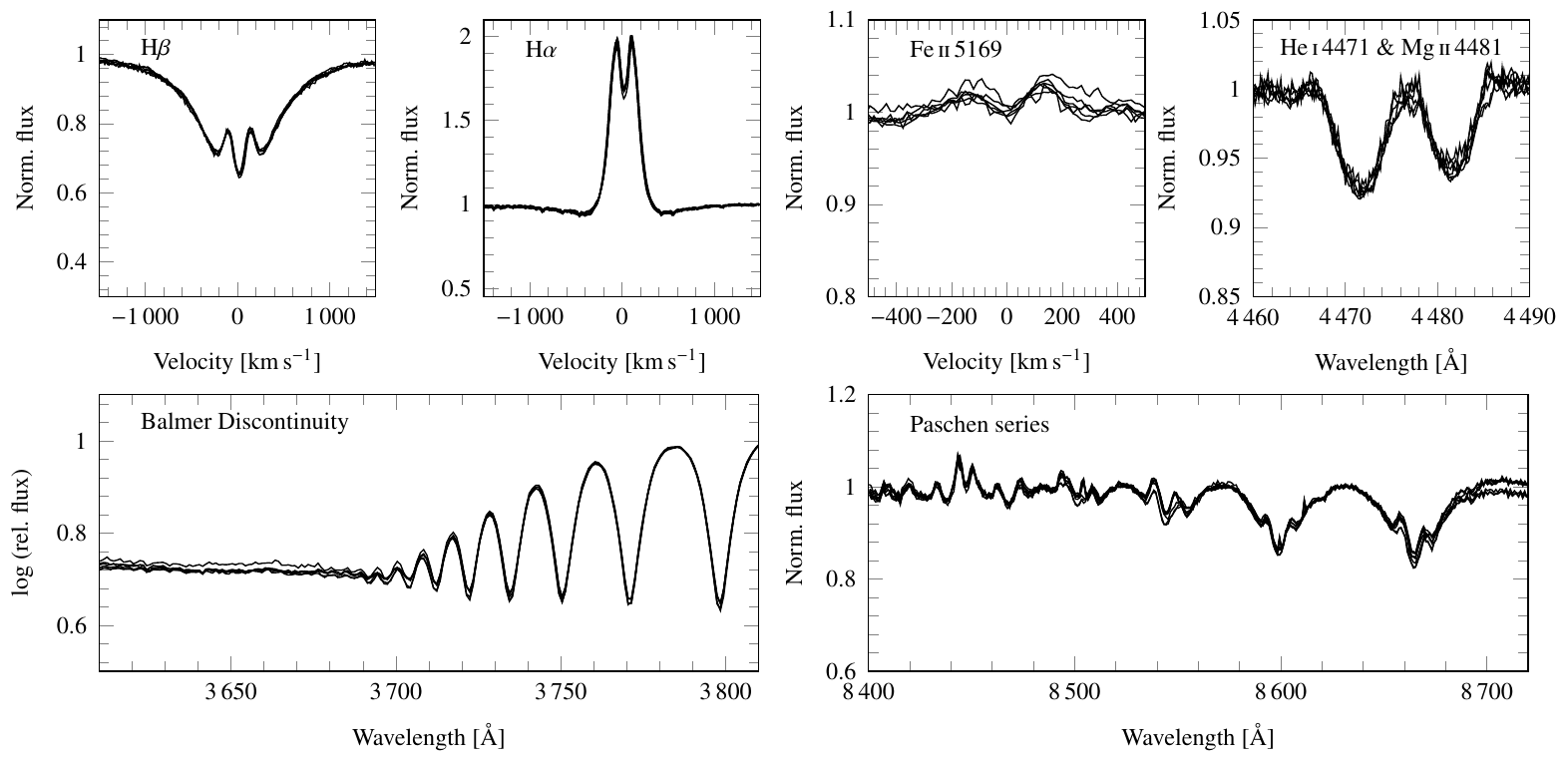}
\end{center}
\caption{Spectrum overview plot for Hip\,37007}
\end{figure*}

\begin{figure*}
\begin{center}
\includegraphics[angle=0,width=14cm,clip]{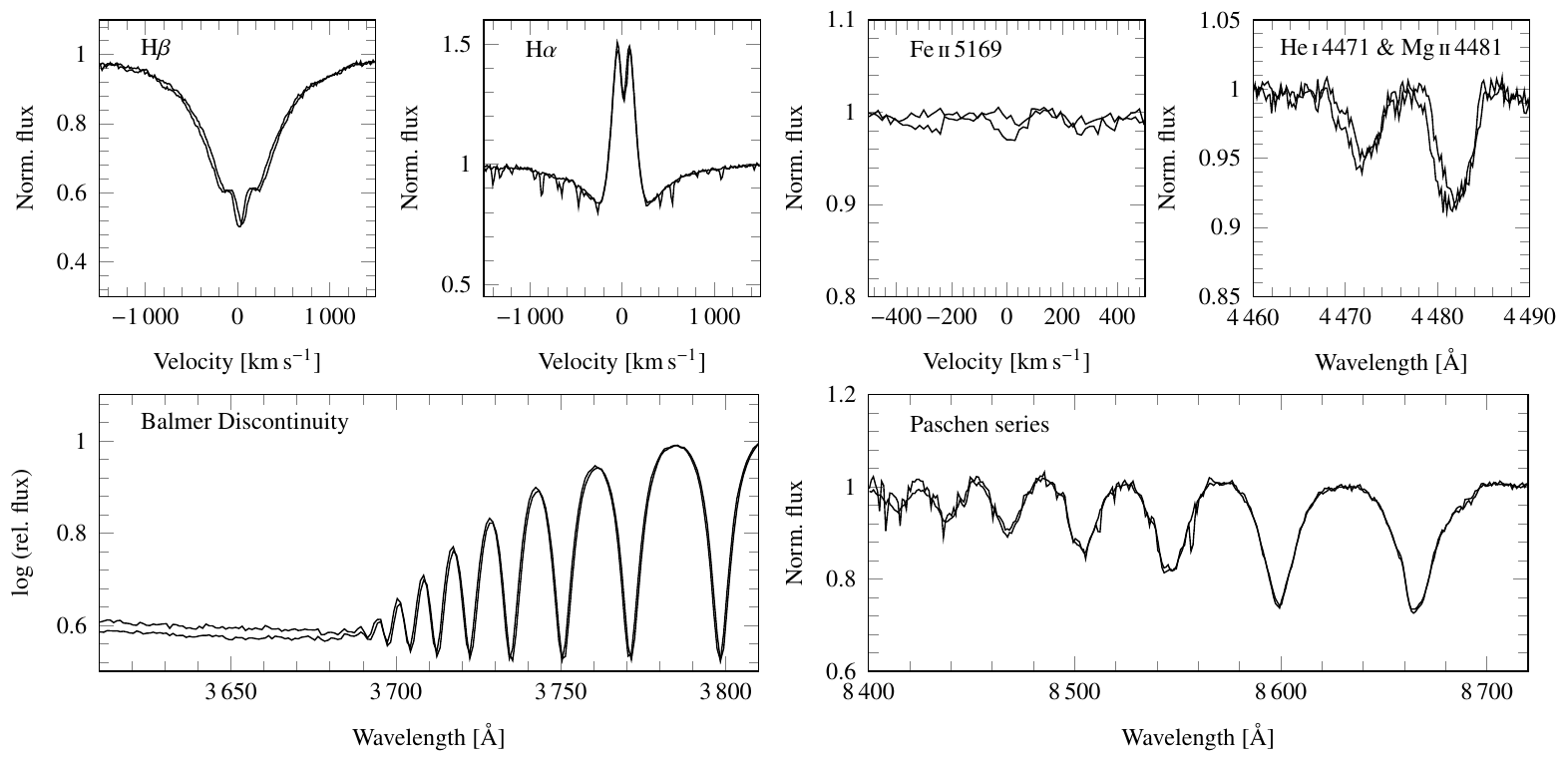}
\end{center}
\caption{Spectrum overview plot for Hip\,39183}
\end{figure*}
\clearpage

\begin{figure*}
\begin{center}
\includegraphics[angle=0,width=14cm,clip]{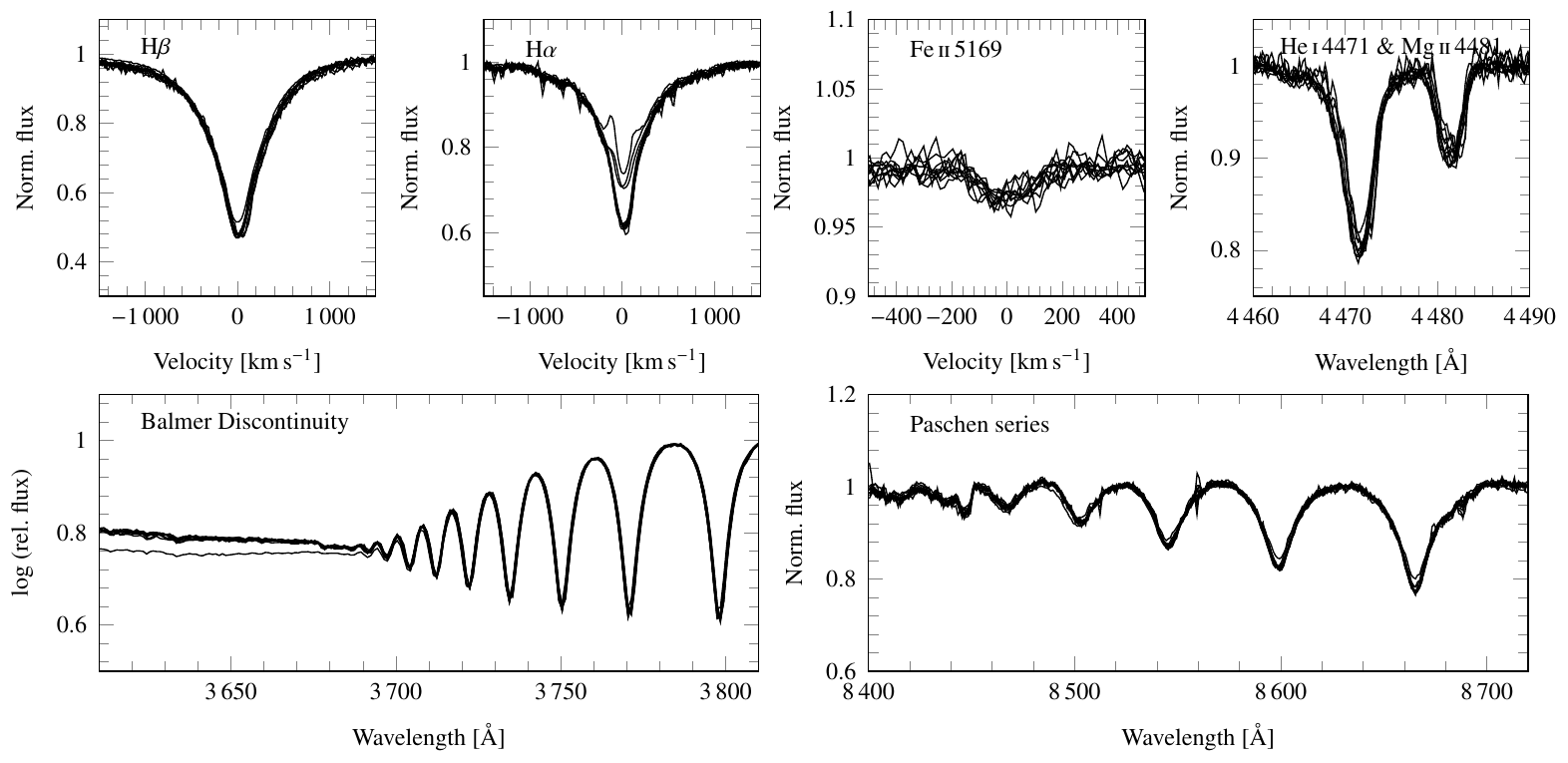}
\end{center}
\caption{Spectrum overview plot for Hip\,39483}
\end{figure*}

\begin{figure*}
\begin{center}
\includegraphics[angle=0,width=14cm,clip]{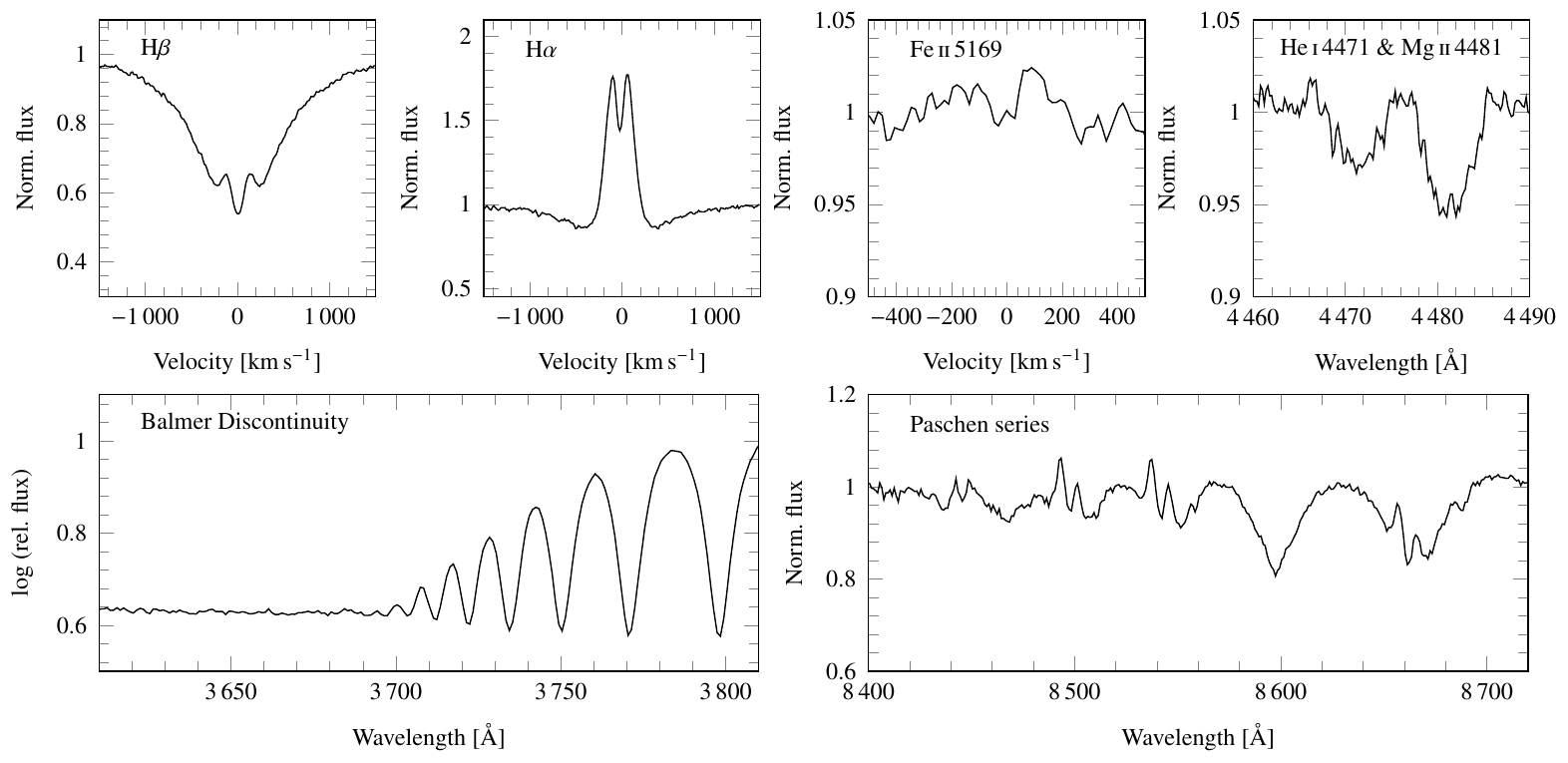}
\end{center}
\caption{Spectrum overview plot for Hip\,39595}
\end{figure*}

\begin{figure*}
\begin{center}
\includegraphics[angle=0,width=14cm,clip]{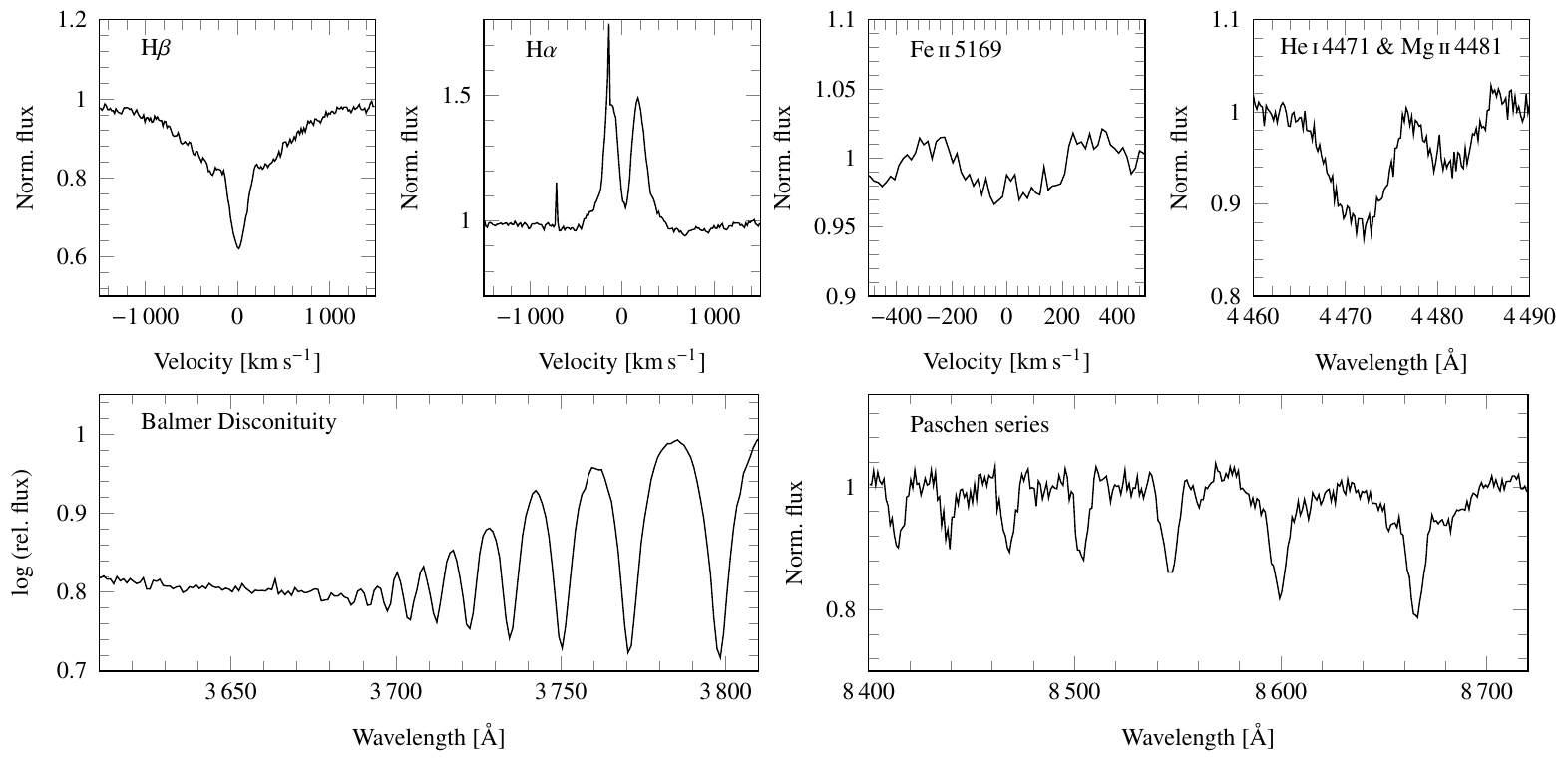}
\end{center}
\caption{Spectrum overview plot for Hip\,41085}
\end{figure*}
\clearpage

\begin{figure*}
\begin{center}
\includegraphics[angle=0,width=14cm,clip]{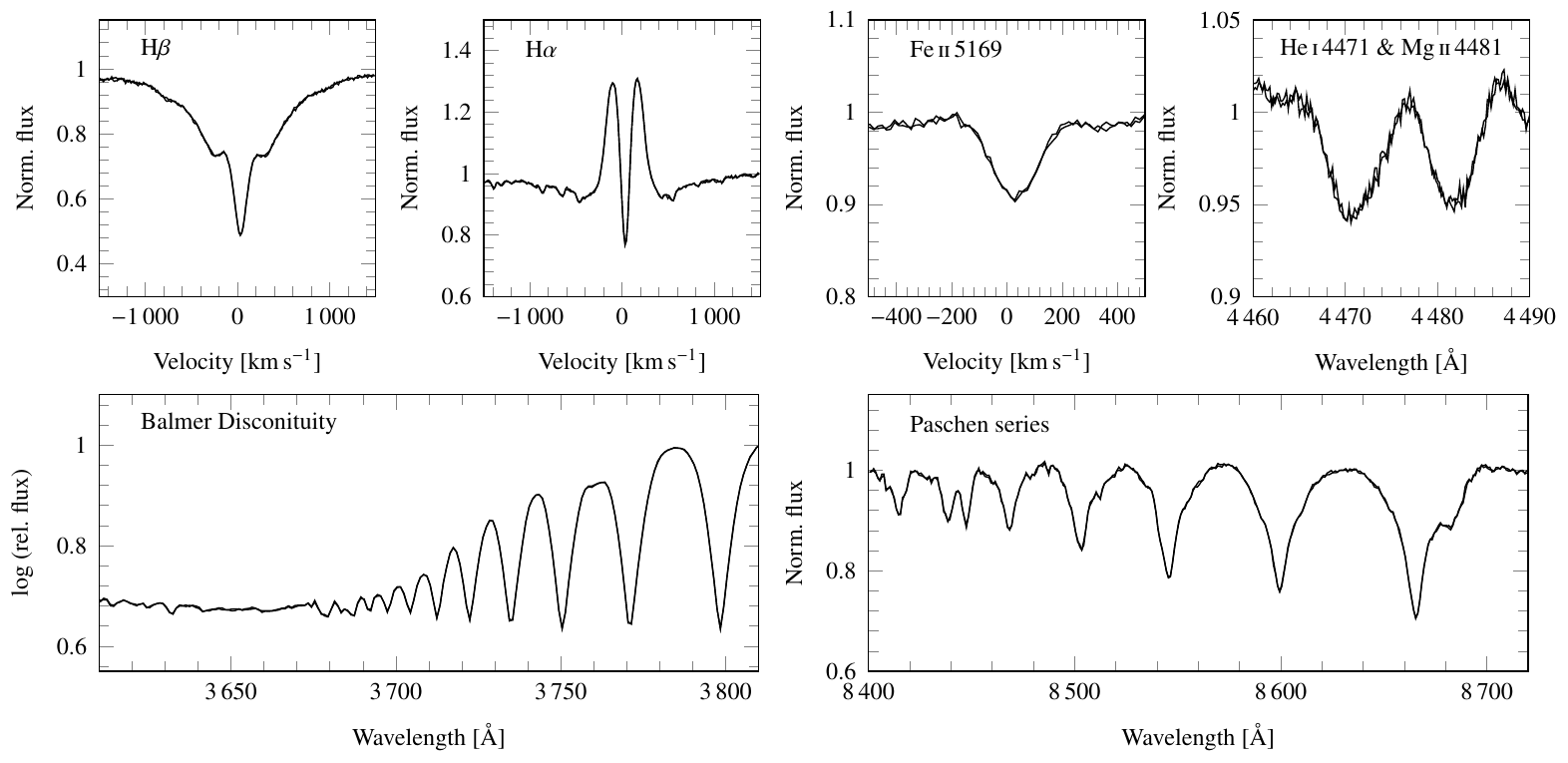}
\end{center}
\caption{Spectrum overview plot for Hip\,41268}
\end{figure*}

\begin{figure*}
\begin{center}
\includegraphics[angle=0,width=14cm,clip]{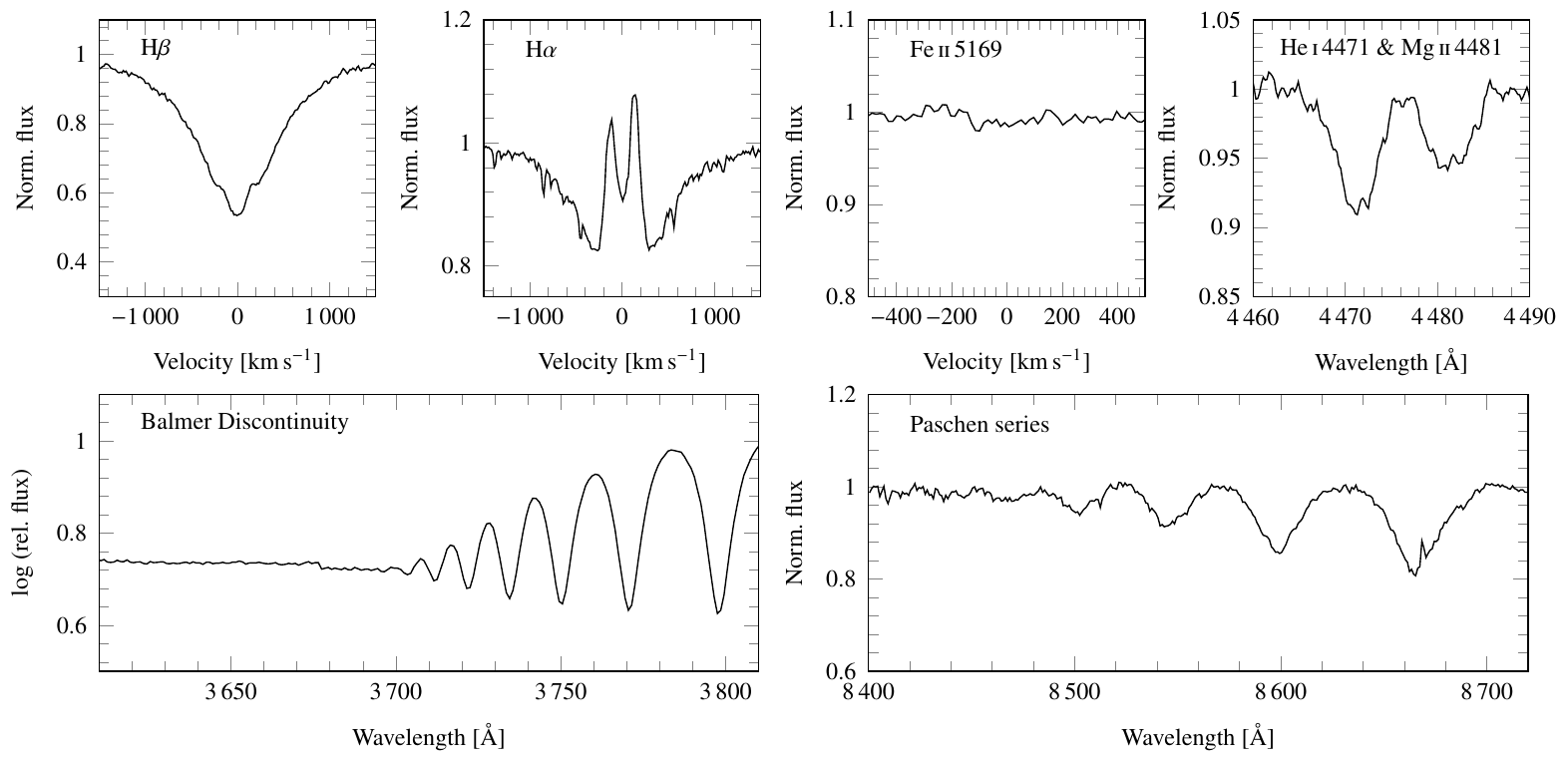}
\end{center}
\caption{Spectrum overview plot for Hip\,42060}
\end{figure*}

\begin{figure*}
\begin{center}
\includegraphics[angle=0,width=14cm,clip]{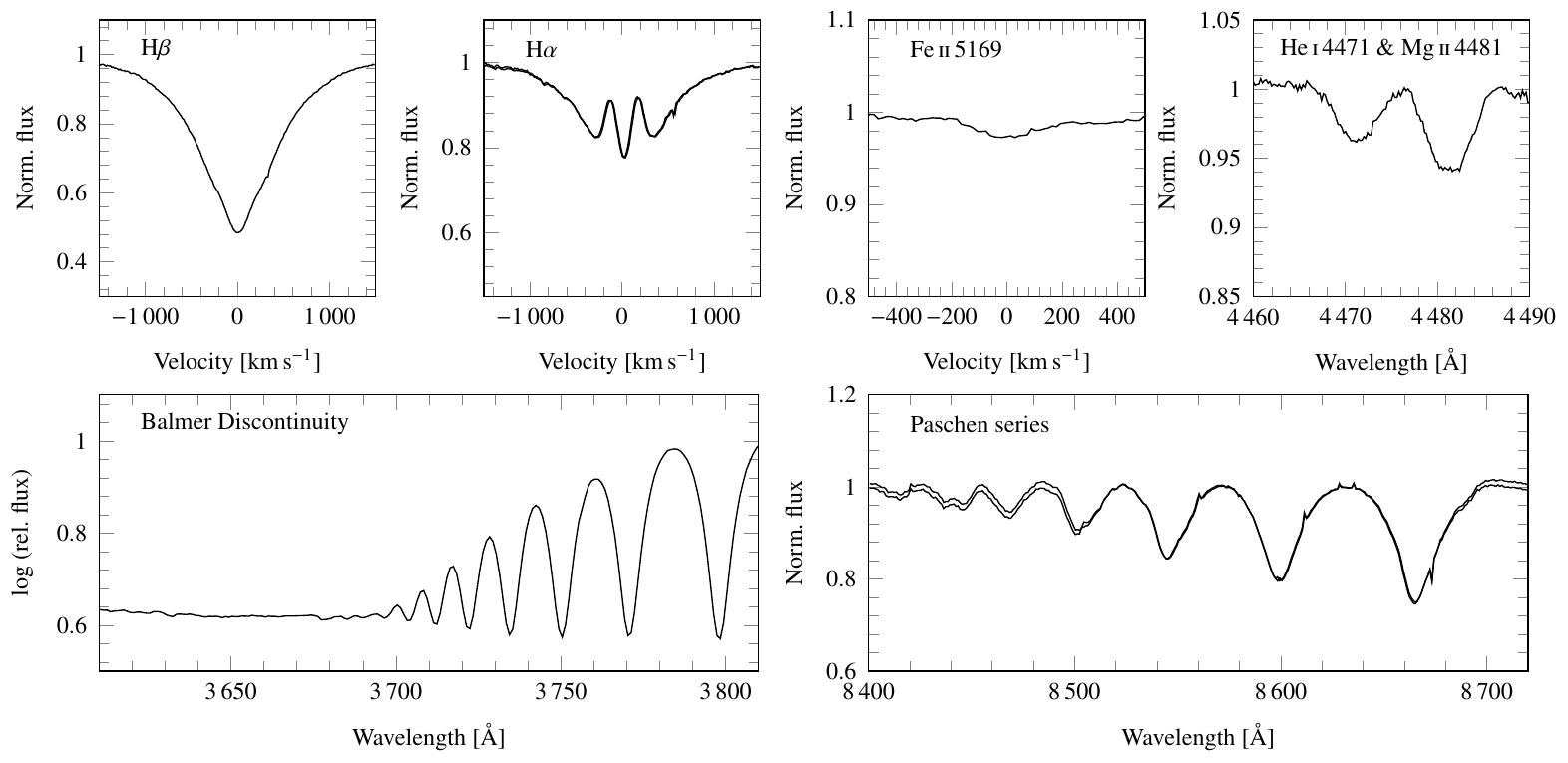}
\end{center}
\caption{Spectrum overview plot for Hip\,43073}
\end{figure*}
\clearpage

\begin{figure*}
\begin{center}
\includegraphics[angle=0,width=14cm,clip]{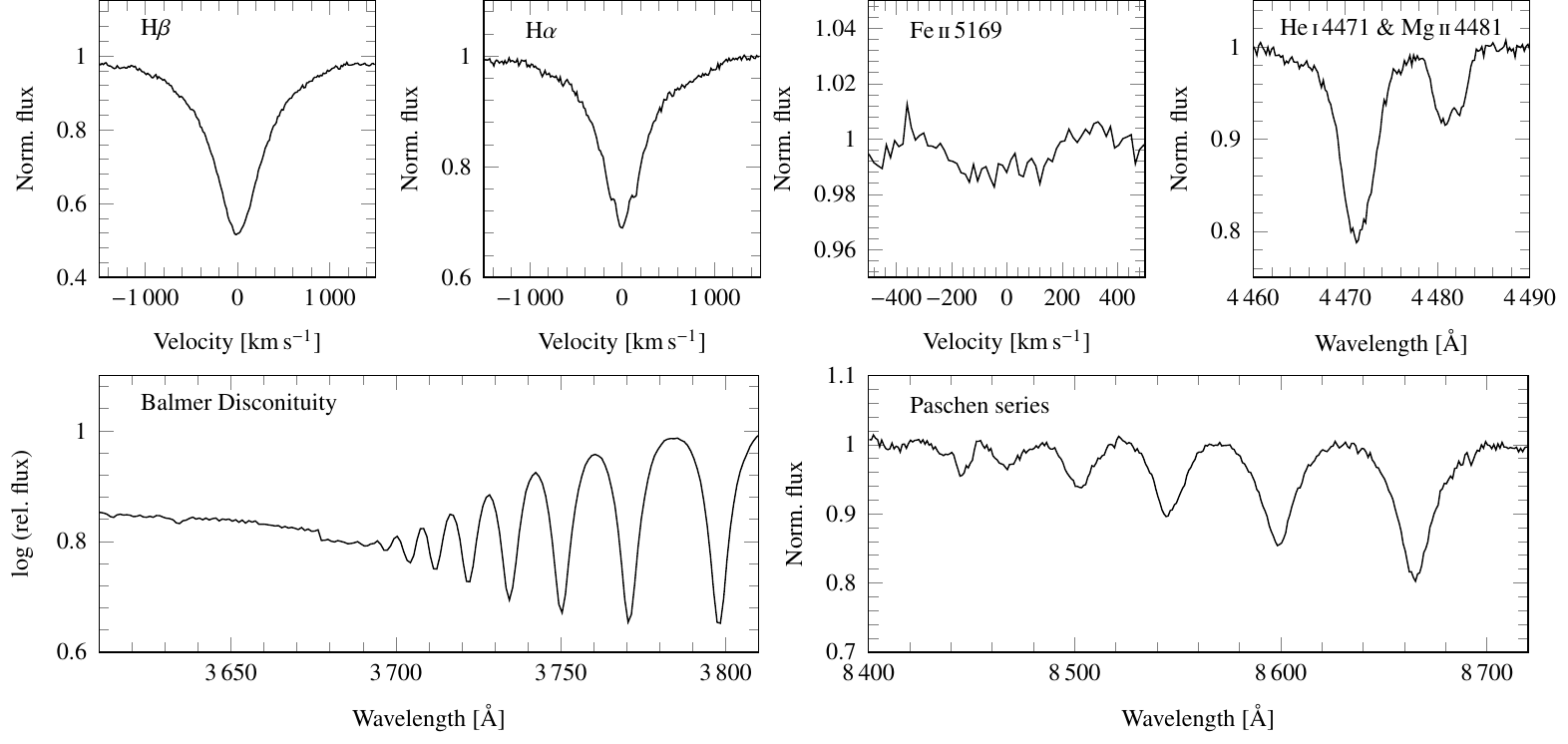}
\end{center}
\caption{Spectrum overview plot for Hip\,43114}
\end{figure*}

\begin{figure*}
\begin{center}
\includegraphics[angle=0,width=14cm,clip]{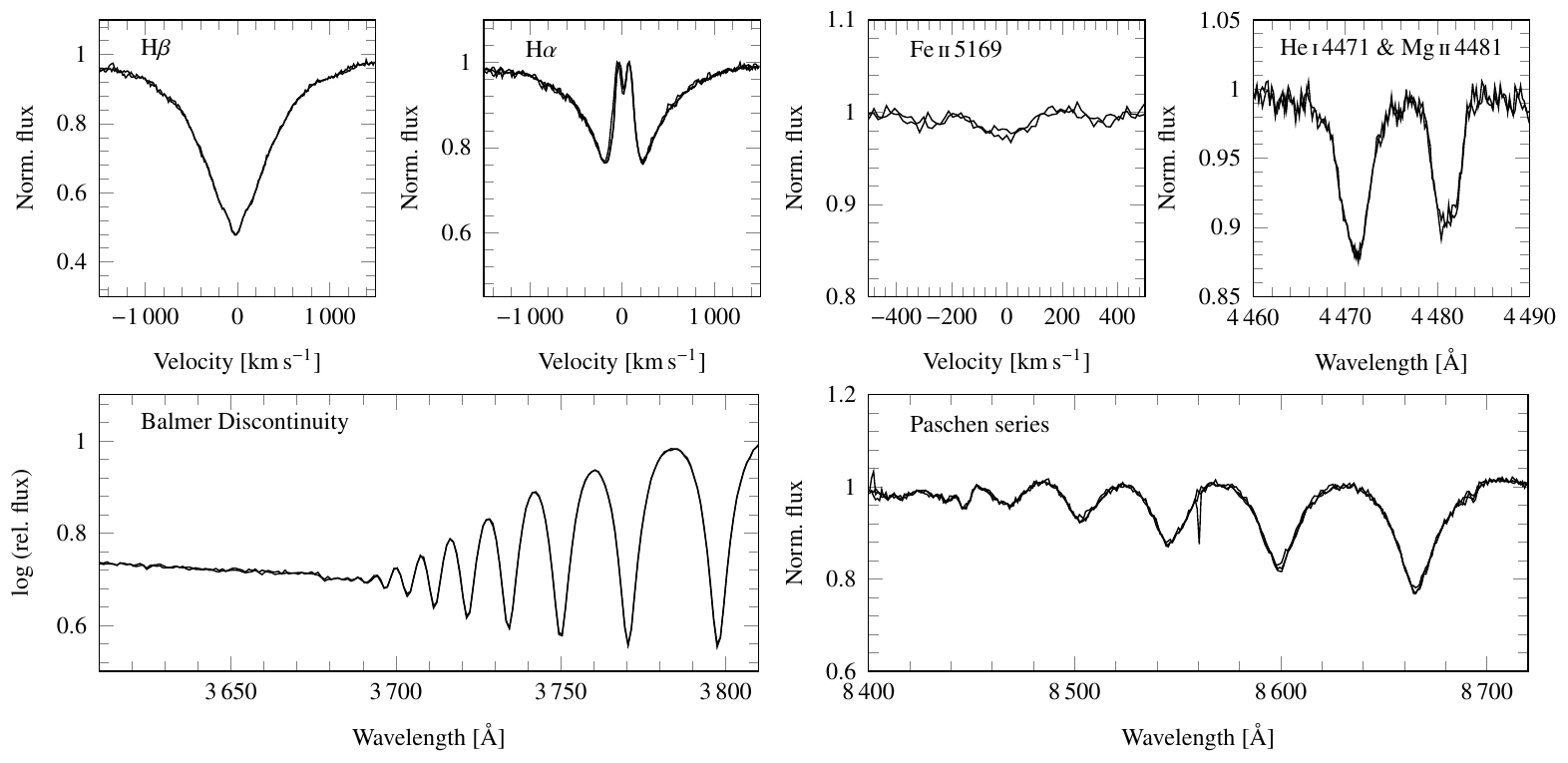}
\end{center}
\caption{Spectrum overview plot for Hip\,44423}
\end{figure*}

\begin{figure*}
\begin{center}
\includegraphics[angle=0,width=14cm,clip]{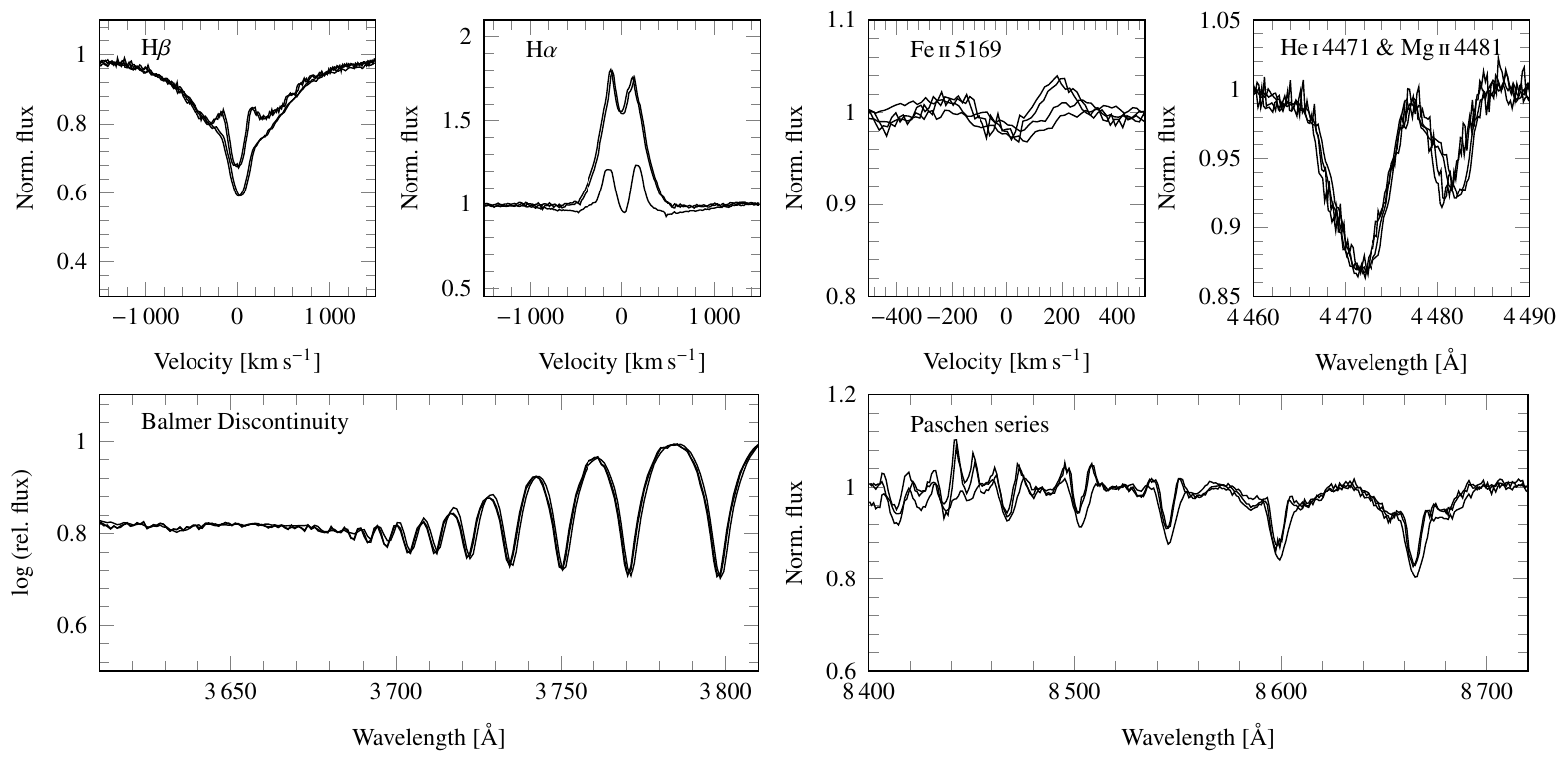}
\end{center}
\caption{Spectrum overview plot for Hip\,46329}
\end{figure*}
\clearpage

\begin{figure*}
\begin{center}
\includegraphics[angle=0,width=14cm,clip]{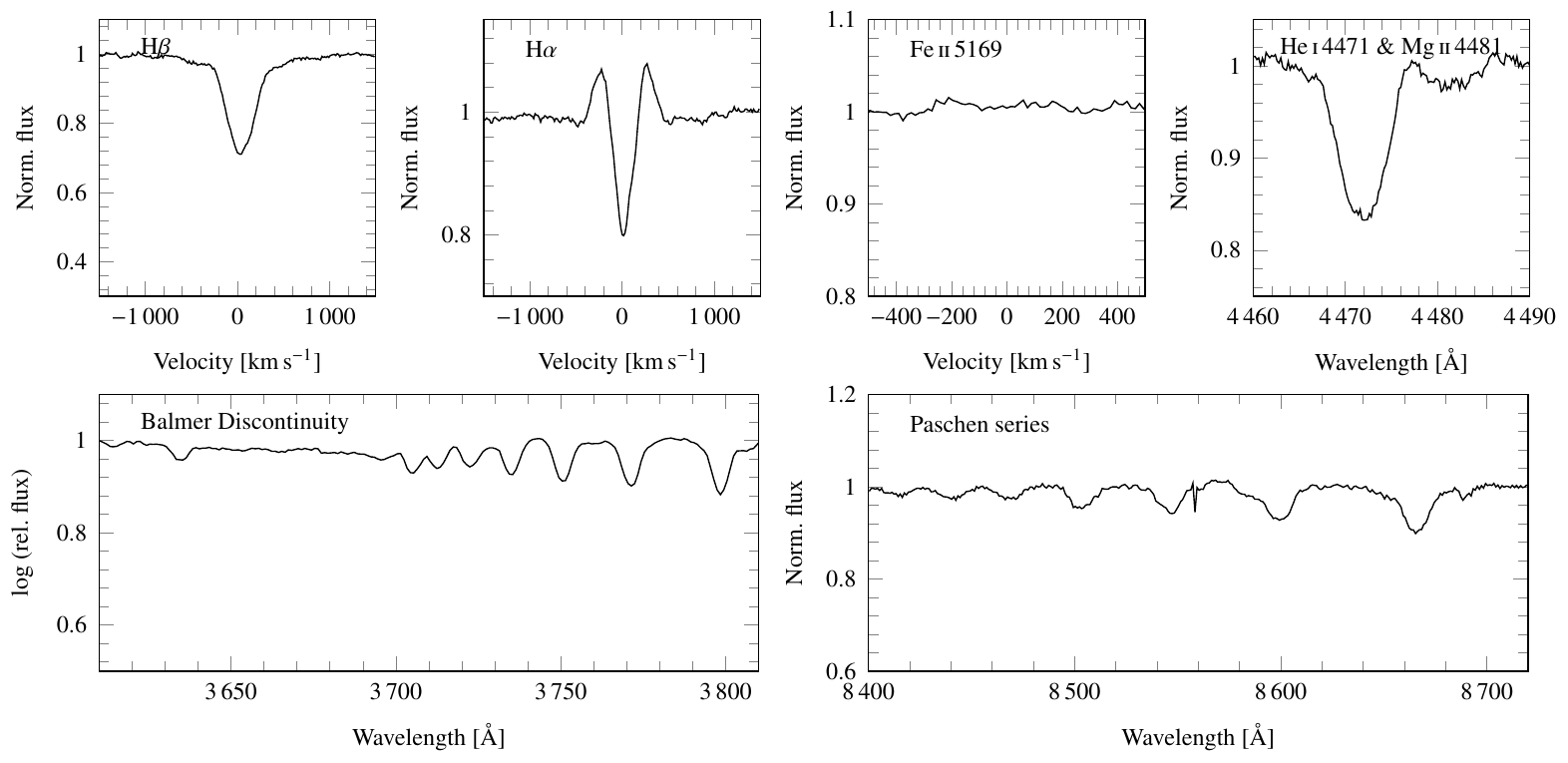}
\end{center}
\caption{Spectrum overview plot for Hip\,47868}
\end{figure*}

\begin{figure*}
\begin{center}
\includegraphics[angle=0,width=14cm,clip]{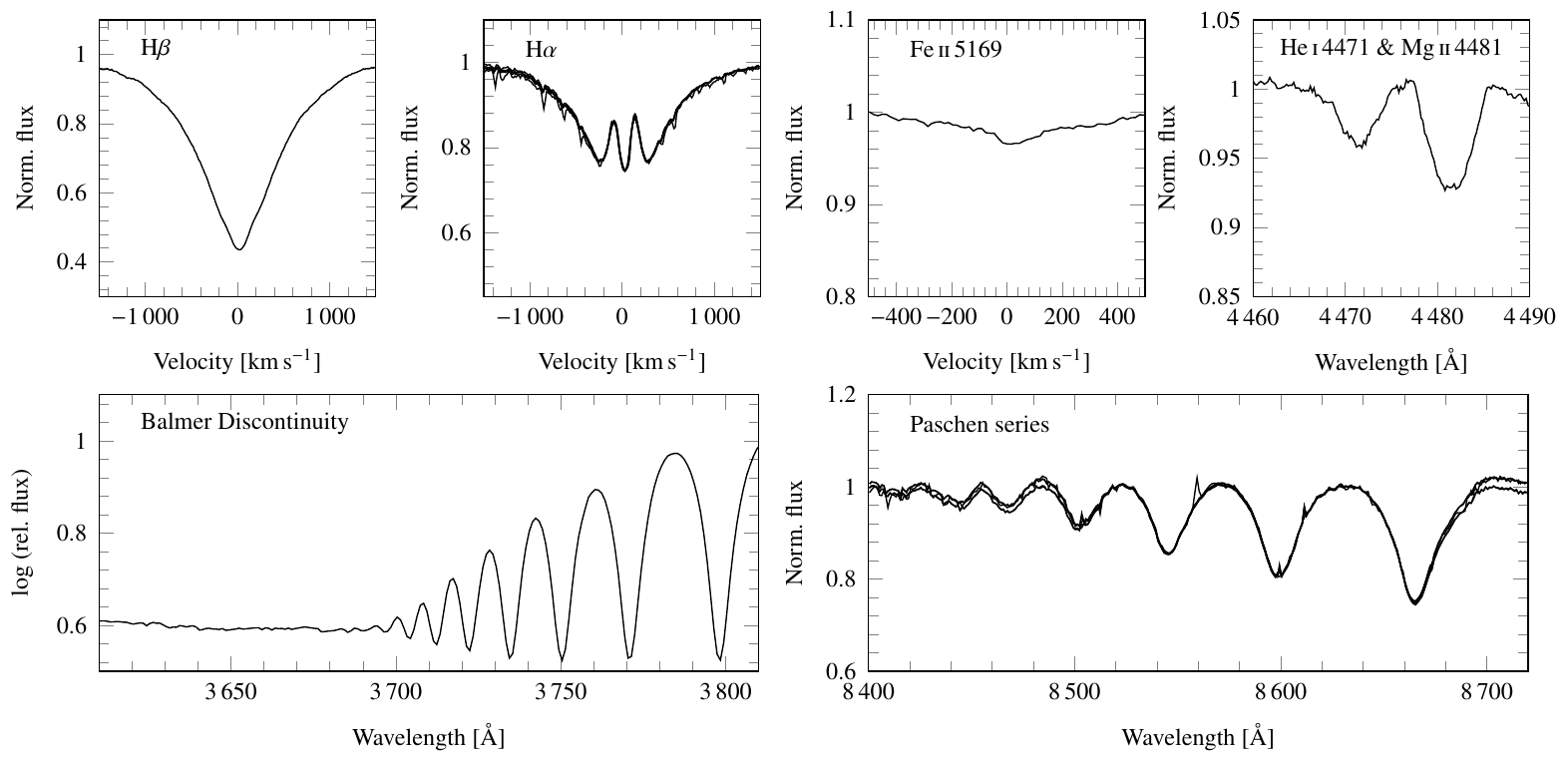}
\end{center}
\caption{Spectrum overview plot for Hip\,47962}
\end{figure*}

\begin{figure*}
\begin{center}
\includegraphics[angle=0,width=14cm,clip]{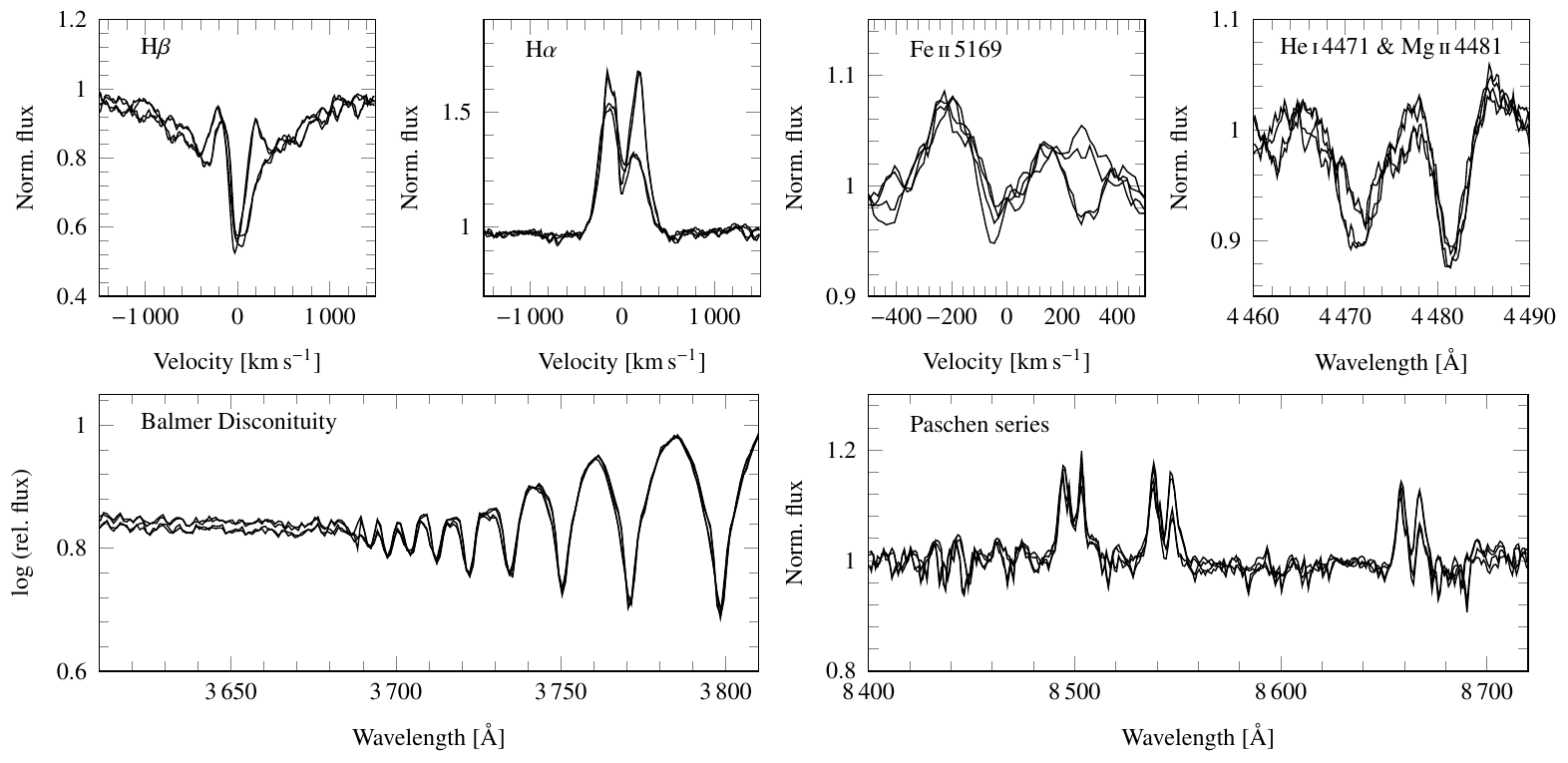}
\end{center}
\caption{Spectrum overview plot for Hip\,48582}
\end{figure*}

\clearpage

\begin{figure*}
\begin{center}
\includegraphics[angle=0,width=14cm,clip]{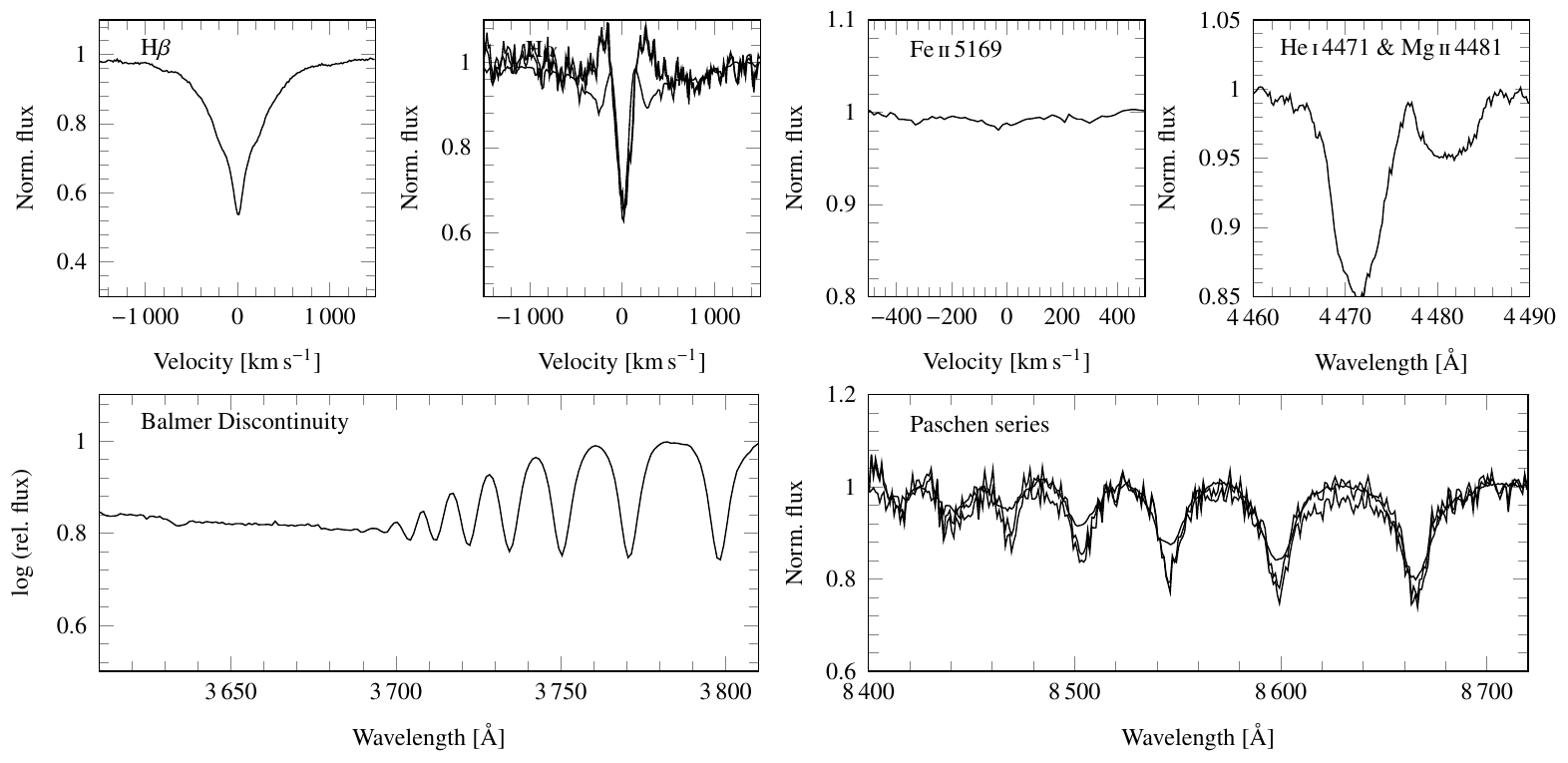}
\end{center}
\caption{Spectrum overview plot for Hip\,51444}
\end{figure*}

\begin{figure*}
\begin{center}
\includegraphics[angle=0,width=14cm,clip]{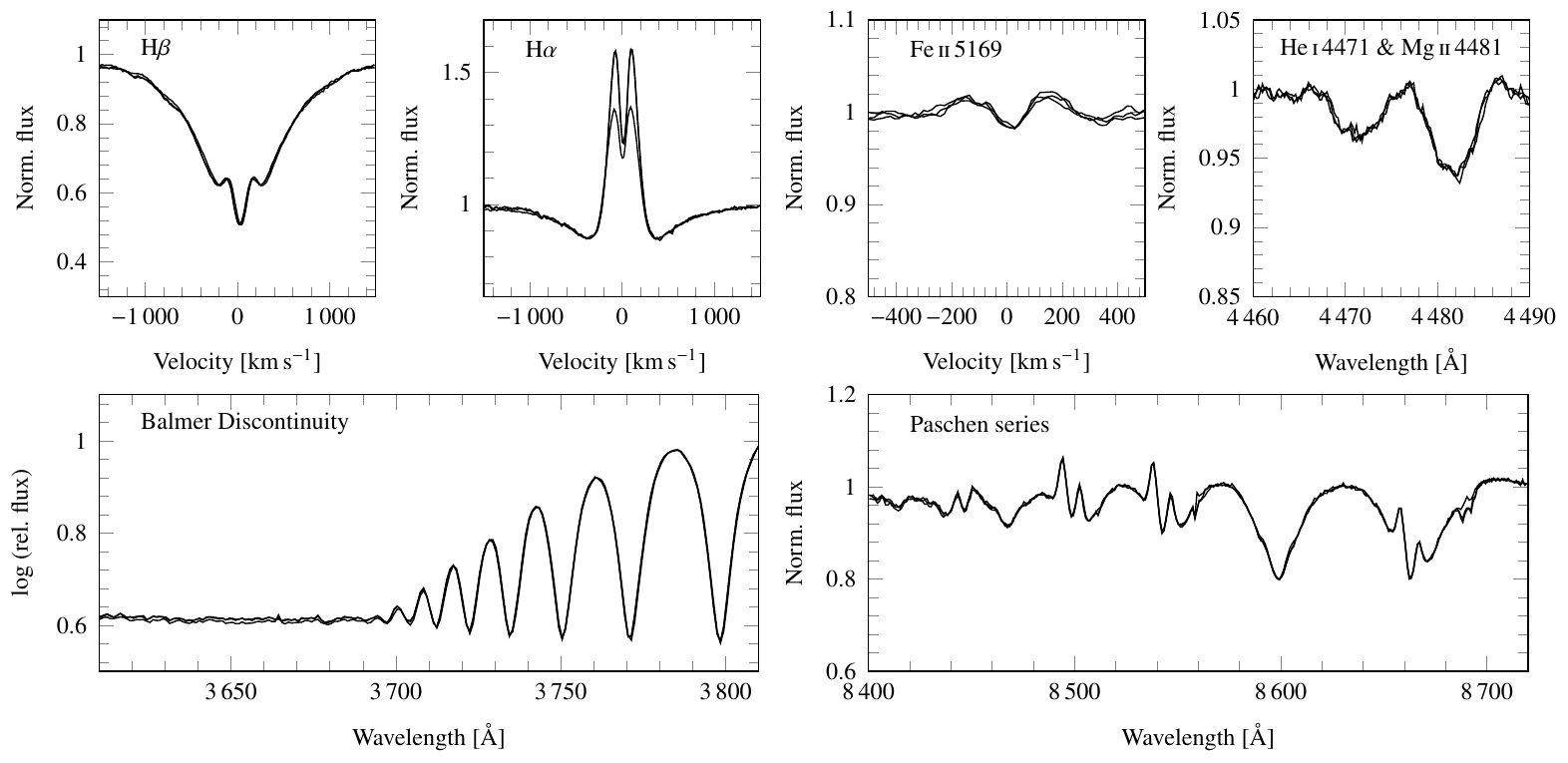}
\end{center}
\caption{Spectrum overview plot for Hip\,51491}
\end{figure*}

\begin{figure*}
\begin{center}
\includegraphics[angle=0,width=14cm,clip]{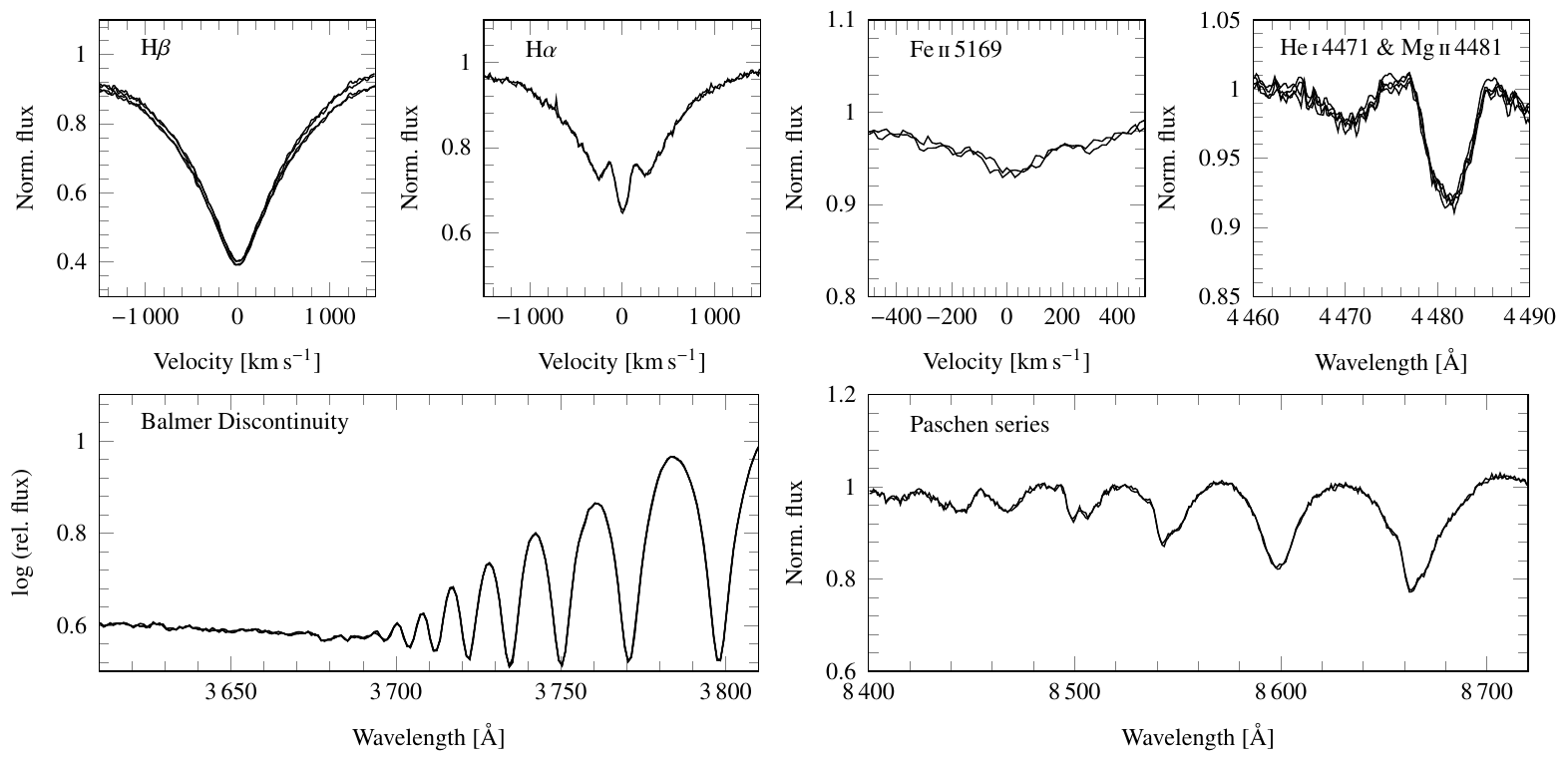}
\end{center}
\caption{Spectrum overview plot for Hip\,51546}
\end{figure*}

\clearpage

\begin{figure*}
\begin{center}
\includegraphics[angle=0,width=14cm,clip]{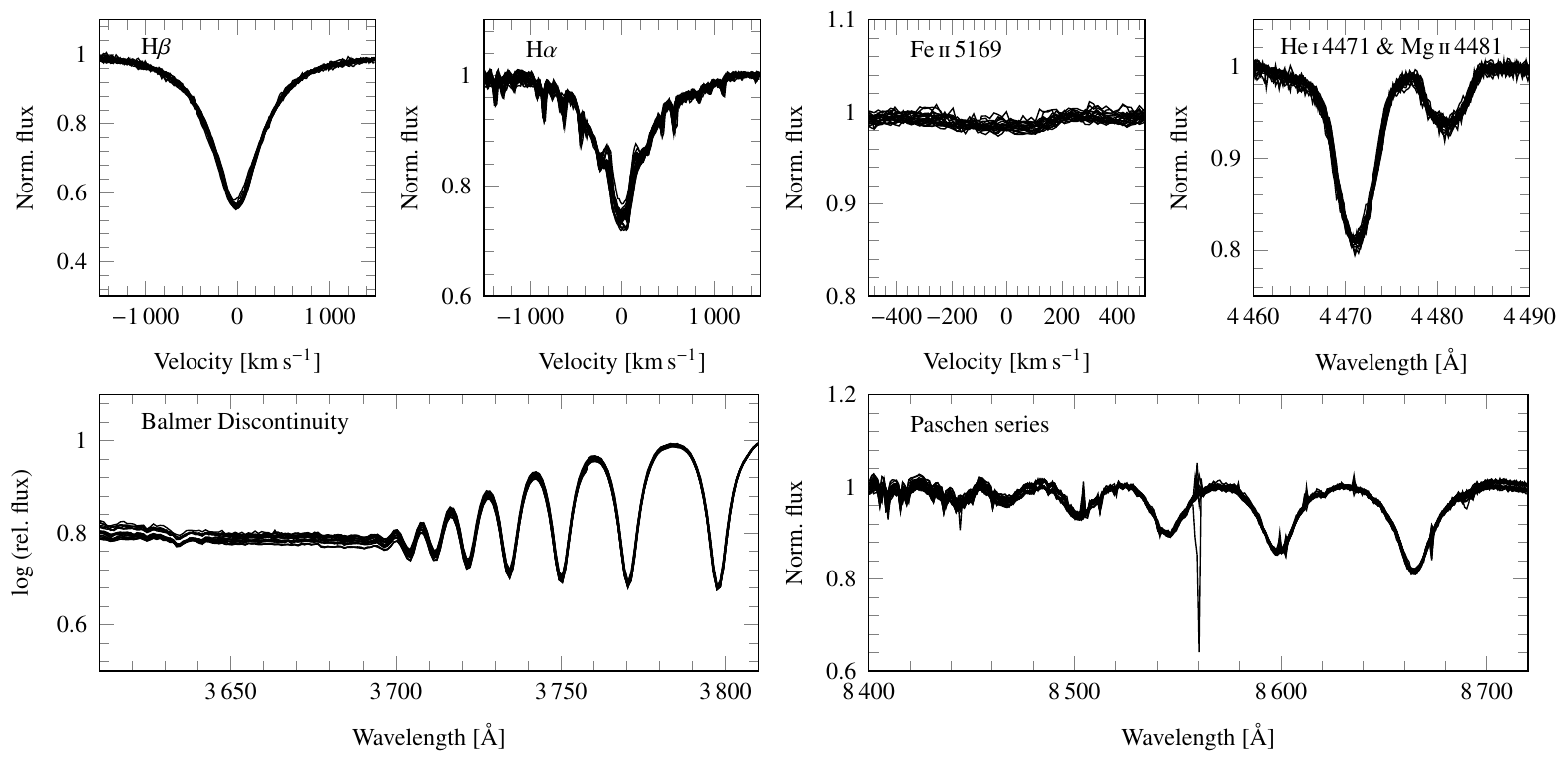}
\end{center}
\caption{Spectrum overview plot for Hip\,52977}
\end{figure*}

\begin{figure*}
\begin{center}
\includegraphics[angle=0,width=14cm,clip]{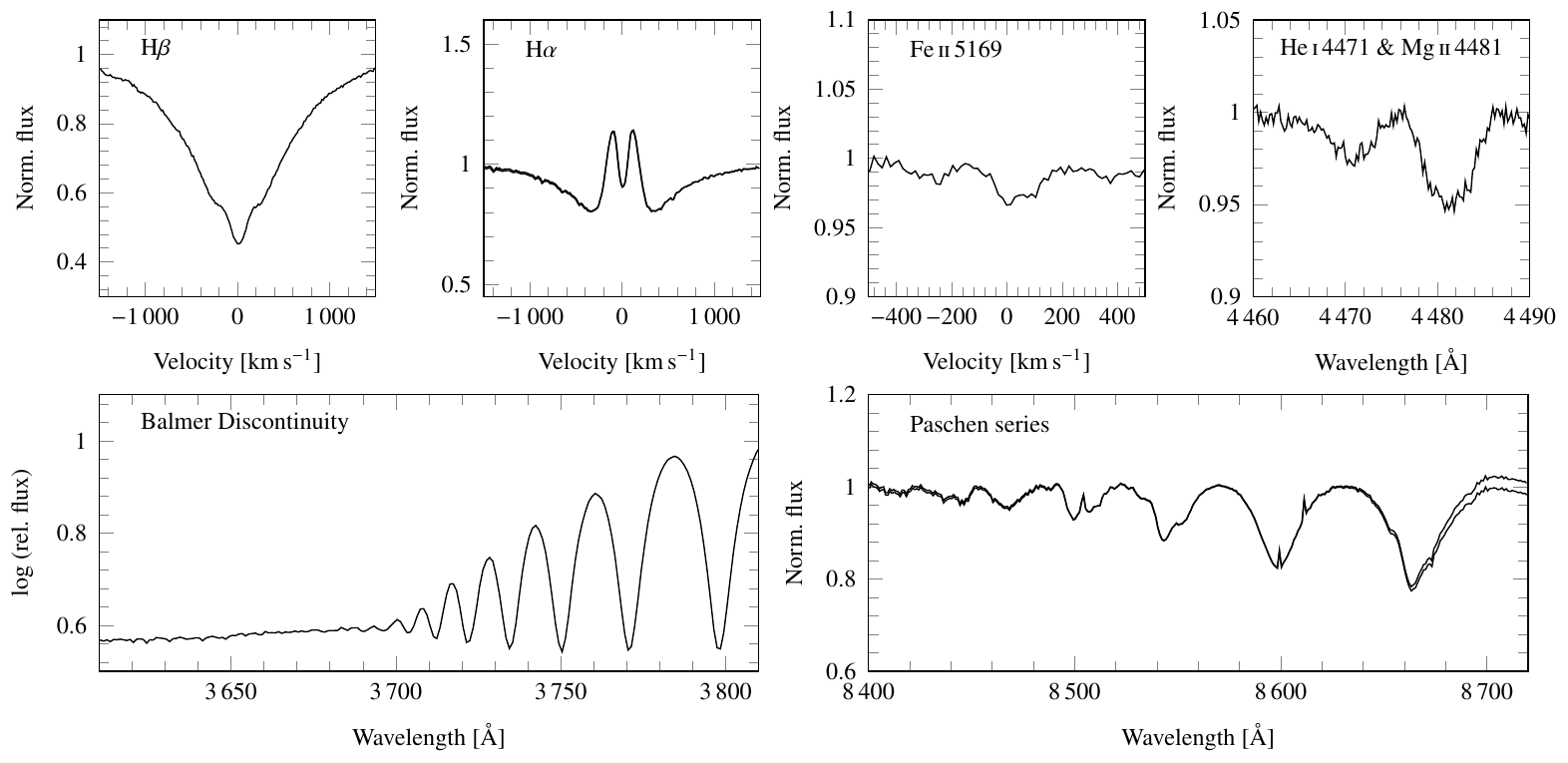}
\end{center}
\caption{Spectrum overview plot for Hip\,56393}
\end{figure*}

\begin{figure*}
\begin{center}
\includegraphics[angle=0,width=14cm,clip]{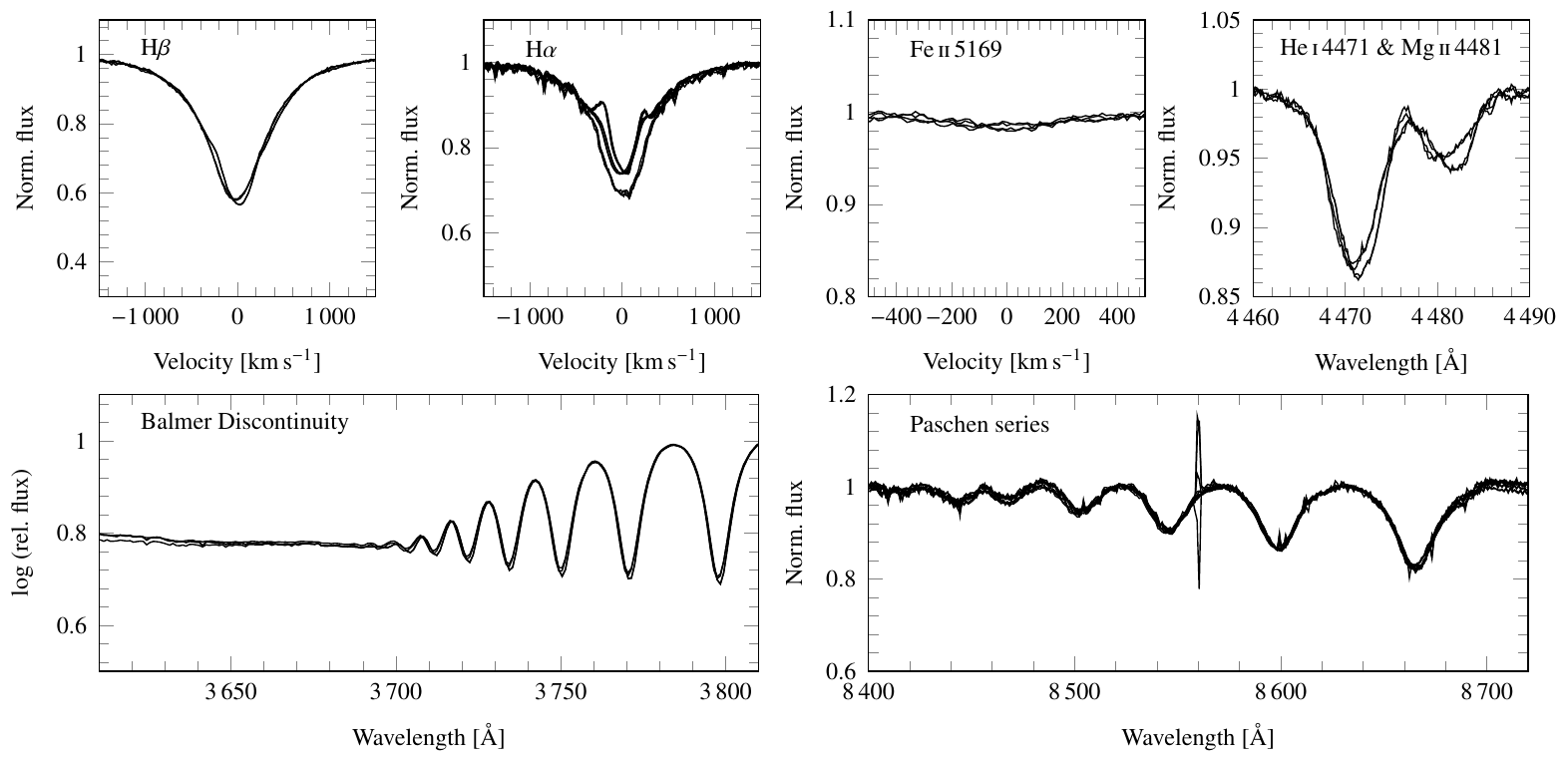}
\end{center}
\caption{Spectrum overview plot for Hip\,57861}
\end{figure*}

\clearpage

\begin{figure*}
\begin{center}
\includegraphics[angle=0,width=14cm,clip]{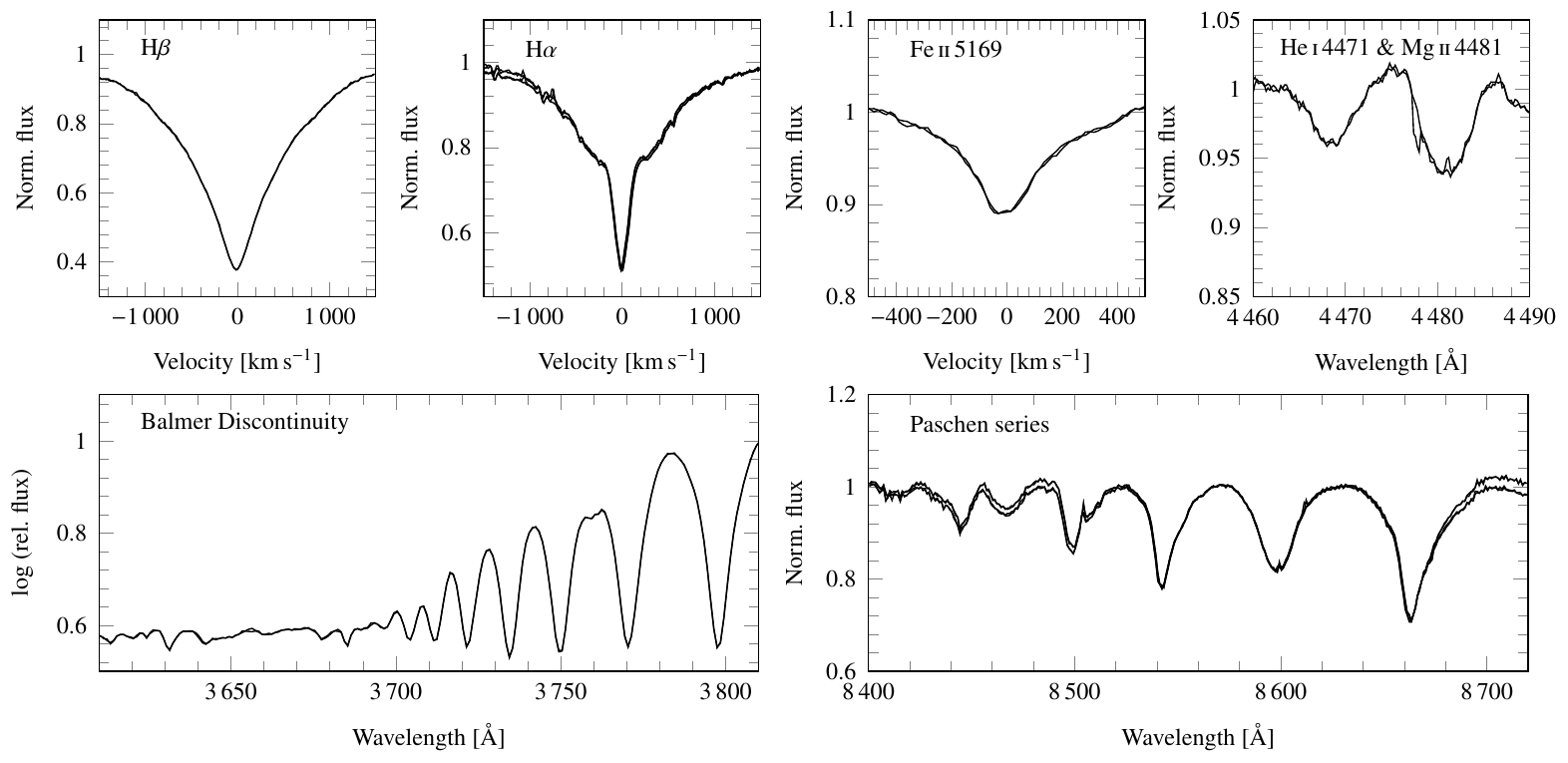}
\end{center}
\caption{Spectrum overview plot for Hip\,59970}
\end{figure*}

\begin{figure*}
\begin{center}
\includegraphics[angle=0,width=14cm,clip]{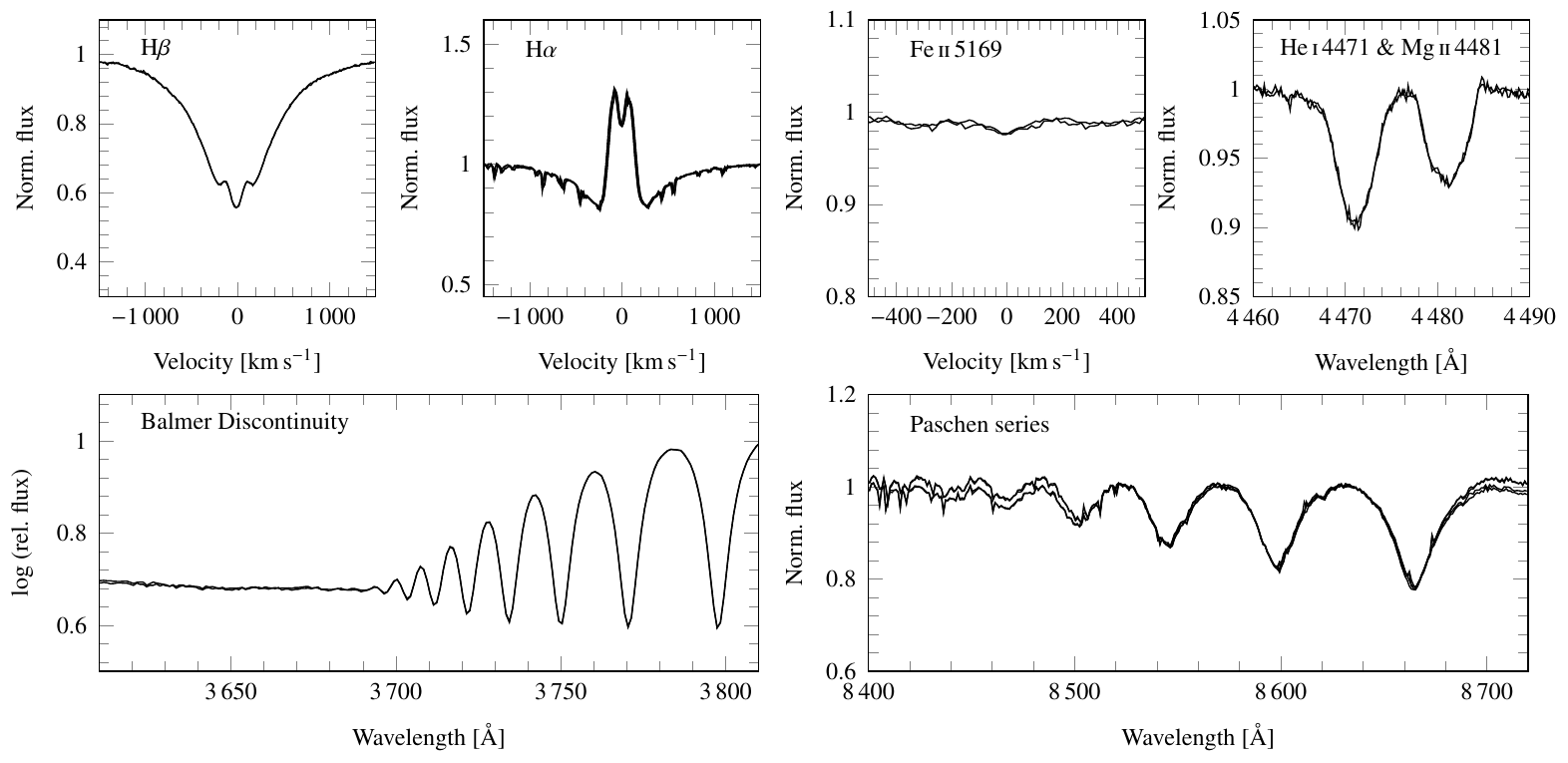}
\end{center}
\caption{Spectrum overview plot for Hip\,64501}
\end{figure*}

\begin{figure*}
\begin{center}
\includegraphics[angle=0,width=14cm,clip]{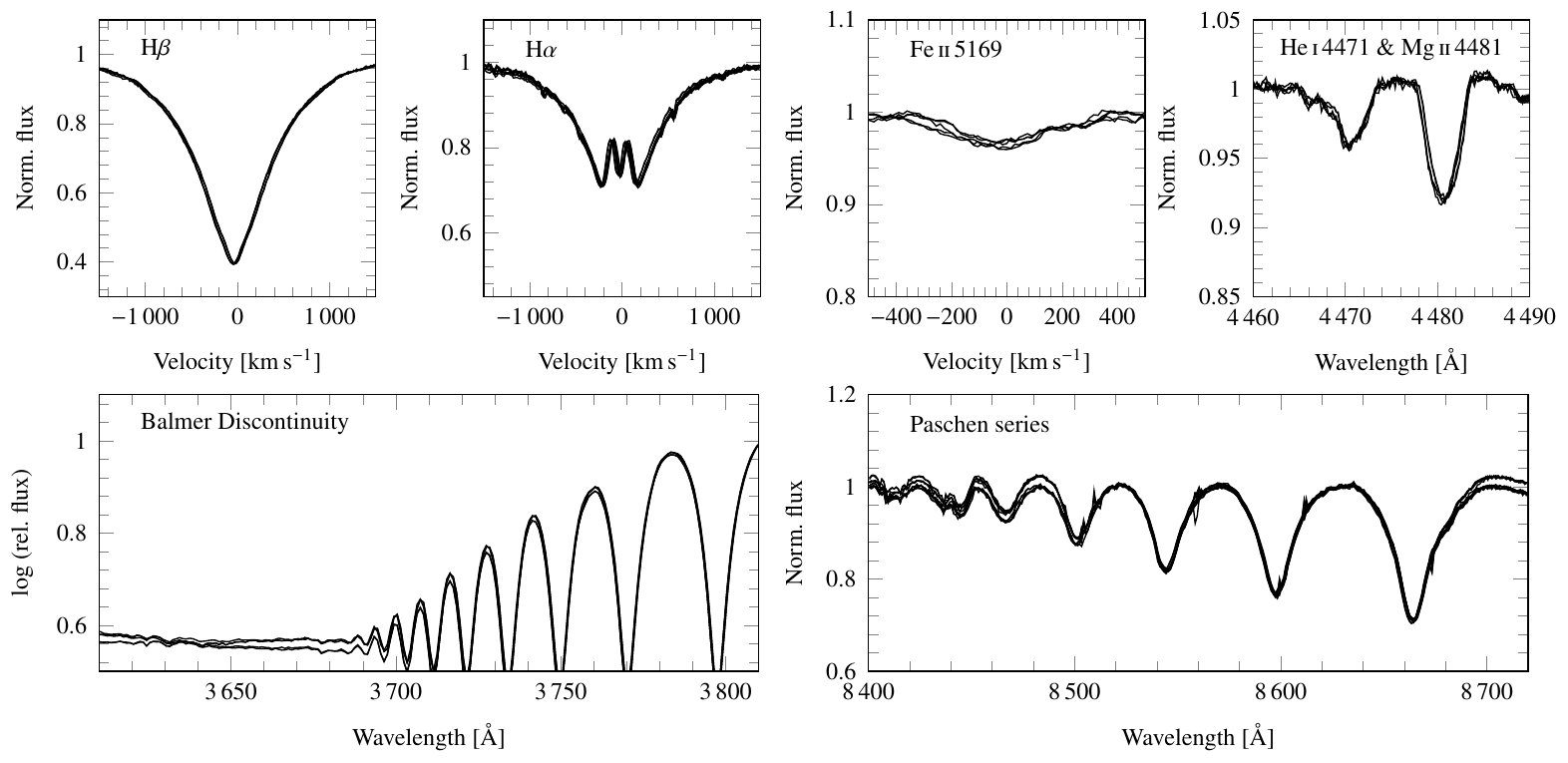}
\end{center}
\caption{Spectrum overview plot for Hip\,64867}
\end{figure*}

\clearpage

\begin{figure*}
\begin{center}
\includegraphics[angle=0,width=14cm,clip]{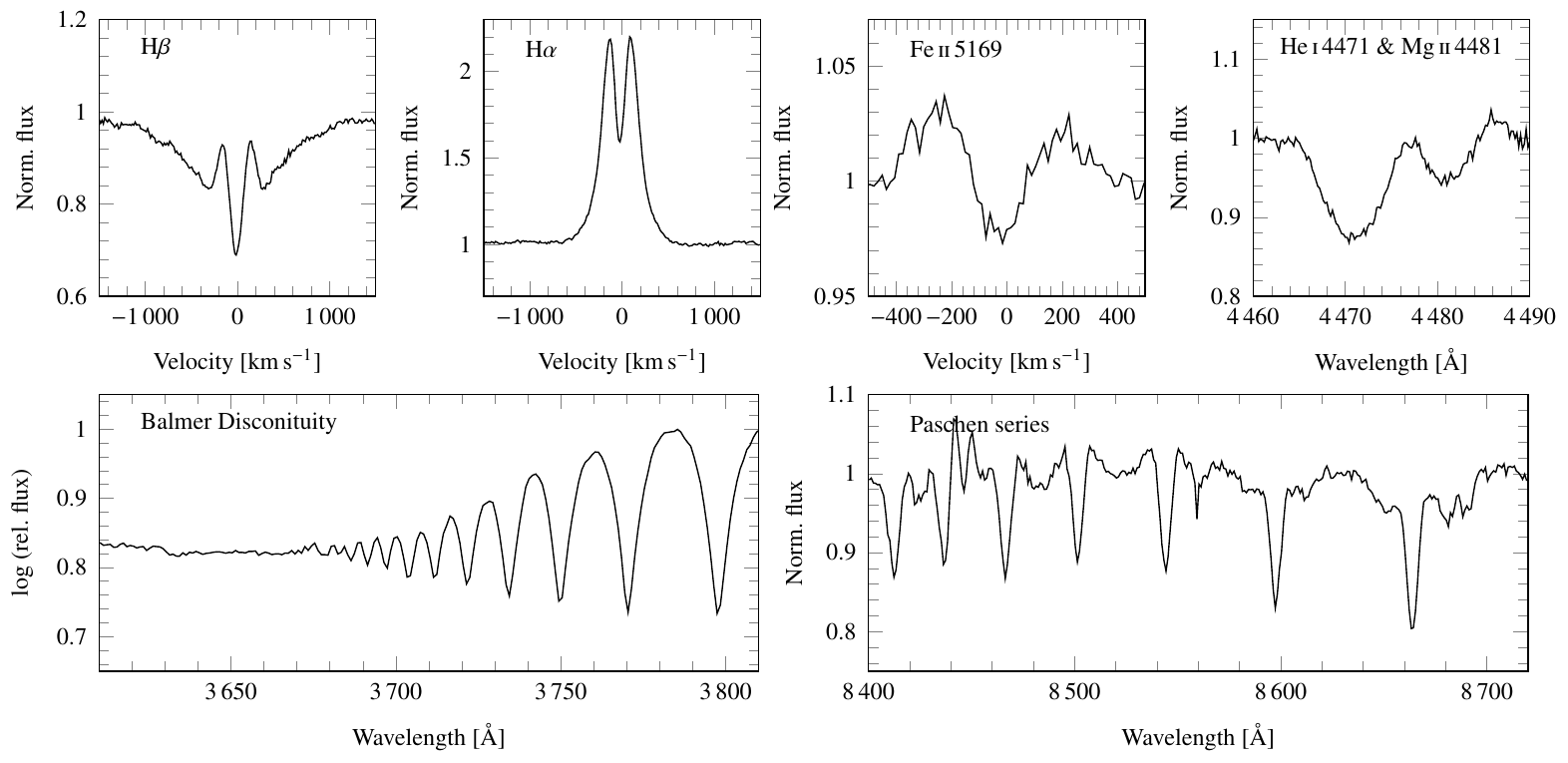}
\end{center}
\caption{Spectrum overview plot for Hip\,66339}
\end{figure*}

\begin{figure*}
\begin{center}
\includegraphics[angle=0,width=14cm,clip]{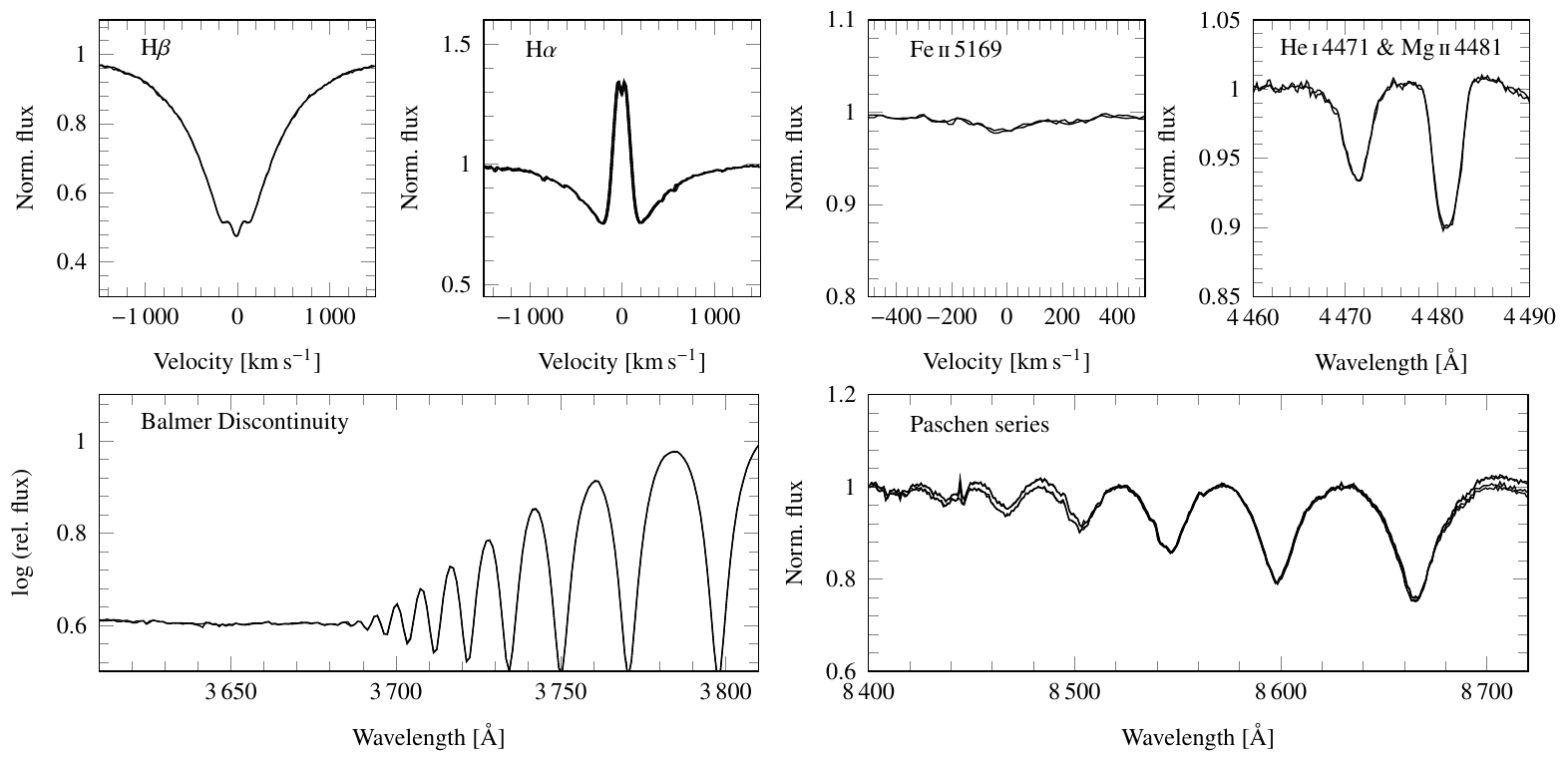}
\end{center}
\caption{Spectrum overview plot for Hip\,66351}
\end{figure*}

\begin{figure*}
\begin{center}
\includegraphics[angle=0,width=14cm,clip]{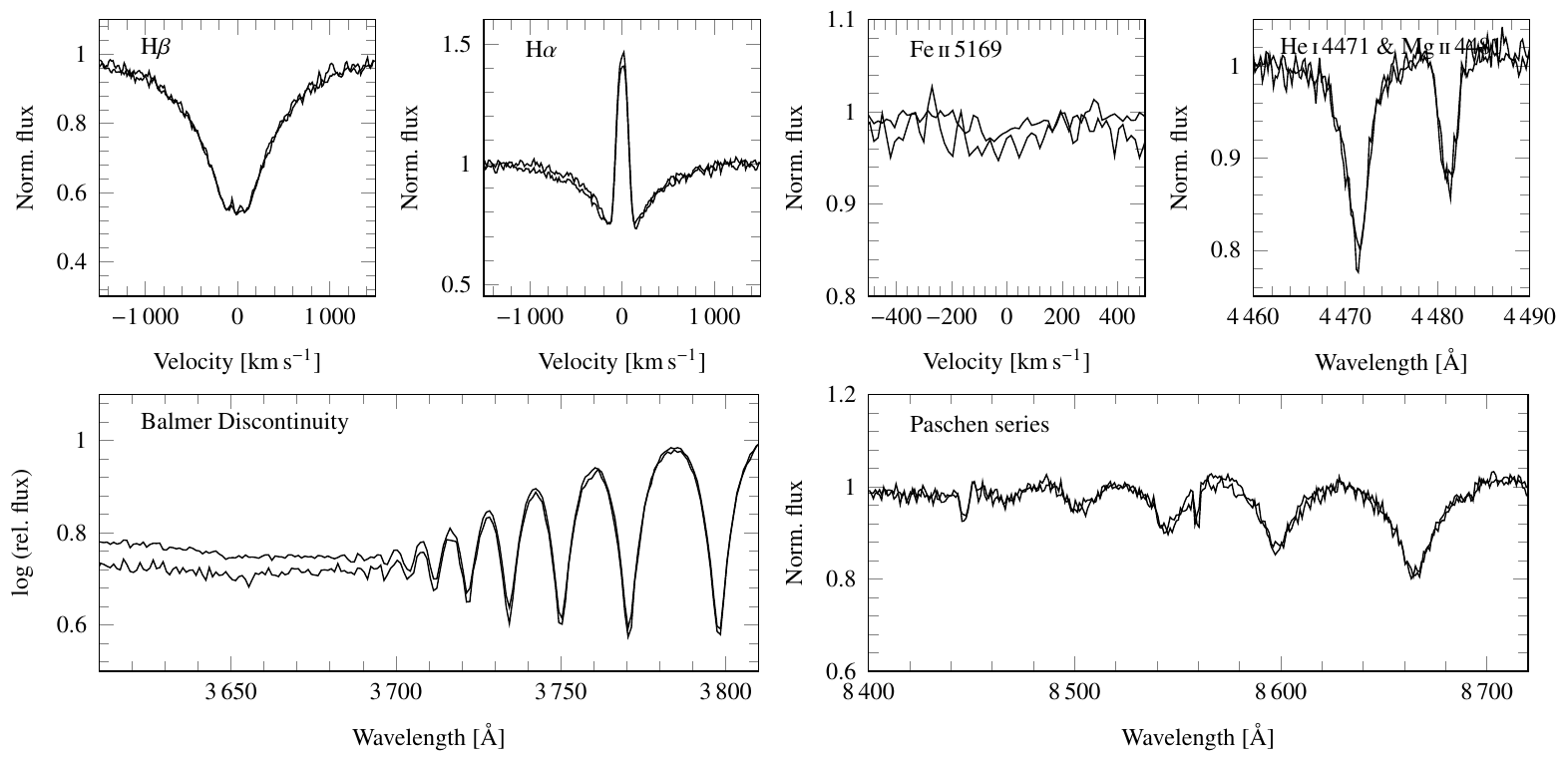}
\end{center}
\caption{Spectrum overview plot for Hip\,68100}
\end{figure*}

\clearpage

\begin{figure*}
\begin{center}
\includegraphics[angle=0,width=14cm,clip]{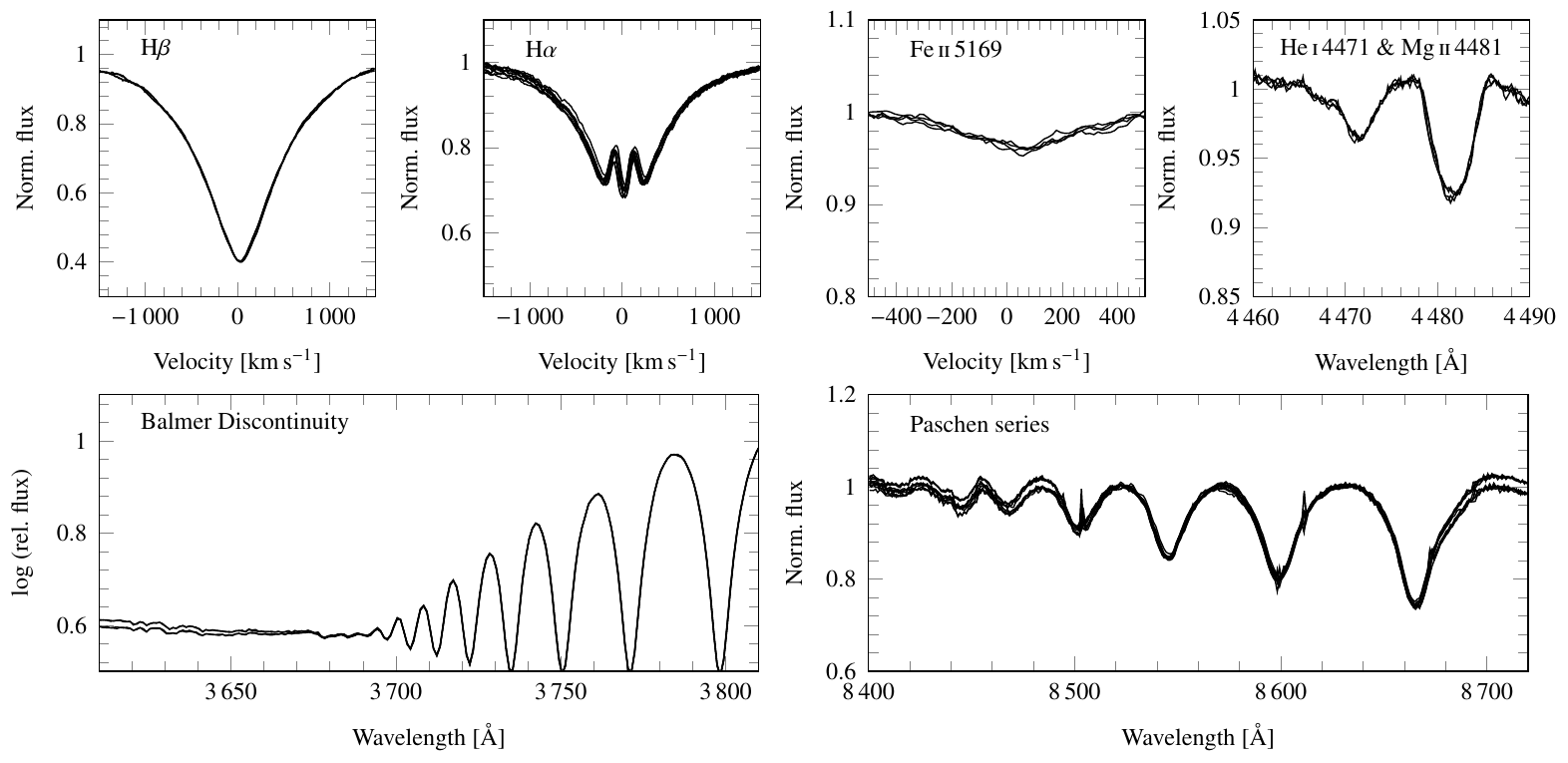}
\end{center}
\caption{Spectrum overview plot for Hip\,69429}
\end{figure*}

\begin{figure*}
\begin{center}
\includegraphics[angle=0,width=14cm,clip]{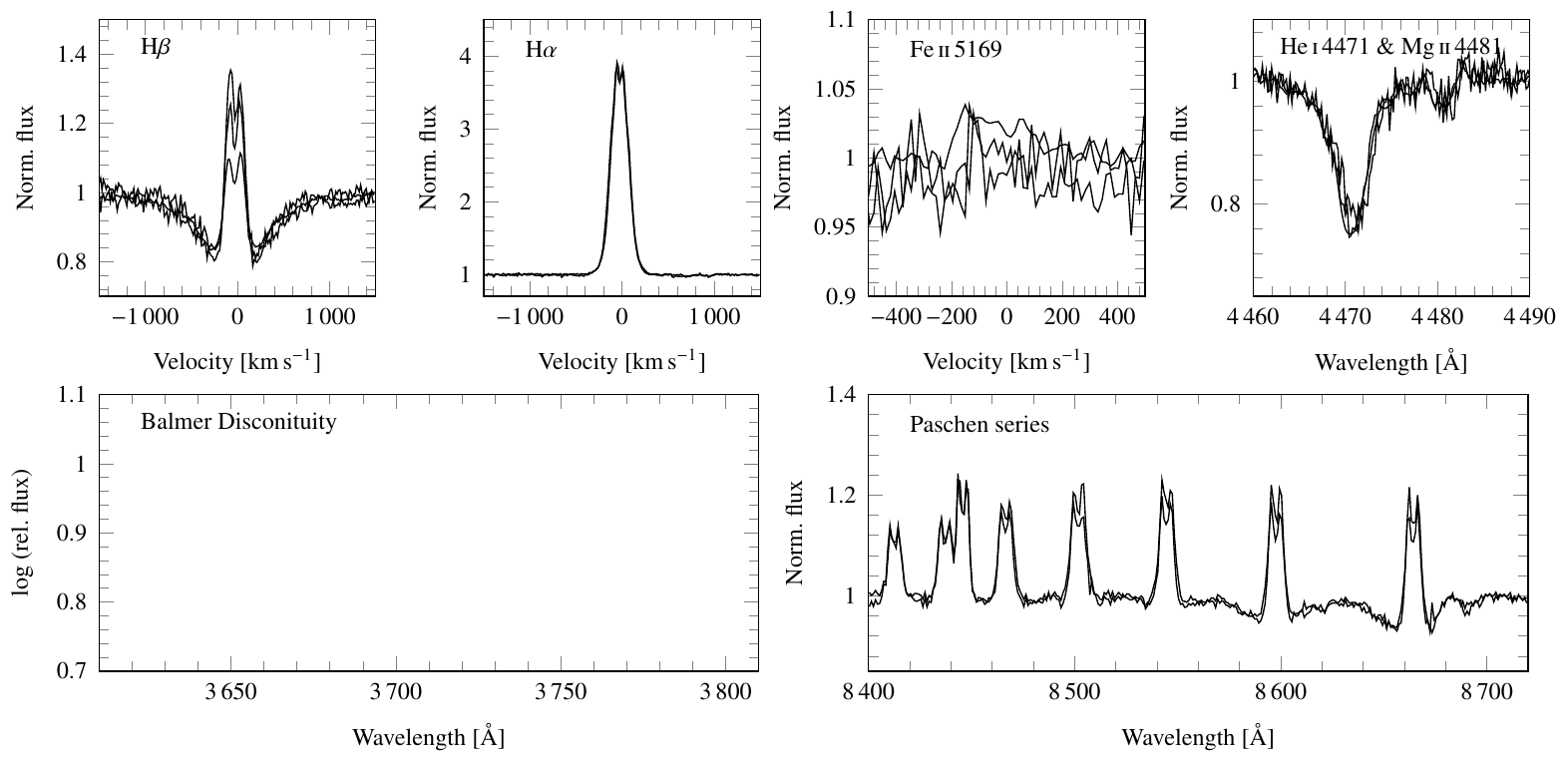}
\end{center}
\caption{Spectrum overview plot for Hip\,71668}
\end{figure*}

\begin{figure*}
\begin{center}
\includegraphics[angle=0,width=14cm,clip]{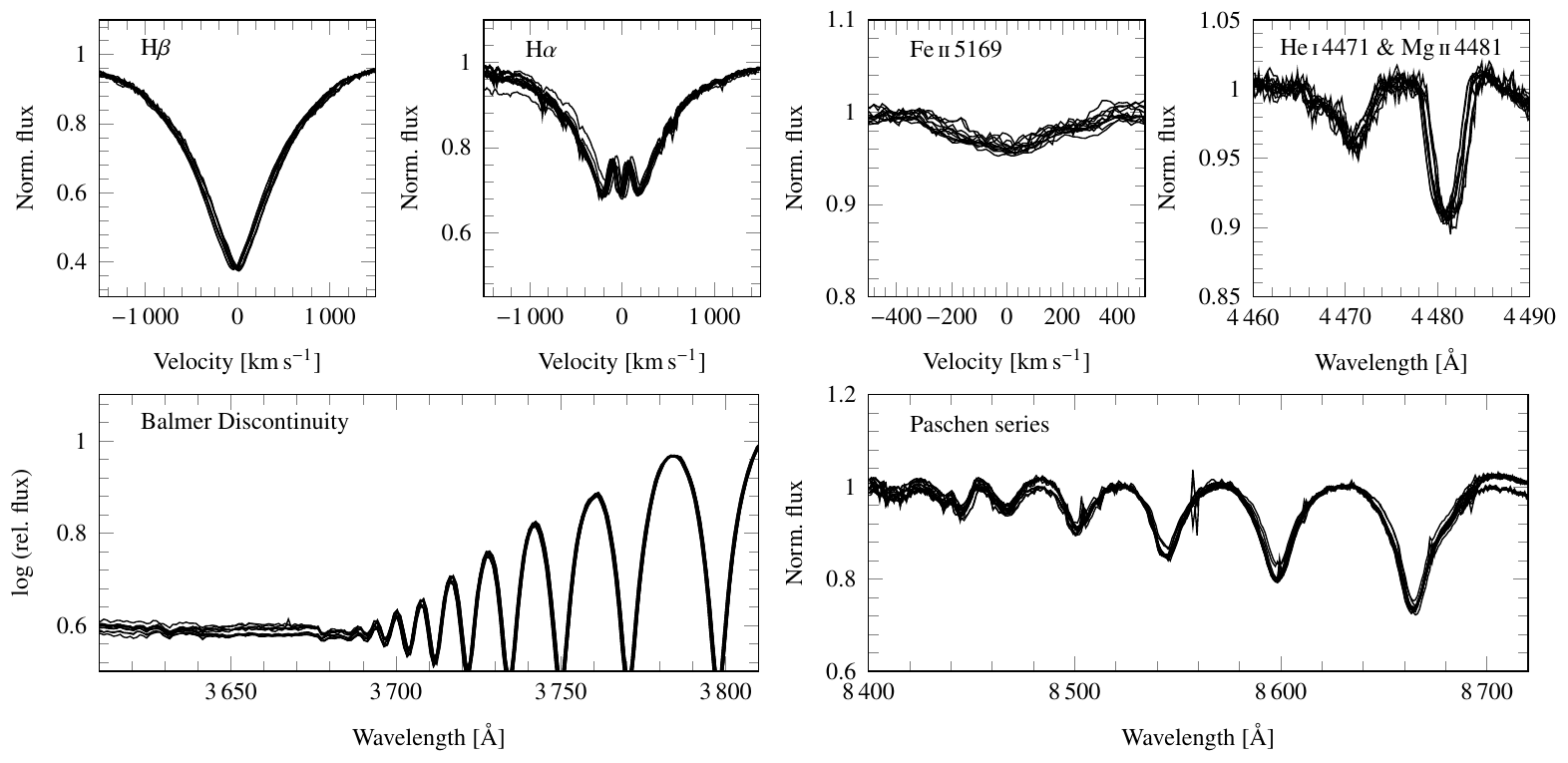}
\end{center}
\caption{Spectrum overview plot for Hip\,71974}
\end{figure*}

\clearpage

\begin{figure*}
\begin{center}
\includegraphics[angle=0,width=14cm,clip]{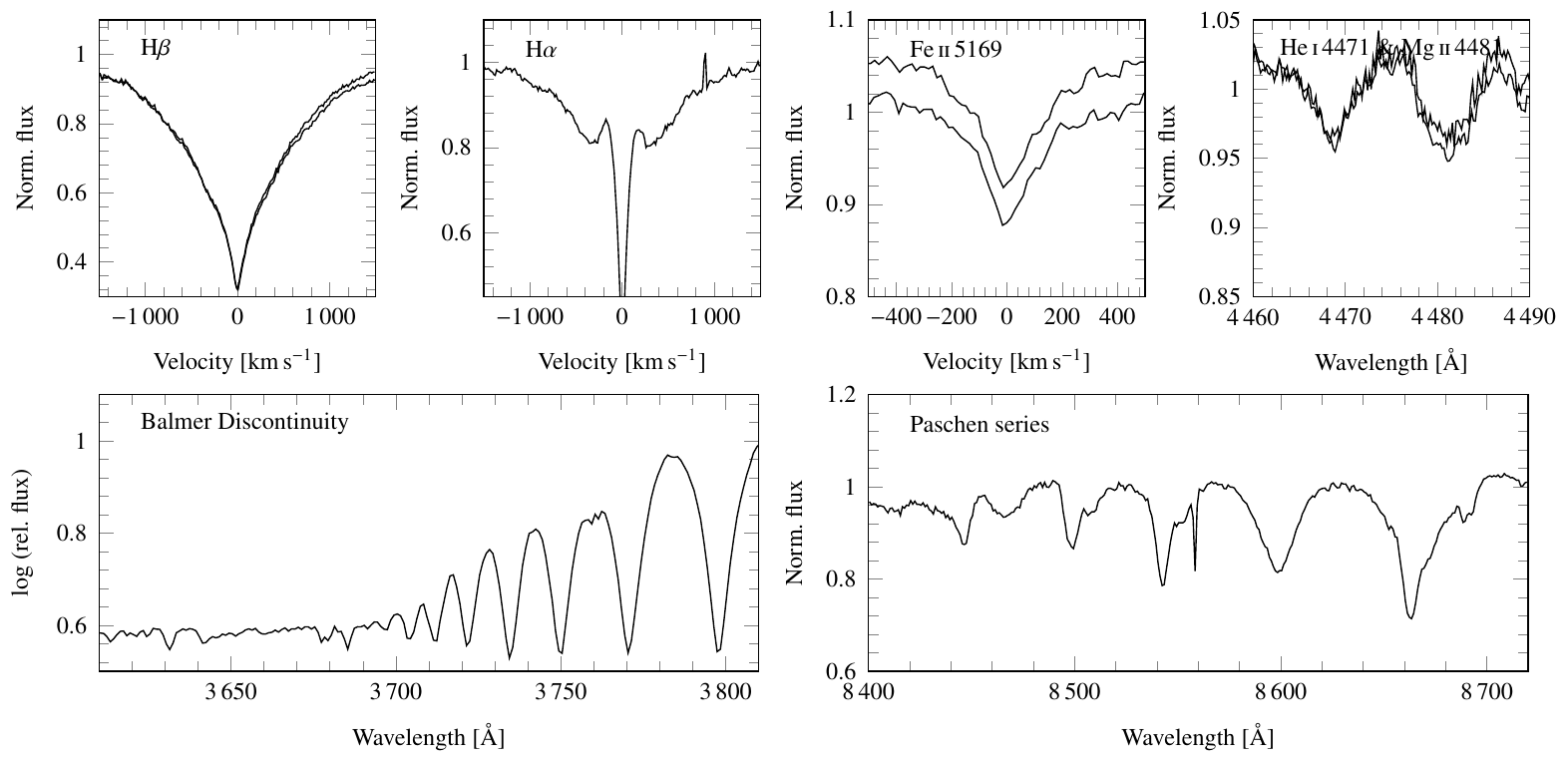}
\end{center}
\caption{Spectrum overview plot for Hip\,78375}
\end{figure*}

\begin{figure*}
\begin{center}
\includegraphics[angle=0,width=14cm,clip]{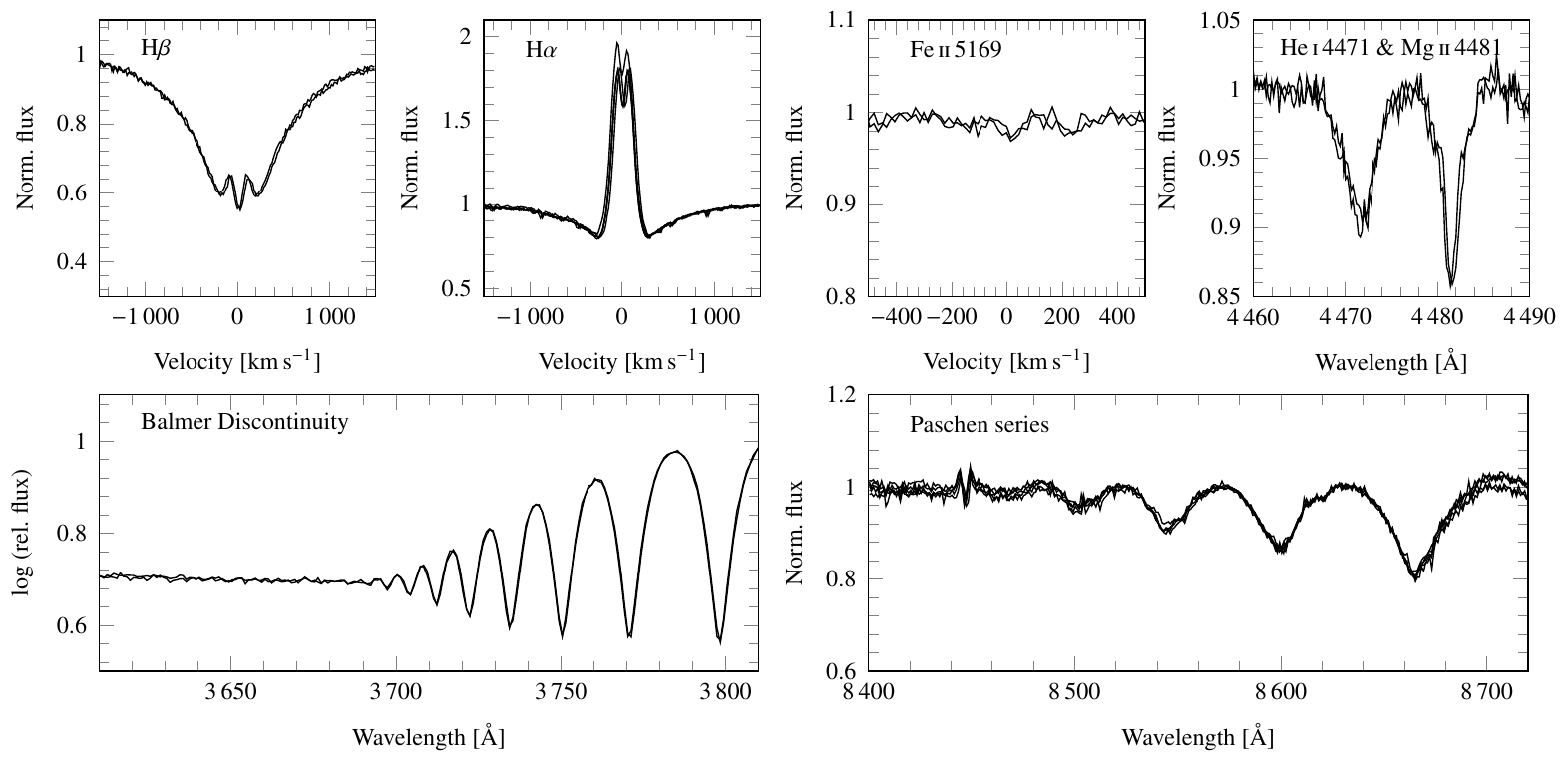}
\end{center}
\caption{Spectrum overview plot for Hip\,80577}
\end{figure*}

\begin{figure*}
\begin{center}
\includegraphics[angle=0,width=14cm,clip]{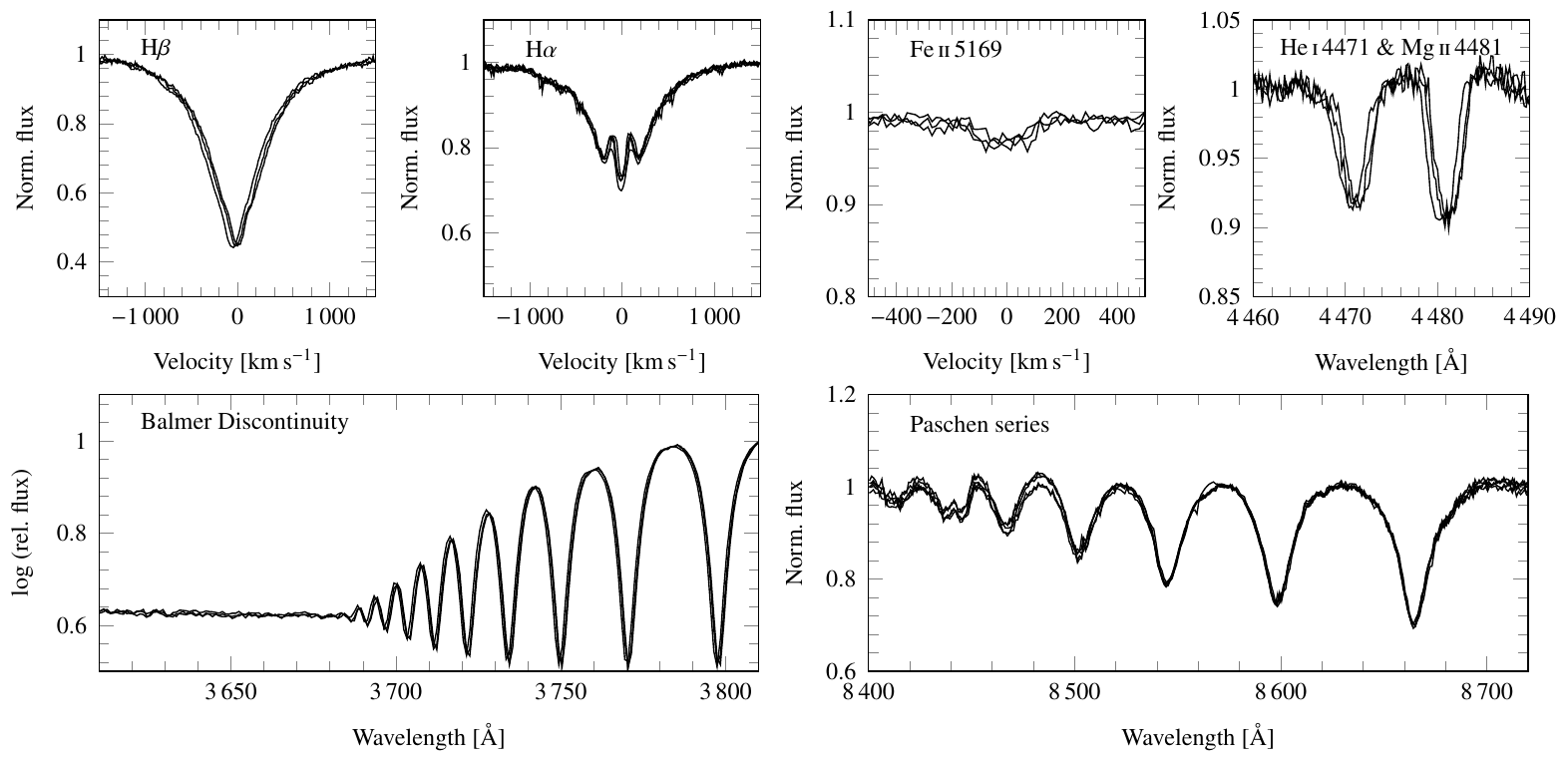}
\end{center}
\caption{Spectrum overview plot for Hip\,80820}
\end{figure*}

\clearpage

\begin{figure*}
\begin{center}
\includegraphics[angle=0,width=14cm,clip]{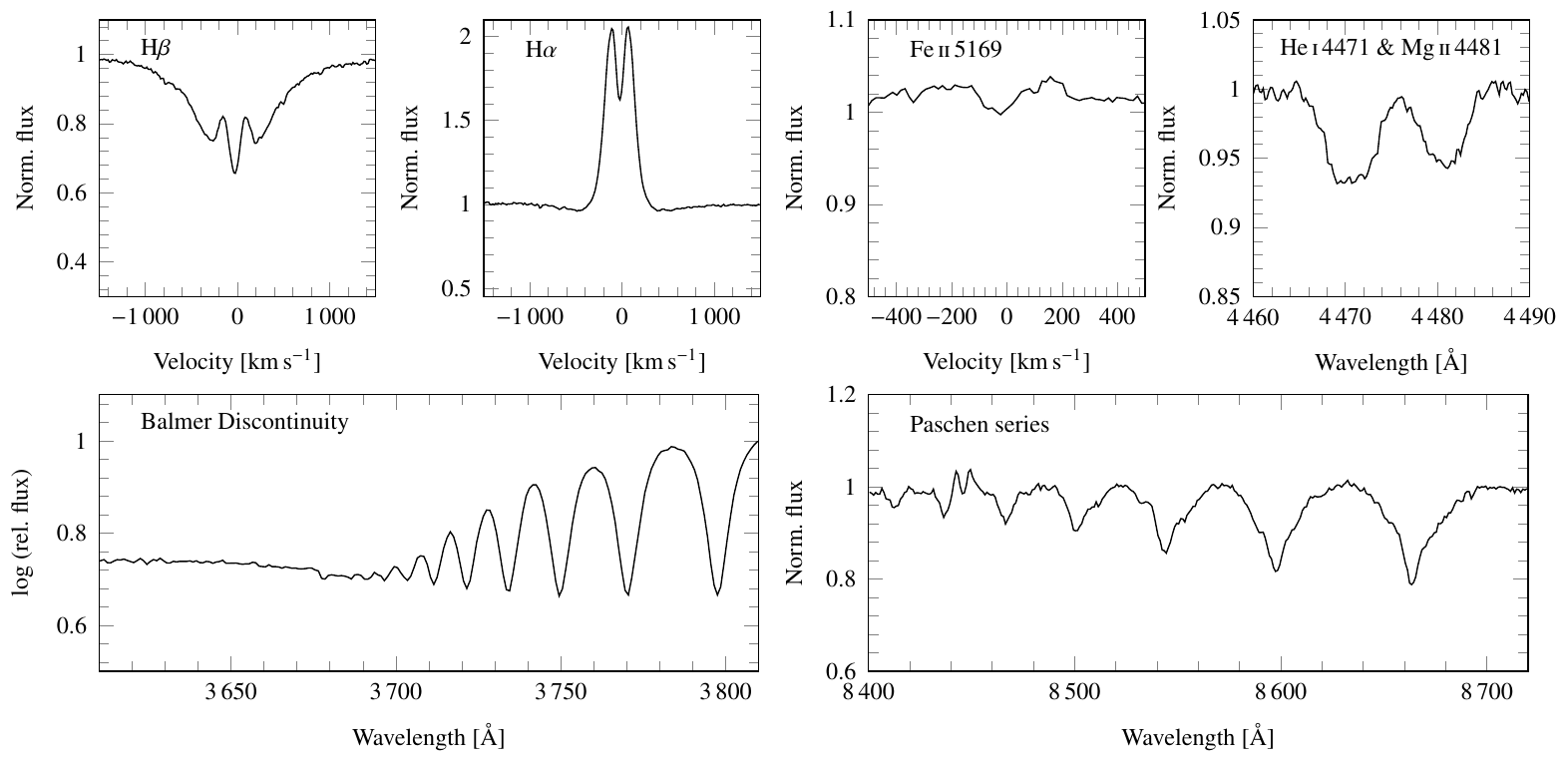}
\end{center}
\caption{Spectrum overview plot for Hip\,82874}
\end{figure*}

\begin{figure*}
\begin{center}
\includegraphics[angle=0,width=14cm,clip]{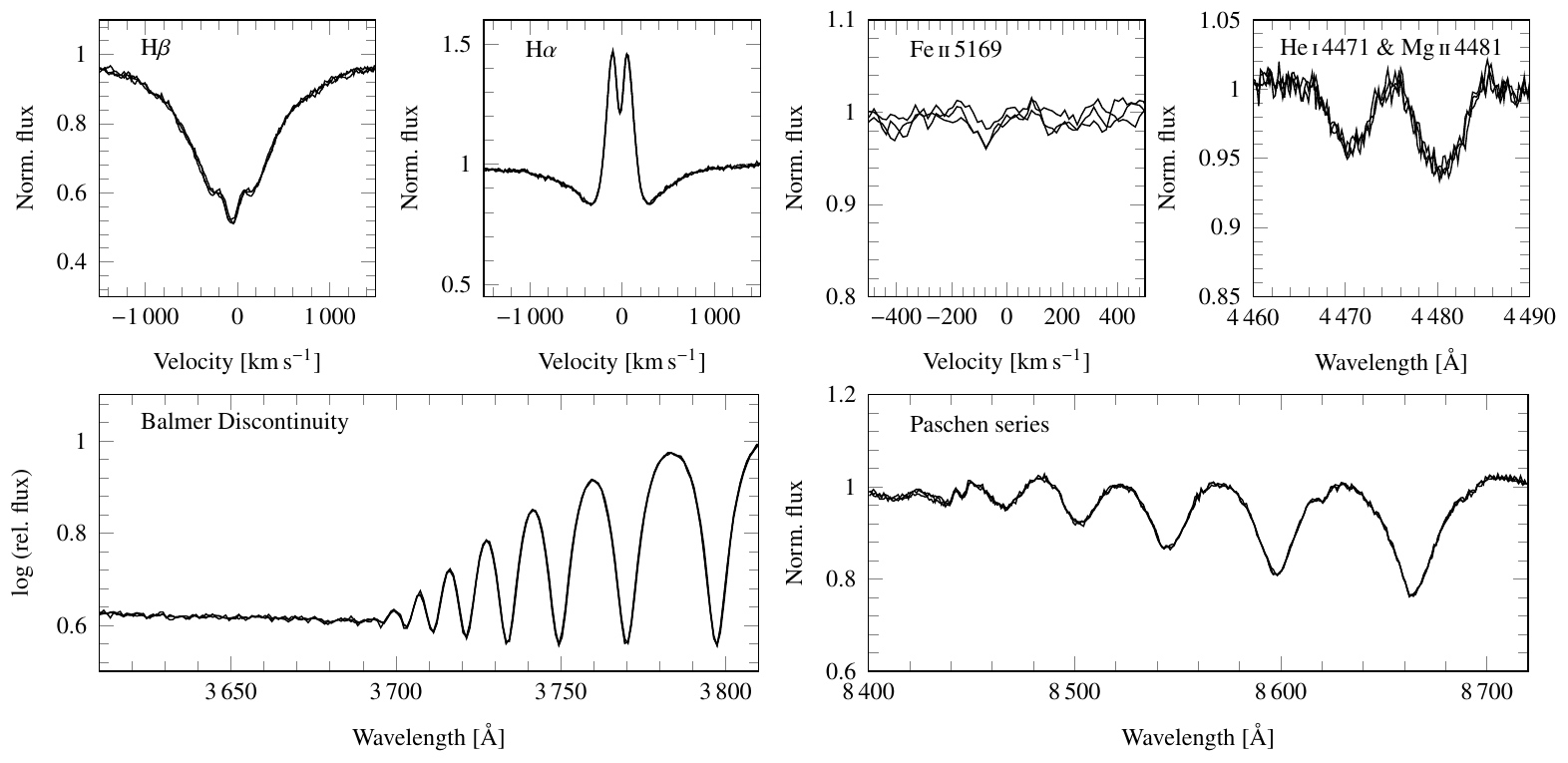}
\end{center}
\caption{Spectrum overview plot for Hip\,83278}
\end{figure*}

\begin{figure*}
\begin{center}
\includegraphics[angle=0,width=14cm,clip]{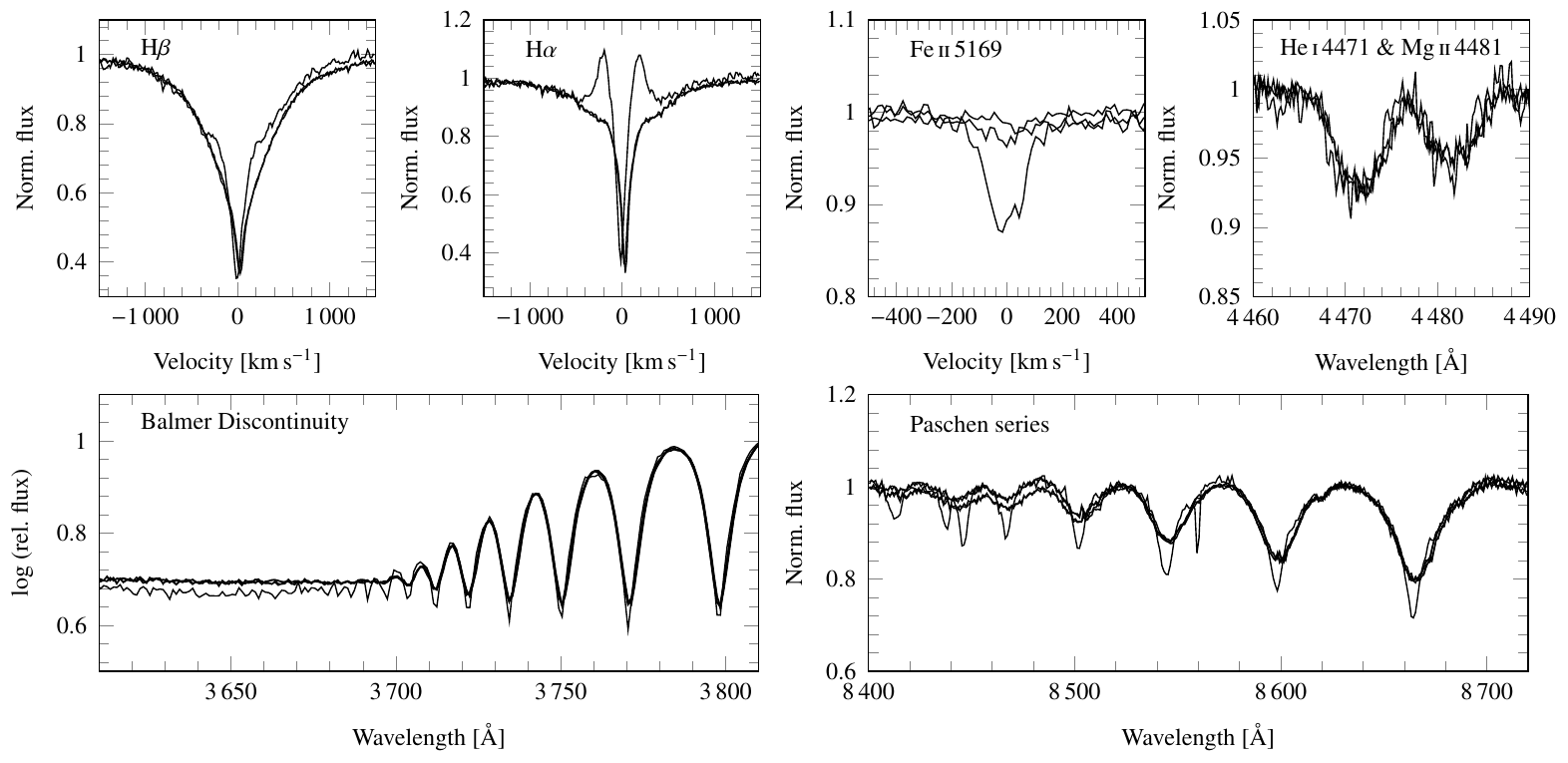}
\end{center}
\caption{Spectrum overview plot for Hip\,84184}
\end{figure*}

\clearpage

\begin{figure*}
\begin{center}
\includegraphics[angle=0,width=14cm,clip]{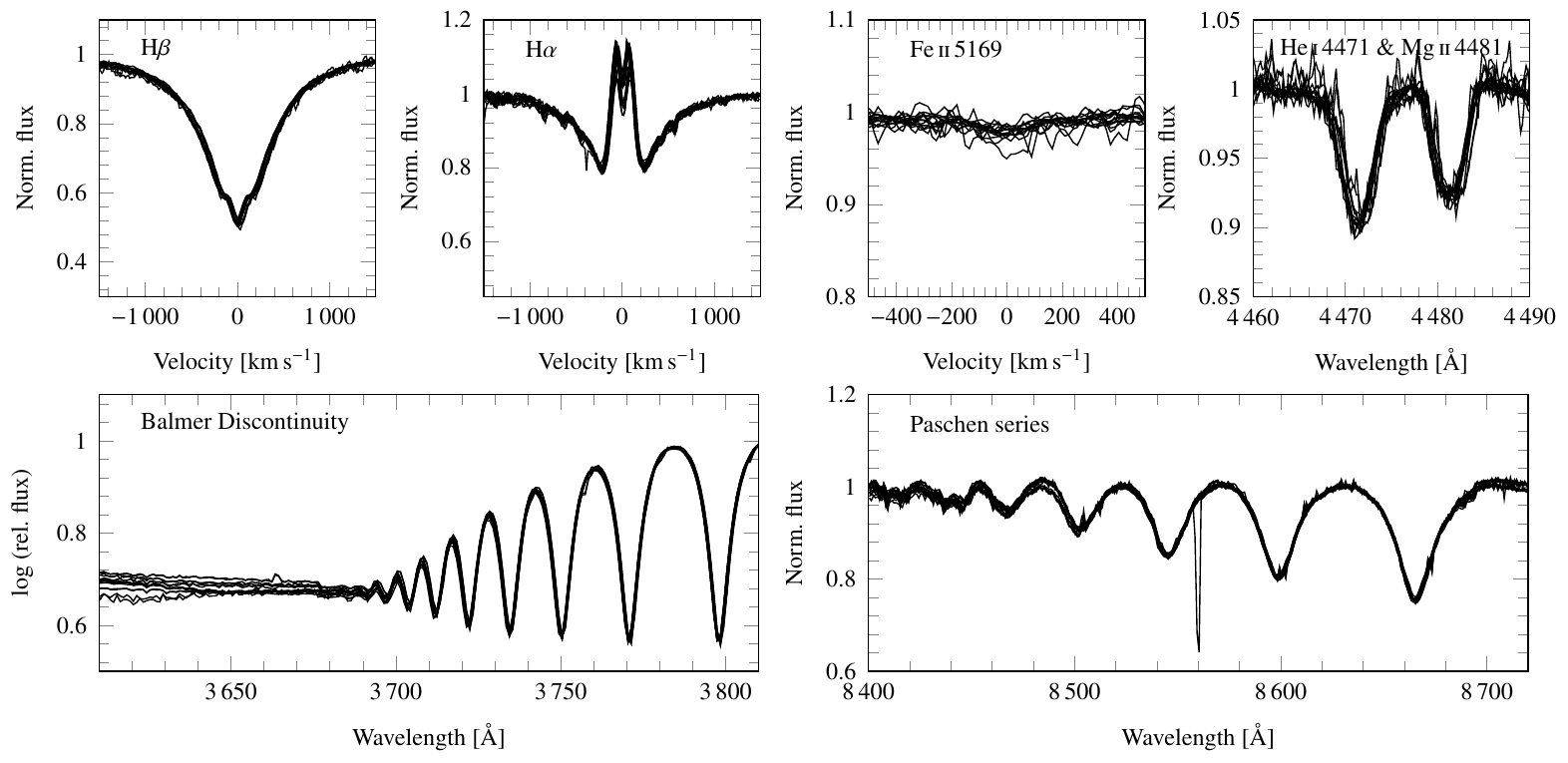}
\end{center}
\caption{Spectrum overview plot for Hip\,85138}
\end{figure*}

\begin{figure*}
\begin{center}
\includegraphics[angle=0,width=14cm,clip]{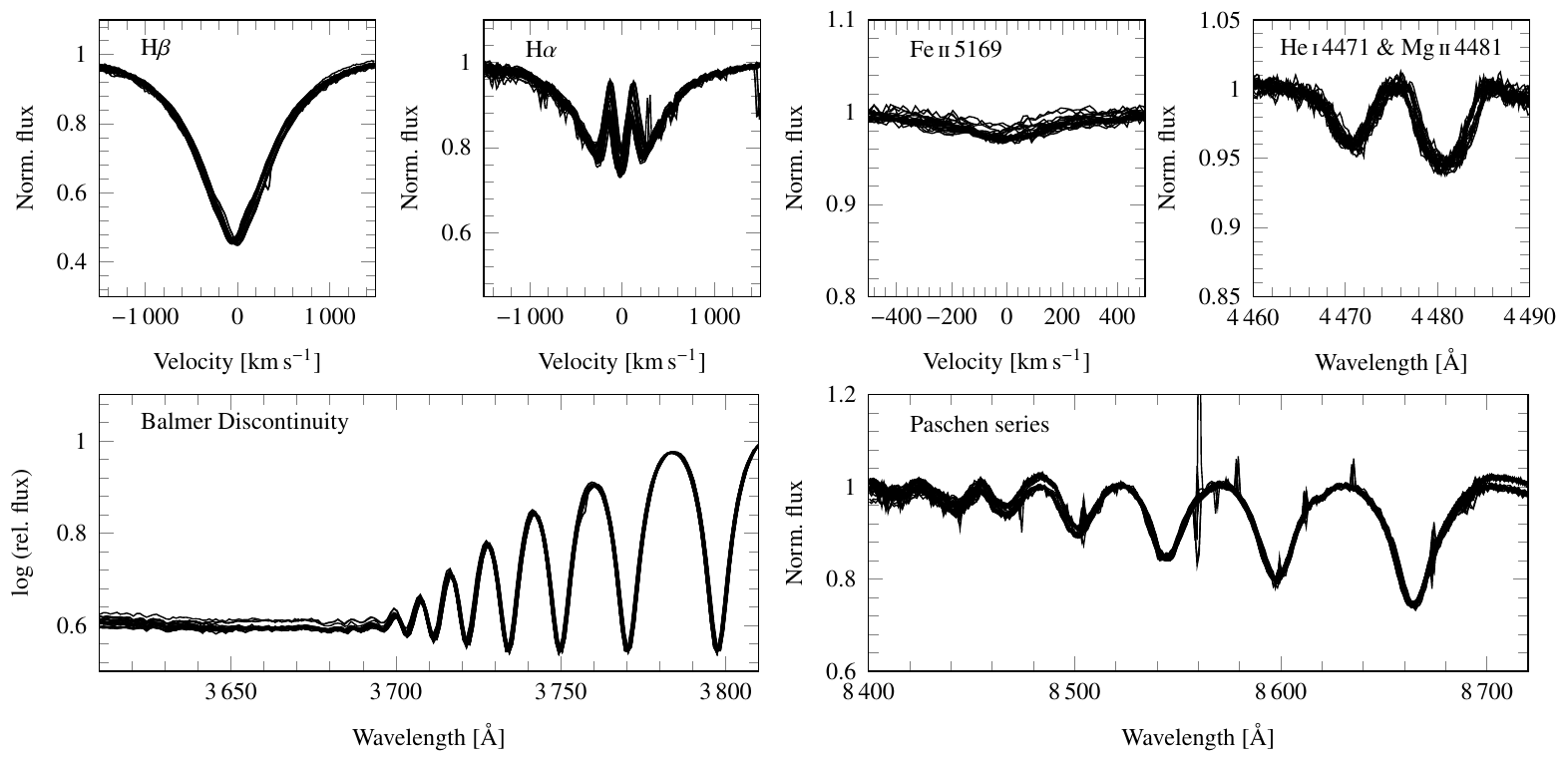}
\end{center}
\caption{Spectrum overview plot for Hip\,85195}
\end{figure*}

\begin{figure*}
\begin{center}
\includegraphics[angle=0,width=14cm,clip]{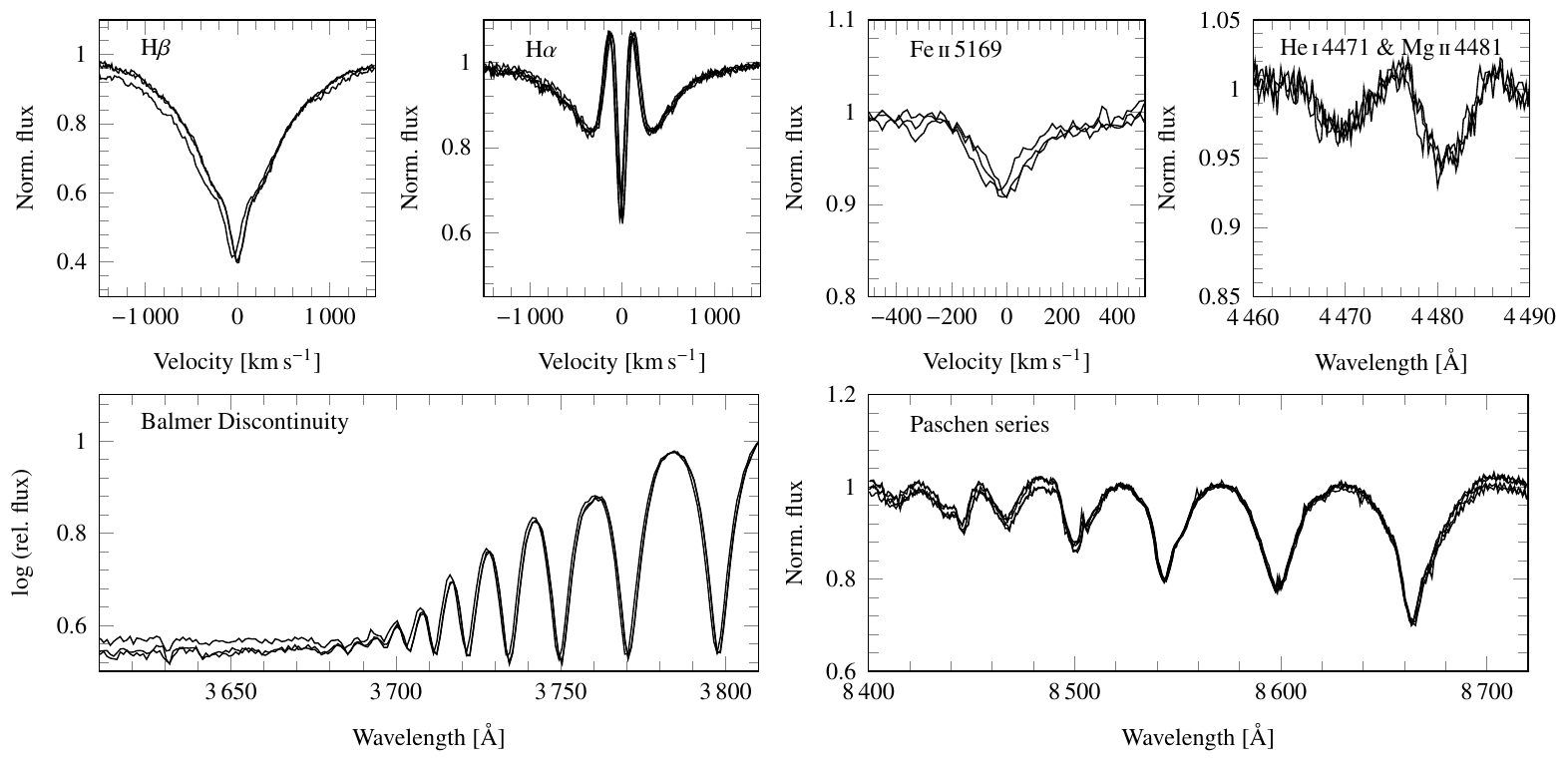}
\end{center}
\caption{Spectrum overview plot for Hip\,85566}
\end{figure*}

\clearpage

\begin{figure*}
\begin{center}
\includegraphics[angle=0,width=14cm,clip]{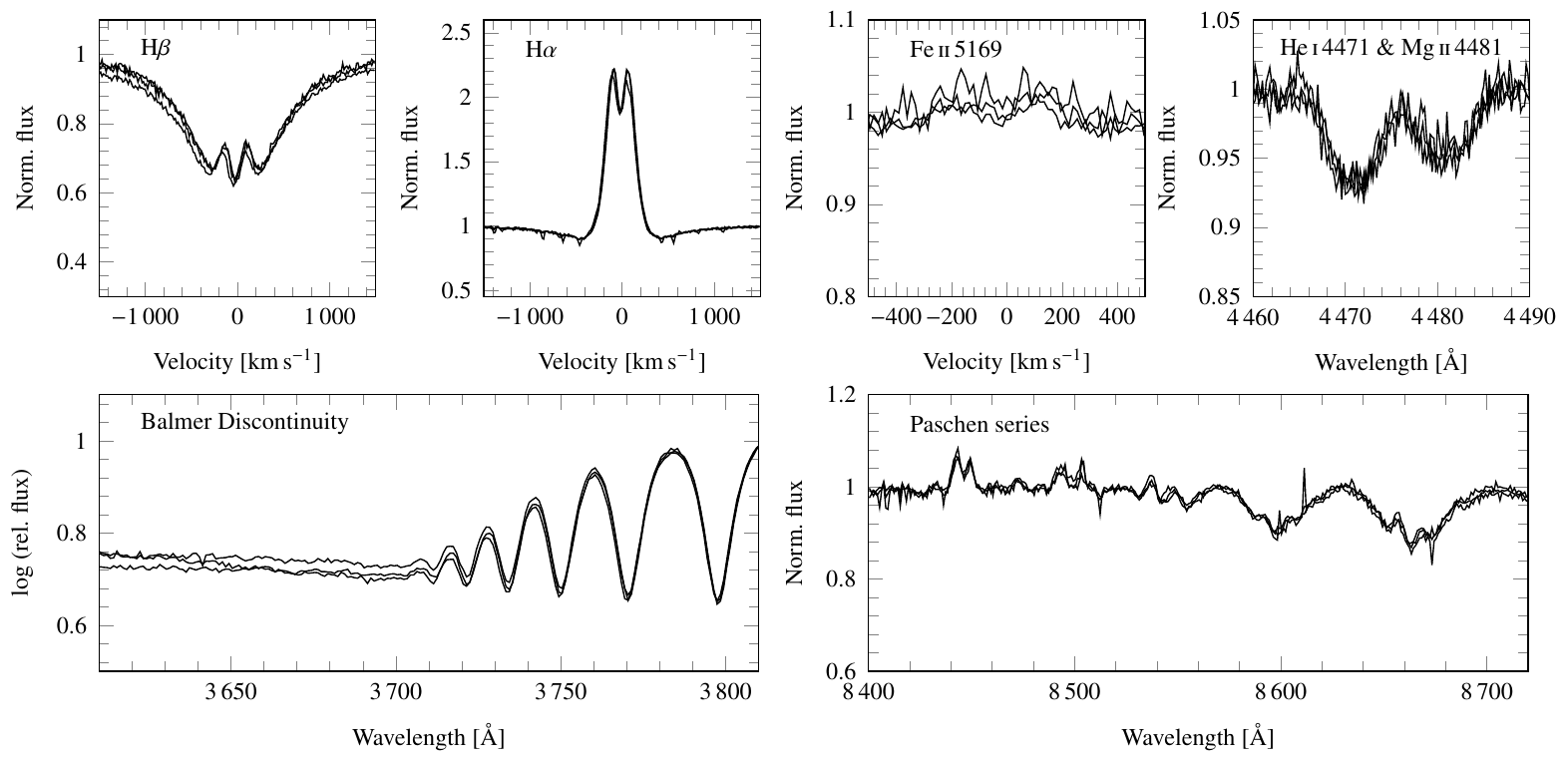}
\end{center}
\caption{Spectrum overview plot for Hip\,87032}
\end{figure*}

\begin{figure*}
\begin{center}
\includegraphics[angle=0,width=14cm,clip]{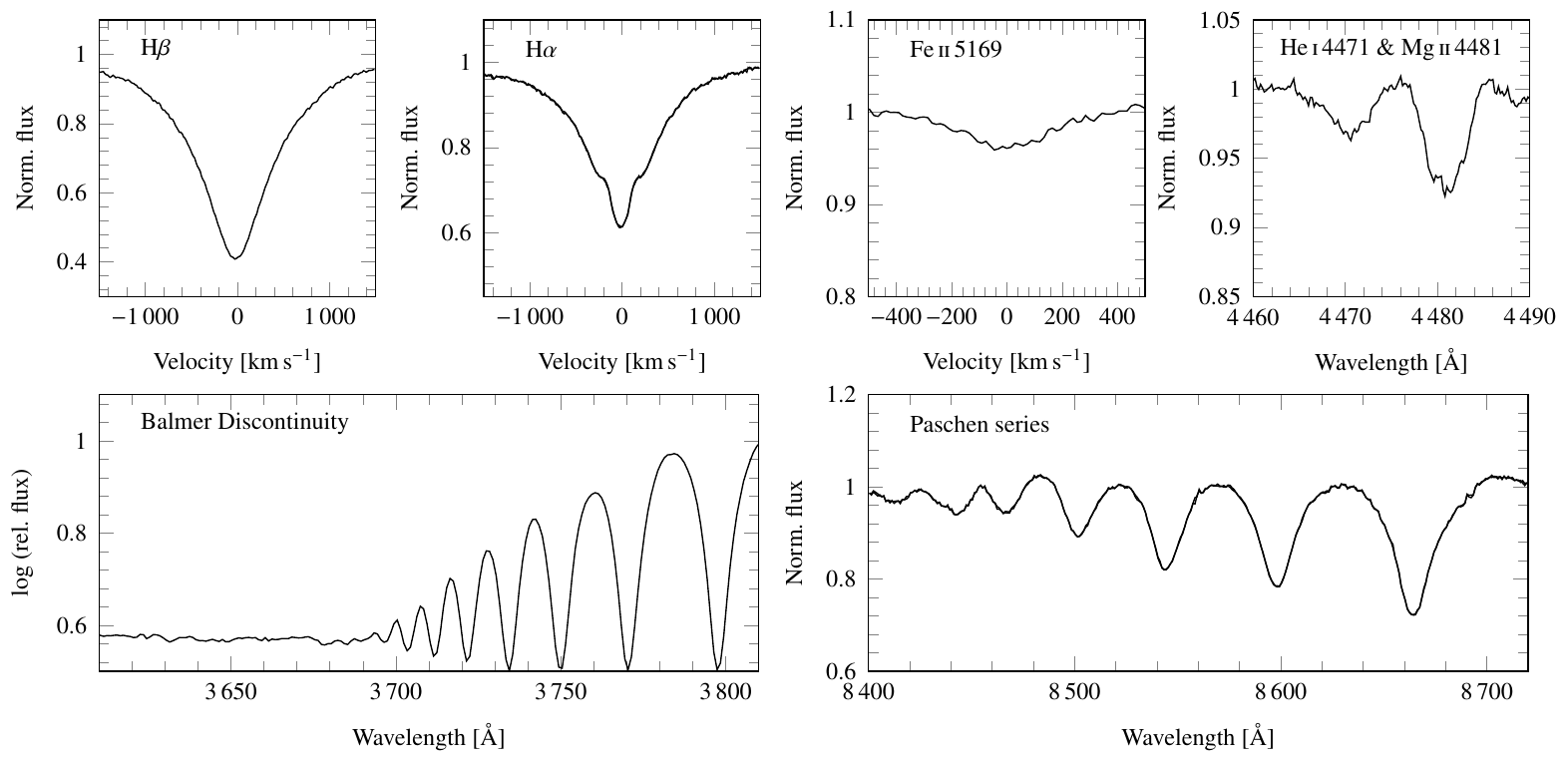}
\end{center}
\caption{Spectrum overview plot for Hip\,87698}
\end{figure*}

\begin{figure*}
\begin{center}
\includegraphics[angle=0,width=14cm,clip]{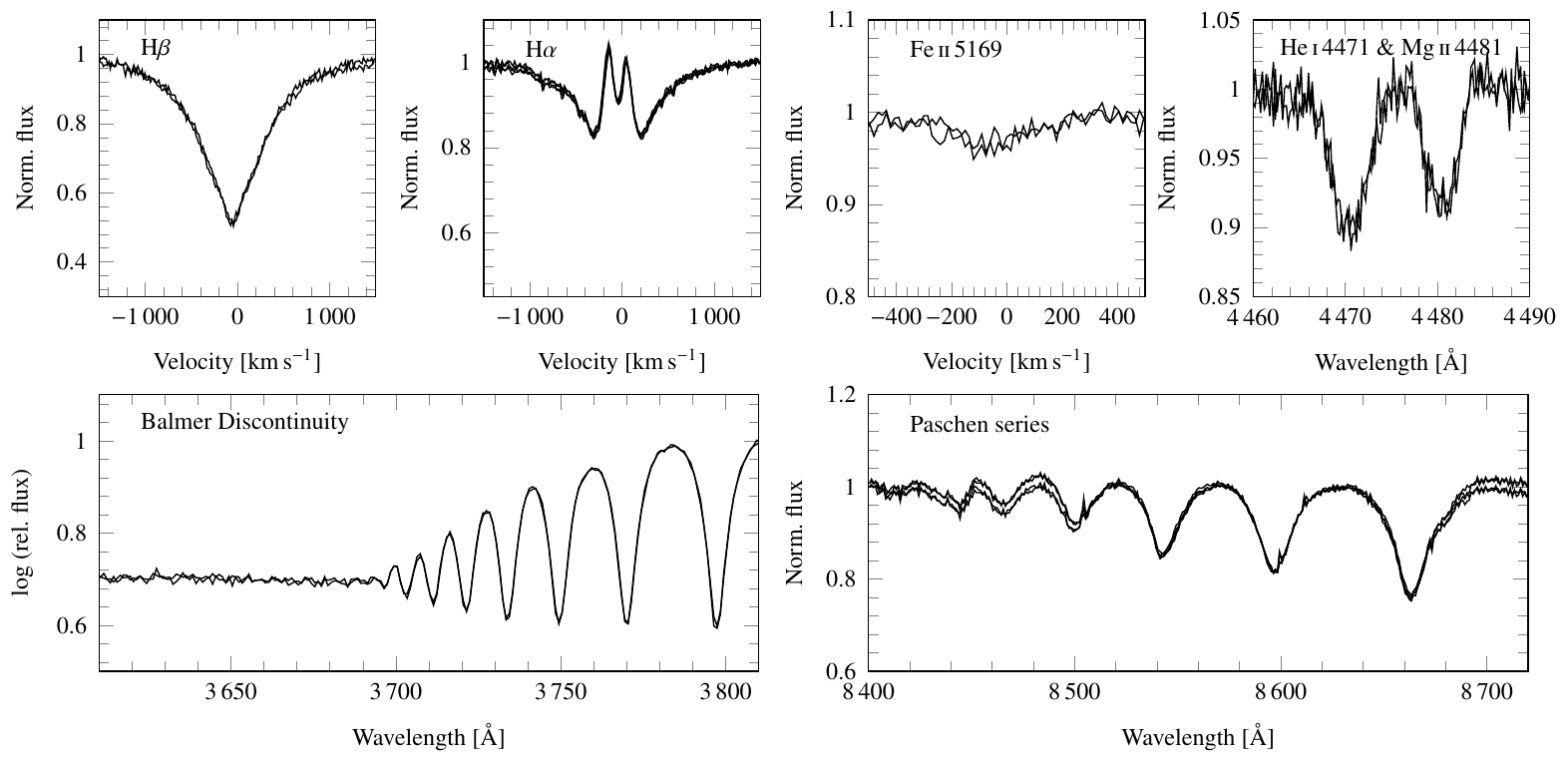}
\end{center}
\caption{Spectrum overview plot for Hip\,88172}
\end{figure*}
\clearpage

\begin{figure*}
\begin{center}
\includegraphics[angle=0,width=14cm,clip]{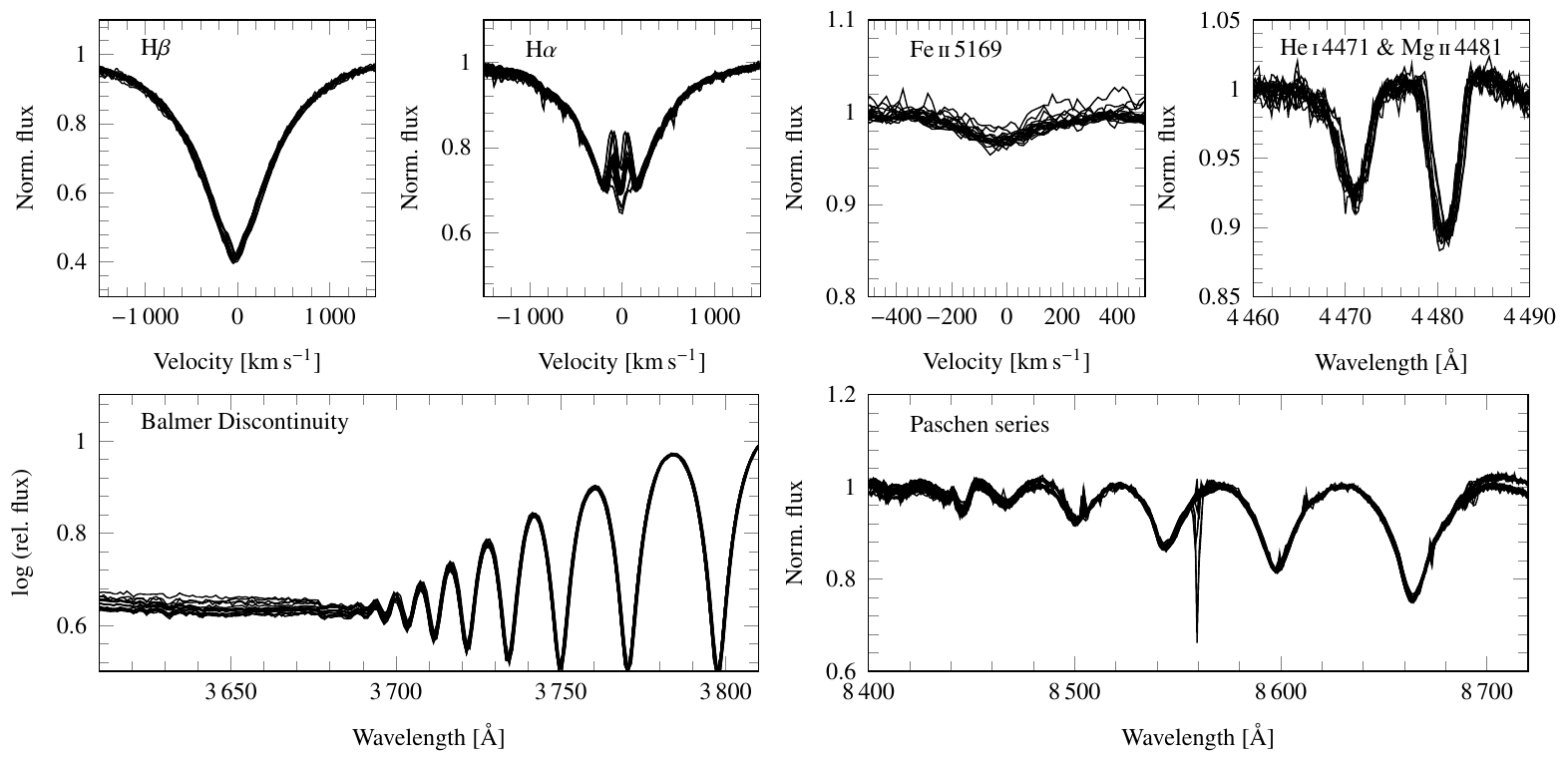}
\end{center}
\caption{Spectrum overview plot for Hip\,88374}
\end{figure*}

\begin{figure*}
\begin{center}
\includegraphics[angle=0,width=14cm,clip]{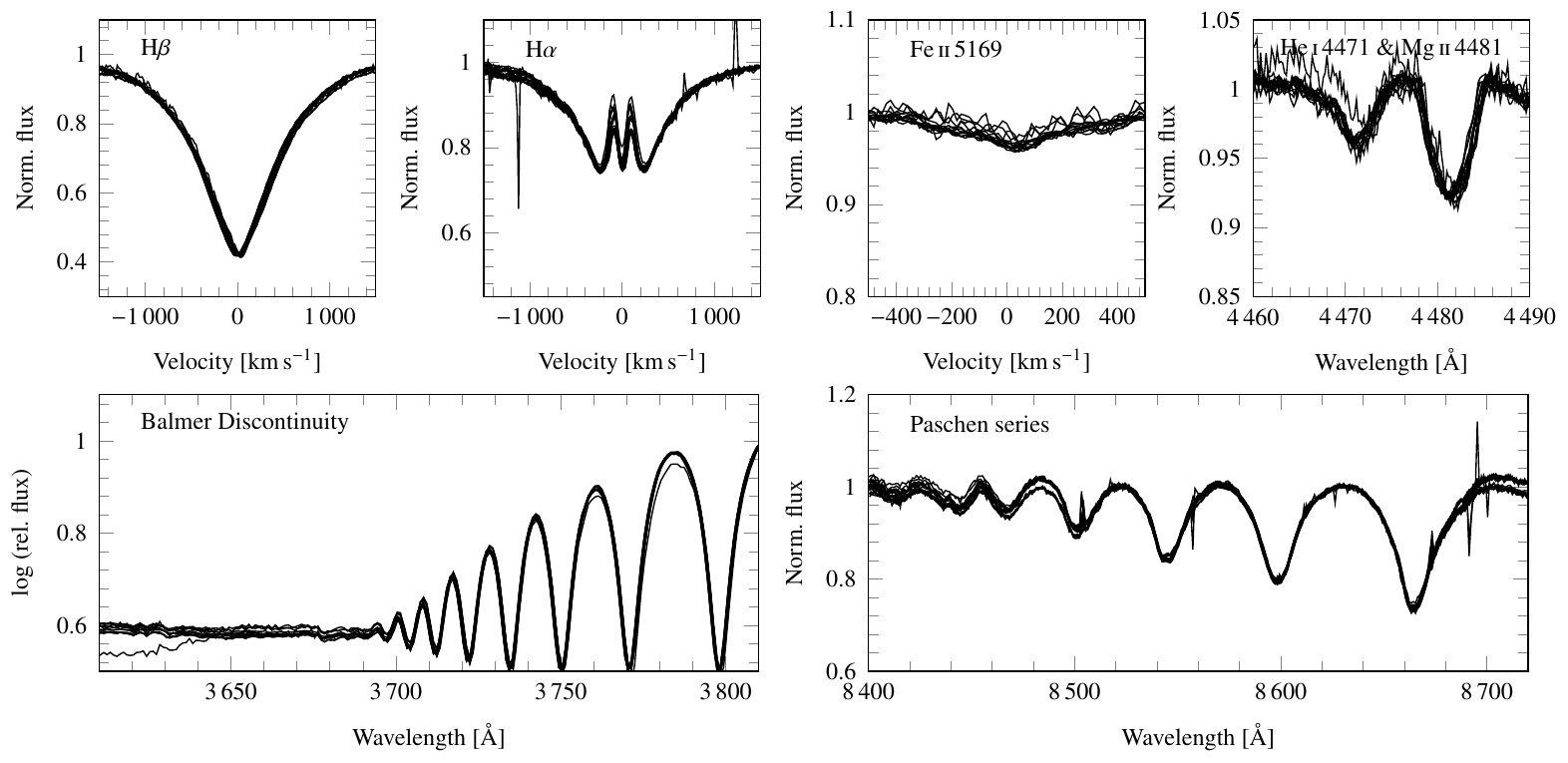}
\end{center}
\caption{Spectrum overview plot for Hip\,89486}
\end{figure*}

\begin{figure*}
\begin{center}
\includegraphics[angle=0,width=14cm,clip]{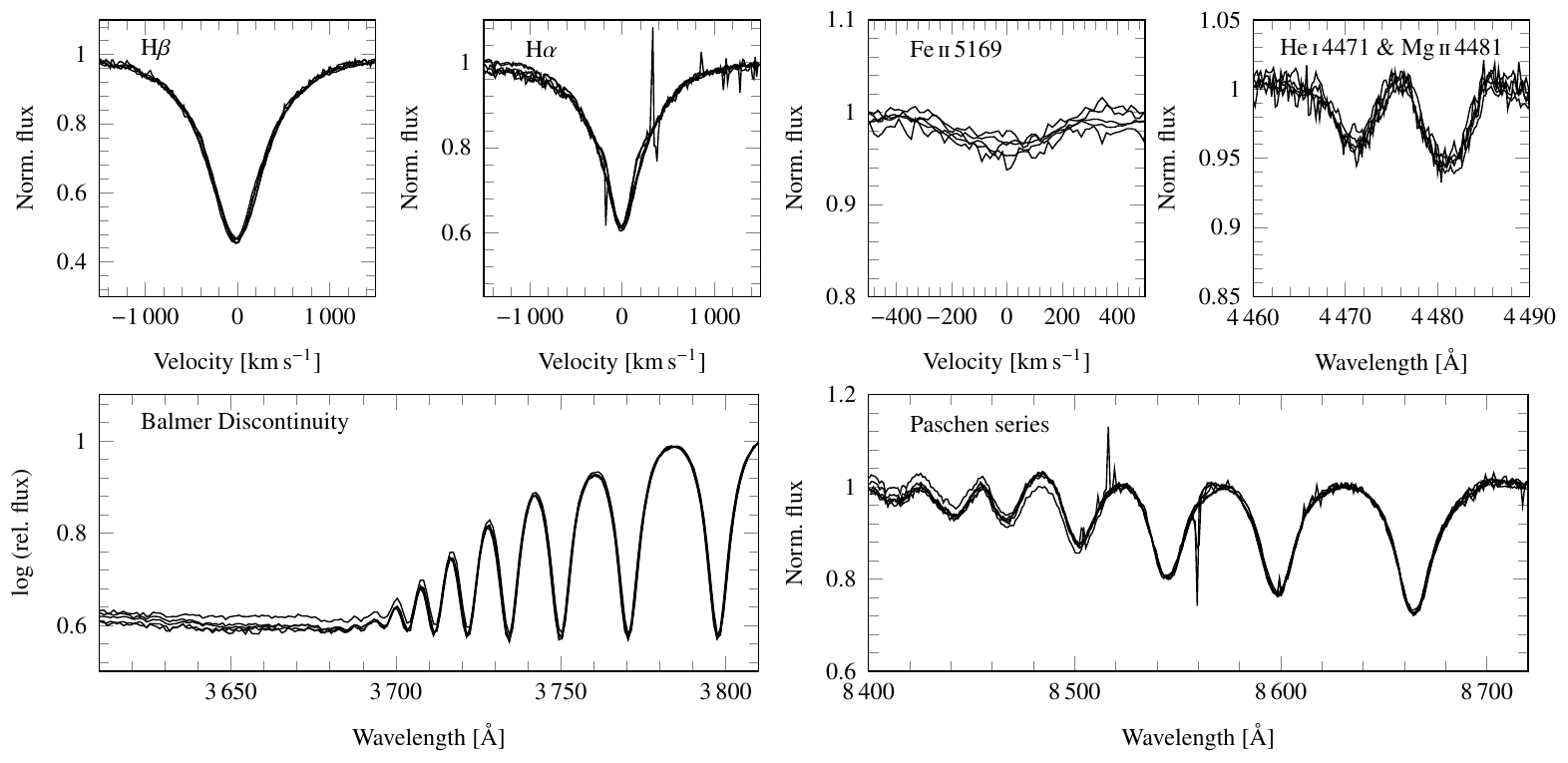}
\end{center}
\caption{Spectrum overview plot for Hip\,89500}
\end{figure*}
\clearpage

\begin{figure*}
\begin{center}
\includegraphics[angle=0,width=14cm,clip]{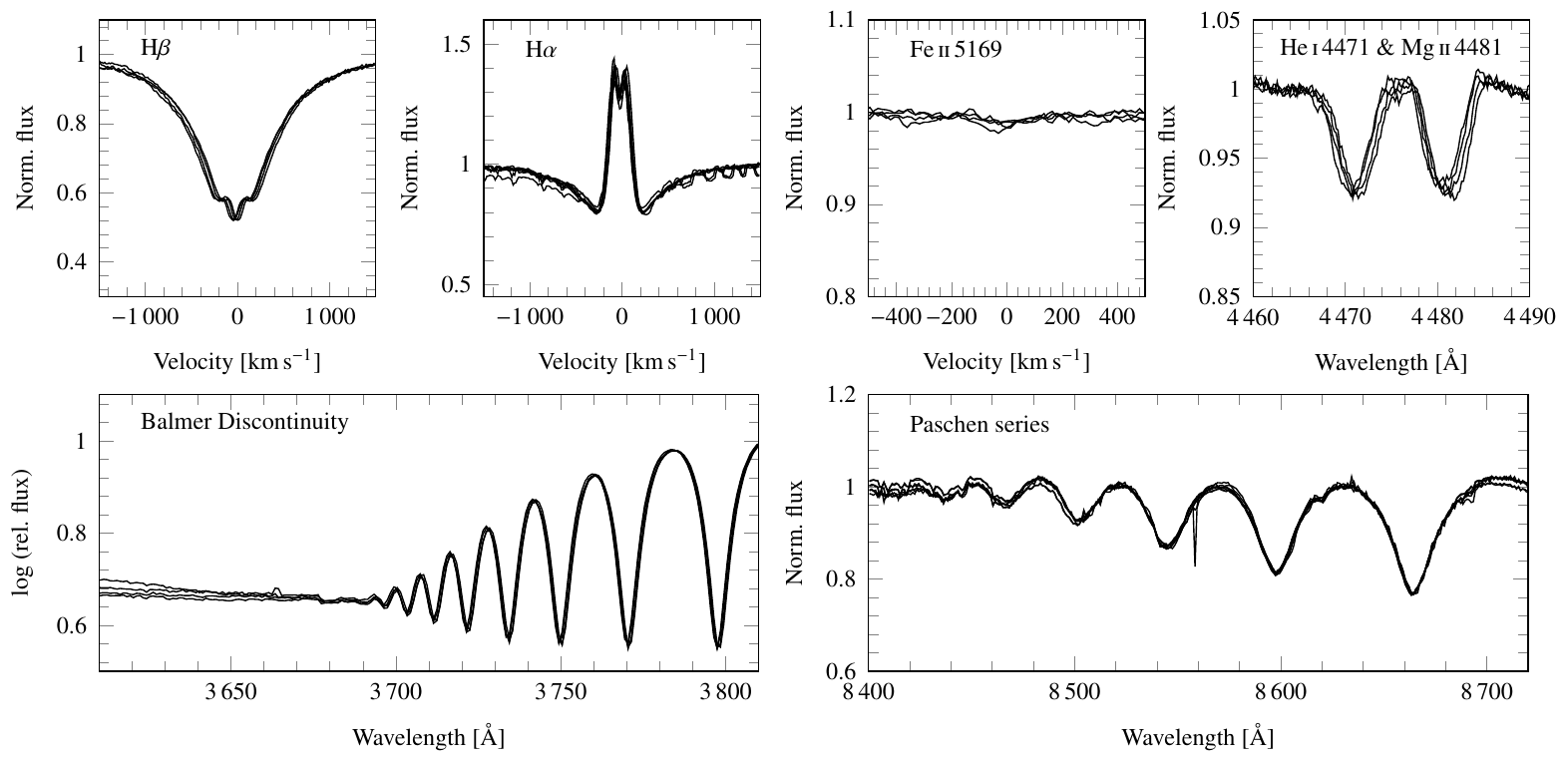}
\end{center}
\caption{Spectrum overview plot for Hip\,90096}
\end{figure*}

\begin{figure*}
\begin{center}
\includegraphics[angle=0,width=14cm,clip]{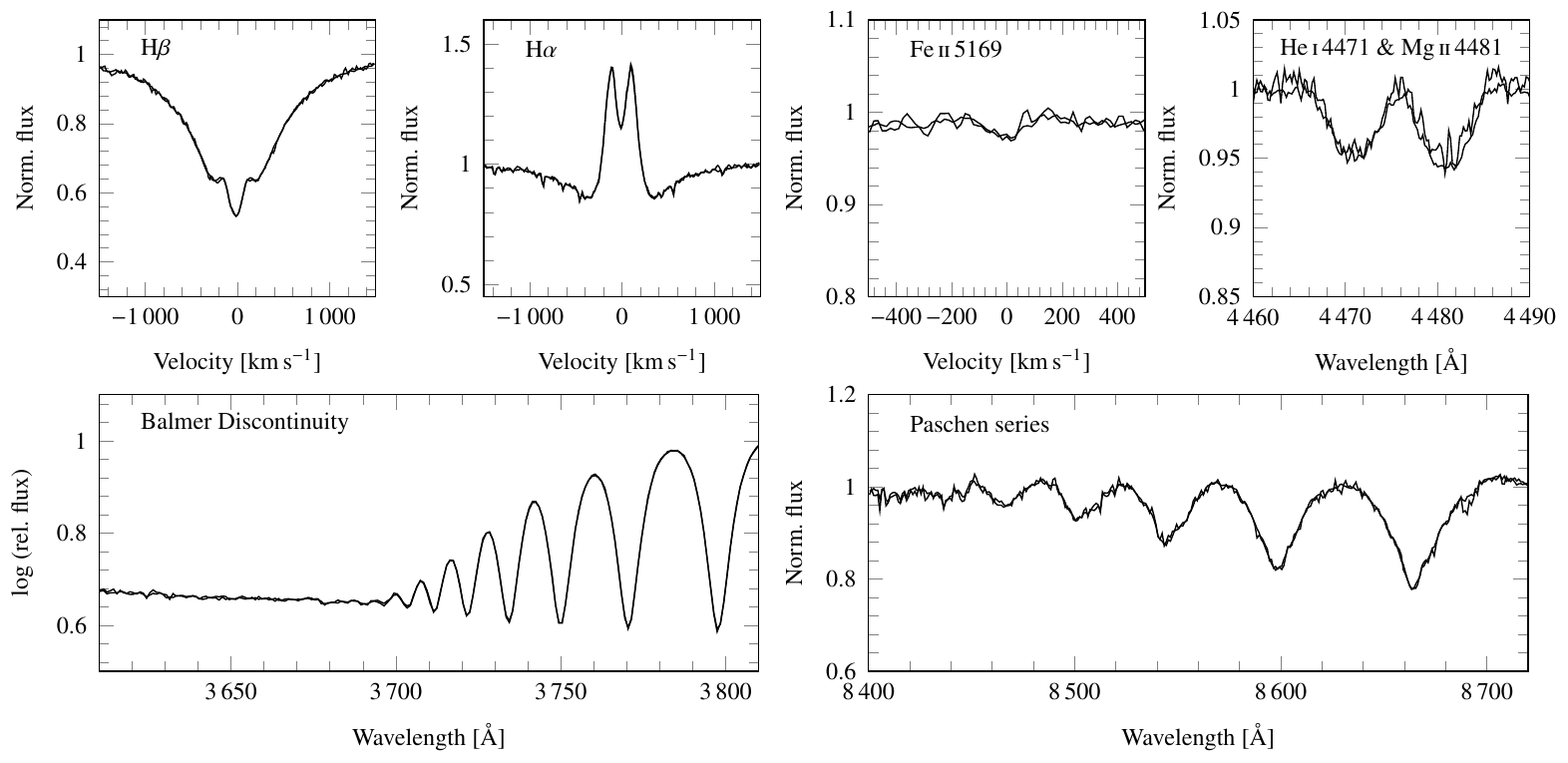}
\end{center}
\caption{Spectrum overview plot for Hip\,90509}
\end{figure*}

\begin{figure*}
\begin{center}
\includegraphics[angle=0,width=14cm,clip]{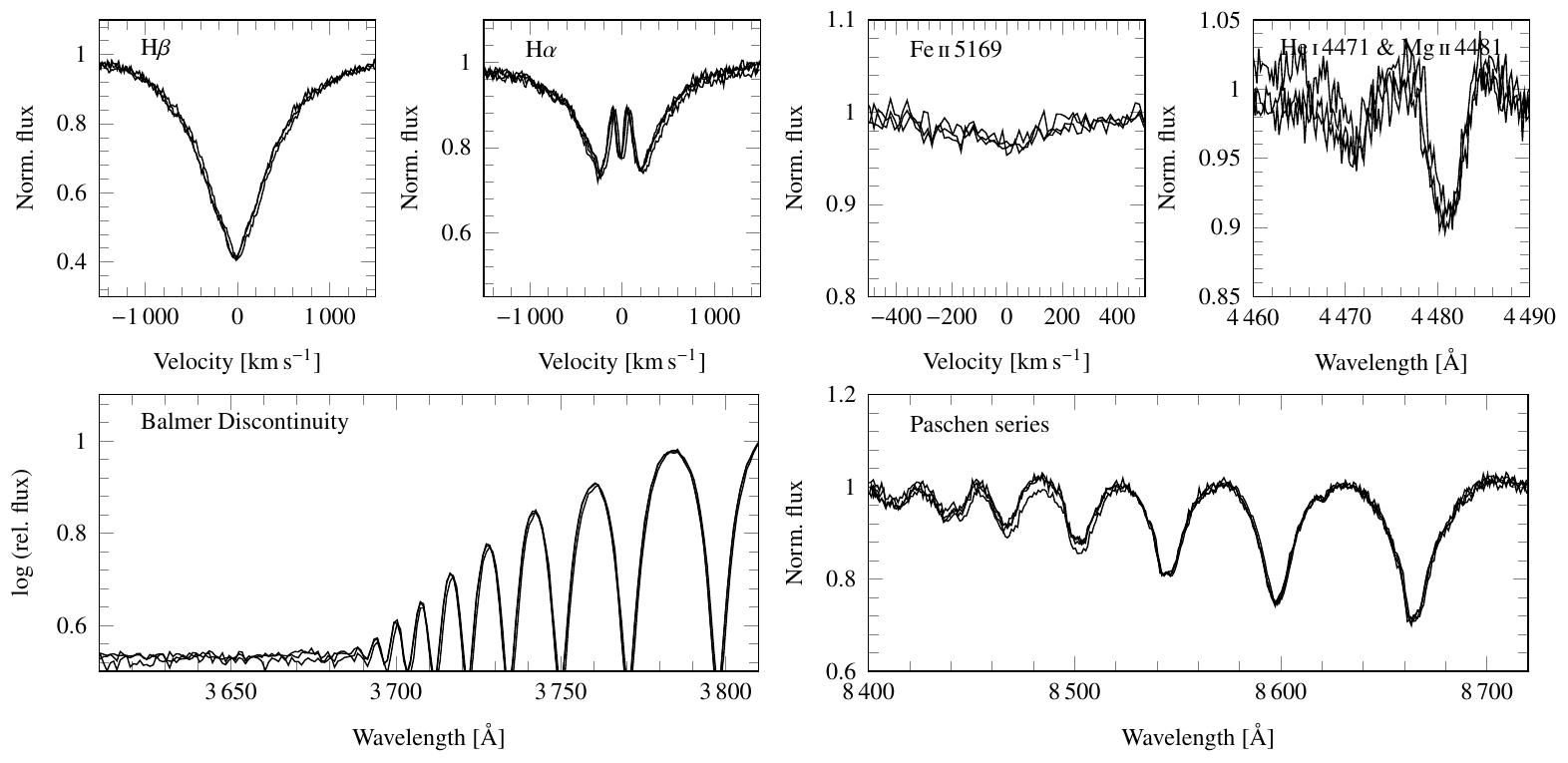}
\end{center}
\caption{Spectrum overview plot for Hip\,91460}
\end{figure*}
\clearpage

\begin{figure*}
\begin{center}
\includegraphics[angle=0,width=14cm,clip]{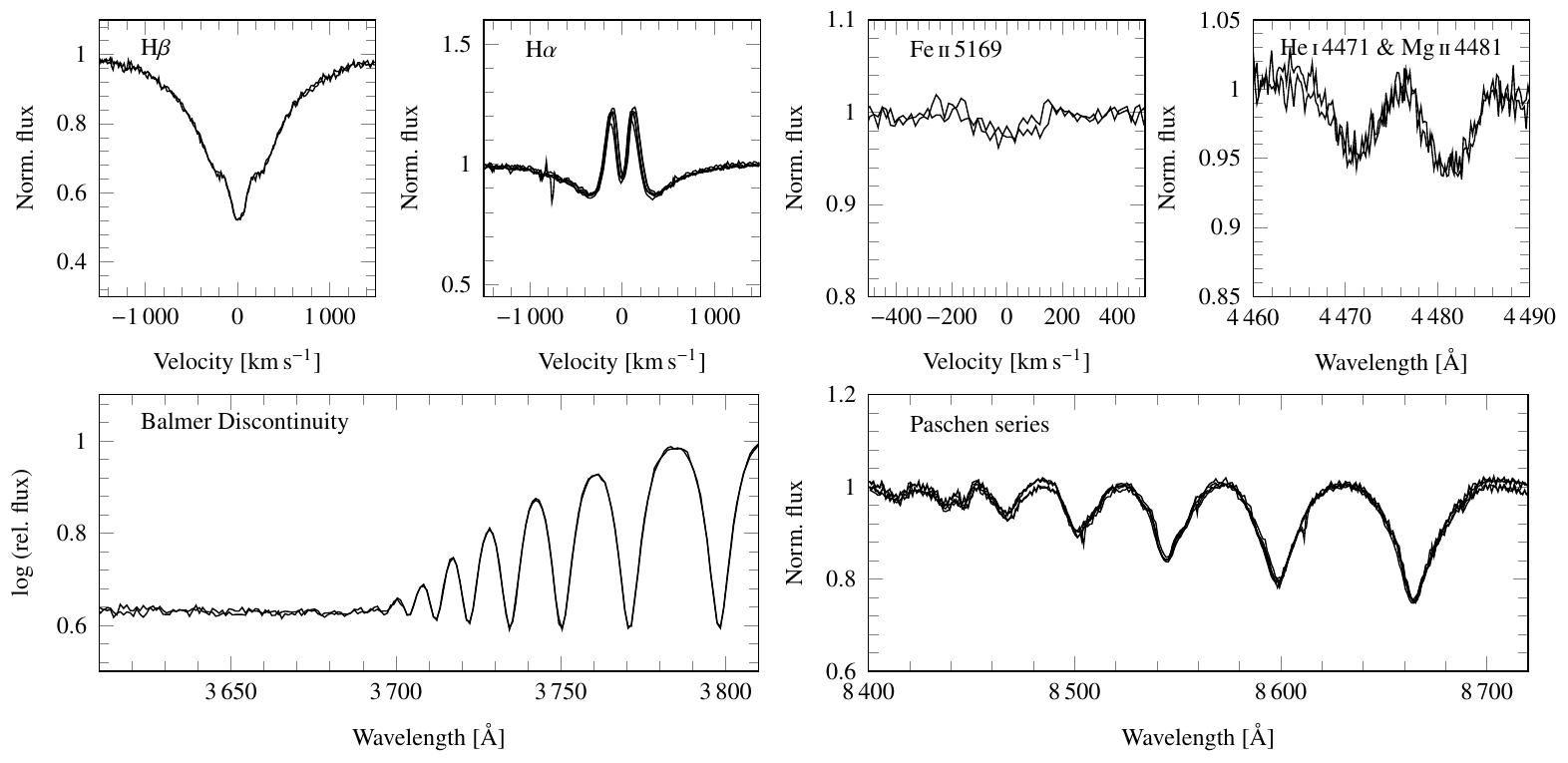}
\end{center}
\caption{Spectrum overview plot for Hip\,91975}
\end{figure*}

\begin{figure*}
\begin{center}
\includegraphics[angle=0,width=14cm,clip]{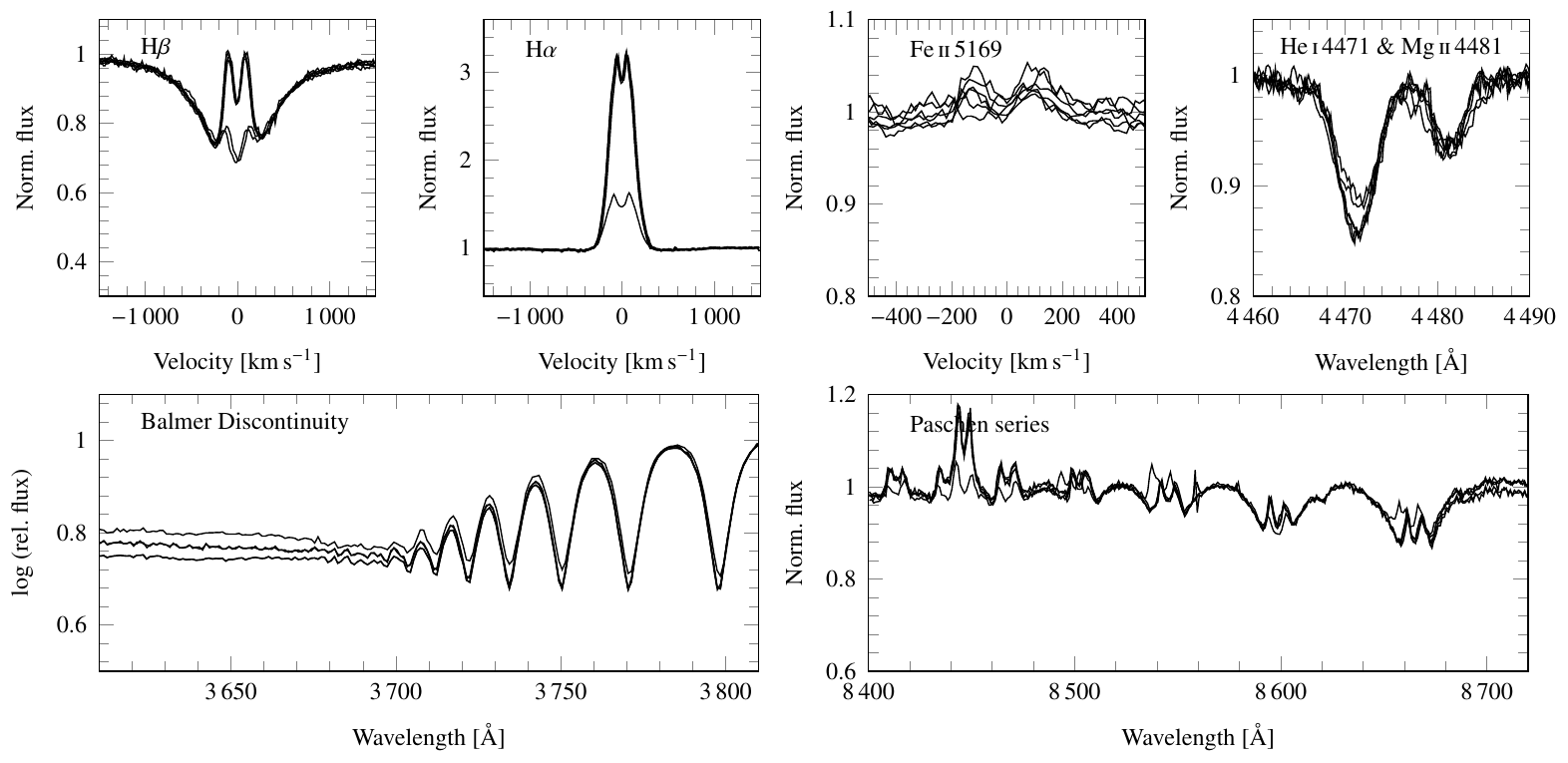}
\end{center}
\caption{Spectrum overview plot for Hip\,92038}
\end{figure*}

\begin{figure*}
\begin{center}
\includegraphics[angle=0,width=14cm,clip]{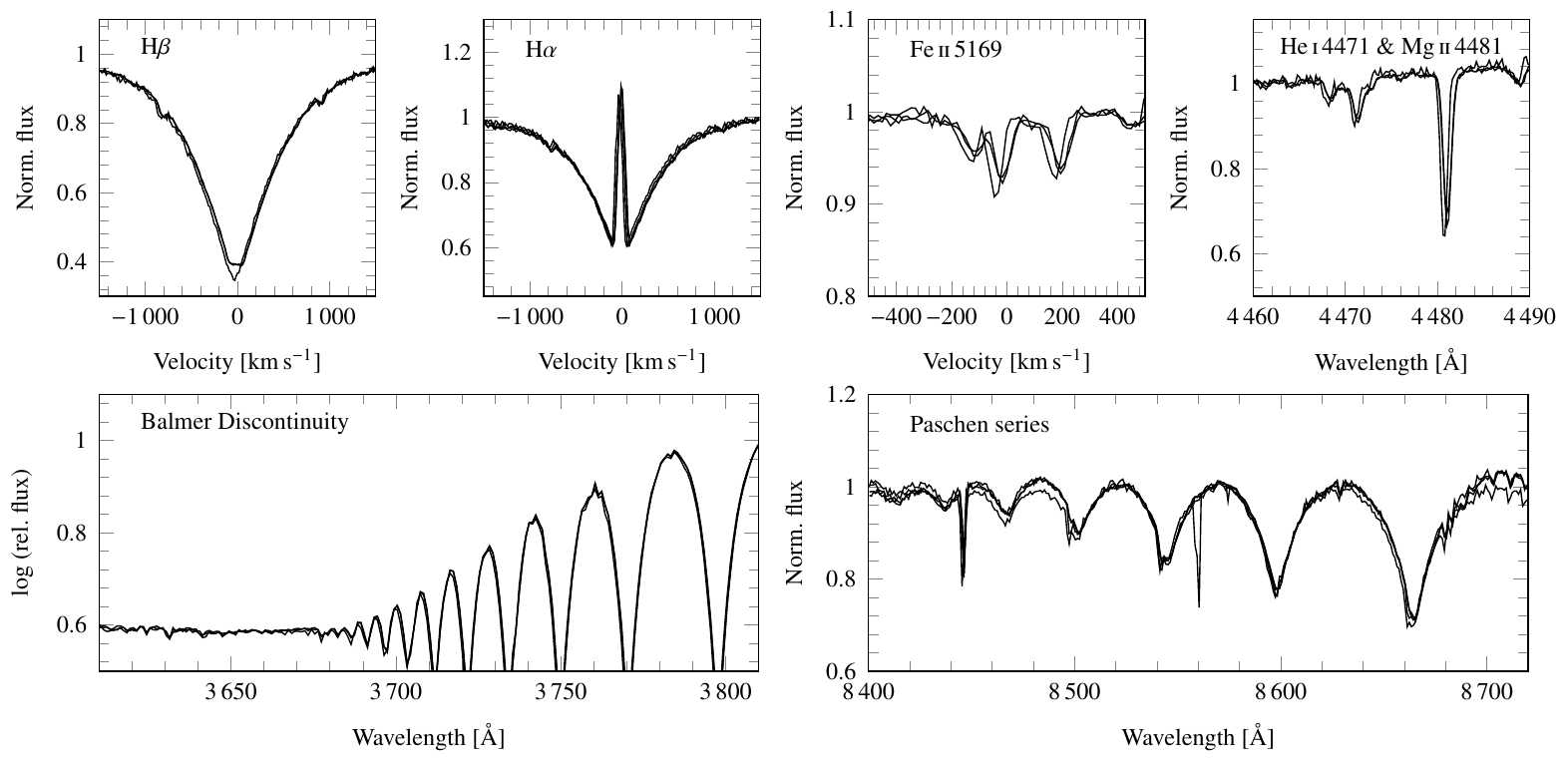}
\end{center}
\caption{Spectrum overview plot for Hip\,93993}
\end{figure*}
\clearpage

\begin{figure*}
\begin{center}
\includegraphics[angle=0,width=14cm,clip]{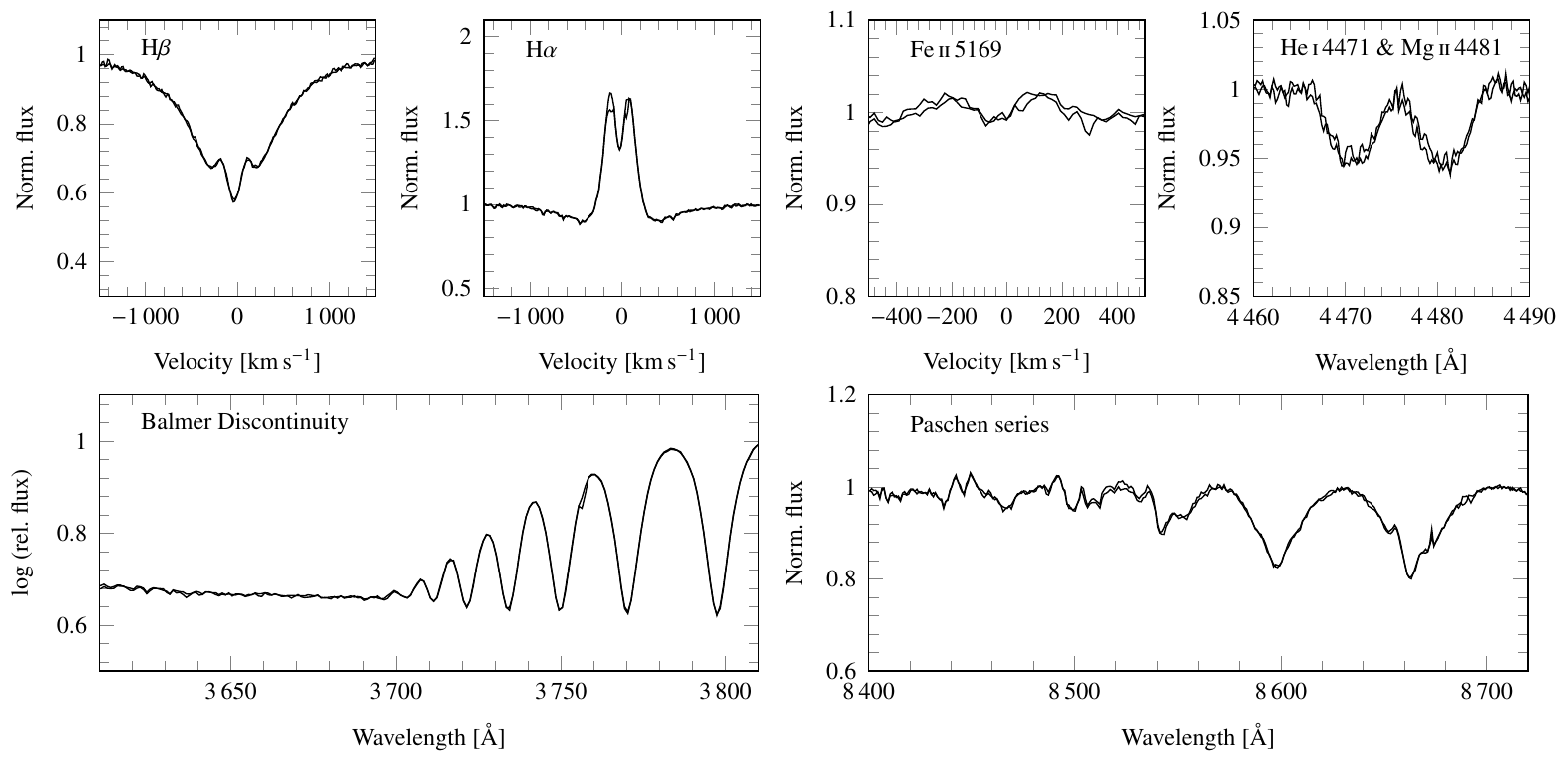}
\end{center}
\caption{Spectrum overview plot for Hip\,94770}
\end{figure*}

\begin{figure*}
\begin{center}
\includegraphics[angle=0,width=14cm,clip]{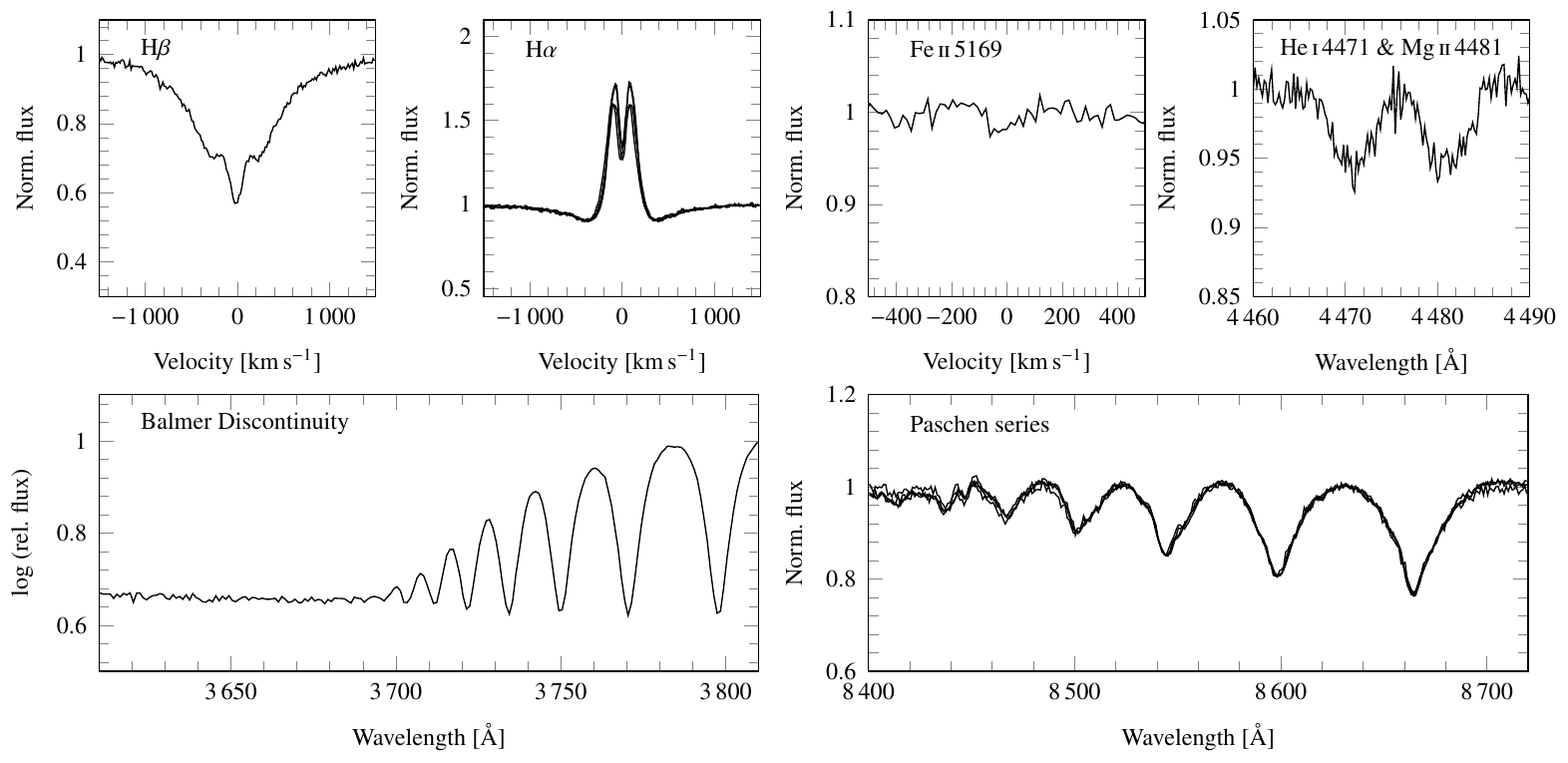}
\end{center}
\caption{Spectrum overview plot for Hip\,94859}
\end{figure*}

\begin{figure*}
\begin{center}
\includegraphics[angle=0,width=14cm,clip]{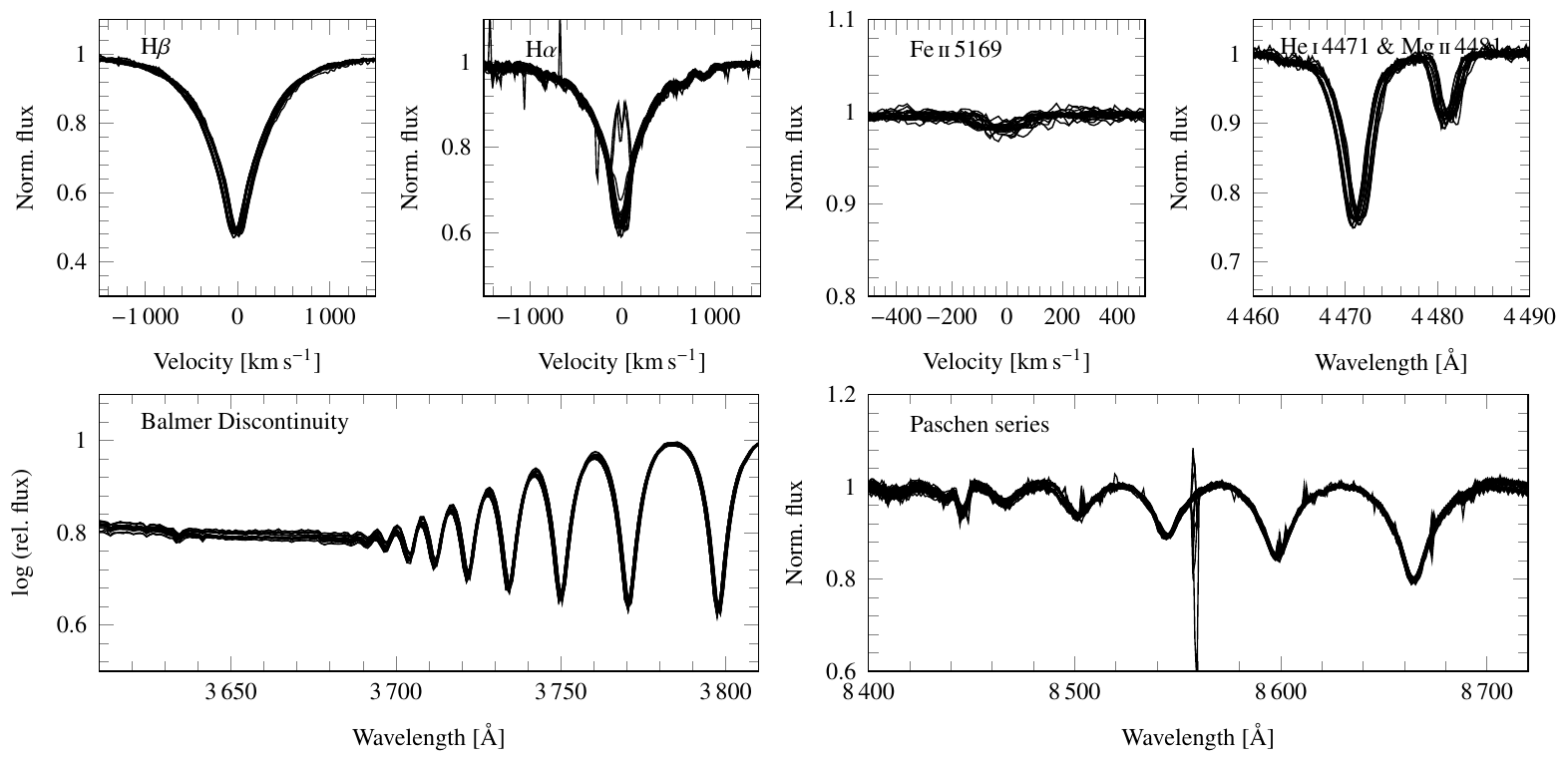}
\end{center}
\caption{Spectrum overview plot for Hip\,94986}
\end{figure*}
\clearpage

\begin{figure*}
\begin{center}
\includegraphics[angle=0,width=14cm,clip]{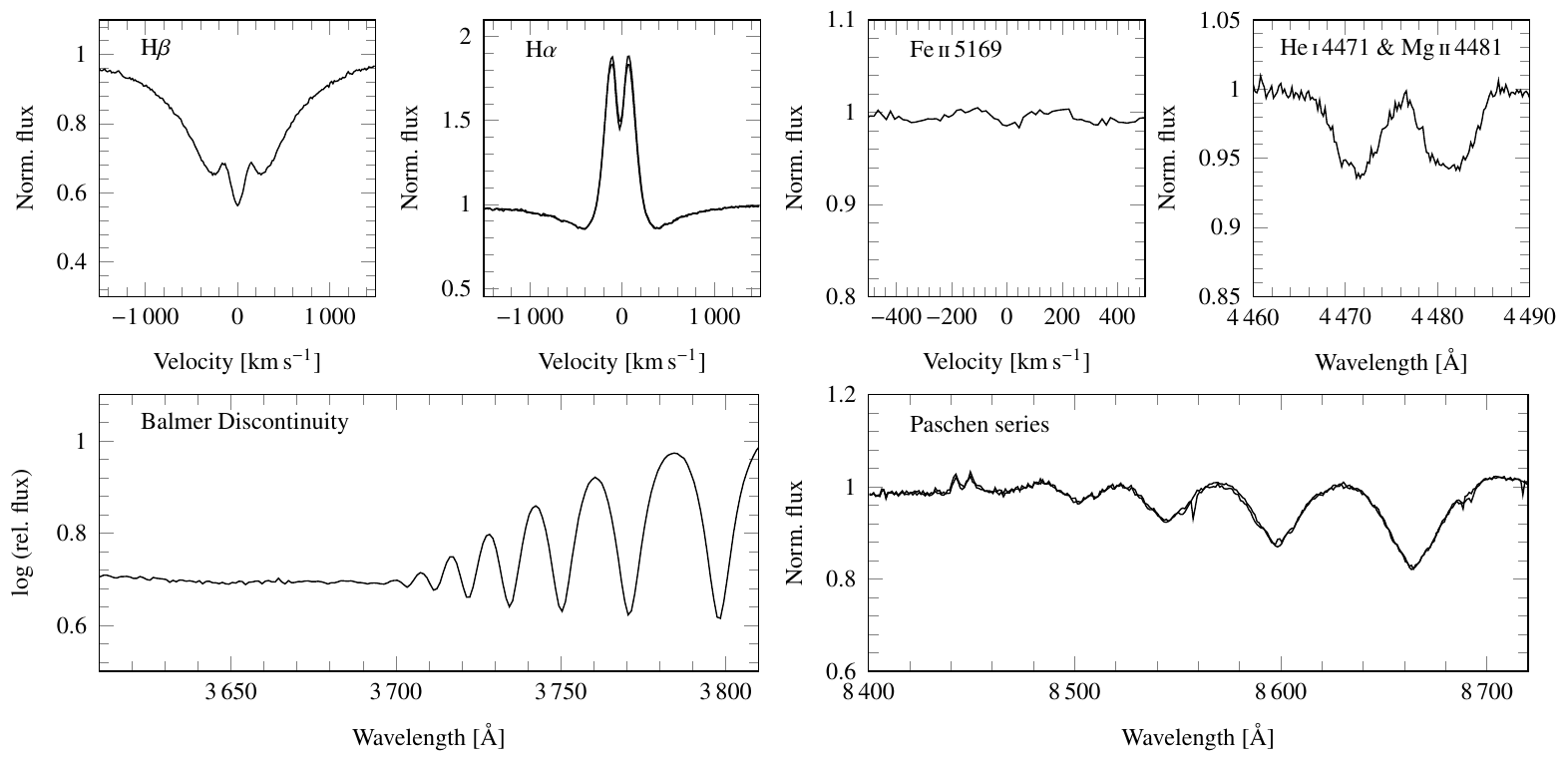}
\end{center}
\caption{Spectrum overview plot for Hip\,95109}
\end{figure*}

\begin{figure*}
\begin{center}
\includegraphics[angle=0,width=14cm,clip]{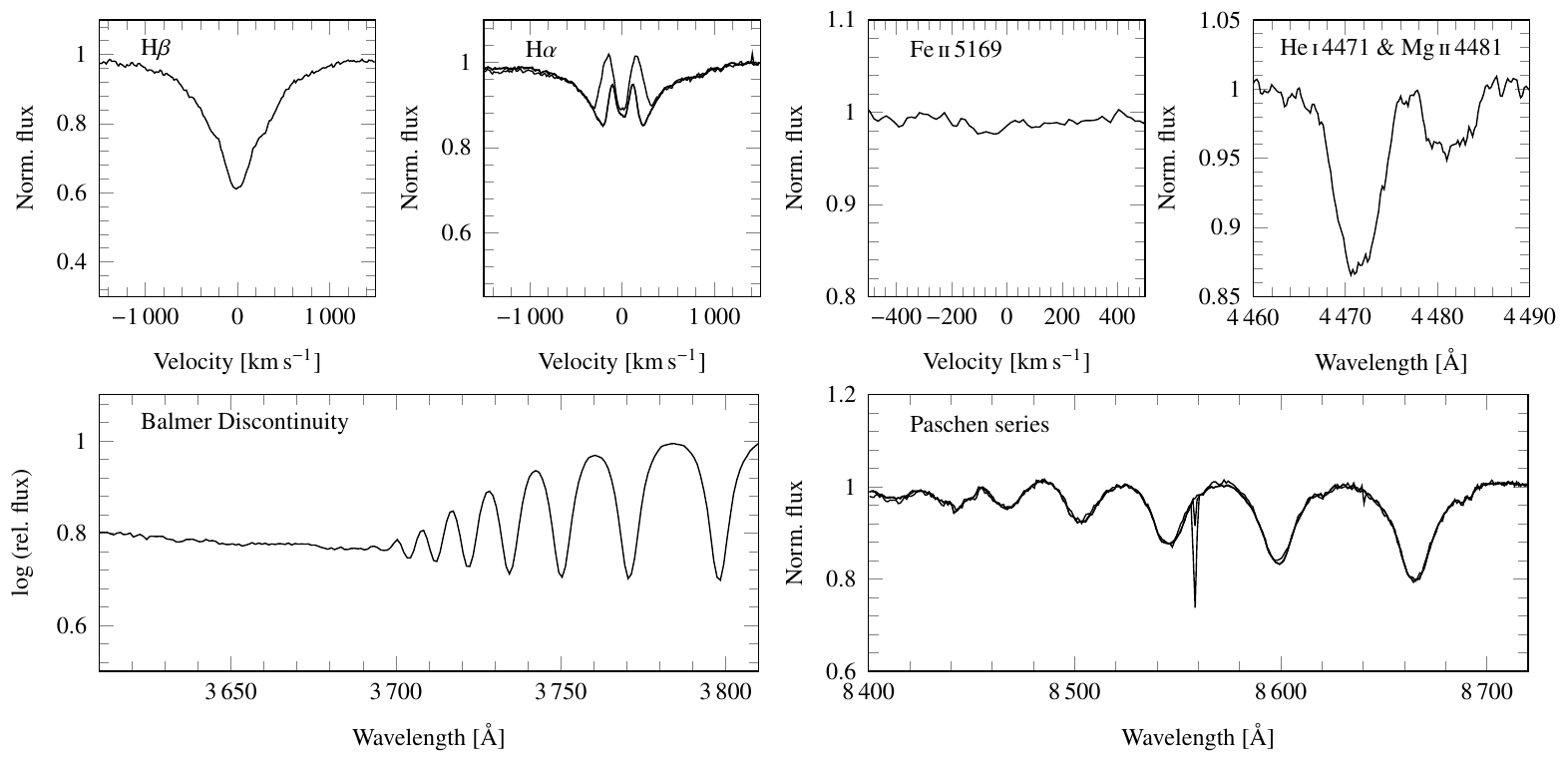}
\end{center}
\caption{Spectrum overview plot for Hip\,96453}
\end{figure*}

\begin{figure*}
\begin{center}
\includegraphics[angle=0,width=14cm,clip]{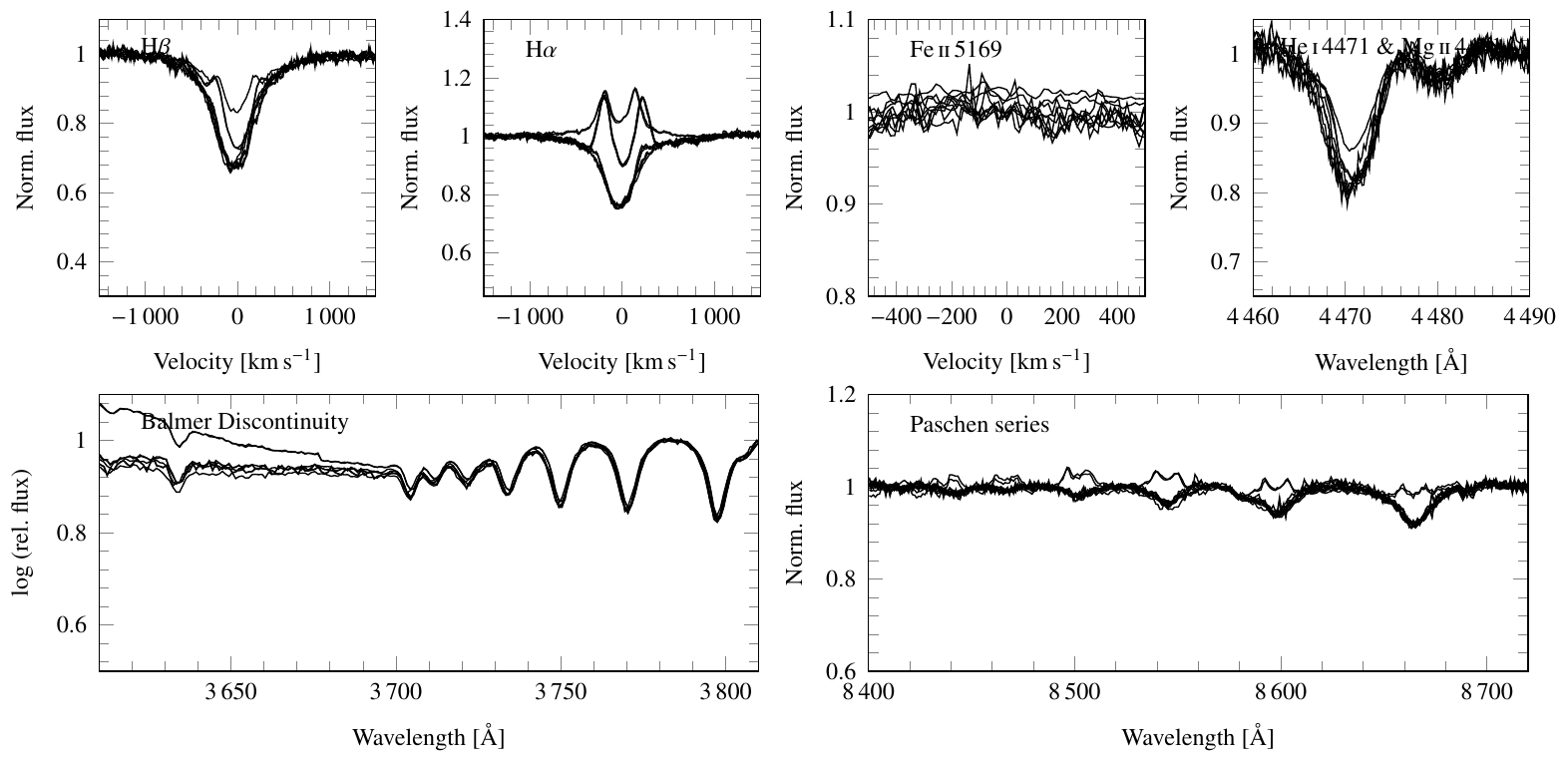}
\end{center}
\caption{Spectrum overview plot for Hip\,99457}
\end{figure*}
\clearpage

\begin{figure*}
\begin{center}
\includegraphics[angle=0,width=14cm,clip]{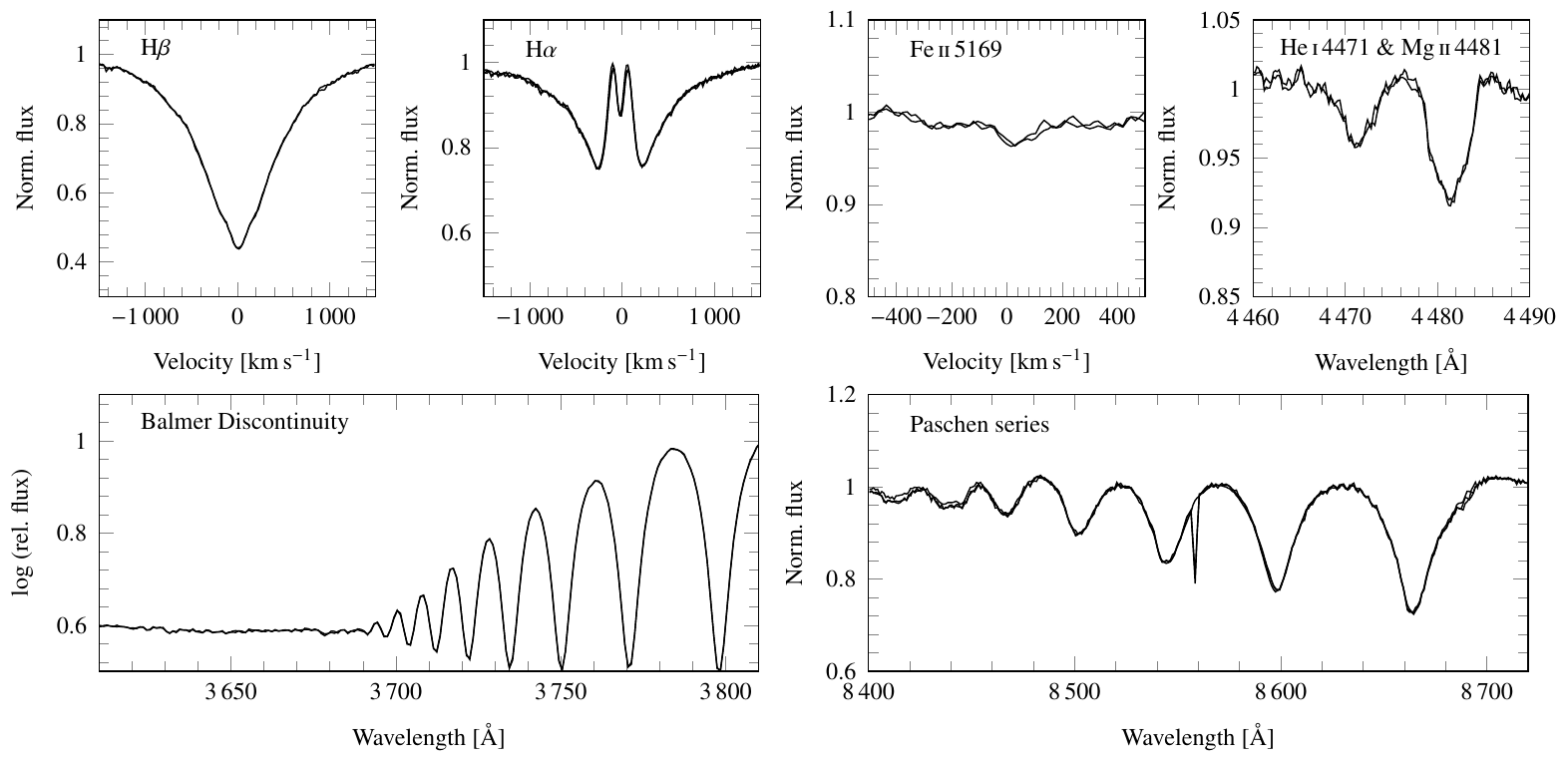}
\end{center}
\caption{Spectrum overview plot for Hip\,100664}
\end{figure*}

\begin{figure*}
\begin{center}
\includegraphics[angle=0,width=14cm,clip]{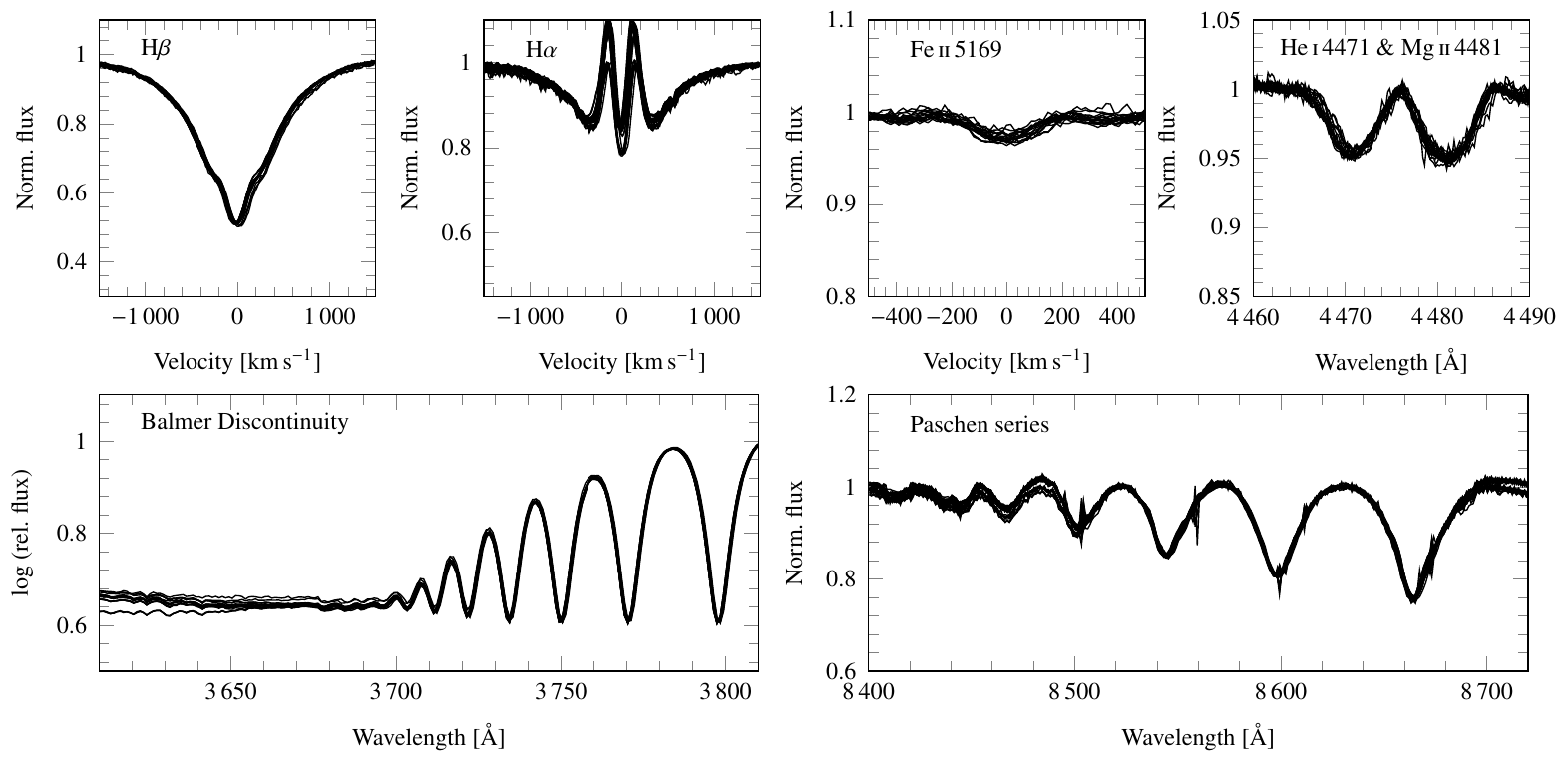}
\end{center}
\caption{Spectrum overview plot for Hip\,104508}
\end{figure*}

\begin{figure*}
\begin{center}
\includegraphics[angle=0,width=14cm,clip]{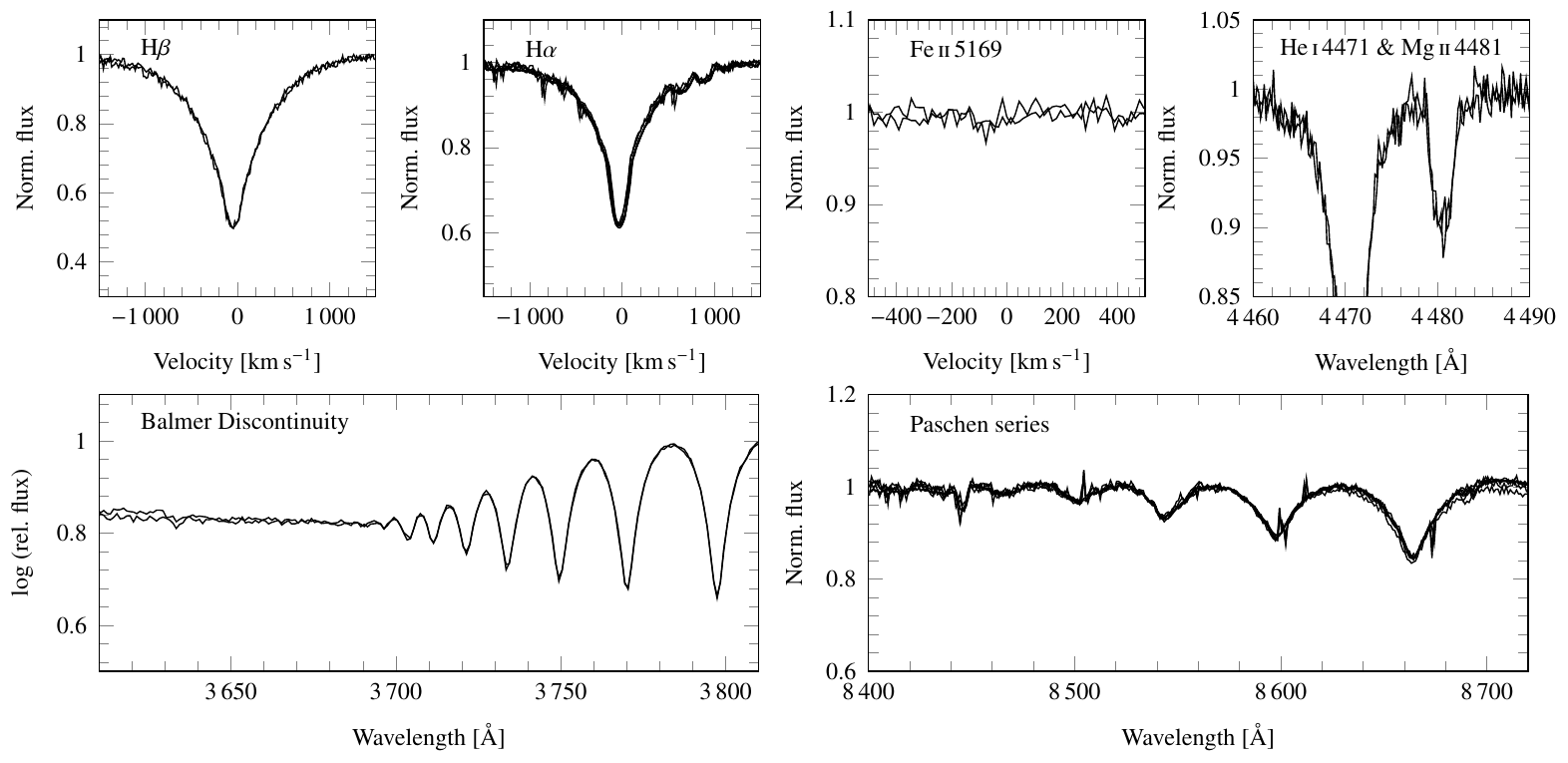}
\end{center}
\caption{Spectrum overview plot for Hip\,108022}
\end{figure*}
\clearpage

\begin{figure*}
\begin{center}
\includegraphics[angle=0,width=14cm,clip]{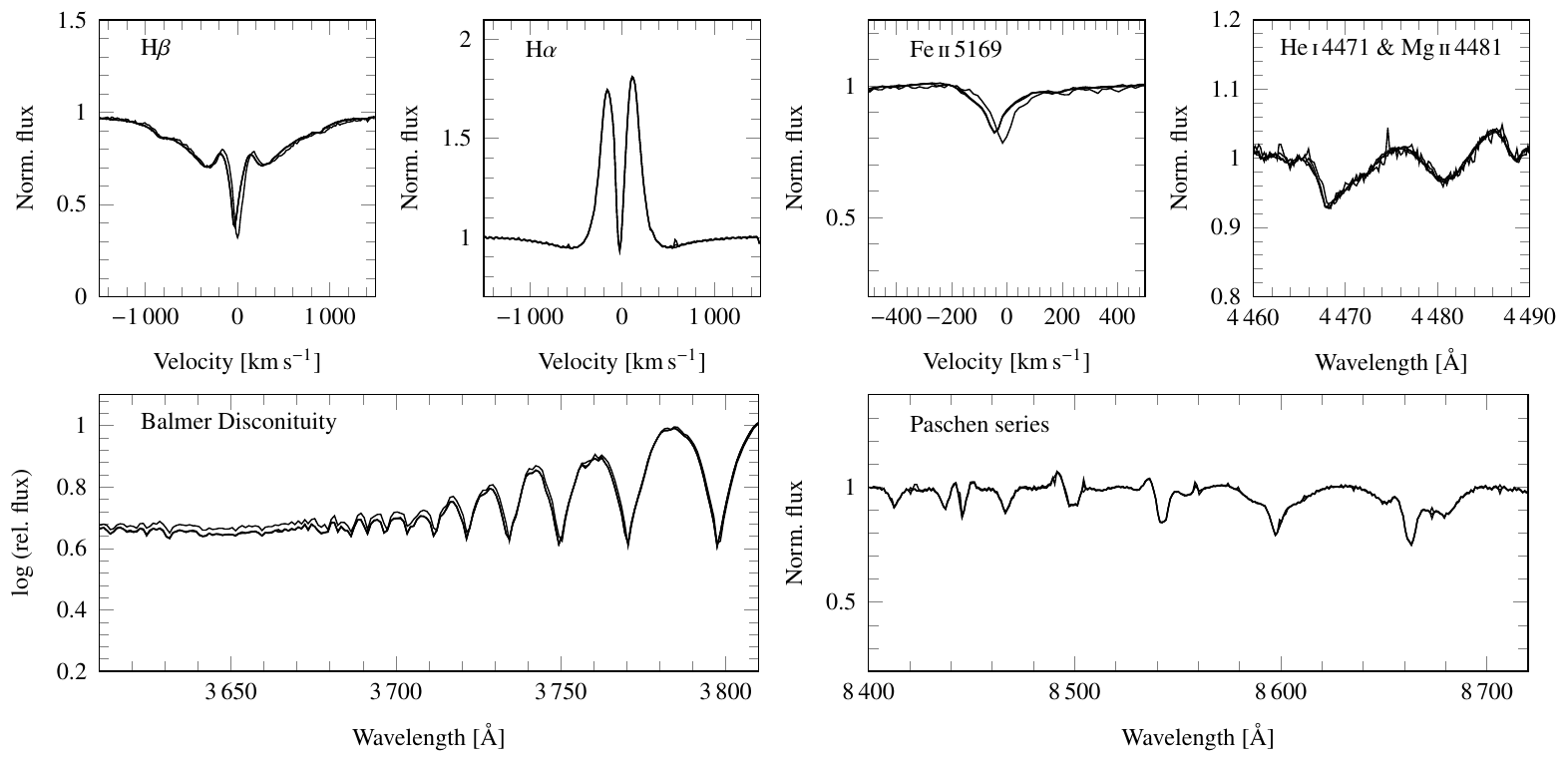}
\end{center}
\caption{Spectrum overview plot for Hip\,108402}
\end{figure*}

\begin{figure*}
\begin{center}
\includegraphics[angle=0,width=14cm,clip]{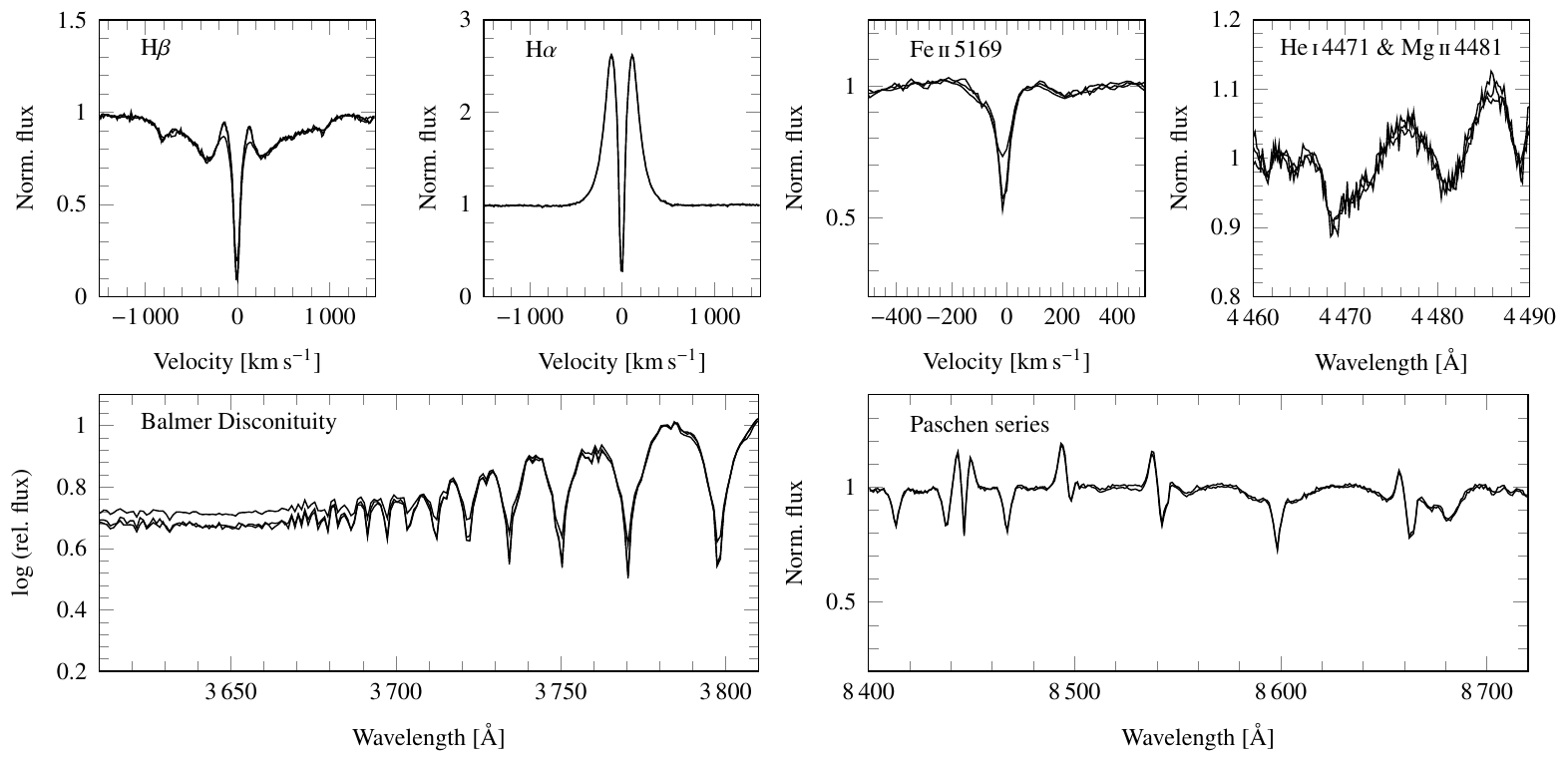}
\end{center}
\caption{Spectrum overview plot for Hip\,108597}
\end{figure*}

\begin{figure*}
\begin{center}
\includegraphics[angle=0,width=14cm,clip]{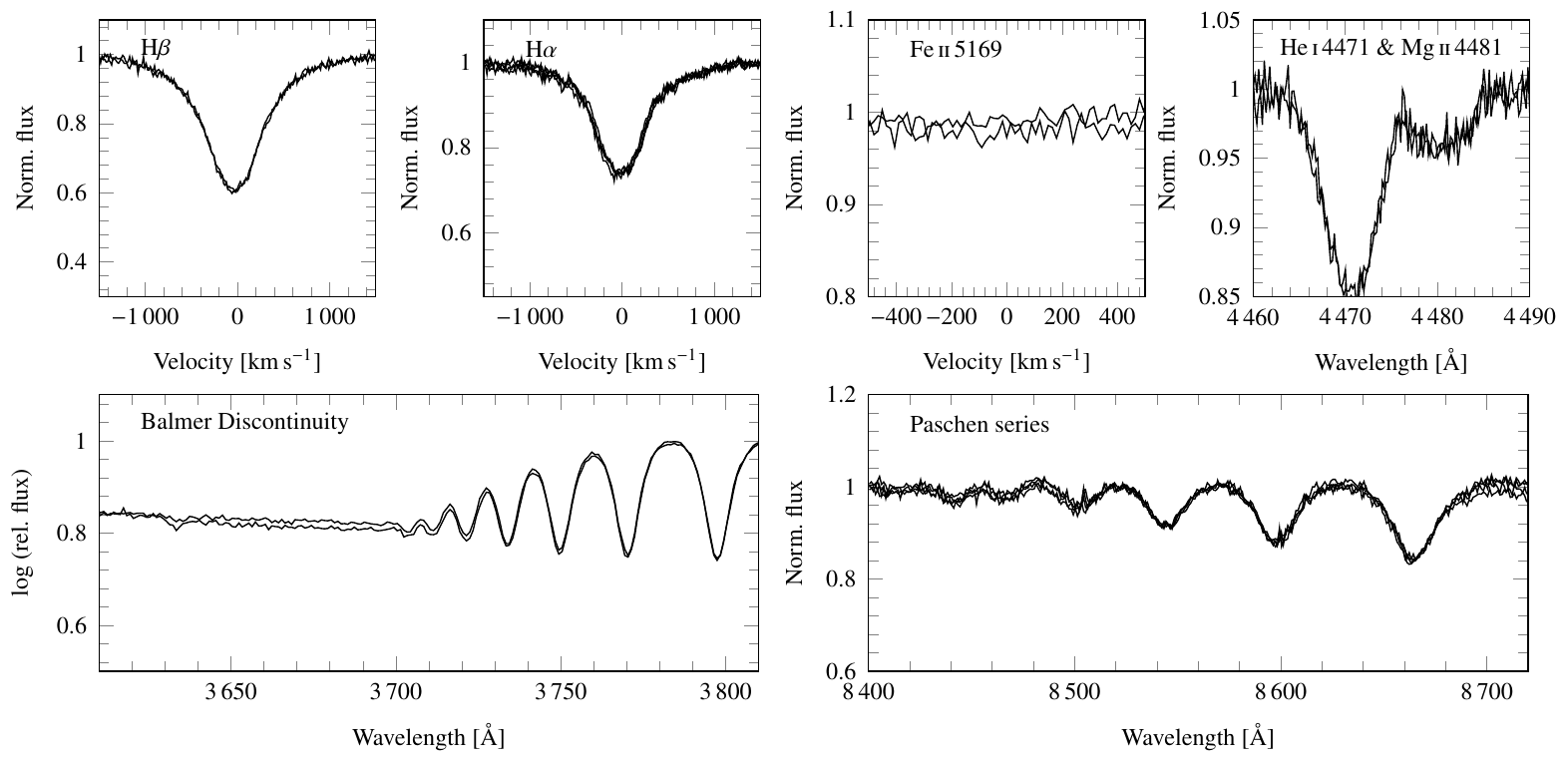}
\end{center}
\caption{Spectrum overview plot for Hip\,108975}
\end{figure*}

\end{appendix}

\end{document}